\documentclass[12pt,titlepage]{article}
\usepackage{latexsym}
\usepackage{amsfonts}
\newcommand{\vu}[1]{\rule[#1 mm]{0mm}{#1 mm}}

\newcommand{\vt}[1]{\rule{0mm}{#1mm}}
\newcommand{\hs}{\hspace}
\newcommand{\cem}{\hspace{1cm}}
\newcommand{\vs}[1]{\rule[- #1 mm]{0mm}{#1 mm}}

\newcommand{\bi}{{\bar{\imath}}}
\newcommand{\bj}{{\bar{\jmath}}}
\newcommand{\bk}{{\bar{k}}}
\newcommand{\bl}{{\bar{l}}}

\newcommand{\br}{{\bar{r}}}

\newcommand{\by}{{\bar{y}}}

\newcommand{\bA}{{\bar{A}}}
\newcommand{\bD}{{\bar{D}}}
\newcommand{\bF}{{\bar{F}}}

\newcommand{\bM}{{\bar{M}}}
\newcommand{\bS}{{\bar{S}}}
\newcommand{\bT }{{\bar{T}}}

\newcommand{\bV}{{\bar{V}}}

\newcommand{\bX}{{\bar{X}}}
\newcommand{\bY}{{\bar{Y}}}

\newcommand{\ovM}{{\overline{M}}}
\newcommand{\ovP}{{\overline{P}}}
\newcommand{\ovW}{{\overline{W}}}

\newcommand{\ovY}{{\overline{Y}}}

\newcommand{\undB}{{\underline{B}}}
\newcommand{\undA}{{\underline{A}}}
\newcommand{\undC}{{\underline{C}}}
\newcommand{\hC}{\hat{C}}
\newcommand{\hW}{\hat{W}}
\newcommand{\hhW}{\widehat{W}}

\newcommand{\hhK}{\widehat{K}}

\newcommand{\ca}{{\cal A}}
\newcommand{\cb}{{\cal B}}
\newcommand{\cc}{{\cal C}}
\newcommand{\cd}{{\cal D}}

\newcommand{\cf}{{\cal F}}
\newcommand{\cah}{{\cal H}}
\newcommand{\ck}{{\cal K}}
\newcommand{\cl}{{\cal L}}
\newcommand{\cm}{{\cal M}}

\newcommand{\cq}{{\cal Q}}
\newcommand{\car}{{\cal R}}
\newcommand{\cs}{{\cal S}}
\newcommand{\ct}{{\cal T}}
\newcommand{\cu}{{\cal U}}
\newcommand{\cv}{{\cal V}}
\newcommand{\cw}{{\cal W}}
\newcommand{\cx}{{\cal X}}
\newcommand{\cy}{{\cal Y}}
\newcommand{\cz}{{\cal Z}}

\newcommand{\cab}{{\bar{\ca}}}
\newcommand{\cbb}{{\bar{\cb}}}
\newcommand{\ccb}{{\bar{\cc}}}
\newcommand{\cdb}{{\bar{\cd}}}

\newcommand{\cwb}{{\bar{\cw}}}
\newcommand{\ovch}{\overline{\cah}}
\newcommand{\ovcf}{{\overline{\cf}}}

\newcommand{\al}{{\alpha}}
\newcommand{\bt}{{\beta}}
\newcommand{\gm}{{\gamma}}
\newcommand{\dt}{{\delta}}
\newcommand{\eps}{{\epsilon}}
\newcommand{\vep}{{\varepsilon}}
\newcommand{\la}{{\lambda}}
\newcommand{\si}{{\sigma}}
\renewcommand{\th}{\theta}
\newcommand{\vp}{\varphi}
\newcommand{\om}{{\omega}}

\newcommand{\da}{{\dot{\alpha}}}
\newcommand{\db}{{\dot{\beta}}}
\newcommand{\dg}{{\dot{\gamma}}}
\newcommand{\dd}{{\dot{\delta}}}

\newcommand{\dmu}{{\dot{\mu}}}
\newcommand{\dv}{{\dot{\varphi}}}

\newcommand{\thb}{\bar{\theta}}
\newcommand{\etab}{\bar{\eta}}
\newcommand{\vpb}{{\bar{\vp}}}
\newcommand{\sib}{{\bar{\sigma}}}
\newcommand{\lab}{{\bar{\lambda}}}
\newcommand{\xib}{{\bar{\xi}}}
\newcommand{\psib}{{\bar{\psi}}}
\newcommand{\chib}{{\bar{\chi}}}
\newcommand{\phib}{{\bar{\phi}}}

\newcommand{\Dt}{\Delta}
\newcommand{\Gm}{\Gamma}
\newcommand{\La}{\Lambda}
\newcommand{\Si}{\Sigma}

\newcommand{\Om}{{\Omega}}

\newcommand{\Phib}{\bar{\Phi}}

\newcommand{\Psib}{{\bar{\Psi}}}
\newcommand{\Lab}{\bar{\Lambda}}
\newcommand{\Gmb}{\bar{\Gamma}}
\newcommand{\oGm}{{\overline{\Gm}}}
\newcommand{\oPi}{{\overline{\Pi}}}

\newcommand{\oSi}{\overline{\Sigma}}

\newcommand{\undal}{{\underline{\alpha}}}
\newcommand{\undbt}{{\underline{\beta}}}
\newcommand{\undgm}{{\underline{\gamma}}}
\newcommand{\unddt}{\underline{\delta}}

\newcommand{\undSi}{{\underline{\Sigma}}}
\newcommand{\undOm}{{\underline{\Omega}}}

\newfont{\twelvemsb}{msbm10 scaled\magstep1}
\newfont{\eightmsb}{msbm8}
\newfont{\sixmsb}{msbm6}
\newfam\msbfam
\textfont\msbfam=\twelvemsb
\scriptfont\msbfam=\eightmsb
\scriptscriptfont\msbfam=\sixmsb
\catcode`\@=11
\def\Bbb{\ifmmode\let\next\Bbb@\else
  \def\next{\errmessage{Use \string\Bbb\space only in math mode}}\fi\next}
\def\Bbb@#1{{\Bbb@@{#1}}}
\def\Bbb@@#1{\fam\msbfam#1}
\newfont{\twelvegoth}{eufm10 scaled\magstep1}
\newfont{\fourteengoth}{eufm10 scaled\magstep2}
\newfont{\sixteengoth}{eufm10 scaled\magstep3}
\newfont{\tengoth}{eufm10}
\newfont{\eightgoth}{eufm8}
\newfam\gothfam
\textfont\gothfam=\twelvegoth
\scriptfont\gothfam=\eightgoth
\def\frak{\ifmmode\let\next\frak@\else
  \def\next{\errmessage{Use \string\frak\space only in math mode}}\fi\next}
\def\frak@#1{{\fam\gothfam{{#1}}}}
\catcode`@=12

\newcommand{\ima}{\mathop{\mathrm{Im}}}
\newcommand{\rea}{\mathop{\mathrm{Re}}}
\newcommand{\gom}{{\frak{M}}}

\newcommand{\lgoa}{\mbox{\sixteengoth a}}
\newcommand{\lgoab}{{\bar{\lgoa}}}

\newcommand{\sect}[1]{\setcounter{equation}{0}\section{#1}}
\newcommand{\subsec}[1]{\setcounter{equation}{0}\subsection{#1}}
\renewcommand{\theequation}{\thesection.\arabic{equation}}
\renewcommand{\theequation}{\thesubsection.\arabic{equation}}
\renewcommand{\thesection}{\Roman{section}}
\renewcommand{\thesubsection}{\thesection-\arabic{subsection}}
\newcommand{\ba}{\begin{array}}
\newcommand{\ea}{\end{array}}
\newcommand{\nn}{\nonumber}

\def\lp{\left(}
\def\rp{\right)}
\def\l[{\left[}
\def\r]{\right]}
\newcommand{\f}[2]{{\displaystyle\frac{#1}{#2}}}
\newcommand{\tdem} {{\scriptstyle\frac{3}{2}}} 
\newcommand{\dem} {{\scriptstyle\frac{1}{2}}}  
\def\2#1{\mbox{#1}}
\def\s2{\sqrt{2}}
\def\kl  {\f{k}{L}}
\newcommand{\ie}{{\em {i.e. }}}
\newcommand{\eg}{{\em {e.g. }}}
\newcommand{\prt}{\partial}
\newcommand{\tr}{\mathop{\mathrm{tr}}} 
\def\dtdu#1#2{{\delta_{#1}}^{#2}} 
\def\dtud#1#2{{\delta^{#1}}_{#2}} 

\newcommand{\aym}{{\mbox{\boldmath $\mathit{a}$}}} 
\newcommand{\fym}{{\mbox{\boldmath $\mathit{f}$}}} 
\newcommand{\gym}{{\mbox{\boldmath $\mathit{g}$}}} 
\newcommand{\alym}{{\mbox{\boldmath $\mathit{\al}$}}} 
\def\ym{\scriptscriptstyle{\cy  \cm}}
\def\WZ{\scriptscriptstyle{\mathrm{WZ }}}

\newcommand{\LSM}{{{\cl}_{\mathrm{supergravity+matter}}}}
\newcommand{\LYM}{{{\cl}_{\mathrm{Yang-Mills}}}}
\newcommand{\LSPOT}{{{\cl}_{\mathrm{superpotential}}}}
\newcommand{\ASM}{{{\ca}_{\mathrm{supergravity+matter}}}}
\newcommand{\AYM}{{{\ca}_{\mathrm{Yang-Mills}}}}
\newcommand{\ASPOT}{{{\ca}_{\mathrm{superpotential}}}}
\newcommand{\rSM}{{{\bf r}_{\mathrm{supergravity+matter}}}}
\newcommand{\rYM}{{{\bf r}_{\mathrm{Yang-Mills}}}}
\newcommand{\rgen}{{{\bf r}}}
\newcommand{\rSPOT}{{{\bf r}_{\mathrm{superpotential}}}}

\newcommand{\DM}{{\bf D}_{\mathrm{matter}}}

\newcommand{\ka}{{K\"{a}hler\ }}
\newcommand{\lsym}[1]{\stackrel{\scriptstyle#1}{\mbox{$\smile$}}}
\newcommand{\sym}[1]{\stackrel{\scriptstyle#1}{\mbox{\tiny $\smile$}}}
\newcommand{\ci}[1]{\raise5pt\hbox{$\scriptstyle#1$}} 

\newcommand{\rd}{{R^\dagger}}
\newcommand{\prokib}{{(\cd^2 - 8 \rd)}}
\newcommand{\proki}{{(\cdb^2 - 8 R)}}
\newcommand{\loco}{{\mathop{ \, \rule[-.06in]{.2mm}{3.8mm}\,}}} 
\newcommand{\doubar}{{{\loco}\!{\loco}}}
\newcommand{\kil}{{  \left[ K_k \lp{{\bf
T}_{(r)}}{\phi} \rp^k  + K_\bk \lp \phib{{{\bf T}}_{(r)}}\rp^{\bk}
\right]}}
\newcommand{\kilc}{{  \left[ K_k \lp{{\bf
T}_{(r)}}{A} \rp^k  + K_\bk \lp {\bA}{{{\bf T}}_{(r)}}\rp^{\bk}
\right]}}

\newcommand{\uti}{{\raisebox{.3ex}{\mbox{\tiny{I}}}}}
\newcommand{\dti}{{\mbox{\tiny{I}}}}
\newcommand{\utj}{{\raisebox{.3ex}{\mbox{\tiny{J}}}}}
\newcommand{\dtj}{{\mbox{\tiny{J}}}}

\newcommand{\bti}{\mbox{\tiny{(I)}}}
\newcommand{\mod}{{\mathop{ \, \rule[-.1in]{.2mm}{8.3mm}\,}}}

\begin{document}

\begin{titlepage}
\begin{center}
{\LARGE {\bf  SUPERGRAVITY COUPLINGS:
\\[3.mm]
A GEOMETRIC FORMULATION }}
\vskip 1cm

{\large P. Bin\'etruy } \\
{\em Laboratoire de Physique Th\'eorique} \\
{\em LPT \footnote{UMR 8627 du CNRS}, Universit\'e Paris-XI,
B\^atiment 210, F-91405 Orsay Cedex, France}
\\[.7cm]

{\large G. Girardi} \\
{\em Laboratoire d'Annecy-le-Vieux de Physique Th\'eorique} \\
{\em LAPTH \footnote{ UMR 5108 du CNRS associ\'ee au L.A.P.P.},
Chemin de Bellevue, BP 110, F-74941 Annecy-le-Vieux Cedex,
France}
\\[.7cm]

{\large R. Grimm} \\
{\em Centre de Physique Th\'eorique} \\
{\em CPT \footnote{UPR 7061 du CNRS}, C.N.R.S. Luminy, Case 907,
F-13288 Marseille Cedex 09, France} \\[1.2cm]
\end{center}

\centerline{ {\bf Abstract}}

\indent This report provides a pedagogical introduction to the
description of the general Poincar\'e
supergravity/matter/Yang-Mills couplings using methods of K\"ahler
superspace geometry. At a more advanced level this approach is
generalized to include tensor field and Chern-Simons couplings in
supersymmetry and supergravity, relevant in the context of weakly
and strongly coupled string theories.

\vfill \rightline{LPTHE-Orsay \ 99/90} \rightline{LAPTH \ 755/99}
\rightline{CPT-99/P.3906} \vskip .2cm \rightline{\today}
\end{titlepage}

\newpage
\pagenumbering{roman}
\tableofcontents

\newpage
\pagenumbering{arabic}
\section{  INTRODUCTION}

\noindent Since its appearance in string theory \cite{NS71,
NST71,Ram71,CS74,SS74d}, in elementary particle physics
\cite{GL71,VA73} and in quantum field theory
\cite{WZ74c,WZ74a,WZ74b,HLS75}, supersymmetry has become a
central issue in the quest for unification of the fundamental
forces of Nature.

Mathematically, supersymmetry transformations fall in the category
of graded Lie groups, with commuting and anticommuting parameters
\cite{BK70,CNS75}. In addition to the generators of Lorentz
transformations and translations in a $D$-dimensional space-time,
the supersymmetry algebra contains one or more spinor supercharges
("simple" or "$N$-extended" supersymmetry). As a consequence of
the particular algebraic structure, Wigner's analysis of unitary
representations \cite{Wig39} can be generalized to the
supersymmetric case \cite{SS74c,Nah77,Fre79,FSZ81}, giving rise to
the notion of supermultiplets which combine bosons and fermions.

Although theoretically very appealing, no explicit sign of such a
Bose-Fermi symmetry has been observed experimentally. This does
not prevent experimental physicists to put supersymmetric versions
of the standard model \cite{Nil84,HK85} to the test
\cite{Par98a,Par98b}. So far they turn out to be compatible with
data.

On a more fundamental level, in the context of recent
developments in string/brane theory \cite{Sch75,
Sch85,GSW87,Pol98}, supergravity in eleven dimensions
\cite{Nah77,CJS78a} seems to play an important role.  Such a
string, or membrane theory is expected to manifest itself in a
four dimensional point particle limit as some locally
supersymmetric effective theory.

The basic structure of a generic $D=4$, $N=1$ effective theory is
provided by supergravity \cite{DZ76,FvNF76} coupled to various
lower spin multiplets. The off-shell supergravity multiplet is
usually taken to be the one with minimal auxiliary field content
\cite{SW78,FvN78}, the so called {\em minimal supergravity
multiplet}\footnote{Other possibilities, such as the {\em new
minimal supergravity multiplet} \cite{AVS77,SW81} and the {\em
non-minimal supergravity multiplet} \cite{Bre77,SG79} are less
popular \cite{Ovr88,OR90,OR91} in this context.}.

{\em Chiral multiplets} are expected to appear in the form of some
nonlinear sigma model. Supersymmetry requires a K\"ahler structure
\cite{Zum79b}:
the complex scalar fields of the chiral multiplets are
coordinates of a K\"ahler manifold \cite{FG80,BW82,AGF83,BW87}. At
the same time they may be subject to Yang-Mills gauge
transformations, requiring the coupling to supersymmetric {\em
Yang-Mills multiplets} \cite{FZ74,SS74b}.

The general theory, combining minimal supergravity, chiral
matter and supersymmetric Yang-Mills theory has been worked out
in \cite{CJS78a,CJS+79,CFGvP82,CFGvP83}. In this construction,
generalized rescalings, compatible with supersymmetry, had to be
carried out to establish the canonical normalization of the
Einstein term. In its final form, this theory exhibits chiral
K\"ahler phase transformations. Alternatively, using conformal
tensor calculus and particular gauge conditions
\cite{KU83b,KU83a}, the cumbersome Weyl rescalings could be
avoided.

But string/membrane theory requires more
fields and more structures - {\em linear
multiplets} \cite{FZW74,Sie79} and {\em 3-form multiplets}
\cite{Gat81}, together with Chern-Simons terms of the gauge and
gravitational types should be included. They
are relevant for string corrections to gauge couplings
\cite{DKL91,ANT91,CO91,Lou91,SV91,DFKZ92,DQQ94,CLO95,Der95}, in
particular non-holomorphic gauge coupling functions, and for
effective descriptions of gaugino condensation \cite{Wu97}, as
well as for a supersymmetric implementation of the consequences
of the Green-Schwarz mechanism \cite{GS84} in an effective theory
\cite{CFGP85,FV87}.

It is clear that a systematic approach should be employed to
cope with such complex structures. This report provides
a presentation of the geometric superspace approach.

The notion of superspace is based on the concept of superfields
\cite{SS74a,FZW74,SS78}: space-time is promoted to superspace in
adding anticommuting parameters and superfields are functions of
space-time coordinates and the anticommuting coordinates.
Supersymmetry transformations are realized as differential
operations involving spinor derivatives.

Implementing the machinery of differential geometry, like
differential forms, exterior derivatives, interior product, etc.
on superspace gives rise to superspace geometry. In this framework
supersymmetry and general coordinate transformations are described
in a unified way as certain diffeomorphisms. Both the graviton and
its superpartner, the gravitino, are identified in the frame
differential form of superspace.

The superspace formulation of supergravity
\cite{WZ77,GWZ77,WZ78a,WZ78b,GWZ79,Zum79a} and supersymmetric
gauge theory \cite{Wes78,Wes83} is by now standard textbook
knowledge \cite{GGRS83,WB83}. A characteristic feature of this
formulation is that the structure group in superspace is
represented by the vector and spinor representations of the
Lorentz group.

This superspace geometry may be modified by adding a chiral
$U(1)$ to the structure group transformations, accompanied by the
corresponding gauge potential differential form. Associated with
this Abelian gauge group is an unconstrained pre-potential
superfield. By itself, this structure is called $U(1)$ superspace
\cite{How82}, it allows to obtain the known supergravity
multiplets mentioned above: minimal, new minimal and non-minimal,
upon applying suitable restrictions \cite{Mul86}.

The superspace description of the supergravity-matter coupling is
obtained from $U(1)$ superspace as well: in this case the chiral
$U(1)$ is replaced by superfield K\"ahler transformations. At the
same time the unconstrained pre-potential is identified with the
superfield K\"ahler potential \cite{BGGM87b,BGG90,Gri90a,Gri90b}.
In this formulation, called {\em K\"ahler superspace} geometry,
or $U_K(1)$ superspace geometry,
the K\"ahler phase transformations are implemented {\em{ab
initio}} at a geometric level, the K\"ahler weights of all the
super- and component fields are given intrinsically and no
rescalings are needed in the construction of the supersymmetric
action. The K\"ahler superspace formulation is related to the
K\"ahler-Weyl formalism \cite{Wes87} in a straightforward way
\cite{BGG90}.

The construction of the general super\-gravity/matter/Yang-Mills
system using the K\"ahler superspace formulation is the central
issue of this report.

In section \ref{RS} we review rigid superspace geometry in some
detail, including supersymmetric gauge theory. Notational details
are presented in appendix \ref{appA}. Section \ref{GRA} contains a
detailed account of the K\"ahler superspace construction. A
collection of elements of $U(1)$ superspace can be found in
appendix \ref{appB}. A more general setting which includes \ka
gauged isometries is treated in appendix \ref{appC}. Derivation
of the superfield equations of motion is reviewed in appendix
\ref{appD}

In section \ref{CPN} we define component fields, their
supersymmetry transformations and construct the complete
component field action.
The K\"ahler superspace formulation is
particularly convenient when the
super\-gravity/\-matter/Yang-Mills system is to be extended to
contain linear multiplets, Chern-Simons forms and 3-form
multiplets, as explained in detail in  sections \ref{2F} and
\ref{3F}. Appendices \ref{appE} and \ref{appF} contain
complements to these sections.

This report is not intended to provide a review of supersymmetry
and its applications. It is rather focused on a quite special
issue, the description of $D=4$, $N=1$ supergravity couplings in
geometric terms, more precisely in terms of superspace geometry.
We have made an effort to furnish a self-contained and exhaustive
presentation of this highly technical subject.

Even when restricted to $D=4$, $N=1$, there are many
topics we have not mentioned, among them supersymmetry
breaking, quantization, anomalies and their cohomological BRS
construction, conformal supergravity or
gravitational Chern-Simons forms.

Similar remarks apply to the bibliography.
The references cited are rather restricted to those directly related to
the technical aspects of differential geometry in superspace applied to
supergravity couplings. 
Even though we cannot claim to have a complete
bibliographical list and apologize in advance for any undue omissions.

\newpage
\sect{ RIGID SUPERSPACE GEOMETRY}
\label{RS}


We gather, here, some of the basic features of superspace geometry
which will be useful later on. In section \ref{RS1} we begin with
a list of the known off-shell multiplets in $D=4, N=1$
supersymmetry, recall the properties of rigid superspace endowed
with constant torsion, and define supersymmetry transformations
in this geometric framework. Next supersymmetric Abelian gauge
theory is reviewed in detail in section \ref{RS2} as an
illustration of the methods of superspace geometry and also in
view of its important role in the context of supergravity/matter
coupling. Although very similar in structure, the non-Abelian
case is presented separately in section \ref{RS3}. In section
\ref{RS4} we emphasize the similarity of \ka transformations with
the Abelian gauge structure, in particular the interpretation of
the kinetic matter action as a composite D-term.

\subsec{ Prolegomena \label{RS1}}

\subsubsection{ $D=4,N=1$ supermultiplet catalogue \label{RS11}}

Since the supersymmetry algebra is an extension of the Poincar\'e
algebra, Wigner's analysis \cite{Wig39} can be generalized to
classify unitary representations \cite{SS74c,Nah77,Fre79,FSZ81} in
terms of physical states. On the other hand, field theories are
usually described in terms of local fields. As on-shell
representations of supersymmetry combine different spins, resp.
helicities, supermultiplets of local fields will contain
components in different representations of the Lorentz group. A
multiplet of a given helicity content can have several
incarnations in terms of local fields. In the simplest case, the
massless helicity $(1/2,0)$ multiplet may be realized in three
different ways, the {\em{chiral multiplet}}, sometimes also called
scalar multiplet \cite{WZ74c}, the {\em{linear multiplet}}
\cite{FZW74,Sie79} or the {\em{3-form multiplet}} \cite{Gat81},
which will be displayed below. At helicity $(1,1/2)$ only one
realization is known: the usual gauge multiplet \cite{WZ74b}. The
$(3/2,1)$ multiplet has a number of avatars as well
\cite{OS76,OS77,Gat77,GS80,GG85}. Finally the $(2,3/2)$
multiplet, which contains the graviton, is known in three versions: the
minimal multiplet
\cite{SW78,FvN78}, the new minimal multiplet \cite{AVS77,SW81} and
the non-minimal multiplet \cite{Bre77,SG79}. This exhausts the
list of massless multiplets in $D=4,N=1$ supersymmetry in the
sense of irreducible multiplets. The massive multiplet of spin content
$(1,1/2,1/2,0)$ which will be presented below may be understood
as a combination of a gauge and a chiral multiplet.
 We just display
the content of some of the off-shell supermultiplets that we shall
use in the sequel, indicating the number of bosonic ($\bf b$) and
fermionic ($\bf f$) degrees of freedom (the vertical bar
separates auxiliary fields from physical ones) \vskip1cm
\begin{itemize}

\item{The chiral/scalar multiplet}
 \[
{\begin{array}{ccc} \phi &\sim& \lp A\ , \ \chi_\al ~|~ F \, \rp
\end{array}
 \left\{\begin{array}{ccl} A    & \ 2 \, {\bf b}
& \ \mbox{complex scalar}
\\[2mm] \chi_\al & \ 4
\, {\bf f} &
         \ \mbox{Weyl spinor}
\\[1.8mm]
F & \ 2 \, {\bf b} & \ \mbox{complex scalar}
\end{array}\right.}
\]
The conjugate multiplet
$\phib \sim \lp \bA \ , \ \chib^\da ~|~ \bar F \, \rp$, consists of the
complex conjugate component fields. It has the same number of degree of
freedom.    \vskip1cm

\item{The generic vector multiplet}   
\[
V \ \sim \ \lp C \ , \
\raisebox{-1.5ex}{$\stackrel{ \raise4pt\hbox{$ {\textstyle
\vp_\al} $}} {\vpb^\da}$}
\ , \
H%
\ , \ V_m \ , \
\raisebox{-1.5ex}{$\stackrel{ \raise4pt\hbox{$ {\textstyle
\la_\al} $}} {\lab^\da}$}
\ , \ D \rp  \left\{\begin{array}{ccl} C & \ 1 \, {\bf b} & \cem
\mbox{real scalar}\\[1.5mm] \vp_\al , \vpb^\da & \ 4 \, {\bf f}
& \cem \mbox{Majorana spinor} \\[1.5mm] H & \ 2 \, {\bf b} & \cem
\mbox{complex  scalar} \\[1.5mm] V_m & \ 4 \, {\bf b} & \cem
\mbox{real vector} \\[1.5mm] \la_\al , \lab^\da & \ 4 \, {\bf f} &
\cem \mbox{Majorana spinor} \\[1.5mm]
    D   & \ 1 \, {\bf b} & \cem \mbox{real scalar}
\end{array}\right.
\]
This vector multiplet can occur in two ways in physical models: as a massive
vector field and its supersymmetric partners or as a gauge multiplet. In the
massive vector case all dynamical fields have the same mass, the Majorana
spinors,
$\vp_\al$, $\vpb^\da$ and $\la_\al$, $\lab^\da$
combine into a Dirac spinor;
the auxiliary sector contains one real and one complex scalar.

\[V_{\mbox{massive}} \ \sim \ \lp C \ , \ V_m \ , \ \Psi ~|~ H \ , \
D \rp
\left\{\begin{array}{ccl} C & \ 1 \, {\bf b} & \ \mbox{real
scalar}
\\[1.5mm] \Psi & \ 8 \, {\bf f} & \
\mbox{Dirac spinor}
\\[1.5mm]
V_m & \ 4 \, {\bf b} & \ \mbox{real vector}
\\[1.5mm] H & \ 2 \, {\bf b} & \ \mbox{complex scalar}
\\[1.5mm]
D   & \ 1 \, {\bf b} & \ \mbox{real scalar}
\end{array}\right.
\]
\vskip.5cm

The gauge multiplet contains less dynamical degrees of freedom due
to gauge transformations which have the structure of scalar
multiplets. One is left with a massless vector, a Majorana spinor
(the gaugino) and an auxiliary scalar
\[
V_{\mbox{gauge}} \ \sim \ \lp V_m \ , \
\raisebox{-1.5ex}{$\stackrel{ \raise4pt\hbox{$ {\textstyle
\la_\al} $}} {\lab^\da}$}
 ~|~ D \rp
 \left\{\begin{array}{ccl} V_m & \ 3 \, {\bf b} & \
\mbox{gauge vector}
\\[1.5mm] \la_\al ,
\lab^\da & \ 4 \, {\bf f} & \ \mbox{Majorana spinor}
\\[1.5mm]
    D   & \ 1 \, {\bf b} & \ \mbox{real scalar}
\end{array}\right.
\]
\item{The 2-form (or linear) multiplet}
\[L_{\mbox{linear}} \ \sim \ \lp L \ , \
\raisebox{-1.5ex}{$\stackrel{ \raise4pt\hbox{$ {\textstyle
\La_\al} $}} {\Lab^\da}$}
 \ , \ b_{mn} \rp
 \left\{\begin{array}{ccl} L & \ 1 \, {\bf b} & \
\mbox{real scalar}
\\[1.5mm] \La_\al ,
\Lab^\da & \ 4 \, {\bf f} & \ \mbox{Majorana spinor}
\\[1.5mm]
    b_{mn}   & \ 3 \, {\bf b} & \ \mbox{antisym. tensor}
\end{array}\right.
\]
The number of physical degrees of freedom of $b_{mn}$ is
$3\,=\,6-4+1$. This multiplet contains no auxiliary field.
\item{The 3-form (or constrained chiral) multiplet}
\[
C_{(3)} \ \sim \ \lp Y \ , \
\raisebox{-1.5ex}{$\stackrel{ \raise4pt\hbox{$ {\textstyle
\eta_\al} $}} {\etab^\da}$}\ , \ C_{lmn} \,
 ~|~H \rp
 \left\{\begin{array}{ccll} Y & \ 2 \, {\bf b} & \
\mbox{complex scalar}
\\[1.5mm] \eta_\al ,
\etab^\da & \ 4 \, {\bf f} & \, \mbox{Majorana spinor}
\\[1.5mm]
C_{lmn}& \ 1 \, {\bf b} & \, \mbox{antisym. tensor}
\\[1.5mm]
    H   & \ 1 \, {\bf b} & \, \mbox{real scalar}
\end{array}\right.
\]
The number of physical degrees of freedom of $C_{lmn}$ is
$1\,=\,4-6+4-1$.
\end{itemize}
Although this section is devoted to rigid superspace, to be
complete, we include here the list of multiplets appearing in
supergravity :
\vskip1cm
\begin{itemize}
\item{The minimal multiplet $(12\, +\, 12)$}  \[  \lp {e_m}^a \ , \
\raisebox{-1.5ex}{$\stackrel{ \raise4pt\hbox{$ {\textstyle
{\psi_m}^\al} $}}{ {\psib}_{m\, \da}}$}
\  | \ b_a \
 , \ M \rp  \left\{\begin{array}{ccl} {e_m}^a & \ 6 \, {\bf b} & \cem
\mbox{graviton}\\[1.5mm] {\psi_m}^\al , {\psib}_{m\, \da} & \ 12 \,
{\bf f} & \cem \mbox{gravitino}
\\[1.5mm] b_a & \ 4 \, {\bf b} & \cem \mbox{real vector}
\\[1.5mm]
    M   & \ 2 \, {\bf b} & \cem \mbox{complex scalar}
\end{array}\right.
\]  \vskip.5cm
\item{The new minimal multiplet $(12\, +\, 12)$}  \[  \lp {e_m}^a \ , \
\raisebox{-1.5ex}{$\stackrel{ \raise4pt\hbox{$ {\textstyle
{\psi_m}^\al} $}}{ {\psib}_{m\, \da}}$}
\  | \ V_m \
 , \ b_{mn} \rp  \left\{\begin{array}{ccl} {e_m}^a & \ 6 \, {\bf b} & \cem
\mbox{graviton}\\[1.5mm] {\psi_m}^\al , {\psib}_{m\, \da} & \ 12 \,
{\bf f} & \cem \mbox{gravitino}
\\[1.5mm] V_m & \ 3 \, {\bf b} & \cem \mbox{gauge vector}
\\[1.5mm]
    b_{mn}   & \ 3 \, {\bf b} & \cem \mbox{antisym. tensor}
\end{array}\right.
\]    \vskip.5cm
\item{The non-minimal multiplet $(20\, +\, 20)$}  \[  \lp {e_m}^a \, ,
\,
\raisebox{-1.5ex}{$\stackrel{ \raise4pt\hbox{$ {\textstyle
{\psi_m}^\al} $}}{ {\psib}_{m\, \da}}$} \,  | \, b_a \,
 , \, c_a\, ,\,
 \raisebox{-1.5ex}{$\stackrel{ \raise4pt\hbox{$ {\textstyle
{\chi}_\al} $}}{ {\chib}^{\da}}$}\, , \,
\raisebox{-1.5ex}{$\stackrel{ \raise4pt\hbox{$ {\textstyle
{T}_\al} $}}{ {\bT}^{\da}}$}\, , \, S
 \rp  \left\{\begin{array}{ccl} {e_m}^a & \ 6 \, {\bf b} & \cem
\mbox{graviton}\\[1.5mm] {\psi_m}^\al , {\psib}_{m\, \da} & \ 12 \,
{\bf f} & \cem \mbox{gravitino}
\\[1.5mm] b_a & \ 4 \, {\bf b} & \cem \mbox{real vector}
\\[1.5mm] c_a & \ 4 \, {\bf b} & \cem \mbox{real vector}
\\[1.5mm] \chi_\al, \chib^\da & \ 4 \, {\bf f} & \cem \mbox{Majorana}
\\[1.5mm] T_\al, \bT^\da & \ 4 \, {\bf f} & \cem \mbox{Majorana}
\\[1.5mm]S   & \ 2 \, {\bf b} & \cem \mbox{complex scalar}
\end{array}\right.
\]
\end{itemize}
In this report we will only be concerned with the minimal supergravity
multiplet.

We conclude the list of known $N=1$ supermuliplets with the
$(3/2, 1)$ multiplet \cite{OS76,dWvH79,FvN76,FV79}. It describes
physical states of helicities $3/2$ and $1$, its off-shell
realization contains 20 bosonic and 20 fermionic component fields.

\begin{itemize}
\item{The (3/2,1) multiplet $(20\, +\, 20)$}  \[  \lp
\,  B_m  \, , \,
\raisebox{-1.5ex}{$\stackrel{ \raise4pt\hbox{$ {\textstyle
{{\Gamma}_m}^\al} $}}{ {\bar \Gamma}_{m\, \da}}$} \,
 | \
 \raisebox{-1.5ex}{$\stackrel{ \raise4pt\hbox{$ {\textstyle
{\rho}_\al} $}}{ {\bar \rho}^{\da}}$}\, , \,
P \, , \, J \, , \, Y_a \, , \, T_{ba} \, , \,
\raisebox{-1.5ex}{$\stackrel{ \raise4pt\hbox{$ {\textstyle
{\Sigma}^\al} $}}{ {\bar \Sigma}_{\da}}$}\, \rp
\left\{\begin{array}{ccl}
B_m & \ 3 \, {\bf b} & \hs{.8cm}
\mbox{gauge vector}
\\[1.5mm]
{\Gamma_m}^\al , {\bar \Gamma}_{m\, \da} &
\ 12 \, {\bf f} & \hs{.8cm} \mbox{Rarita-Schwinger}
\\[1.5mm]
\rho_\al, \bar \rho{}^\da & \ 4 \, {\bf f} &
\hs{.8cm} \mbox{Majorana}
\\[1.5mm]
P   & \ 1 \, {\bf b} & \hs{.8cm} \mbox{real scalar}
\\[1.5mm]
J   & \ 2 \, {\bf b} & \hs{.8cm} \mbox{complex scalar}
\\[1.5mm]
Y_a & \ 4 \, {\bf b} & \hs{.8cm} \mbox{complex vector}
\\[1.5mm]
T_{ba} & \ 6 \, {\bf b} & \hs{.8cm} \mbox{antisym. tensor}
\\[1.5mm]
{\Sigma}^\al, {\bar \Sigma}_{\da} & \ 4 \, {\bf f} &
\hs{.8cm} \mbox{Majorana}
\end{array}\right.
\]
\end{itemize}
The component field content displayed here corresponds to the de
Wit-van Holten multiplet \cite{dWvH79}. It is related to the
Ogievetsky-Sokatchev multiplet \cite{OS76}
by a duality relation \cite{LR83, GK84},
similar to that between chiral and linear multiplet. Superspace
descriptions are discussed in \cite{GS80,GG85,GK84}.

\subsubsection{ Superfields and multiplets \label{RS12}}

The anticommutation relation,
\begin{equation}
\{Q_{\al},\bar{Q}^{\da}\} \, = \, 2{(\si^{a}\eps)_{\al}} ^{\da}P_a
, \label{RS.1}
\end{equation}
 which relates the generators
$Q_\al$ and $\bar{Q}^{\da}$ of supersymmetry transformations to
translations $P_a$ in space-time is at the heart of the
supersymmetry algebra. Superspace geometry, on the other hand, is
based on the notion of superfields which are functions depending
on space-time coordinates $x^m$ as well as on spinor,
anticommuting variables $\theta^{\al}$ and $\bar\theta_{\da}$. Due to
the anticommutativity, superfields are polynomials of finite degree
in the spinor variables. Coefficients of the monomials in
$\theta^{\al}, \,\bar\theta_{\da}$ are called {\em component
fields}.

\indent

Supersymmetry transformations of superfields are generated by the
differential operators
\begin{equation}
Q_{\al} \, = \, \frac{\partial}{\partial\theta^{\al}} -
i\bar\theta_{\da} {{(\sib^{m}\eps)}^{\da}}_{\al}
\frac{\partial}{\partial x^{m}}, \label{RS.2} \end{equation}
\begin{equation}
{\bar{Q}}^{\da} \, = \,
\frac{\partial}{\partial\bar{\theta}_{\da}} - i{{\theta}
^\al{{(\si^{m}\eps)}_\al}^{\da}} \frac{\partial}{\partial{x}^m}
\label{RS.3} \end{equation}
 which,
of course, together with $P_a = -i{\partial}/{\partial{x}^a}$
satisfy (\ref{RS.1}) as well. A general superfield, however, does not
necessarily provide an irreducible representation of supersymmetry.

The differential operators
\begin{equation}
D_\al \, = \, \frac{\partial}{{\partial\theta}^\al} +
i{{\thb}_{\da}} {{({\sib}^{m}\eps)} ^{\da}}_{\al} \frac{\partial}
{\partial{x}^m}, \label{RS.4} \end{equation}
\begin{equation}
 D^{\da} \, = \, \frac{\partial}{{\partial}{\thb}_{\da}} + i{\theta^{\al}}
{{({\si}^{m}\eps)}_\al}^{\da} \frac{\partial}{\partial{x}^m},
\label{RS.5} \end{equation} anticommute with the supersymmetry
generators, \ie they are  covariant with respect to supersymmetry
transformations and satisfy, by definition, the anticommutation
relations
\begin{equation}
\{D_\al, D^{\da}\} \, = \, 2i{{(\si^{m}\eps)}_\al}^{\da}
\frac{\partial}{\partial{x}^m}, \label{RS.6} \end{equation}
\begin{equation}
\{D_{\al},D_{\beta}\} \, = \, 0, \cem \{D^{\da}, D^{\db}\} \, = \,
0. \label{RS.7} \end{equation}
 \noindent
These spinor covariant derivatives can be employed to define
{\em{constrained superfields}} which may be used to define
irreducible field representations of the supersymmetry algebra.

The most important ones are
\begin{itemize} \item{The chiral superfields $\phi,\;\phib$}
are complex superfields, subject to the constraints
\begin{equation}
 D^{\da}\phi \,=\,0,\cem
 D_\al{\phib} \,=\,0.    \label{RS.8} \end{equation}
 They are usually employed to describe supersymmetric matter
multiplets.

 \item{The superfields $W^\al,\;W_{\da}$}, subject to the constraints
\begin{equation}
 D_{\al}W_{\da} \,=\,0,  \cem
 D^{\da} W^{\al} \,=\,0,   \label{RS.12}
\end{equation}
\begin{equation}
 D^{\al}W_{\al} \,=\,D_{\da}W^{\da},     \label{RS.14}
\end{equation}
are related to the field strength tensor and play a key role in
the description of supersymmetric gauge theories.
\item{The linear superfield $L$}, subject to the linearity
constraints\footnote{With the usual notations $D^2=D^\al D_\al$
and $\bD^2= D_\da D^\da$, which will be used throughout this
paper.}
\begin{equation}
 D^2 L \,= \, 0,  \cem {\bD}^2 L \,= \, 0.     \label{RS.10}
\end{equation}
 As explained above, it describes the supermultiplet of an
antisymmetric tensor or 2-form gauge potential, as such it plays a
key role in describing moduli fields in superstring effective
theories.
 \item{The 3-form superfields $Y, \ovY$},
are chiral superfields ($D_\da Y\,=\,0,\,D_\al \ovY\,=\,0 $)
with a further constraint
\begin{equation}
D^2 Y - \bD^2 \ovY \, = \, \f{8i}{3} \vep^{klmn}
\Si_{klmn},\label{chicont}
\end{equation}
with  $\Si_{klmn}$, the field strength of the 3-form. These
superfields are relevant in the context of gaugino condensation
and of Chern-Simons forms couplings.
\end{itemize}
 The superfields $L$ and $W^\al,\;
W_{\da}$ are invariant under the respective gauge transformations,
they can be viewed as some kind of invariant field strengths. As
is well known, geometric formulations of 1-, 2- and 3-form gauge
theories in superspace exist such that indeed
$W^\al$, ${W}_{\da},\,L$ and $Y$, $\ovY$ are properly identified
as field strength superfields with (\ref{RS.12}) - (\ref{chicont})
constituting the corresponding Bianchi identities.

\subsubsection{ Geometry and supersymmetry \label{RS13}}

In order to prepare the ground for a geometric superspace
formulation of such theories one introduces a local frame for
rigid superspace. It is suggestive to re-express
(\ref{RS.4}) - (\ref{RS.7}) in terms of supervielbein (a generalization of
Cartan's local frame) and torsion in a superspace of
coordinates
$z^{M}\sim(x^m,\theta^\mu,{{\bar\theta}_{\dot{\mu}}})$,
derivatives
$\partial_{M}\sim({\partial}/{\partial{x}^m},
{\partial}/{\partial\theta^\mu},
{\partial}/{\partial{\bar\theta}_{\dot\mu}})$
and differentials
$dz^M\sim (dx^m,d{\theta}^{\mu},d\bar{\theta}_{\dot\mu})$.
The latter may be viewed as the tangent and cotangent frames of superspace,
respectively.
 The supervielbein 1-form of rigid superspace is
\begin{equation}
 E^A \,=\,dz^M {{E_M}^A},      \label{RS.17}
\end{equation}
with
\begin{equation}
{E_{M}}^{A} \,=\,\left(
\begin{array}{ccc}
\vspace{4 mm}{{\delta_{m}}^{a}} & 0 & 0\\ \vspace{5
mm}-i{({\thb}{\sib}^{a}{\eps})}_{\mu} & {\dtdu{\mu}{\al}} & 0\\
-i{({\theta}{\si}^{a}{\eps})}^{\dot{\mu}} & 0 &
{\delta^{\dot{\mu}}}_{\da}
\end{array}
\right).
\label{RS.18}
\end{equation}
The inverse vielbein ${E_{A}}^{M}$, defined by the relations
 $${E_M}^A(z)\,{E_A}^N(z)\,=\,{\delta_M}^N,\cem
{E_A}^M(z)\,{E_M}^B(z)\,=\, {\delta_A}^B,$$ reads
\begin{equation}
{E_{A}}^{M} \,=\,\left(
\begin{array}{ccc}
\vspace{4 mm}{{\delta_{a}}^{m}} & 0 & 0\\ \vspace{5
mm}i{({\thb}{\sib}^{m}{\eps})}_{\al} & {\dtdu{\al}{\mu}} & 0\\
i{({\theta}{\si}^{m}{\eps})}^{\da} & 0 & {\dtud{\da} {\dot{\mu}}}
\end{array}
\right).      \label{RS.19}  \end{equation} \vskip.6cm \noindent
The torsion 2-form in rigid superspace is defined as the exterior
derivative of the vielbein 1-form:
\begin{equation}
 dE^A \,=\,T^A\,=\,\f{1}{2} E^B E^C {T_{CB}}^A.
\label{RS.20} \end{equation}
 Now, for the differential operators
 $D_A \,=\, ({\partial}/{\partial{x}^a}, D_\al,{D^{\da}})$ we
have
\begin{equation}
D_A \,=\,{E_A}^{M}\partial_{M},     \label{RS.15} \end{equation}
\begin{equation}
 (D_C,D_B) \,=\,- {T_{CB}}^{A} D_{A},     \label{RS.16}
\end{equation}
 with the graded commutator defined as $(D_{C},D_{B}) = D_C D_B
 - (-)^{bc} D_B D_C$ with $b=0$ for a vector and $b=1$ for a spinor index.
The fact that the same torsion coefficient appears in
(\ref{RS.16}) and in (\ref{RS.16}) reflects the fact that $dd =
0$ in superspace. To be more precise consider the action of $dd$
on some generic $0$-form superfield ${\Phi}$. Application of $d$
to the expression $d \Phi = E^B \! D_B \Phi$, in combination with
the rules of superspace exterior calculus, \ie $dd \Phi \, = \, d
E^B \! D_B \Phi \, = \, E^B E^C D_C D_B \Phi + (d E^A) D_A \Phi$,
and the definitions introduced so far gives immediately
\begin{equation}
dd \Phi \, = \,
\frac{1}{2} E^B E^C
\lp  ( D_B , D_A ) \Phi + T_{C B }{}^A D_A \Phi \rp,
\end{equation}
establishing the assertion. A glance at the differential algebra
of the $D_A$'s, in particular (\ref{RS.6}), shows then that the
only non-vanishing torsion component is
\begin{equation}
{T_{\gamma}}^{\db a} \,=\,- 2i {{(\si^{a}\eps)}_\gamma} ^{\db}.
\label{RS.22}
\end{equation}

Given the relation between supersymmetry transformations and the
"square root" of space-time translations (\ref{RS.1}), we would
like to interpret them as diffeomorphisms in superspace. The
action of diffeomorphisms on geometric objects such as vector and
tensor fields or differential forms is encoded in the Lie
derivative, which can be defined in terms of basic operations of a
differential algebra (suitably extended to superspace), \ie the
exterior derivative, $d$, and the interior product, $\iota_\zeta$,
such that
\begin{equation}
L_{\zeta} \,=\,\iota_{\zeta}d + {d\iota}_{\zeta}. \label{RS.24}
\end{equation}
The interior product, for instance, of a vector field $\zeta$ with
the vielbein 1-form is
\begin{equation}
 \iota_{\zeta}E^A \,=\,{\zeta}^{M} {E_{M}}^A \,=\,{\zeta}^{A}.
\label{RS.23} \end{equation}
 The definition of
differential forms in superspace (or superforms) and the
conventions for the differential calculus are those of Wess and
Bagger \cite{WB83} -cf. appendix \ref{appA0} below for a summary.
Then, on superforms $d$ acts as an anti-derivation of degree $+1$,
the exterior derivative of a p-form is a (p$+1$)-form. Likewise,
$\iota_{\zeta}$ acts as an anti-derivation of degree $-1$ so that
the Lie derivative $L_{\zeta}$, defined by ({\ref{RS.24}),
 does not change the degree of differential forms. This
 geometric formulation will prove to be very efficient to construct more
 general
 supersymmetric or supergravity theories involving p-form fields.

For the vielbein itself, combination of (\ref{RS.20}) and
(\ref{RS.23}) yields
\begin{equation}
L_{\zeta} E^A \,=\,{d\zeta}^{A} + \iota_{\zeta} T^{A}.
\label{RS.25}
\end{equation}
 On a $0$-form superfield, ${\Phi}$, the Lie derivative acts
according to
\begin{equation}
L_{\zeta}{\Phi} \,=\,{\iota}_{\zeta}d{\Phi} \,=\,{\zeta}^{A}
D_{A}{\Phi}\,=\,{\zeta}^{M}\partial_{M}{\Phi}. \label{RS.26}
\end{equation}
The Lie derivative $L_{\xi}$ with respect to the particular vector
field
\begin{equation}
\xi^{M} \, = \,
(i\theta^{\al}{(\si^{m}\eps)}_{\al}{}^{\da}{\xib}_{\da} +
i{\bar{\theta}}_{\da}{(\sib^{m}{\eps})}^{\da}{}_{\al} \xi^{\al} \
, \ \xi^{\mu} \ , \ {\xib}_{\dot{\mu}}), \label{RS.27}
\end{equation}
 leaves the vielbein $1$-form
(\ref{RS.17}), (\ref{RS.18}) invariant, \ie
\begin{equation}
L_{\xi} E^A \,=\,0. \label{RS.28} \end{equation}
This is most easily seen in terms of $\xi^A = \iota_\xi E^A$, which is
explicitly given as
\begin{equation}
\xi^A \,=\,(2i(\theta\si^{a}{\xib}) + 2i
({\bar{\theta}}{\sib}^{a}{\xi}) \ , \ \xi^{\al} \ , \
{\xib}_{\da}). \label{RS.29}
\end{equation}
Recall that $L_{\xi} E^A \,=\, d \xi^A + \iota_\xi T^A$. This
shows immediately that for the spinor components the equation is
satisfied, because $\xi^\undal\,$ is constant and $T^\undal\,$
vanishes. As to the vector part one keeps in mind that in $d
\xi^a \, = \, E^B D_B \, \xi^a$ only the derivatives with respect
to $\th$, $\thb$ contribute and compare the result
\[ d \xi^a \, = \,
2i E^{\al}{(\si^{a}\eps)}_{\al}{}^{\da}{\xib}_{\da} +
2i E_{\da}{(\sib^{a}{\eps})}^{\da}{}_{\al} \xi^{\al}
\]
to the expression for the interior product acting on
$T^a = 2i E_\db E^\gm {{(\si^{a}\eps)}_{\gm}}^{\db}$, \ie
\[ \iota_\xi T^A \, = \,
2i E^{\gm}{(\si^{a}\eps)}_{\al}{}^{\da}{\xib}_{\da} +
2i E_{\da}{(\sib^{a}{\eps})}^{\da}{}_{\al} \xi^{\al}.
\]
The Lie derivative of a generic superfield $\Phi$ in terms of the particular
vector field $\xi^A$ defined in (\ref{RS.29}) is given as
\begin{equation}
L_{\xi}{\Phi} \,=\,\xi^{A} D_{A} \Phi \,=\,(\xi^{\al}Q_{\al} +
{\xib}_{\da}{\bar{Q}^{\da}}){\Phi},
\label{RS.30}
\end{equation}
reproducing the infinitesimal supersymmetry
transformation with $Q_{\al}$ and ${\bar{Q}}^{\da}$
as defined in (\ref{RS.2}) and (\ref{RS.3}).

\begin{itemize}
\item{{\em{Supersymmetry transformations can be
 identified as diffeomorphisms of parameters $\xi^\al,
  {\xib}_\da$ which leave $E^A$ invariant.}}}
  \end{itemize}
Combining such a supersymmetry transformation with a translation of parameter
$\vep^a$, we obtain
\begin{eqnarray}
 L_\vep \Phi +L_\xi \Phi &=& \vep^a \prt_a \Phi +\xi^A D_A
 \Phi \nonumber \\ &=& (\vep^a +\xi^a)\prt_a \Phi+\xi^\al
 D_\al \Phi +{\xib}_\da D^\da \Phi.
\end{eqnarray}
The transformations with the particular choice $\vep^a =-\xi^a$ of a $\xi$
dependent space-time translation, will be called {\em super-translations}.
They are given as
 \begin{equation} \delta \Phi =\lp\xi^\al
 D_\al  +{\xib}_\da D^\da \rp \Phi.
 \end{equation}
  These special
transformations will be used in the formulation of supersymmetric
theories (and in particular in supergravity \cite{Zum79a}). Let us
stress that for $\theta = {\bar{\theta}} = 0$, supersymmetry
transformations and super-translations coincide. The components of
a superfield are traditionally defined as coefficients in an
expansion with respect to $\theta$ and ${\bar{\theta}}$. In the
geometric approach presented here, component fields are defined
as lowest components of superfields. Higher components are
obtained by successive applications of {\em{covariant}}
derivatives and subsequent projection to $\theta = {\bar{\theta}}
= 0$. Component fields defined this way are naturally related by
super-translations. The basic operational structure is the algebra
of covariant derivatives.

\subsec{ Abelian Gauge Structure \label{RS2}}

\subsubsection{ Abelian gauge potential \label{RS21}}

In analogy to usual gauge theory, gauge potentials in
supersymmetric gauge theories are defined as 1-forms in
superspace
\begin{equation}
A \,=\,E^{A}A_{A} \,=\,E^{a}A_{a}+E^{\al}A_{\al}+E_{\da}A^{\da}.
\label{RS.31}
\end{equation}
 The coefficients $A_{a},\; A_{\al},\; A^{\da}$
are, by themselves, superfields. Since we consider here an Abelian
gauge theory, $A$ transforms under gauge transformations as
\begin{equation}
{A} \, \mapsto \,A-g^{-1}dg.    \label{RS.32} \end{equation}
 The gauge transformation parameters $g$ are 0-form superfields and the
invariant field strength is a 2-form,
\begin{equation}
F \,=\,dA\,=\,\f{1}{2} E^{A}E^{B}F_{BA}. \label{RS.34}
\end{equation} Observe that, following (\ref{RS.20})
a torsion term appears
 in its explicit expression:
\begin{equation}
F_{BA} \,=\,D_{B}A_{A} - (-)^{ab}D_{A}A_{B}+{T_{BA}}^CA_{C}.
\label{RS.35} \end{equation}
 By definition, (\ref{RS.34}), the
field strength satisfies the Bianchi identity
\begin{equation}
dF \,=\,0.    \label{RS.36} \end{equation}

Consider next a covariant (0-form) superfield ${\Phi}$ of weight
$w( {\Phi})$ under Abelian superfield gauge transformations, \ie
\begin{equation}
{\Phi} \,\stackrel{g}{\mapsto}\,g^{w({\Phi})}{\Phi}. \label{RS.37}
\end{equation} Its covariant (exterior) derivative,
\begin{equation}
{\cd}{\Phi} \,=\,E^{A} {\cd}_{A} {\Phi},   \label{RS.38}
\end{equation} is defined as{\footnote{If $\Phi_p$ is a p-form, we
define it as ${\cd}{\Phi}_p \,=\,d{\Phi}_p +(-)^p
w({\Phi}_p)\,A\,{\Phi}_p$.}}
\begin{equation}
{\cd}{\Phi} \,=\,d{\Phi} + w({\Phi})\,A\,{\Phi}. \label{RS.39}
\end{equation} Covariant differentiation of (\ref{RS.38}) yields
in turn $( w({\cd}{\Phi}) \,=\,w({\Phi}))$
\begin{equation}
{\cd}{\cd}{\Phi} \,=\,w({\Phi})\, F\,{\Phi}, \label{RS.40}
\end{equation} leading to the graded commutator
\begin{equation}
({\cd}_{B},{\cd}_{A}){\Phi} \,=\,w({\Phi})F_{BA}\,{\Phi} -
{T_{BA}}^{C} {\cd}_{C} {\Phi}. \label{RS.41} \end{equation}
Super-translations in superspace {\em{and}} infinitesimal
superfield gauge transformations, $g\approx 1 + \al$, with $\al$ a real
superfield, change
$A$ and ${\Phi}$ into ${A'}=A+\delta A$ and ${\Phi}' = {\Phi} + \delta
{\Phi}$ such that
\begin{equation} {\delta} A \, =\,{{\iota}_{\xi}}F - d({\al} -
{\iota}_{\xi}A) \label{RS.42}
\end{equation}
 and
\begin{equation}
\delta {\Phi} \,=\,\iota_{\xi}{\cd}{\Phi} + w({\Phi})(\al -
{\iota}_{\xi}A){\Phi}. \label{RS.43}
\end{equation}
 The combination
of a super-translation and of a {\em compensating} gauge
transformation of superfield parameter $\al = {\iota}_{\xi}A$
gives rise to remarkably simple transformation laws. This
parametrization is particularly useful for the definition of
component fields and their supersymmetry transformations. We shall
call these special transformations:{\em{ Wess-Zumino
transformations}}, they are given as
 \begin{equation}
  \delta_{\WZ} \Phi \,=\,
\iota_{\xi}{\cd}{\Phi} , \cem \delta_{\WZ}A \,= \,
{{\iota}_{\xi}}F. \label{WZTfs} \end{equation} Let us stress that
the formalism developed here is well adapted to describe
supersymmetry transformations of differential forms.

So far, $\Phi$ was considered as some generic superfield. Matter fields are
described in terms of chiral superfields. In the context of a gauge structure
the chirality conditions are most conveniently defined in terms of covariant
derivatives. A superfield $\phi$ is called {\em covariantly chiral} and a
superfield $\phib$ is called {\em covariantly antichiral}, if they satisfy
the conditions
\begin{equation}
{\cal D}^{\da} \! {\phi} \,=\,0, \cem {\cal D}_{\al}{\phib} \,=\,0.
\label{RS.44} \end{equation}
Observe that usually they are supposed to have opposite weights
$w(\phib) = -w(\phi)$.
Consistency of the covariant chirality constraints (\ref{RS.44}) with the
graded commutation relations (\ref{RS.41}) implies then
\begin{equation}
F^{{\db}{\da}} \,=\,0, \cem F_{\beta\al} \,=\,0. \label{RS.46}
\end{equation}
   Moreover, due to the (constant)
torsion term in (\ref{RS.35}), \ie
\begin{equation}
{F_{\beta}}^{\da} \,=\,D_{\beta}A^{\da}+D^{\da}A_{\beta}-
2i{{({\si}^{a} {\eps})}_{\beta}}^{\da}A_{a}, \label{RS.48}
\end{equation}
the condition
\begin{equation}
{F_{\beta}}^{\da} \,=\,0    \label{RS.49} \end{equation}
 amounts
to a mere covariant redefinition of the vector superfield gauge
potential $A_a$. Given the constraints (\ref{RS.46}) on $
F_{{\beta}{\al}}$ and $F^{{\db}{\da}}$, the properties of the
remaining components $F_{{\beta}{a}}$, ${F^{\db}}_{a}$ and $F_{ba}$
of the superfield strength $F_{BA}$ are easily derived from the
Bianchi identities (\ref{RS.36}) which
read\footnote{${\oint}_{(CBA)}$ stands for the graded cyclic
permutation on the super-indices $CBA$, explicitly defined as: \
$\oint_{(CBA)} CBA \, = \,
CBA+{(-)}^{a(c+b)}ACB+{(-)}^{(b+a)c}BAC$.}:
\begin{equation}
{\oint}_{(CBA)}(D_{C}F_{BA}+{{T_{CB}}^{D}}F_{DA}) \,=\,0.
\label{RS.50} \end{equation}
 It turns out that the whole geometric
structure which describes supersymmetric gauge theories can be
formulated only in terms of the superfields $W_{\al}$ and
$W^{\da}$ such that
 \begin{eqnarray} F_{{\beta}{a}} &=&
+i{{\si}_{a}}_{\beta \db}\,W^{\db}, \label{RS.51}
\\ {F^{\db}}_{a} &=&
-i{{\sib}_a}^{\db \beta}W_{\beta}, \label{RS.52}
\\ \hspace{-1.5cm} F_{ba} \ &=& \f{1}{2}
{{({\sib}_{ba})}^{\db}}_{\da} D^{\da}
W_{\db}-\frac{1}{2}{(\si_{ba})_{\beta}}^{\al} D_{\al} W^{\beta}.
\label{RS.53} \end{eqnarray}
 Furthermore the Bianchi identities imply the
restrictions (\ref{RS.12}), (\ref{RS.14}). In this sense these
equations have an interpretation as Bianchi identities, providing
a condensed version of (\ref{RS.50}).

\subsubsection{ Solution of constraints and pre-potentials \label{RS22}}

Equation (\ref{RS.50}) is the supersymmetric analogue of the
geometric part of Maxwell's equations
\begin{equation}
  \prt_c F_{ba} +\prt_a F_{cb} + \prt_b F_{ac}\,=\,0, \nonumber
\end{equation}
which are solved in terms of a vector potential, $A_a$, such that
$F_{ba} = \prt_b A_a - \prt_a A_b $. In the supersymmetric case a
similar mechanism takes place, via the explicit solution of the
constraints (\ref{RS.12}), (\ref{RS.14}). To be more precise these
solutions can be written in terms of superfields $T$ and $U$ as
\begin{equation}
A_{\al} \,=\,- T^{-1}D_{\al}T\, =\, -D_\al \log T,\label{RS.54}
\end{equation}
\begin{equation}
A^{\da} \,=\,- U^{-1} D^{\da}U\,=\,-D^\da \log U. \label{RS.55}
\end{equation}
Indeed one obtains from (\ref{RS.51}), (\ref{RS.52})
 \begin{equation}
 W_\al \,=\, +\f{1}{8} {\bD}^2 D_\al \log(T U^{-1}), \ \ \
  W^\da \,=\, +\f{1}{8} {D}^2 D^\da \log(T U^{-1}), \end{equation}
  which is easily seen to satisfy (\ref{RS.12}), (\ref{RS.14}).
 The superfields $T$ and $U$ are called pre-potentials; they are subject
to gauge transformations which have to be consistent with the
gauge transformations (\ref{RS.32}) of the potentials. However due
to the special form of the solutions (\ref{RS.54}), (\ref{RS.55}),
we have the freedom to make extra chiral, resp. antichiral,
transformations, explicitly
\begin{equation}
T \ \mapsto \ {\overline{{\bf P}}}\ T\ g,   \label{RS.57}
\end{equation}
\begin{equation}
U \ \mapsto \ {\bf Q}\ U\ g. \label{RS.58} \end{equation}
 The
new superfields $\overline{{\bf P}}$ and ${\bf Q} $ parametrize so
called pre-gauge transformations which do not show up in the
transformation laws of the potentials themselves due to their
chirality properties
\begin{equation}
D_{\al}{\overline{{\bf P}}} \,=\,0,   \cem  D^{\da} {\bf Q} \, =
\, 0. \label{RS.60} \end{equation} The terminology originates from
the fact that, due to the covariant constraints, the gauge
potentials can be expressed in terms of more fundamental
{\em{unconstrained}} quantities, the pre-potentials, which in turn
give rise to new gauge structures, the  pre-gauge transformations.

 The pre-potentials serve to mediate between quantities subject to
different types of gauge (pre-gauge) transformations $g$
($\overline{{\bf P}}$ and  ${\bf Q}$ ) and we can build
combinations of these which are sensitive to all these
transformations. For instance, the composite field $T^a\, U^b$
transforms under gauge and pre-gauge transformations as follows
\begin{equation}\label{RStfab}
  (T^a \, U^b) \mapsto (T^a \, U^b)\ {\overline{{\bf P}}}^{a}\,
  {\bf Q}^{b}\, g^{a+b}
\end{equation}

Now if we consider a generic superfield ${\Phi}$ of weight $w({
{\Phi}})$ as in (\ref{RS.37}) and define
\begin{equation}
{\Phi}(a,b) \,=\,{(T^{a}\,U^{b})}^{-w(\Phi)}\,{\Phi},
\label{RS.61}
\end{equation}
 this new superfield ${\Phi}(a,b)$ is
inert under $g$ superfield gauge transformations if $ a+b \,=\,1$,
but still transforms under chiral and antichiral superfield gauge
transformations ${\bf Q}$ and $\overline{{\bf P}}$ as
\begin{equation}
{\Phi}(a,b) \ \mapsto \left[ g^{(a+b-1)}\,{\overline{\bf P}}^a
\,{\bf Q}^b \right]^{-w( \Phi)}\, {\Phi}(a,b). \label{RS.63}
\end{equation} ${\Phi}(a,b)$ will be said to be in the $(a,b)$-basis with respect
to $\overline{{\bf P}}$ and ${\bf Q}$ superfield pre-gauge
transformations. It is convenient to introduce the corresponding
definitions for the gauge potential as well
\begin{eqnarray}
A(a,b) \ &=& \  A+ \lp T^a U^b \rp^{-1} d \lp T^a U^b \rp\nonumber
\\  &=&\ A + a\ d \log T + b\ d \log U  . \label{RS.64}
\end{eqnarray}
It should be clear that $F(a,b) \equiv dA(a,b)\,=\,F\,=\,dA$, in
any basis and thus that the superfields $W^\al, W_\da$ are basis
independent. It is interesting to note that we can write
\begin{eqnarray}
{A_\al}(a,b)\,=\, (a-\frac{1}{2})\,D_\al \log T +
(b-\frac{1}{2})\, D_\al \log U - \frac{1}{2} D_\al \log W
\nonumber \\ {A^\da}(a,b)\,=\, (a-\frac{1}{2}) D^\da \log T +
(b-\frac{1}{2}) D^\da \log U + \frac{1}{2} D^\da \log W,
\end{eqnarray}
 where the superfield $W \,=\,\lp T\ U^{-1} \rp$ is
 inert under $g$ gauge transformations (\ref{RStfab}), basis independent
  and transforms as
\begin{equation}\label{RSWtf}
  W \mapsto {\overline{{\bf P}}}\, W \, {\bf Q}^{-1}.
\end{equation}
Therefore, we can gauge away the $T$ and $U$ terms in the
expressions for $A_\al (a,b)$ and $A^\da (a,b)$, but not the $W$
one.
 The covariant derivative in the $(a,b)$-basis is then defined as
\begin{equation}
{\cd}\,{\Phi}(a,b) \,=\,d\,{\Phi}(a,b) + w({\Phi}) \ A(a,b)\,
{\Phi}(a,b), \label{RS.66} \end{equation} and transforms in
accordance with (\ref{RS.61}):
\begin{equation}
{\cd}{\Phi}(a,b) \, = \, {(T^{a}U^{b})}^{-w({\Phi})}\,{\cd}{\Phi}.
\label{RS.67}
\end{equation} Again ${\cd}{\Phi}(a,b)$ is inert under $g$ gauge
transformations if $a+b=1$, so hereafter we will stick to this
case and omit the label $b$, unless specified. Observe now that
\begin{eqnarray}(a,b) &=& (\dem,\dem)\ \Rightarrow\ \
A_\al (\dem) = -\frac{1}{2} D_\al \log W, \cem A^\da (\dem) = +
\frac{1}{2} D^\da \log W,\nonumber \\ (a,b) &=& (1,0)\
\Rightarrow\ \ A_\al(1) =0,\cem \cem \cem A^\da(1) =+ D^\da \log
W, \nonumber
\\ (a,b) &=& (0,1)\ \Rightarrow\ \ A_\al(0) = - D_\al \log W,\cem
\ \  A^\da(0)= 0\label{RS3bas}.
\end{eqnarray}
The three particular bases presented in (\ref{RS3bas}) are useful
in different situations.
 Later on, in the discussion of \ka
transformations and in the construction of supergravity/matter
couplings, we shall identify spinor components of the \ka $U(1)$
connection with spinor derivatives of the \ka potential, namely
\begin{equation} \label{RSKalcon}
  A_\al \,=\,\frac{1}{4} D_\al K, \cem
  A^\da \,=\, - \frac{1}{4} D^\da K.
\end{equation}
Such an identification is easily made in the $(\dem , \dem)$
base, called the vector basis: setting
\begin{equation}\label{RSWdemi}
  W\  \equiv\  \exp(-K\,/\,2),
\end{equation}
we obtain (\ref{RSKalcon}). Moreover, if we parametrize
${\overline{{\bf P}}}\,=\, \exp (-{\bF}/2)$ and ${\bf Q} \,=\,\exp
(F/2)$ (we take $\bF$ and $F$ since $K$ is real) we obtain, given
(\ref{RSWtf}),
\begin{equation}\label{Kael}
  K \mapsto K \ + \ F\ +\ \bF,
\end{equation} the usual form of \ka transformations. A generic superfield
$\Phi$, in this base, transforms as
 \begin{equation} \Phi(\dem)\mapsto
  e^{-\frac{i}{2}w(\Phi)\, \ima F} \, \Phi(\dem).
  \end{equation}
 In addition for the connection we obtain
\begin{equation}
  A(\dem) \mapsto A(\dem) +\f{i}{2}\, d\, \ima F,
\end{equation}
where the vector component is, using (\ref{RS.49}),
\begin{equation}\label{RSAaK}
  A(\dem)_a \,=\, \f{i}{16} {\sib_a}^{\da \al}
   \left[ D_\al, D_\da \right]\ K.
\end{equation}

In other contexts (anomalies and Chern-Simons forms study) the
$(0,1)$ and $(1,0)$ bases are relevant; we name them respectively
chiral and antichiral bases. Indeed, let us consider the covariant
chiral superfield $\phi$, with $w(\phi)=+w$, in the $(0,1)$-basis
the superfield ${{\phi}}(0)=U^{-w} \phi $ transforms under ${\bf
Q}$-transformations only,
\begin{equation}
{{\phi}}(0) \ \mapsto \ {{\bf Q}}^{-w}{{\phi}}(0), \label{RS.70}
\end{equation}
 whereas the gauge potential
has the property $A^{\da}(0) \,=\,0$. Then, in this basis, the
covariant chirality constraint for $\phi$, (\ref{RS.44}), takes a
very simple form for $\phi(0)$: $D^{\da}{\phi}(0)  =  0$.
Analogous arguments hold for $\phib$, with weight $w(\phib)= -w$,
in the $(0,1)$-basis, \ie $D_\al {\phib}(1) =0$. So it is
$\phi(0)$ and $\phib(1)$ which are actually the "traditional"
chiral superfields, our $\phi$ and $\phib$ are different objects,
they are {\em{covariant}} (anti)chiral superfields. We emphasize
this point because to build the matter action coupled to gauge
fields we shall simply use the density
\begin{equation}
\phib \phi \,=\, \phib(1)\, W^w \phi(0) \,=\,  \phib(1)\,
e^{2wV} \phi(0), \end{equation} where we have defined
\begin{equation}
 W\  \equiv \  e^{2\,V}.\end{equation}
We thus recover the standard formulation of the textbooks in terms
of non-covariantly chiral superfields $\phi (0),\, \phib (1)$,
with $V$ the usual vector superfield; this is illustrated in
section \ref{RS24}.
 The chiral and the
antichiral bases are related among themselves by means of the
superfield $W$, $\phi(0)=W^w \,\phi(1)$.

Similarly, $A(1)$ and $A(0)$ are related by a gauge-like
transformation
\begin{equation}
A(0) \,=\, A(1)\,-\, W^{-1}dW. \label{RS.77}
\end{equation}
Finally, the basis independent superfields $W^{\al}$ and $W_{\da}$
are easily obtained as
\begin{equation}
W_{\al} \,=\,{\f{1}{4}} {\bar{D}}^{2} D_{\al} V, \cem
W^{\da} \,=\,\f{1}{4} D^{2} D^{\da} V,
\label{RS.82}
\end{equation}
which is nothing but the solution to the reduced Bianchi
identities (\ref{RS.12}), (\ref{RS.14}).

\subsubsection{ Components and Wess-Zumino transformations}
\label{RS23}

Component fields are systematically defined as lowest components
of superfields, expansion in terms of anticommuting parameters is
replaced by successive application of covariant derivatives. In
this approach the component fields of a chiral multiplet $\phi$
of weight $w$ are defined as
\begin{equation}
 \phi\,{\loco}\,=\,A(x),\cem
\cd_\al\, \phi\,{\loco}\,=\, \sqrt{2}\chi_\al(x),\cem
\cd^\al \cd_\al\, \phi\,{\loco}\,=\,-4 F(x),
\end{equation}
whereas those of the gauge supermultiplet are identified as
\begin{equation}
{A}_m {\loco} \,=\, i {a}_m, \cem {W}^{\db} {\loco} \,=\, i
{\lab}^{\db},
 \cem {W}_\beta {\loco} \,=\, - i\la_\beta,
\cem \cd^\al {W}_\al {\loco} \,=\, -2 {\bf D}.
\label{Gaucomp}
\end{equation}
Their Wess-Zumino transformations are obtained from (\ref{WZTfs})
in identifying $\Phi$ successively with $\phi$, $\cd_\al \phi$
and $\cd^\al \cd_\al \phi$. We obtain
 \begin{eqnarray}
\dt_{\WZ}\, A &=& \sqrt{2} \, \xi \chi~, \label{dtA}
\\[1mm] \dt_{\WZ}\, \chi_\al &=& +i\sqrt{2} \, (
\xib \sib^m \eps)_\al \, \cd_m A +
              \sqrt2\,\xi_\al F~,
\label{dtchi} \\[2mm] \dt_{\WZ}\, F &=& i\sqrt{2} \, ( \xib
\sib^m)^\al \cd_m \chi_\al + 2i w \lp \xib \lab\rp A~.
\label{ts0}
\end{eqnarray}
The covariant derivatives arise in a very natural way due to our
geometric construction; they are given as
\begin{equation}
\cd_m A \,=\,( \prt_m + iw a_m) A, \cem \cd_m \chi_\al \,=\,(
\prt_m + iw a_m) \chi_\al, \end{equation}
\begin{equation}
\cd_m \bA \,=\,( \prt_m - iw a_m) \bA, \cem \cd_m  \chib^\da \, =
\, ( \prt_m - iw a_m)  \chib^\da.
\end{equation}
As to the gauge
supermultiplet, the supersymmetry transformation of the component field
gauge potential $A_m$ is obtained from the Wess-Zumino transformation of the
1-form $A$ in (\ref{WZTfs}), projected to the lowest vector component, with
the result
\begin{equation}
\delta_{\WZ}\, {a}_m \, = \, i (\xi \si_m \lab) + i (\xib \sib_m \la).
\label{Wzpot}
\end{equation}
The corresponding equations of the gaugino component fields are obtained
replacing $\Phi$ with $W_\al$ and $W^\da$
\begin{eqnarray}
\delta_{\WZ}\, \la^\al & = & -(\xi \si^{mn})^\al {f}_{mn}
 + i \xi^\al {\bf D}, \\ \delta_{\WZ}\,
{\lab}_{\da} & = & -(\xib \sib^{mn})_{\da}  {f}_{mn}-i \xib_\da
{\bf D},
\end{eqnarray}
where ${f}_{mn} = \prt_m a_n -\prt_n a_m=-i F_{mn}{\loco}$ and we
used the Abelian versions of (\ref{B.121}), (\ref{B.122}).
Finally, for the auxiliary component we have
\begin{equation}
 \delta_{\WZ}\, {\bf D} \,=\, -\xi
\si^m\prt_m\lab +\xib \sib^m\prt_m\la .
\end{equation}
Observe that these are
the supersymmetry transformations which would have been obtained
in the Wess-Zumino gauge of the traditional approach.
This is due to the definition of Wess-Zumino transformation in terms of
particular compensating gauge transformation. In this way
the Wess-Zumino gauge is realized in a geometric manner.

We should like to comment briefly on the implementation of
R-transformations \cite{SS75}, \cite{Fay75}, \cite{Fay77},
\cite{FF77}, related to a phase freedom on the superspace
anticommuting coordinates, in the language employed here. As the
role of $\theta, \bar \theta$ is now taken by the covariant
spinor derivatives, we assign to the latters R-parity charges of
opposite sign to those of the corresponding $\theta$'s. This way
it is easy to recover the usual arguments in the discussion of
properties and consequences of R-transformations in
supersymmetric theories.

\subsubsection{ Component field actions}
\label{RS24}

We have seen how component fields and their Wess-Zumino transformations are
obtained from the algebra of covariant superspace derivatives and projections
to lowest superfield components. This kind of mechanism is applied to the
construction of supersymmetric component field actions as well.

Let us explain this with the example of the kinetic action of the
chiral matter multiplet. The key idea is to consider the D-term
of the gauge invariant superfield $\phi \phib$, given as the
lowest component of the superfield $D^2 \bar D^2  \phib \phi$. To
be exact, this definition differs from the earlier one by a total
space-time derivative, irrelevant in the construction of
invariant actions. The explicit component field action is
obtained expanding the product of spinor derivatives and using
the Leibniz rule. When acting on $\phi$ or $\phib$ individually
the ordinary covariant derivatives, $D_A$, transmute into gauge
covariant derivatives, $\cd_A$, giving rise to the expansion
 \begin{equation}
D^2 \bar D^2 ( \phib \phi) \,=\,\phi\, \cd^2 \bar \cd^2  \phib + 2
(\cd^\al \phi) \cd_\al \bar \cd^2 \phib + (\cd^2 \phi) \bar \cd^2
\phib~.
\label{exp1}
\end{equation}
At this point the algebra of covariant derivatives intervenes.
The relations
\begin{eqnarray} \cd_\al \cdb^2  \phib &=&
   -4i \si^a_{\al \da} \cd_a \cd^\da  \phib - 8w \, W_\al  \phib~,
\label{exp2}
\\[2mm]
 \cd^2 \bar \cd^2  \phib &=& 16 \cd^a \cd_a  \phib
   - 16w \, W_\da \bar \cd^\da \! \phib
   - 8w \, \phib \, D^\al W_\al~,
\label{exp3}
\end{eqnarray}
illustrate how gauge covariant derivatives will appear in the component field
way in a completely natural way by construction. This should be contrasted
with the method using explicit expansions in the anticommuting coordinates of
superspace. In the approach pursued here, the component field action is simply
obtained from combining (\ref{exp2}) and (\ref{exp3}) with (\ref{exp1})
and projecting to lowest components, with the result
\begin{eqnarray}
 \frac{1}{16} D^2 \bar D^2 ( \phib \phi){\loco} &=&
-\cd^m \! A \, \cd_m \bA
- \f{i}{2} \lp \chi \si^m \cd_m  \chib
+ \chib \sib^m \cd_m \chi \rp
\nonumber \\[1mm] &&
+ F \bar F + w \, {\bf D} A \bA +
i w \sqrt2 \lp \bA \la \chi - A \lab \chib \rp~.
\label{kinchir}
\end{eqnarray}
In this approach $D^2 {\bar D}^2$ plays the role of the volume
element of superspace. Again, as in the derivation of the
Wess-Zumino transformations (\ref{dtchi}), (\ref{ts0}), the
covariant spacetime derivatives appear in a very natural way as a
consequence of use of covariant differential calculus, without
recourse to the introduction of the vector superfield $V$. The
relation between the present formulation and the traditional one
is established in section \ref{RS22}.

The kinetic terms of the gauge multiplet are derived from the superfield
$W^\al W_\al$ and its complex conjugate $W_\da W^\da$. As $W^\al W_\al$ is
chiral, and $W_\da W^\da$ antichiral, this will be achieved  by a F-term
construction. The relevant superfields we have to consider are therefore
$D^2 (W^\al W_\al)$ and ${\bar D}^2 (W_\da W^\da)$. In the explicit evaluation
we will make use of certain superfield building blocks, which are the Abelian
flat superspace versions of (\ref{B.121}),
(\ref{B.122}) and (\ref{B.129}), (\ref{B.130}).
Simple spinor derivatives of the gaugino superfields are given as
\begin{eqnarray}
 D_{\bt} W_{\al} &=&
-(\si^{ba}\eps)_{\bt \al} F_{ba} -\eps_{\bt \al}{\bf D},
\label{BB.121} \\
D_{\db}W_{\da} &=&
-(\eps\sib^{ba})_{\db \da}F_{ba} +\eps_{\db \da}{\bf D},
\label{BB.122}
\end{eqnarray}
with
\[ {\bf D} \, = \, - \frac{1}{2} D^\al W_\al, \]
the D-term superfield. Double spinor
derivatives arising in the construction are
\begin{equation}
D^2 W_\al \, = \,
4i \si^m_{\al \da} \prt_m W^{\da}, \cem
{\bar D}^2 W^\da \, = \,
4i \sib^{m \, \da \al} \prt_m W_\al.
\label{BB.130}
\end{equation}
It is then straightforward to derive
\begin{equation}
D^2 (W^\al W_\al) \,=\,-2 F^{ba} F_{ba}
+ 8i W^\al \si^a_{\al \da} \prt_a \bar W^\da
- 4{\bf D}^2 - i \varepsilon^{dcba} F_{dc} F_{ba}
\end{equation}
\begin{equation}
\bar D^2 (W_\da W^\da) \,=\,-2 F^{ba} F_{ba}
+ 8i \bar W_\da \sib^{a \da \al} \prt_a W_\al
- 4{\bf D}^2
+ i \varepsilon^{dcba} F_{dc} F_{ba}.
\end{equation}
Projection to lowest components identifies the component field kinetic terms
of the gauge multiplet in{\footnote{The gauge coupling
$g$ may be restored explicitly: rescaling the
components of the gauge multiplet by $g$ and the pure gauge action
by $g^{-2}$.}}
\begin{equation}
-\f{1}{16} ( D^2 \, W^\al W_\al + \bar D^2  \, W_\da W^\da ){\loco} \,=\,
-\f{1}{4} f^{mn} f_{mn} - \f{i}{2} \la \si^m \prt_m \lab - \f{i}{2}
\lab \sib^m \prt_m \la + \f{1}{2}{\bf D}^2,
\end{equation}
whereas the orthogonal combination yields a total space-time derivative.

So far, we have illustrated the construction of the component
field Lagrangian for a chiral matter multiplet with an Abelian
gauge multiplet. The discussion of the F-term construction of
mass term and self-interactions of the matter multiplet, arising
from the chiral superpotential and its complex conjugate will be
postponed to more interesting situations.

As is clear from its supersymmetry transformation law, the component field
${\bf D}$ may be added to the supersymmetric action - this is the genuine
Fayet-Iliopoulos D-term. In the terminology employed here, it arises from
projection to the lowest component of the D-term superfield
\begin{equation}
{\bf D} \,=\, - \f{1}{8} D^\al {\bar D}^2 D_\al V.
\end{equation}
From this point of view, gauge invariance
\begin{equation}
V \mapsto V + i \lp \Lambda - \Lab \rp,
\label{gaugeinvar}\end{equation}
is ensured due to the fact that chiral and antichiral superfields
are annihilated by the superspace volume element
$D^\al {\bar D}^2 D_\al \,=\, {\bar D}_\da { D}^2 {\bar D}^\da$.

As we have noted above,
the kinetic term of the chiral matter multiplet
may be viewed as a D-term as well,
identifying $V$ with $\phi \phib$.
In this case the gauge invariance (\ref{gaugeinvar}) indicates that the
addition of holomorphic or anti-holomorphic superfield functions $F(\phi)$
or $\bF(\phib)$ will not change the Lagrangian.

We have described here the simplest case of a supersymmetric gauge theory, a
single chiral multiplet interacting with an Abelian gauge multiplet.

Mass terms and self-interactions of the chiral multiplet, on
the other hand,
would arise from a F-term construction
applied to $\phi^2$ and $\phi^3$ and
their complex conjugates, for power-counting renormalizable theories,
or to
holomorphic and antiholomorphic functions $W(\phi)$ and
$\overline W (\phib)$ in more general situations. In the simplest
case of a
single chiral superfield with non-vanishing Abelian charge, as discussed
here, this kind of superpotential terms are incompatible with gauge
invariance. The construction of a non-trivial invariant superpotential
requires several chiral superfields with suitably adjusted weights
under gauge
transformations.

For the sake of pedagogical simplicity, we will now describe the
superpotential term for a single chiral superfield, restricting ourselves to
the case of a self-interacting scalar multiplet in the absence of gauge
couplings.

The F-term construction amounts to evaluate $D^2 W$ and project to lowest
superfield components, resulting in
\begin{equation}\label{D2W}
  -\frac{1}{4}  D^2 W(\phi) \loco
  = -\f{1}{2} \f{\prt^2 W}{\prt A^2} \lp \chi \chi \rp
+ \f{\prt W}{\prt A} F,
\end{equation}
for $W$ and
\begin{equation}\label{bD2bW}
  -\frac{1}{4}  \bD^2 \overline W(\phib) \loco
  =
  -\f{1}{2} \f{\prt^2 \overline W}{\prt {\bA}^2} \lp \chib \chib \rp
 + \f{\prt \overline W}{ \prt \bA} \bF.
\end{equation}
In the component field expressions, the holomorphic function $W$
is to be considered as a function of the complex scalar $A$ and
correspondingly $\overline W$ as a function of $\bar A$. Combining
the superpotential terms with the kinetic terms (\ref{kinchir}),
for $w=0$, and eliminating the auxiliary fields, $F,\bF$, through
their algebraic equations of motion, $F=- \prt \overline W / \prt
\bA$, we obtain the on-shell Lagrangian
\begin{equation}\label{Lagshell}
-\prt^m \! A \, \prt_m \bA
- \f{i}{2} \lp \chi \si^m \prt_m  \chib + \chib \sib^m \prt_m \chi \rp
 -\f{1}{2} \f{\prt^2 W}{\prt A^2} \lp \chi \chi \rp
-\f{1}{2} \f{\prt^2 \overline W}{\prt {\bA}^2} \lp \chib \chib \rp -
\mod \f{\prt W}{\prt A}{\mod}^2,
\end{equation}
for a single self-interacting scalar multiplet,
the last term being just the usual scalar potential contribution.

\subsec{ Supersymmetric Yang-Mills Theories \label{RS3}}

The interplay between chiral, antichiral and real gauge
transformation formulations, as encountered in the Abelian case,
persists in the case of supersymmetric Yang-Mills theory. These
properties are not only of academic interest, but quite useful,
if not indispensable in contexts like Chern-Simons couplings or
supersymmetric chiral anomalies.

The 1-form Yang-Mills gauge potential is now Lie algebra
valued,
\begin{equation}
{\ca}\,=\,E^A {\ca}_A \,=\,E^A {\ca}_A^{(r)}{\bf T}_{(r)},
\label{RS.84}
\end{equation}
the generators ${\bf T}_{(r)}$ fulfill the commutation relations
\begin{equation}
\left[ {\bf T}_{(r)},{\bf T}_{(s)} \right] \, = \,
i{c_{(r)(s)}}^{(t)} {\bf T}_{(t)}.
\label{RS.85}
\end{equation}
Under a gauge transformation, parametrized by a matrix superfield ${\gym}$,
the gauge potential $\ca$ transforms as
\begin{equation}
{\ca} \, \mapsto \, {\gym}^{-1} {\ca} {\gym} -{\gym}^{-1} d
{\gym}. \label{RS.86} \end{equation} Observe that this
corresponds to a gauge transformation in the {\em real basis},
\ie the parameters of the gauge transformations are real
unconstrained superfields. The covariant field strength is
defined by
\begin{equation}
 {\cf} \,=\, d{\ca}+ {\ca}{\ca},
\label{RS.87}
\end{equation}
 and transforms covariantly,
 \begin{equation}
  \cf \, \mapsto \, {\gym}^{-1} \cf {\gym}.
\end{equation}
   Its components are given by
\begin{equation}
{\cf}_{BA} \,=\,D_B {\ca}_A - (-)^{ab} D_A {\ca}_B
-({\ca}_B,{\ca}_A) +{T_{BA}}^C {\ca}_C, \label{RS.89}
\end{equation}
exhibiting now, in addition to the derivative terms and the
torsion term, the graded commutator $({\ca}_B,{\ca}_A)$.

Due to its definition, the field strength, ${\cf}$, satisfies
Bianchi identities
\begin{equation}
{\cd}{\cf} \, = \, d{\cf} -{\ca}{\cf} + {\cf}{\ca} \,=\,0.
\label{RS.90}
\end{equation}

Consider next generic superfields $\Phi$ and $\Phib$ of gauge transformation
\begin{equation}
{\Phib} \, \mapsto \, {\Phib}{\gym}, \cem
 {\Phi} \, \mapsto \, {\gym}^{-1} {\Phi},
\label{YMtrans}
\end{equation}
so that $\Phib \Phi$ is invariant. Covariant exterior derivatives
${\cd}{\Phi} \,=\,E^A{\cd}_A {\Phi}$ are defined as
\begin{equation}
{\cd}{\Phib}  \,=\,d{\Phib} + {\Phib}{\ca}, \cem
{\cd}{\Phi}  \,=\,d{\Phi} - {\ca}{\Phi}.
\label{RS.91}
\end{equation}
Double exterior covariant derivatives
\[ \cd \cd \Phib = + \Phib \cf, \cem \cd \cd \Phi = - \cf \Phi,\]
give rise to
\begin{equation}
({\cd}_B,{\cd}_A) {\Phi} \,=\,-{\cf}_{BA}{\Phi} -{T_{BA}}^C
{\cd}_C {\Phi}.
\label{RS.92}
\end{equation}
\begin{equation}
({\cd}_B,{\cd}_A) {\Phib} \,=\,+\,{\Phib}{\cf}_{BA} -{T_{BA}}^C
{\cd}_C {\Phib}.
\label{RS.94}
\end{equation}

\indent

In this framework, matter fields are described by covariantly chiral
superfields, \ie we specialize the generic superfields $\Phi$ and $\Phib$
to matter superfields $\phi$ and ${\phib}$, which still transform under
(\ref{YMtrans}), but are required to be covariantly chiral and antichiral,
respectively,
\ie
\begin{equation}
{\cd}^{\da}{\phi} \,=\,0, \cem {\cd}_{\! \al} {\phib} \,=\,0.
\label{RS.95}
\end{equation}
Compatibility of these conditions with the graded commutation relations
(\ref{RS.92}) and (\ref{RS.94}) above
suggest to impose the constraints
\begin{equation}
{\cf}^{{\db}{\da}} \,=\,0, \cem {\cf}_{\beta \al} \,=\,0,
\label{RS.96} \end{equation}
called {\em representation preserving constraints}.
Furthermore, in view of the explicit expression
\begin{equation}
{{\cf}_{\beta}}^{\da} \,=\,D_{\beta} {\ca}^{\da} +D^{\da}
{\ca}_{\beta} -\left\{ {\ca}_{\beta},{\ca}^{\da} \right\}
-2i{({\si}^c {\eps}) _{\beta}} ^{\da} {\ca}_c, \label{RS.97}
\end{equation}
the constraint
\begin{equation}
{{\cf}_{\beta}}^{\da} \,=\,0,
\label{RS.98}
\end{equation}
just corresponds to a linear covariant redefinition of the vector component,
${\ca}_a$, of the connection superfield. For this reason it is called a
{\em conventional constraint}.

\indent

As in the Abelian case, the constraints are solved in terms of pre-potentials.
The representation preserving constraints (\ref{RS.96}) suggest to
express the spinor components of ${\ca}$ as
\begin{equation}
{\ca}_{\al} \,=\, -{\ct}^{-1} D_{\al}{\ct}, \cem {\ca}^{\da} \,=\,
- {\cu}^{-1} D^{\da} {\cu},
\label{RS.99}
\end{equation}
in terms of pre-potential superfields ${\cu}$ and ${\ct}$.
Their gauge transformations should be adjusted such that they reproduce
those of the gauge potentials themselves, that is
\begin{equation}
{\ct} \, \mapsto \, {\overline{\cal{P}}} \, {\ct}{\gym},\cem
{\cu} \, \mapsto \, {\cal Q} \; {\cu} \, {\gym}.
\label{RS.102}
\end{equation}
Here, ${\overline{\cal P}}$ and $\cal Q$ denote the pre-gauge
transformations and are respectively antichiral and chiral
superfields.

Recall that $\ca$ is the gauge potential in the real basis of gauge
transformations; by construction, it is inert under the chiral and
antichiral pre-gauge transformations.
On the other hand,
pre-potential dependent redefinitions of $\ca$, which have the form of gauge
transformations,
 \begin{eqnarray}
 {\ca}(1) &=& {\ct}\,{\ca}\,{\ct}^{-1}-{\ct}d \,{\ct}^{-1},
 \\
{\ca}(0)&=& {\cu}\,{\ca}\,{\cu}^{-1} -{\cu}d \,{\cu}^{-1},
\label{RS.105}
\end{eqnarray}
give rise to new gauge potentials which are inert under the original $\gym$
gauge transformations and transform under chiral, resp. antichiral gauge
transformations, \ie
 \begin{eqnarray}
  \ca(1) &\mapsto& {\overline{\cal P}}\,
\ca(1)\,{\overline{\cal P}}^{-1} -{\overline{\cal P}}d\,
{\overline{\cal P}}^{-1},  \\
  \ca(0) &\mapsto&
{\cal Q}\, \ca(0)\,{\cal Q}^{-1} -{\cal Q}d\,{\cal Q}^{-1}.
\label{RS.106}  \end{eqnarray}
The connections ${\ca}(1)=E^A {\ca}_A(1)$
and ${{\ca}}(0)=E^A {\ca}_A(0)$ take a particularly simple form
\begin{eqnarray}
&&{{\ca}}^{\da}(1)  \, = \,  -{\cw} D^{\da}{\cw}^{-1}, \cem
{{\ca}}_{\al}(1) \, =\,  0,\cem {{\ca}}_{{\al}{\da}}(1)  \,=\,  \f
{i}{2} D_{\al} {{\ca}}_{\da}(1),
\cem \label{RS.114}\\
&& {{\ca}}_{\al}(0)  \, = \,
-{\cw}^{-1}D_{\al} {\cw},\cem {\ca}^{\da}(0)  \,=\,  0, \cem  {
\ca}_{{\al}{\da}}(0)  \,=\, \f {i}{2} D_{\da}{\ca}_{\al}(0), \cem
\label{RS.113}
\end{eqnarray}
expressed in terms of the combination
\begin{equation}
{\cw} \,=\,{\ct}\ {\cu}^{-1} \label{RS.110}
\end{equation}
 with gauge transformations
\begin{equation}
{\cw} \ \mapsto \ {\overline{\cal P}} {\cw}{\cal Q}^{-1}.
\label{RS.111}
\end{equation}

 The corresponding change of basis on the covariant chiral and
antichiral superfields ${\phi}$ and ${\phib}$ is achieved via the
redefinitions which have the form of gauge transformations as well, such that
\begin{eqnarray}
{\phi}(1) \ &=& \ct\, \phi, \cem \phib(1)\, =\,
{\phib}\,{\ct}^{-1}, \\  {\phi}(0) \ &=& \ {\cu}\,{\phi},
\cem \phib(0)\, =\, \phib \, \cu^{-1} . \label{RS.108}
\end{eqnarray}
 In this case, we also obtain particularly simple
chirality conditions for ${\phi}(0)$ and ${\phib}(1)$.
The
invariant combination ${\phib}{\phi}$ behaves under this change of
bases as
\begin{equation}
{\phib}{\phi} \,=\,{\phib}(1)\, {\cw}\, {\phi}(0). \label{RS.109}
\end{equation}
The right hand side of this equation corresponds to the
traditional formulation in terms of simply chiral, resp.
antichiral fields, explicitly
\begin{equation}
D^\da {\phi}(0) \, = \, 0, \cem D_\al {\phib}(1) \, = \, 0.
\end{equation}
The superfield ${\cw}$ provides the bridge
between the chiral and antichiral bases. Setting ${\cw}= \exp 2V$,
we recover the usual description of supersymmetric Yang-Mills
theories.

 As before, the
components of the field strength ${\cf}_{{\beta}a},\;
{{\cf}^{\db}}_a$ and ${\cf}_{ba}$ can be expressed in terms of two
superfields ${\cw}_{\al}$, ${\cw}^{\da}$ and their spinor
derivatives, namely
 \begin{eqnarray}
{\cf}_{{\beta} a} &=&+i{\si}_{a\,\beta \db}\,{\cw}^{\db},
\label{RS.117} \\ {{\cf}^{\db}}_a &=&-i {{\sib}_a}^{\db \beta}
\,{\cw}_{\beta}, \label{RS.118} \\ \hspace{-.5cm} {\cf}_{ba} \,
&=&{\f {1}{2}}({\eps}{\si}_{ba})^{{\beta}{\al}}
{\cd}_{\al}{\cw}_{\beta} +{\f {1}{2}}({\sib}_{ba}
{\eps})^{{\db}{\da}} {\cd}_{\da} {\cw}_{\db}. \label{RS.119}
\end{eqnarray}
The gaugino superfields ${\cw}_{\al}$ and ${\cw}^{\da}$ fulfill
\begin{equation}
{\cd}_{\al}{\cw}^{\da} \,=\,0, \cem {\cd}^{\da} {\cw}_{\al} \, =
\, 0, \label{RS.120} \end{equation}
\begin{equation}
{\cd}^{\al} {\cw}_{\al} \,=\,{\cd}_{\da} {\cw}^{\da},
\label{RS.121} \end{equation}
  as a result of Bianchi identities.

The superfields
${\cw}_{\al}$ and ${\cw}^{\da}$ are the building blocks of the kinetic
terms for the supersymmetric Yang-Mills action. Recall the field
content
  of the Yang-Mills gauge multiplet: it consists of the gauge
potentials ${\aym}_m(x)$, the gauginos $\la(x), \lab(x)$, which
are Majorana spinors, and the auxiliary scalars ${\bf D} (x)$.
All these component fields are Lie-algebra valued, they are
identified in the gaugino superfields $\cw^\al$ and $\cw_\da$,
subject to the constraint  conditions (\ref{RS.120}),
(\ref{RS.121}).

The component fields are defined as lowest components of
superfields; for the gauge potential we have
\begin{equation}
\ca_m \loco \, = \, i \aym_m,
\end{equation}
whereas the gaugino component fields are
defined as the lowest components of the gaugino superfields themselves,
\begin{equation}
\cw_\al{\loco} \,=\,-i \la_\al, \cem
\cw^\da{\loco} \,=\,i \lab^\da.
\end{equation}
The Yang-Mills
field strength ${\fym}_{mn} \,=\,\prt_m {\aym}_n - \prt_n
{\aym}_m - i[{\aym}_m, {\aym}_n]$ and the auxiliary field ${\bf D}
(x)$ appear at the linear level in the superfield
expansion
\begin{eqnarray}
\cd_\bt \cw_\al {\loco}
&=& - i (\si^{mn} \eps)_{\bt \al} \, {\fym}_{mn}
                                  - \eps_{\bt \al} \, {\bf D} (x), \nn \\
\cd_\db \cw_\da {\loco} &=& - i (\eps
\sib^{mn})_{\db \da} \, {\fym}_{mn}
                                  + \eps_{\db \da} \, {\bf D} (x),
\end{eqnarray}
this means that the auxiliary field is identified as
\begin{equation}
\cd^\al \cw_\al {\loco}=\cd_\da \cw^\da
{\loco}=-2 {\bf D} (x). \end{equation}
 The Lagrangian for pure Yang-Mills
gauge theory is then given by (we often use the shorthand notation
$\cw^2= \cw^\al \cw_\al$ and ${\overline \cw}^2= \cw_\da
\cw^\da$)
\begin{equation}
 \cl \,=\, -\f{1}{16} D^2 \tr \lp \cw^2 \rp
-\f{1}{16}{\bar D}^2 \tr \lp {\overline \cw}^2 \rp. \end{equation}
As in the Abelian case, the gauge invariant product $\phib \phi$
provides both the kinetic terms for matter superfields and their
minimal supersymmetric coupling to Yang-Mills fields.

\subsec{ Supersymmetry and K\"ahler Manifolds}
\label{RS4}

As explained by B. Zumino, supersymmetric non-linear sigma models
have necessarily a \ka structure \cite{Zum79b}. The complex
scalars of the chiral matter multiplets have an interpretation as
complex coordinates of a \ka manifold and the supersymmetric
component field Lagrangian is given as\footnote{For the sake of
clarity, we consider, here, only matter multiplets without gauge
couplings. Couplings to Yang-Mills theory will be constructed
later on, in the context of the complete
super\-gravity/matter/Yang-Mills system in section \ref{CPN}, and
gauged isometries described in appendix \ref{appC}.}
\begin{eqnarray}
{\bf \cl}_{\mathrm{K\ddot{a}hler}} & = &
- \, g_{i \bj} \, \eta^{mn} \, \prt_m A^i \, \prt_n \bA^{\bj}
- \f{i}{2} \, g_{i \bj} \lp \chi^{i}  \si^m \cd_m \chib^{\bj} \rp
+ \f{i}{2} \, g_{i \bj}\, \lp \cd_m \chi^{i} \si^m \chib^{\bj} \rp
\nonumber \\ &&
+ \f{1}{4} R_{i \bi j \bj}
          \lp \chi^i \chi^j \rp \lp \chib^{\bi} \chib^{\bj} \rp
+ \, g_{i \bj} \, F^i {\bF}^{\bj}.
\label{CPN2.59}
\end{eqnarray}
As a function of the scalar fields $A^i$ and $\bA^\bj$,
the \ka metric $g_{i \bj}$ derives from a \ka potential.
The covariant derivatives
\begin{equation}
\cd_m \chi_\al^i \, = \,
        \prt_m \chi_\al^i + \Gamma^i{}_{kl} \, \prt_m A^k \, \chi_\al^l, \cem
\cd_m \chib^{\bj \da} \, = \,
        \prt_m \chib^{\bj \da}
           + \Gamma^\bi{}_{\bk \bl} \, \prt_m A^\bk \, \chib^{\bl \da},
\end{equation}
contain the Levi-Civita symbols
($g_{i \bi,k}$ denotes the derivative of $g_{i \bi}$ with respect to $A^k$)
\begin{equation}
\Gamma^i{}_{kl} \, = \, g^{i \bi} g_{k \bi,l} \, , \cem
\Gamma^\bi{}_{\bk \bl} \, = \, g^{i \bi} g_{i \bk, \bl} \, ,
\label{levicivicom}
\end{equation}
whereas the quartic spinor terms exhibit the curvature tensor of the \ka
manifold,
\begin{equation}
R_{i \bi j \bj} \, = \, g_{i \bi , j \bj}
      - g_{k \bk} \, \Gamma^k{}_{ij} \, \Gamma^\bk{}_{\bi \bj}.
\label{GRA2.165}
\end{equation}
The auxiliary fields, here, correspond to those of the
diagonalized version in \cite{Zum79b};
 more details will be given below. The supersymmetry
transformations of the chiral multiplet, which leave the action invariant
are given as
 \begin{eqnarray}
\dt A^i &=& \sqrt{2} \, \xi \chi^i~,
\\[1mm]
\dt \chi^i_\al &=&
+i\sqrt{2} \, (\xib \sib^m \eps)_\al \, \prt_m A^i
+\sqrt2\,\xi_\al F^i~,
\\[1mm]
\dt F^i &=&
i\sqrt{2} \, ( \xib \sib^m)^\al \cd_m \chi^i_\al.
\end{eqnarray}

As pointed out by B. Zumino in the same paper, the structure of
the supersymmetric non-linear sigma model is most conveniently
understood in the language of superfields. As he explained, the
lowest component of the superfield
\[
{\bf \cl}_{\mathrm{K\ddot{a}hler}} \, = \;
\frac{1}{16} D^\al {\bD}^2 D_\al K(\phi, \phib),
\]
reproduces exactly the component field Lagrangian given above.
In other words, the kinetic Lagrangian may be understood as a
Fayet-Iliopoulos D-term.  The \ka metric, defined as the lowest
superfield component of (using the same symbols for the component and the
superfield)
\begin{equation}
g_{k \bk} \,=\, \frac{\prt^2 K}{\prt \phi^k \prt \phib^{\bk}},
\label{Kamet}
\end{equation}
the Levi-Civita symbol and the \ka curvature appear in the process
of successive application of spinor derivatives and subsequent
projection to lowest components. Chirality of the matter
superfields and the fact that the differential operator $ D^\al
{\bD}^2 D_\al \, = \, {D}_\da D^2 {D}^\da $ annihilates chiral
superfields, imply invariance under the superfield \ka
transformations
\begin{equation}\label{KTF}
K(\phi, \phib) \, \mapsto \, K(\phi, \phib) + F(\phi) + \bF(\phib).
\end{equation}
This shows that, in fact, the \ka manifold is spanned by the
chiral resp. antichiral matter superfields $\phi^i$ and $\phib^\bj$, \ie
a mapping from superspace into the \ka manifold.
Complex structure on the one hand, in \ka geometry and chirality conditions on
the other hand, in supersymmetry, give rise to intriguing analogies
\cite{Gri90a}.

In the following we will elaborate somewhat more on these geometric superspace
aspects, which will be of essential importance later on in the context of
supergravity/matter coupling.
The properties of the pre-potential $V$ in the $( \dem , \dem )$ basis
of Abelian gauge theory -cf. section \ref{RS22}- suggest to interprete $K(\phi,
\phib)$ as a particular, superfield
dependent, pre-potential\footnote{It should however be noted that, in
distinction to section \ref{RS22}, there are no phase transformations on the
matter fields corresponding to \ka transformations. In the language of section
\ref{RS22}, all the matter fields have weight zero. Non trivial \ka phase
transformations will only appear later on in the coupling of matter to
supergravity.}.

Replacing the unconstrained
pre-potential $V$ by the \ka potential $K(\phi, \phib)$
we define{\footnote {Normalizations are chosen for later convenience in
the supergravity/matter system.}
\begin{equation}\label{Aal}
  A_\al\,=\, + \frac{1}{4} D_\al K
        \,=\, + \frac{1}{4} K_k D_\al {\phi}^k,
\end{equation}
\begin{equation}\label{Ada}
  A^\da \, = \, - \frac{1}{4} D^\da K
         \, = \, - \frac{1}{4} K_\bk D^\da
  {\phib}^\bk.
\end{equation}
Here $K_k$ resp. $K_\bk$ denote derivatives of the \ka potential
with respect to the superfield coordinates $\phi^k$ and $\phib^\bk$.
Following the construction in Abelian gauge theory we define
furthermore
\begin{equation}\label{Aalda}
  A_{\al \da}\,=\,\frac{i}{2} \lp D_\al A_\da +D_\da A_\al \rp.
\end{equation}
This corresponds to a conventional constraint.
Substituting for $A_\al$ and $A_\da$ yields
\begin{equation}
A_a \,=\,\f {1}{4}
   \lp K_i \prt_a{\phi}^i -K_{\bj} \prt_a {\phib}^{\bj} \rp
+\f {i}{8} {\sib}_a^{{\da} \al} g_{i \bj}
     D_\al {\phi}^i D_{\da}{\phib}^{\bj}. \label{RSAa}
\end{equation}
The expressions for $A_\al, A^\da$ and $A_a$ can be subsumed
compactly in superform language,
\begin{equation}
A \,=\,\f {1}{4} (K_i d{\phi}^i -K_{\bj}d{\phib}^{\bj})
+\f {i}{8}E^a {\sib}_a^{{\da} \al} g_{i {\bj}} D_\al
{\phi}^i D_{\da}{\phib}^{\bj}. \label{RSA}
\end{equation}
Let us note that this potential, $A$, transforms as it should (\ie
as a connection) under \ka transformations,
\begin{equation}\label{Atf}
  A \, \mapsto A \, +\frac{i}{2}d \ima F.
\end{equation}
We can now apply the machinery of Abelian gauge structure in
superspace to determine the component field action as the
corresponding D-term. First, applying the exterior derivative to
$A$ gives the composite field strength 2-form
\begin{equation}
F \,=\,dA \,=\,
\f {1}{2} g_{i{\bj}} \, d\phi^i d{\phib}^{\bj}
+\f {i}{8} d \left( E^a {\sib}_a^{{\da} \al}
g_{i{\bj}} D_\al {\phi}^i D_{\da}{\phib}^{\bj}
\right). \label{RSFka}
\end{equation}
As in the generic Abelian case, the coefficients of $F$ are
expressed in terms of a single Weyl spinor and its complex
conjugate, in particular
\begin{equation}
{F_{\beta a}} \,=\, + \f {i}{2} \si_{a \beta \db} \bX^{\db},
 \cem {F^\db}_a\,=\,- \f
{i}{2}{\sib_a}^{\db \beta} X_{\beta}. \label{RSFX}
\end{equation}
On the one hand $X_\al$ and $\bX^\da$ are given in terms of the
\ka potential
\begin{equation}\label{RSXK}
  X_\al\,=\,-\frac{1}{8} {\bD}^2 D_\al K, \cem \bX^\da\,=\,-\frac{1}{8}
  D^2 D^\da K,
\end{equation}
on the other hand, identifying (\ref{RSFX}) in (\ref{RSFka}), we
obtain
\begin{eqnarray}
 X_\al  & = &
-\f {i}{2} \, g_{i {\bj}} \, {\si}^a_{\al{\da}} \,
{\prt}_a{\phi}^i \, D^{\da} \! {\phib}^{\bj}
+\f {1}{2} g_{i {\bj}} \, D_{\! \al} {\phi}^i \, {\bF}^{\bj},
\label{RSXphi} \\
{\bX}^{\da} & = &
- \f {i}{2} \, g_{k {\bar{k}}} \, {\sib}^{a \, {\da} \al} \,
{\prt}_a{\phib}^{\bj} \, D_{\! \al}{\phi}^i
+ \f {1}{2}g_{i {\bj}} \, D^{\da} \! {\phib}^{\bj} \, F^i.
\label{RSXBphi}
\end{eqnarray}
Here we used the definitions
\begin{equation}
F^i \,=\,- \f {1}{4}{\cd}^\al D_\al {\phi}^i, \cem
{\bF}^{\bj} \,=\,- \f{1}{4}{\cd}_\da D^\da {\phib}^{\bj}
\label{RSF}
\end{equation}
with second covariant derivatives defined as
\begin{eqnarray}
{\cd}_B D_\al \phi^i &=&
D_B D_\al \phi^i
+ \Gamma^i{}_{k l} D_B \phi^k D_\al \phi^l,
\label{RScoD1}\\[2mm]
{\cd}_B D^{\da} \phib^{\bj} &=&
{D_B} D^{\da} \phib^{\bj}
+ {\Gamma^{\bj}}_{\bk \bl} D_B \phib^{\bk}D^{\da}\phib^{\bl},
  \label{RScoD2}
\end{eqnarray}
 assuring covariance with respect to K\"ahler
transformations and (ungauged) isometries of the K\"ahler
manifold. Observe that, in terms of these definitions, the
component field Lagrangian will come out to be diagonal in the
auxiliary fields \cite{Zum79b}.
 Due to their definition, the
superfields
$X_{\al}, {\bX}^{\da}$ have the following properties
\begin{equation}
D^{\da}X_{\al} \,=\,0 ,\cem D_{\al}{\bX}^{\da} \,=\,0,
\end{equation}
\begin{equation}
D^{\al} \! X_{\al} \,=\,D_{\da}{\bX}^{\da}.
\end{equation}
It is then easy to obtain the superfield expression of the
K\"ahler D-term
\begin{eqnarray}
-\f {1}{2}D^\al \! X_\al&=&
-{\eta}^{ab} g_{i {\bj}} \, {\prt}_b {\phi}^i \, {\prt}_a {\phib}^{\bj}
-\f{i}{4} g_{i {\bj}} \, {\si}^a_{\al{\da}} \,
       D^{\al} \! {\phi}^i \, {\cd}_a D^{\da} \! {\phib}^{\bj}
\nonumber \\ & &
-\f {i}{4}g_{i{\bj}} \, {\si}^a_{\al{\da}} \,
      D^{\da} \! {\phib}^{\bj} \, {\cd}_a D^{\al} \! {\phi}^i
+g_{i {\bj}} F^i {\bF}^{\bj}
\nonumber \\ & &
+\f {1}{16} R_{i{\bi} j {\bj}} \,
D^{\al} \! {\phi}^i \, D_{\al}{\phi}^j \,
D_{\! \da}{\phib}^{\bi} \, D^{\da} \! {\phib}^{\bj},
\end{eqnarray}
with covariant derivatives defined above in (\ref{RScoD1}),
(\ref{RScoD2}) and the curvature tensor given in (\ref{GRA2.165}).
Projection of this last equation to lowest superfield components
results in the component field Lagrangian (\ref{CPN2.59}). The construction
presented here will be generalized later on and applied to the full
supergravity/matter/Yang-Mills system.

\newpage
\sect{ MATTER IN CURVED SUPERSPACE}
\label{GRA}

The formulation of supersymmetry as a local symmetry naturally leads
to supergravity, where the graviton, of helicity 2,
has a fermionic partner, the gravitino, of helicity 3/2.
The corresponding local fields are the
vierbein ${e_m}^a(x)$ and the Rarita-Schwinger field
${\psi_m}^\al(x)$, $\psib_{m \, \da}(x)$.
As mentioned in section \ref{RS11},
the different $D=4, N=1$ supergravity multiplets (minimal,
new minimal and non-minimal) all contain the graviton and the
gravitino, but differ by their systems of auxiliary fields.

In the geometric formulation of supergravity, the vierbein
${e_m}^a(x)$ is generalized to the  frame superfield $E_M{}^A$ in
superspace, describing the graviton and the gravitino in a
unified way. The three different supergravity multiplets, as well
as the coupling of minimal supergravity to matter, which will be
presented here, are then derived from a superspace geometry in
suitably choosing the structure group and torsion constraints.

The choice of a structure group, which we take to be the
product of Lorentz and chiral  $U(1)$ transformations,
already determines the properties of superspace geometry
to a large extent.

Further specification derives from requiring appropriate
covariant constraints on the torsion and curvature tensors,
which, given the extension of the notion of space-time to
superspace, acquire a plethora of new components. One
distinguishes between {\em{geometric}} and {\em{dynamical}}
constraints. Geometric constraints help to restrict the
properties of superspace geometry without leading to any
dynamics, \ie to any equation of motion. Dynamical constraints
may then be imposed as further restrictions which imply equations
of motion.

Geometric constraints come in two categories: first, the
so-called conventional constraints which are used to absorb  part
of the torsion in covariant redefinitions of the Lorentz and
$U(1)$ connection and of the frame of superspace; second, the
so-called representation preserving constraints, which arise from
consistency conditions for covariant chiral superfields
(essential for the description of supergravity/matter couplings)
with their commutation relations.

Different supergravity multiplets (minimal, new minimal or
non-minimal) are obtained from different kinds of geometric
constraints.

As emphasized in the
introduction, we will only consider the {\em{minimal multiplet of
supergravity}}, whose superspace description is briefly recalled in
section \ref{GRA1}.

We will then show in some detail how
supergravity/matter/Yang-Mills couplings are obtained from a unified
geometric setting by including superfield \ka transformations in the
structure group.

In section \ref{GRA2} we show explicitly how
this formulation can be obtained from the conventional one,
\cite{CJS78a,CJS+79,CFGvP82,CFGvP83}, by means of field dependent
superfield rescalings.
This leads in a natural way to the identification of the
supergravity/matter system as a special case of $U(1)$ superspace
geometry whose structure is reviewed in section \ref{GRA23}. In
section \ref{GRA3}, we identify \ka superspace as a special case
of $U(1)$ superspace geometry, define supergravity
transformations and present
invariant actions and equations of motion at the superfield level.

\subsec{ Minimal Supergravity}
{\label{GRA1}

In supergravity, the dynamical degrees of freedom are the graviton
and the gravitino. They are identified as the local frame of
space-time or vierbein, $e_m{}^a(x)$, and the Rarita-Schwinger
\cite{RS41} field $\psi_m{}^\al(x)$, $\psib_{m \, \da}(x)$. The
supergravity action \cite{DZ76,FvNF76} is then defined as a
certain combination of the Einstein and Rarita-Schwinger actions,
invariant under space-time dependent supersymmetry
transformations relating the graviton and the gravitino. The
commutators of these  transformations only close on-shell, \ie
modulo equations of motion. In minimal supergravity
\cite{SW78,FvN78}, a complex scalar $M$, $\bar M$ and a real
vector $b_a$ are added as auxiliary fields to avoid the
appearance of the equations of motion at the geometric level and
to define an off-shell theory.

The formulation of supergravity in superspace  \cite{AVS75,WZ77}
provides a unified description of the vierbein and the
Rarita-Schwinger fields. They are identified in a common geometric
object, the local frame of superspace, \begin{equation} E^A \ = \
dz^M E_M{}^A (z)~, \end{equation} defined as a 1-form over
superspace, with coefficient
superfields $E_M{}^A(z)$,
generalizing the usual frame, $e^a = dx^m e_m{}^a(x)$, which
is a space-time differential form. Vierbein and Rarita-Schwinger
fields are identified as lowest superfield components, such that
\begin{equation}
e_m{}^a(x) \ = \ E_m{}^a{\loco}\, , \cem \frac{1}{2}
\psi_m{}^\al(x) \ = \ E_m{}^\al{\loco}\, , \cem \frac{1}{2}
\psib_{m \, \da}(x) \ = \ E_{m \, \da}{\loco}\, . \end{equation}
Correspondingly, as in ordinary gravity, one introduces
super-coordinate transformations, thus unifying the usual general
coordinate transformations and the local supersymmetry
transformations as their vector and spinor parts, respectively.
Local Lorentz transformations act through their vector and spinor
representations on $E^a$ and $E^\al, E_\da$.

Covariant derivatives with respect to local Lorentz
transformations are constructed by means of the spin connection,
which is a 1-form in superspace as well,
\begin{equation}
{\phi_B}^A  \ = \ dz^M {\phi_{MB}}^A (z)~.
\end{equation} It takes values in the Lie algebra of the Lorentz group such
that its spinor components are given in terms of the vector
ones as
\begin{equation}
{\phi_\beta}^\al \ = \ - \frac{1}{2} {(\si^{ba})_\beta}^\al
\phi_{ba}~, \cem {\phi^{\db}}_{\da} \ = \ - \frac{1}{2}
{(\sib^{ba})^{\db}}_{\da} \, \phi_{ba}~.
\label{spincom}
\end{equation} These are the basic geometric objects in the superspace
description of supergravity. The covariant exterior derivative of
the frame in superspace,
\begin{equation}
T^A \ = \ dE^A + E^B {\phi_B}^A ~,
\end{equation} defines torsion in superspace as a 2-form
\begin{equation}
T^A \ = \ \frac{1}{2} E^B E^C {T_{CB}}^A~.
\end{equation} Likewise, the covariant expression
\begin{equation}
{R_B}^A \ = \ d {\phi_B}^A \ + \ {\phi_B}^C {\phi_C}^A~,
\end{equation} defines the  curvature 2-form in superspace,
\begin{equation}
{R_B}^A \ = \ \frac{1}{2} E^C E^D {R_{DC \ B}}^A~.
\end{equation}
It is a special feature of supergravity that the curvature tensor
is completely expressed in terms of the torsion and its
derivatives \cite{Dra79}. We do not intend here to give a complete
and detailed review of this geometric structure; for a detailed
exposition we refer to \cite{WB83}.

Recall that superspace torsion is subject to covariant constraints
\cite{GSW80,Wes90} which imply that all the coefficients of
torsion are given in terms of the  covariant supergravity
superfields
\begin{equation}
R~, \cem \rd~, \cem G_a~, \cem W_{\lsym{\gm \beta \al}}~, \cem
W_{\lsym{{\dg} {\db}\da}}~,
\end{equation} and their covariant derivatives. To be more explicit, the
non-vanishing components of superspace torsion are
\begin{equation}
{T_\gm}^{{\db} a} \ = \ -2i {(\si^a \eps)_\gm}^{\db}~,
\end{equation}
\begin{equation}
T_{\gm b \da} \ = \ -i \si_{b \gm \da} \rd~, \cem T^{\dg \ \
\al}_{\ \ b} \ = \ -i {\sib_b}^{\dg \al} R~,
\end{equation}
\begin{equation}
{T_{\gm b}}^\al \ = \ \frac{i}{2} G^c{(\si_c \sib_b)_\gm}^\al
                      + \frac{3i}{2} \dtdu{\gm}{\al} G_b~, \cem
{T^\dg}_{b \da} \ = \ -\frac{i}{2}  G^c {({\sib_c
\si_b)}^{\dg}}_\da
                      -\frac{3i}{2} \dtud{\dg}{\da} G_b~,
\end{equation}
As for ${T_{cb}}^\al$ and ${T_{cb}}_\da$, they will be interpreted
later on as   the covariant
Rarita-Schwinger field strength superfields. They involve the superfields
$W_{\lsym{\gm \beta \al}}$ and $W_{\lsym{\dg \db\da}}$  called
Weyl tensor superfields, because they occur in the decomposition
of these Rarita-Schwinger superfields in very much the same way as
the usual Weyl tensor occurs in the decomposition of the covariant
curvature tensor.

The auxiliary component fields mentioned above appear as lowest
components in the basic superfields $R$, $\rd$ and $G_a$ such that
\begin{equation}
M(x) \ = \ -6 R {\loco}\, , \cem \overline M(x) \ = \ -6 \rd {\loco}\,
, \cem b_a \ = \ -3 G_a {\loco}\, . \end{equation}

Consistency of the superspace Bianchi identities with the special
form of the torsion components displayed so far implies the
chirality conditions
\begin{equation}
{\cd}_\al \rd \ = \ 0~, \cem {\cd}_{\da} R \ = \ 0~,
\end{equation}
\begin{equation}
{\cd}_\al {W}_{\lsym{{\dg} {\db} \da}} \ = \ 0~, \cem
{\cd}_{\da}W_{\lsym{\gm \beta \al}} \ = \ 0~,
\end{equation} as well as the relations
\begin{equation}
\cd_\al R \ = \ \cd^\da G_{\al \da}~, \cem \cd^\da \rd \ = \ -
\cd_\al G^{\al \da}~. \label{relrg} \end{equation} Moreover
\begin{equation}
\cd^2 R + \bar \cd^2 \rd \ = \ -\frac{2}{3} \, \car + 4 \, G^a G_a
+ 32 \, \rd R~, \label{curv} \end{equation} where $\car \equiv
{R_{ab}}^{ab}$ is the curvature scalar superfield. This relation
is at the heart of the construction of the supersymmetric
component field action. On the other hand, the orthogonal
combination \begin{equation} \cd^2 R - \bar \cd^2 \rd \ = \ 4i \,
\cd_a G^a~, \end{equation} is a consequence of (\ref{relrg}), it
has an intriguing resemblance with the 3-form constraint in
superspace -cf. (\ref{con3f}).

The component field Lagrangian is obtained from the superspace integral
\cite{WZ78a}
\begin{equation}
{\cal L}_{\mathrm{supergravity}} \, = \, \int \! \! E,
\label{GRA.0}
\end{equation}
where $\int \! \! E$ stands for $\int \! d^2 \th d^2
{\bar{\theta}} E$, and $E$ denotes the superdeterminant of
$E_M{}^A$. Integration over $d^2 \th d^2 {\bar{\theta}}$ yields
the usual curvature scalar term, $-\frac{1}{2} e \car$, together
with all the other terms necessary for the supersymmetric
completion, with the usual canonical normalization.

\subsec{ Superfield Rescaling}
{\label{GRA2}

In the conventional superfield approach \cite{Wes87} to the
coupling of matter fields to supergravity, the superspace action
for the kinetic terms is taken to be
\begin{equation}
{\cal L}_{\mathrm{kin}}
\,=\,-3 \int \! \! E e^{- \frac{1}{3} K(\phi, \phib)}.
\label{GRA.1}
\end{equation}
Given (\ref{GRA.0}) we may hope that,
by a suitable modification of the superspace geometry, the factor
$\exp (-K(\phi, \phib)/3) $ can be absorbed into $E$; however this
will be possible only if there are symmetries which allow such a
modification, so let us analyze the situation in that respect.
Supersymmetry transformations as well as general coordinate
transformations are encoded in the diffeomorphisms of superspace;
precisely the action (\ref{GRA.1}) is invariant under
super-diffeomorphisms and thereby under supersymmetry and general
coordinate transformations. The superspace geometry relevant to
(\ref{GRA.1}) is that of the so-called minimal supergravity
multiplet. The structure group in superspace in this case is the
Lorentz group. By construction, (\ref{GRA.1})  is Lorentz
invariant.

In addition to super-diffeomorphisms and Lorentz transformations,
which are symmetries of the kinetic action (\ref{GRA.1}),
superspace geometry allows also for a generalization of
dilatation transformations to the supersymmetric case, which are
known as super-Weyl or Howe-Tucker transformations \cite{HT78}.
These are defined as transformations of the frame in superspace
and of the Lorentz superfield connection which respect the torsion
constraints and reduce to ordinary dilatations when supersymmetry
is switched off.

As a result, for the minimal supergravity multiplet, they change
the frame of superspace in such a way that
\begin{eqnarray}
{E_M}^{a} & \mapsto & {E_M}^{a} e^{\Sigma + \overline{\Sigma}}\label{GRA.2}
\\[2mm]
{E_M}^{\al} & \mapsto & e^{2 \overline{\Sigma} -
\Sigma} \left( {E_M}^{\al} + \f{i}{2} {E_M}^{b} {{(\eps
\si_{b})}^{\al}}_{\da} {\cal D}^{\da} \overline{\Sigma}\right),
\label{GRA.3} \\ \hspace{-.2cm} {E_{M \da}} & \mapsto & e^{2
\Sigma - \overline{\Sigma}} \left( {E_{M \da}} + \f{i}{2}
{E_M}^{b} {{{(\eps \sib_{b})}}_{\da}}^{\al} {\cal D}_{\al}
\Sigma\right). \label{GRA.4}
\end{eqnarray}
The chirality conditions
\begin{equation}
{\cal D} _{\al} \overline{\Sigma} \,=\,0,\cem {\cal D}^{\da}
\Sigma \,=\,0, \label{GRA.5} \end{equation}
of the superfield parameters $\overline{\Sigma}$ and $\Sigma$  are a
characteristic feature of the superspace geometry of minimal
supergravity \ie of the torsion constraints which model it.

As a consequence of (\ref{GRA.2}) - (\ref{GRA.4}), the
superdeterminant of the frame in superspace is subject to the
following super-Weyl transformations:
\begin{equation}
E \ \mapsto \ E\, e^{2\, (\Sigma + \overline{\Sigma})}.
\label{GRA.6}
\end{equation}
 Since the K\"ahler potential $K(\phi, \phib)$ is inert
under super-Weyl transformations, (\ref{GRA.6}) indicates that the
kinetic action (\ref{GRA.1}) is not invariant.

However, $K(\phi, \phib)$ is subject to K\"ahler transformations
\begin{equation}
K(\phi, \phib) \ \mapsto \ K(\phi, \phib) + F(\phi) +
{\bF}(\phib), \label{GRA.7} \end{equation}
 which by
themselves are not an invariance of (\ref{GRA.1}) either. Then, it
is easy to see that the kinetic superfield action is K\"ahler
invariant, if together with (\ref{GRA.7}), a {\em{compensating}}
super-Weyl transformation \cite{Wes87} of parameters
\begin{equation}
\Sigma \,=\,\f{1}{6} F(\phi), \cem \overline{\Sigma} \, = \,
\f{1}{6} {\bF} (\phib), \label{GRA.8} \end{equation}
is performed.

In this way a K\"ahler invariant action in superspace is obtained
which contains the kinetic terms for supergravity and
matter superfields and leads to the correct result in the
flat superspace limit.

On the other hand, the component field action which derives from
(\ref{GRA.1}) in the conventional approach, yields the correctly
normalized Einstein action only after a field dependent rescaling
of the component fields \cite{WB83}. The correct
K\"ahler transformations of the various component fields are then
identified on the rescaled fields.

These complications can be avoided, however, if one
starts right away from K\"ahler superspace as explained below. In
particular, K\"ahler transformations are then consistently
introduced  at the {\em{superfield level}}. Another way to
understand this is to perform the rescalings directly in terms of
superfields: this will give the explicit relation between the
conventional superfield approach described just above and our
K\"ahler superspace construction.

The aim is therefore to absorb the exponential of the K\"ahler
potential in (\ref{GRA.1}) by means of a superfield rescaling of
the frame in superspace. The first attempt might have been to
employ a super-Weyl transformation. However, this does  not work
because the combination, $\Sigma + \overline{\Sigma}$, of chiral
and antichiral superfield  in (\ref{GRA.6}) is not sufficient to
absorb the more general real superfield $K(\phi, \phib)$ in
(\ref{GRA.1}). On the other hand, the chirality (resp.
anti-chirality) conditions on $\Sigma$ (resp. $\overline{\Sigma}$)
are consequences of the invariance of the torsion constraints
under the transformations (\ref{GRA.2}) - (\ref{GRA.4}). If one
is willing to give up this requirement, more general rescalings
are possible, at the price of changing the torsion constraints
and thus the superspace structure. We are therefore led to study
more general transformations of the frame (and of the Lorentz
connection) together with their consequences for the corresponding
coefficients of the torsion 2-form. To be more precise, note that
the arbitrary transformations of the vielbein ${E_M}^{A}$ and of
the Lorentz connection ${\phi_{MB}}^{A}$
 \begin{eqnarray}
{E'_{M}}^{A} &=& {E_M}^{\undA} {X_{\undA}}^{A},
\label{GRA.9} \\
{\phi '_{MB}}^{A} &=& {\phi _{MB}}^{A} + {\chi_{MB}}^{A},
\label{GRA.10} \end{eqnarray}
 change the torsion coefficients as
\begin{eqnarray}
{{T'}_{CB}}^{A} & = &
  (-)^{\underline{c}(b+\underline{b})} {{X^{-1}}_{C}}^{\undC}
{{X^{-1}}_{B}}^{\undB} ({T_{\undC \undB}}^{\undA} {X_{\undA}}^{A}+
{\cd} _{\undC} {X_{\undB}}^{A} -(-)^{\underline{c} \underline{b}}
{\cd} _{\undB} {X_{\undC}}^{A})
 \nonumber \\
&& +  {{X^{-1}}_{C}}^{\undC}\; {\chi_{\undC B}}^{A} -(-)^{cb}
{{X^{-1}}_{B}} ^{\undB}\; {\chi_{\undB C}}^{A}. \label{GRA.11}
\end{eqnarray}
For our present purpose it is sufficient to consider the
superfield rescalings
\begin{equation}
{X_{B}}^{A} \,=\,\left(
\begin{array}{ccc}
{\delta_{b}}^{a} X {\bX} & {X_{b}}^{\al} & X_{b \, \da}\\ 0 &
{\delta_{\bt}}^{\al}X & 0\\ 0 & 0 & {{\delta}^{\db}}_{\da}\ {\bX}
\end{array}
\right). \label{GRA.12} \end{equation}
 The superfield $X$ and its
complex conjugate ${\bX}$ are arbitrary, furthermore
\begin{equation}
{X_{b}}^{\al} \,=\,\f{i}{2} {{(\eps \si_{b})}^{\al}}_{\da}
{\bX}^{-1} {\cal D}^{\da}(X {\bX}), \label{GRA.13} \end{equation}
\begin{equation}
X_{b \, \da} \,=\,\f{i}{2} {{{(\eps \sib_{b})}_{\da}}^{\al}} X^{-1}
{\cal D} _{\al} (X {\bX}). \label{GRA.14} \end{equation}
 Observe that
(\ref{GRA.12}) - (\ref{GRA.14}) differ from (\ref{GRA.2}) - (\ref{GRA.4})
only by the fact that $X$ and ${\bX}$ are, contrary to $\Sigma$
and $\overline{\Sigma}$, not subjected to any
restrictions\footnote{ (\ref{GRA.2}) - (\ref{GRA.4}) can be obtained
from (\ref{GRA.12}) - (\ref{GRA.14}) by restricting $X$ and ${\bX}$ to
be given as $X= \exp (2 \overline{\Sigma} - \Sigma)$ and
$ \bX = \exp(2 \Sigma - \overline{\Sigma})$.}.
 What are the effects of
the superfield rescalings (\ref{GRA.12}) - (\ref{GRA.14}) on the
various torsion coefficients? First of all, note that these
transformations leave the torsion constraints
\begin{equation}
{T_{\gamma \bt}}^{a} \,=\,0, \cem  T^{\dot{\gamma} \db a} \,=\,0,
\label{GRA.15} \end{equation}
\begin{equation}
T_{\gamma \bt \da} \,=\,0, \cem  T^{\dot{\gamma} \db \al} \,=\,0
\label{GRA.16} \end{equation}
 and
\begin{equation}
{T_{\gamma}}^{\db \ a} \,=\,-2i{(\si^{a} \eps)_{\gamma}}^{\db}\label{GRA.17} \end{equation} unchanged. It is well known that the
torsion constraints
\begin{equation}
{T_{\gamma b}}^{a} \,=\,0,\cem  {{T^{\dg}}_{b}}^{a} \,=\,0
\label{GRA.18} \end{equation}
 and
\begin{equation}
{T_{cb}}^{a} \,=\,0, \label{GRA.19} \end{equation}
 allow to
determine the Lorentz connection in superspace completely in terms
of ${E_{M}}^{A}$. Likewise, the requirement that (\ref{GRA.18})
and (\ref{GRA.19}) are left invariant under
(\ref{GRA.12}) - (\ref{GRA.14}) determines ${\chi_{C \, b}}^{a}$ in
terms of $X$ and ${\bX}$,
\begin{eqnarray}
\chi_{\gamma ba} & = & 2{{(\si_{ba})}_{\gamma}}^{\varphi} (X
{\bX})^{-1} {\cal D}_{\varphi}(X {\bX}),
\label{GRA.20} \\[1mm]
{\chi^{\dot{\gamma}}}_{ba} & = &
2{{({\sib}_{ba})}^{\dot{\gamma}}} _{\dot{\varphi}} (X {\bX})^{-1}
{\cal D}^{\dot{\varphi}}(X {\bX}),
\label{GRA.21} \\[1mm]
\chi_{cba} & = & {\eta}_{ca} (X {\bX})^{-1} {\cal D}_{b} (X
{\bX}) -{\eta}_{cb}(X {\bX})^{-1} {\cal D}_{a} (X {\bX})
\nonumber
\\ && + \f{1}{2} \eps_{dcba} (X {\bX})^{-1} {\cal D}_{\varphi} (X
{\bX}) {{(\eps \si^{d})}^{\varphi}}_{\dot{\varphi}} (X {\bX})^{-1}
{\cal D}^{\dot{\varphi}} (X {\bX}).
\label{GRA.22}
\end{eqnarray}
This means that ${X_B}^A$ and ${\chi_{CB}}^A$ are now completely
fixed in terms of the unconstrained superfields $X$ and ${\bX}$.

However, the remaining torsion constraints,
\begin{equation}
{T_{\gamma \bt}}^\al \,=\,0, \cem {{T^{\dot{\gamma}}}_{\bt}}^{\al}
 \,=\,0 \label{GRA.23}
\end{equation}
 and
\begin{equation}
{{T_{\gamma}}^{\db}}_{\da} \,=\,0, \cem  {T^{\dot{\gamma}
\db}}_{\da} \,=\,0, \label{GRA.24} \end{equation}
 are no
longer conserved by the superfield rescalings
 (\ref{GRA.12}) - (\ref{GRA.14}) and (\ref{GRA.20}) - (\ref{GRA.22}).
The new torsion coefficients take the form
\begin{eqnarray}
{{{T'}}_{\gamma \bt}}^\al & = & - \dtdu{\bt}{\al} {A'}_{\gamma} -
\dtdu{\gamma}{\al} {A'}_{\bt},
\label{GRA.25}
\\ {{{{T'}}_{\gamma}}^{\db}}_{\da} & = &
\dtud{\db}{\da} {A'}_{\gamma},
\label{GRA.26}
\end{eqnarray}
with ${A'}_{\gamma}$ defined as
\begin{equation}
{A'}_{\gamma} \,=\,- X^{-1} (2X^{-1} {\cal D}_{\gamma} X +
{{\bX}}^{-1} {\cal D}_{\gamma} {\bX}). \label{GRA.27}
\end{equation}
 The complex conjugate equations are
\begin{eqnarray}
{{T'}^{{\dg} {\db}}}_{\da} & = & \dtud{\db}{\da} {A'}^{{\dg}}
+ \dtud{\dg}{\da} {A'}^{{\db}},
\label{GRA.29}
 \\[1mm]
{T'}^\dg{}_{\bt}{}^{\al} & = & -
\dtdu{\bt}{\al} {A'}^{{\dg}},
\label{GRA.28}
\\[1.5mm]
{A'}^{{ \ \dg}} & = & \bX^{-1} (2 \bX^{-1} {\cd}^{{\dg}} \bX +
X^{-1} {\cd}^{{\dg}}X) \label{GRA.30}.
\end{eqnarray}
Next we examine the consequences of the superfield rescalings for
the remaining torsion coefficients by solving the Bianchi
identities in the presence of the new constraints or,
equivalently, by explicit calculation from
(\ref{GRA.11})\footnote{Solutions to the Bianchi identities in
terms of $R,\;R^{\dagger}$ and $G_a$ are presented in section
\ref{GRA23} and appendix \ref{appB}.}. In either case the tensor
decompositions of ${{T'}}_{\gamma b \da}$ and ${T^{' {\dg}}}_{b
\al}$ do not change, \ie
\begin{eqnarray}
{T'}_{\gamma b \da} & = &  - i \si_{b \gamma \da}
{R'}^{\dagger},
\label{GRA.31} \\[1mm]
{{{T'}^{{\dg}}}_{b}}{}^{\al}& = & -i
{\sib_{b}}^{{\dg} \al} R'.
\label{GRA.32}
\end{eqnarray}
The rescaled superfields ${R'}^{\dagger}$ and $R'$ are related to
the old ones by
\begin{equation}
{R'}^{\dagger} \,=\,X^{-2} \left\{ R^{\dagger } - \f{1}{4}
\left[(X \bX)^{-1} {\cd}^{\al} {\cd}_{\al} (X \bX) +
Y^{\al}Y_{\al} \right] \right\},
\label{GRA.33}
\end{equation}
\begin{equation} R' \,=\,\bX^{-2} \left\{ R -
\f{1}{4} \left[(X \bX)^{-1} {\cd}_{\da} {\cd}^{\da} (X \bX) +
\bY_{\da} \bY^{\da} \right] \right\},\label{GRA.34} \end{equation}
with the definitions
\begin{equation}
Y_{\al} \,=\,(X \bX)^{-1} {\cd}_{\al} (X \bX), \cem \bY^{\da} \ =
\ (X \bX)^{-1} {\cd}^{\da} (X \bX). \label{GRA.35}
\end{equation}
 The torsion coefficients ${T_{\gamma b}}^{\al}$ and
${T^{{\dg}}}_{b \da}$, however, pick up additional terms under the
superfield rescalings,
\begin{eqnarray}
{{T'}_{\gamma b}}^\al & = & i {{(\si_{cb})}_{\gamma}}^{\al} {G'}^c
+ i \dtdu{\gamma}{\al} G'_{b} \nonumber \\ && -\f{i}{4}
\dtdu{\gamma}{\al}{{\sib}_{b}}^{{\db} \bt} \left\{ (X^{-1}
{\cd}_{\bt} + {A'}_{\bt}){A'}_{{\db}} + (\bX^{-1} {\cd}_{{\db}}
 - {A'}_{{\db}}) {A'}_{\bt} \right\},
\label{GRA.36}  \\[2mm]
{{T'}^{{\dg}}}_{b \da} & = & i {{(\si_{bc})}^{{\dg}}}_{\da} {G'}^c
- i \dtud{\dg}{\da} G'_{b} \nonumber \\ && + \f{i}{4}
\dtud{\dg}{\da}{{\sib}_{c}}^{{\db} \bt} \left\{(X^{-1} {\cd}_{\bt}
+ {A'}_{\bt}){A'}_{{\db}} +
 (\bX^{-1} {\cd}_{{\db}} - {A'}_{{\db}}) {A'}_{\bt} \right\}. \label{GRA.37}
\end{eqnarray}
The rescaled superfield $G'_{\bt {\db}}=\si^{b}_{\bt {\db}}
G'_{b}$ is defined as
\begin{equation}
G'_{\bt {\db}} \,=\,(X \bX)^{-1} \left\{ G_{\bt {\db}} - \f{1}{2}
[{\cd}_{\bt}), (\ {\cd}_{{\db}}] \log (X \bX) + Y_{\bt} \bY_{{\db}}
\right\}.\label{GRA.38} \end{equation}
 The purpose of
this detailed presentation of superfield rescalings and their
consequences for the superspace torsion is twofold. First of all,
in the case ${A'}_{\al} = 0, \ {A'}_{\da} = 0$ the usual
super-Weyl or Howe-Tucker transformations, which leave the torsion
constraints invariant, are reproduced. Second, if $X$ and $\bX$
are kept arbitrary, the supervolume $E$ of the moving frame in
superspace changes as
\begin{equation}
E' \,=\,E(X \bX)^{2}. \label{GRA.39} \end{equation}
 This shows
that for the particular field dependent rescalings of parameters
\begin{equation}
X \,=\,\bX = e^{- \frac{1}{12} K (\phi, \phib)},
\label{GRA.40}
\end{equation}
 the kinetic action (\ref{GRA.1})
takes the form
\begin{equation}
\cl_{\mathrm{kin}} \,=\,-3 \int \! \!E'.
\label{GRA.41}
\end{equation}
 That is,
the kinetic Lagrangian action is the integral over a new
superspace defined with the supervolume $E'$.
In addition, in this case, from (\ref{GRA.40}) and (\ref{GRA.27}),
(\ref{GRA.30}) one obtains
\begin{equation}
{A'}_{\gamma} \,=\,+\f{1}{4} {\cd}'_{\gamma} K(\phi, \ \phib),
\label{GRA.42} \end{equation}
\begin{equation}
{A'}^{{\dg}} \,=\,- \f{1}{4} {{\cd}'}^{{\dg}} K(\phi, \ \phib).
\label{GRA.43} \end{equation}
 The primed spinor derivatives are,
of course, given as
\begin{equation}
{\cd}'_{\gamma} \,=\,X^{-1} {\cd}_{\gamma} \cem
{\cd'}^\dg \, = \, \bX^{-1} {\cd}^{{\dg}}.
\label{GRA.44} \end{equation}
 At
this stage it is very suggestive to interprete the additional
terms in (\ref{GRA.25}), (\ref{GRA.26}) and
(\ref{GRA.29}), (\ref{GRA.28}) not as unfortunate contributions to
the torsion but rather as superfield gauge potentials associated
to the structure group of a modified superspace geometry which
realizes K\"ahler transformations as field dependent chiral
rotations. To see this more clearly observe that the new frame is
related to the old one by
\begin{eqnarray}
{E'_{M}}^{a} & =&  e^{- \frac{1}{6}K(\phi, \phib)}\ {E_M}^{a},
\\[2mm]
\label{GRA.45}
{E'_{M}}^{\al} & = & e^{- \frac{1}{12}K(\phi, \phib)} \
\left(
{E_M}^{\al}
- \f{i}{12} {E_M}^{b} {{(\eps \si_b)}^{\al}}_{\da} \,
  {\cd}^{\da} \, K(\phi, \phib)
\right),
\label{GRA.46}
\\
{E'_{M}}_{\da} \, & = & e^{- \frac{1}{12}K(\phi, \phib)} \
\left(
E_{M \da} - \f{i}{12} {E_M}^{b} {{(\eps \sib_b)}_{\da}}^{\al} \,
{\cd}_{\al} K(\phi, \phib)
\right).
\label{GRA.47}
\end{eqnarray}
It is then easy to see that under the combination of K\"ahler
transformations and compensating super-Weyl transformations these
new variables transform homogeneously
\begin{eqnarray}
{E'_M}^{a} & \mapsto & {E'_M}^{a},
\label{GRA.48} \\[1mm]
{E'_M}^{\al} & \mapsto & e^{- \frac{i}{2} \ima F} {E'_M}^{\al},
\label{GRA.49} \\[1mm]
E'_{M \da} & \mapsto & e^{+ \frac{i}{2} \ima F} E'_{M \da}.
\label{GRA.50}
\end{eqnarray}
Indeed, these transformations represent chiral rotations of
parameter $- {i}/{2} \ima F$ and chiral weights $w(E_M{}^a) = 0$,
$w({E_M}^{\al}) = 1$, $w(E_{M \da}) = -1$. Likewise, by the same
mechanism, the superfields $R'$, ${R'}^{\dagger}$ and $G'_b$
undergo chiral rotations of weights
$w(R') = 2$, $w({R'}^{\dagger}) = -2$ and $w(G'_b)=0$.

The corresponding gauge potential 1-form in superspace is then
identified to be
\begin{equation}
{A'} \,=\, {E'}^a {A'}_a + {E'}^{\al} {A'}_{\al} + {E'}_{\da} {A'}^\da,
\label{GRA.51}
\end{equation}
 with field strength $F'=d{A'}$. The spinor coefficients
${A'}_{\al}$ and $A^{' \da}$ are given by (\ref{GRA.42}) and
(\ref{GRA.43}) and give rise to
\begin{equation}
F'_{\bt \al} \,=\,0, \cem  {F'}^{{\db} \da} \,=\,0. \label{GRA.52}
\end{equation}
 The equation for the field strength
${{F'}_{\bt}}^{\da}$ allows to determine the vector component
\begin{equation}
{A'}_{\al \da} \,=\,\f{i}{2} ({\cd}'_{\al} + {A'}_{\al}){A'}_{\da}
+ \f{i}{2} ({\cd}'_{\da}-{A'}_{\da}){A'}_{\al} - \f{i}{2} F'_{\al
\da}.
 \label{GRA.53}
\end{equation}
 Comparing (\ref{GRA.53}) to (\ref{GRA.36}), (\ref{GRA.37}) and
substituting appropriately yields
\begin{eqnarray}
{{T'}_{\gamma b}}^{\al}
&=&
i {(\si_{cb})_{\gamma}}^{\al} {G'}^c
+ i \dtdu{\gamma}{\al}  (G'_b + \f{1}{2} F'_b)
+ \dtdu{\gamma}{\al} {A'}_b,
\label{GRA.54}
\\
T'{}^\dg{}_{b \da}
&=& i {{(\sib_{bc})^{\dot{\gamma}}}_{\da}} {G'}^c
- i \dtud{\dg}{\da} (G'_b + \f{1}{2} F'_b)
- \dtud{\dg}{\da} {A'}_b.
\label{GRA.55}
\end{eqnarray}
Note that in this construction, ${A'}_b$ and $F'_b$ always appear in the
combination ${A'}_b + {i}/{2} F'_b$.

As a consequence of their definition, the coefficients of the
connection 1-form ${A'}$ change  under transformations
(\ref{GRA.7}), (\ref{GRA.8}) as
\begin{eqnarray}
{A'}_{\al} &\mapsto& e^{+\frac{i}{2} \ima F} ({A'}_{\al}
+ \f{i}{2} {\cd}'_{\al} \ima F),
\label{GRA.56}
\\[1mm]
{A'}^{\da} &\mapsto&
e^{- \frac{i}{2} \ima F} (A^{' \da} + \f{i}{2} {\cd}^{' \da} \ima F),
\label{GRA.57}
\\[2mm]
{A'}_a &\mapsto& ({A'}_a  + \f{i}{2} {\cd}'_a \ima F).
\label{GRA.58}
\end{eqnarray}
Taking into account the properties
of the rescaled frame, the transformation law for the 1-form
${A'}$ in superspace becomes simply
\begin{equation}
{A'} \, \mapsto \, {A'} + \f{i}{2} d\ima F. \label{GRA.59}
\end{equation}
 \indent
To summarize, the matter field dependent superfield rescalings of
frame and Lorentz connection, which might have appeared
embarrassing in the first place, because they changed the
geometric structure, actually led to a very elegant and powerful
description of matter fields in the presence of supergravity. The
most remarkable feature is that, in the supersymmetric case,
matter and gravitation lend themselves concisely to a unified
geometric description. Due to the close analogy between the
K\"ahler potential and the pre-potential of supersymmetric gauge
theory it is possible to include K\"ahler transformations in the
structure group of superspace geometry. They are realized by
chiral rotations as explained in detail above and the K\"ahler
potential takes the place of the corresponding pre-potential. The
superspace potentials can then be used to
construct K\"ahler covariant spinor and vector derivatives,
K\"ahler transformations are thus defined from the beginning at
the full superfield level and imbedded in the geometry of
superspace.

Furthermore, we have seen in (\ref{GRA.41}), that
the kinetic action for both supergravity and matter fields is
given by minus three times the volume of superspace. Its expansion
in terms of component fields gives immediately the correctly
normalized kinetic terms for all the component fields without any
need for rescalings or complicated integrations by parts at the
component field level.

\subsec{ $U(1)$ Superspace Geometry}
{\label{GRA23}

The result of the construction in the preceding section has a
natural explanation in the framework of $U(1)$ superspace
geometry, which will be reviewed in this section. In this
approach, the conventional superspace geometry is enlarged to
include a chiral $U(1)$ factor in the structure group. As a
consequence, the basic superfields of the new geometry are the
supervielbein ${E_M}^A(z)$ and the Lorentz gauge connection
${\phi_{MB}}^A (z)$ together with a gauge potential $A_M (z)$ for
chiral $U(1)$ transformations. These superfields define
coefficients of 1-forms in superspace such that
\begin{equation}
E^A \,=\,dz^M {E_M}^A (z), \label{GRA.60} \end{equation}
\begin{equation}
{\phi_B}^A  \,=\,dz^M {\phi_{MB}}^A (z), \label{GRA.61}
\end{equation}
\begin{equation}
A \,=\,dz^M A_M(z). \label{GRA.62} \end{equation}
 Torsion and
field strengths are then defined with the help of the exterior
derivative $d$ in superspace
 \begin{eqnarray}
  T^A &=& dE^A + E^B
{\phi_B}^A + w(E^A) E^A A, \label{GRA.63} \\ {R_B}^A &=& d
{\phi_B}^A \ + \ {\phi_B}^C {\phi_C}^A, \label{GRA.64} \\ F &=&
dA. \label{GRA.65} \end{eqnarray}
 The chiral $U(1)$ weights
$w(E^A)$ are defined as
\begin{equation}
w(E^a) \,=\,0, \cem w(E^\al) \,=\,1, \cem w(E_{\da}) \,=\,-1.
\label{GRA.66} \end{equation}
 \indent The non-vanishing parts ${\phi_b}^a, \
{\phi_\bt}^\al, \ {\phi^{\db}}_{\da}$ of ${\phi_B}^A$ (the Lorentz
connection) are related among each other as usual,
\begin{equation}
{\phi_\bt}^\al \,=\,
- \f{1}{2} {(\si^{ba})_\bt}^\al \phi_{ba}, \cem
{\phi^{\db}}_{\da}\,=\,-
\f{1}{2} {(\sib^{ba})^{\db}}_{\da} \phi_{ba}.
\label{GRA.68}
\end{equation}
As is well-known \cite{Dra79}, for this choice of
structure group, the Lorentz curvature and $U(1)$ field  strength,
\begin{equation}
{R_B}^A \,=\,\f{1}{2} E^C E^D {R_{DC \ B}}^A, \label{GRA.69}
\end{equation}
\begin{equation}
F \,=\,\f{1}{2} E^C E^D F_{DC},
\label{GRA.70} \end{equation}
 are
completely defined in terms of the coefficients of the torsion
2-form,
\begin{equation}
T^A \,=\,\f{1}{2} E^B E^C {T_{CB}}^A, \label{GRA.71}
\end{equation}
 and covariant derivatives thereof as a consequence
of the superspace Bianchi identities,
\begin{equation}
{\cd}T^A - E^B {R_B}^A - w(E^A) E^A F \,=\,0. \label{GRA.72}
\end{equation}

In the present case, covariant derivatives are understood to be
covariant with respect to both Lorentz and $U(1)$
transformations. The covariant derivative of a generic
superfield $\cx_A$ of chiral weight $w(\cx_A)$ is defined as
\begin{equation}
{\cd}_B \cx_A \,=\,{E_B}^M \prt_M \cx_A - {\phi_{BA}}^C \cx_C +
                    w(\cx_A) A_B \cx_A,
\label{GRA.73} \end{equation}
 with (graded) commutator
\begin{equation}
({\cd}_{C}, {\cd}_{B}) \cx_A \,=\,- {T_{CB}}^F {\cd}_F \cx_A  -
{R_{CB \ A}}^F \cx_F + w(\cx_A) F_{CB} \cx_A. \label{GRA.74}
\end{equation}

The chiral weights of the various objects are related to that of
the vielbein, $E^A$, in a simple way, \eg
\begin{eqnarray}
w({\cd}_A) &=& -w(E^A),
\nn \\
w({T_{CB}}^A) &=& w(E^A) - w(E^B) - w(E^C),
\nn \\
w({R_{CB\ A}}^F) &=& - w(E^B) - w(E^C).
\label{GRA.75}
\end{eqnarray}

Finally, the vielbein $E^A$, the covariant derivative ${\cd}_A$
and the $U(1)$ gauge potential $A_A$ change under chiral $U(1)$
structure group transformations $g$ as
 \begin{eqnarray}
  E^{A}
&\mapsto& E^A g^{w(E^A)},
\label{GRA.77} \\[1mm]
 {\cd}_A &\mapsto&
g^{-w(E^A)} {\cd}_A,
\label{GRA.78} \\[1mm]
 A_A &\mapsto& g^{-w(E^A)}
\left( A_A - g^{-1} {E_A}^M \prt_M g \right). \label{GRA.79}
\end{eqnarray}
 As said in the introduction, the choice of
structure group largely determines the $U(1)$ superspace geometry,
which is further specified by appropriate covariant torsion
constraints. For instance, combination of the covariant chirality
conditions with the commutation relation (\ref{GRA.74}) suggests
\begin{equation}
{T_{\gamma \bt}}^a \,=\,0 , \cem T^{{\dg} {\db} a} \,=\,0.
\label{GRA.80} \end{equation}
For a more complete presentation, we refer to \cite{GGMW84b}, and
references therein. Here, we content ourselves to sketch out the essential
features of the resulting structure in superspace.

First of all,
we note that all the coefficients of torsion and of Lorentz and
 $U(1)$ field strengths are given in terms of the covariant superfields $R,
R^{\dagger}$ (respectively chiral and antichiral) and $G_a$ (real)
of canonical dimension $1$ and of the Weyl spinor
superfields $W_{\lsym{\gamma \bt \al}}$ and $W_{\lsym{{\dg} {\db}
\da}}$ of canonical dimension ${3}/{2}$.

Moreover the only non-vanishing component at dimension zero is
the constant torsion already present in rigid superspace,
\begin{equation}
{T_\gamma}^{{\db} a} \,=\,-2i {(\si^a \eps)_\gamma}^{\db}.
\label{GRA.81} \end{equation}
We then proceed in the order of increasing canonical dimension.
At dimension${1}/{2}$, all the torsion coefficients vanish
whereas at dimension $1$ the above mentioned superfields $R,
R^{\dagger}$ and $G_a$ are identified as
 \begin{eqnarray}
  T_{\gamma b \da}
\,=\,-i \si_{b \gamma \da} R^{\dagger} \  ,&& \cem {T_{\gamma
b}}^\al \,=\,\f{i}{2} {(\si_c \sib_b)_\gamma}^{\al} G^c ,
\label{GRA.82} \\ {{T^{\dg}}_b}^\al \,=\,-i {\sib_b}^{{\dg} \al} R
\ ,&& \cem {T^{\dg}}_{b \da} \,=\,-\f{i}{2} {({\sib_c
\si_b)}^{{\dg}}}_{\da} G^c. \label{GRA.83} \end{eqnarray}
 The purely vector torsion is taken to vanish
\begin{equation}
{T_{cb}}^a \,=\,0. \label{GRA.84} \end{equation}

At dimension ${3}/{2}$, the super-covariant Rarita-Schwinger
(super)field strengths ${T_{cb}}^\al$ and $T_{cb \da}$ are most
conveniently displayed in spinor notation
\begin{equation}
{T_{\gamma {\dg}\  \bt {\db}}}^A \,=\,
\si^c_{\gamma {\dg}} \,
\si^b_{\bt {\db}}\, {T_{cb}}^A . \label{GRA.85} \end{equation}
Together with $G_{\al \da} = \si^a_{\al \da} G_a$ we obtain
 \begin{eqnarray}
 T_{\gamma {\dg}\ \bt {\db}\ \al}
&= &
+2 \eps_{{\dg} {\db}} W_{\lsym{\gamma \bt \al}}
+ \f{2}{3} \eps_{{\dg} {\db}} (\eps_{\al \bt} S_\gamma
+ \eps_{\al \gamma} S_\bt )
- 2 \eps_{\gamma \bt} T_{\sym{{{\dg}}{{\db}}}}{}_{\vt{3}\al}, \\
T_{\sym{{{\dg}}{{\db}}}}{}_{\vt{3}\al}
& = &
- \f{1}{4} ({\cd}_{\dg} G_{\al {\db}}
+ {\cd}_{\db} G_{\al {\dg}}),
\\
S_\gamma
& = &
- {\cd}_\gamma R + \f{1}{4} {\cd}^{\dg} G_{\gamma {\dg}}
\label{GRA.86}
\end{eqnarray}
and
\begin{eqnarray}
T_{\gamma {\dg}\ \bt {\db}\ \da}
& = &
-2 \eps_{\gamma \bt} W_{\lsym{{\dg} {\db} \da}}
- \f{2}{3} \eps_{\gamma \bt} (\eps_{\da {\db}} S_{\dg}
+ \eps_{\da {\dg}} S_{\db})
+ 2 \eps_{{\dg} {\db}} T_{\sym{\gamma \bt}}{}_{\da}, \\
T_{\sym{\gamma \bt}}{}_{\da}
& = &
+\f{1}{4} ({\cd}_\gamma G_{\bt \da}
 + {\cd}_\bt G_{\gamma \da}),
\\
S_{\dg} & = &
+ {\cd}_{\dg} R^{\dagger} - \f{1}{4} {\cd}^\gamma G_{\gamma {\dg}}.
\label{GRA.87}
\end{eqnarray}

The $U(1)$ weights of the basic superfields appearing in
(\ref{GRA.82}), (\ref{GRA.83}) and (\ref{GRA.86}), (\ref{GRA.87}) are
\begin{eqnarray}
w(R) \,=\,2, \cem & & w(R^{\dagger}) \,=\,-2,
\nonumber \\
w(G_a) &=& 0, \\
w(W_{\lsym{\gamma \bt \al}}) \,=\,1, \cem
& &
w(W_{\lsym{{\dg} {\db}\da}}) \,=\, -1. \nonumber
\label{GRA.88}
\end{eqnarray}

As already mentioned above, the coefficients of Lorentz curvatures
and $U(1)$ field strengths are expressed in terms of these few
superfields. At dimension one we obtain
\begin{eqnarray}
R_{\delta \gamma\ ba}
& = &
8 (\si_{ba} \eps)_{\delta \gamma}
R^{\dagger},
\label{GRA.89} \\[1.5mm]
{R^{\dd {\dg}}}_{\ ba} & = & 8
(\sib_{ba} \eps)^{\dd {\dg}} R,
\label{GRA.90} \\[1.5mm]
{{R_\delta}^{{\dg}}}_{\ ba} & = & -2i G^d {(\si^c
\eps)_\delta}^{\dg} \eps_{dcba},
\label{GRA.91}
\end{eqnarray}
 for the Lorentz curvatures whereas the chiral
$U(1)$ field strengths are given by
\begin{equation}
F_{\bt \al} \,=\,0, \cem F^{{\db} \da} \,=\,0, \label{GRA.92}
\end{equation}
\begin{equation}
{F_\bt}^{\da} \,=\,3{(\si^a \eps)_\bt}^{\da} G_a. \label{GRA.93}
\end{equation}

At dimension ${3/}{2}$, we find
\begin{eqnarray}
 R_{\delta c \
ba} & = & i \si_{c \delta \dd} {T_{ba}}^\dd +
                        i \si_{b \delta \dd} {T_{ca}}^\dd +
                        i \si_{a \delta \dd} {T_{bc}}^\dd,
\label{GRA.94} \\[2mm]
{R^{\dd}}_{c\ ba} & = & i \sib_{c}^{\dd \delta} {T_{ba \delta}}+ i
\sib_b^{\dd \delta} {T_{ca \delta}} +i \sib_a^{\dd \delta} T_{bc
\delta} \label{GRA.95}
\end{eqnarray}
 and
\begin{equation}
 F_{\delta c} \, = \, \f{3i}{2} {\cd}_\delta G_c
                + \f{i}{2} \si_{c \delta \dd} {\bX}^{\dd}, \cem
{F^{\dd}}_{c} \, = \, \f{3i}{2}
{\cd}^\dd G_c - \f{i}{2} \sib_{c} ^{\dd {\delta}} {X}_\delta,
\label{GRA.97}
\end{equation}
with the definitions
\begin{equation}
X_\delta \, = \, {\cd}_\delta R - {\cd}^\dd G_{\delta \dd}, \cem
{\bX}^\dd \, = \, {\cd}^\dd R^{\dagger} +
{\cd}_\delta G^{\delta \dd}.
\label{GRA.99}
\end{equation}

Finally, having expressed torsions, curvatures and $U(1)$ field
strengths in terms of few covariant superfields, the Bianchi
identities themselves are now represented by a small set of rather
simple conditions, such as
\begin{equation}
{\cd}_\al \bar{W}_{\lsym{{\dg} {\db} \da}} \,=\,0 , \cem
{\cd}_{\da}W_{\lsym{\gamma \bt \al}}  \,=\,  0, \label{GRA.101}
\end{equation}
or
\begin{equation}
{\cd}^\al T_{cb \al} + {\cd}_{\da} {T_{cb}}^{\da} \,=\,0,
\label{GRA.104} \end{equation}
for these superfields. A detailed account of
these relations is given in \ref{appB2}.

Let us stress, that the complex superfield $R$, subject to
chirality conditions
\begin{equation}
{\cd}_\al R^{\dagger} \,=\,0, \cem {\cd}^{\da} R \,=\,0,
\label{GRA.100}
\end{equation}
plays a particularly important role, it contains the curvature scalar in
its superfield expansion. As in our language superfield
expansions are replaced by successive applications of spinor
derivatives, the relevant relation is
\begin{equation}
{\cd}^2R + {\cdb}^2 R^{\dagger} \, = \, - \f{2}{3} {R_{ba}}^{ \
ba} - \f{2}{3} {\cd}^{\al}X_{\al} + 4G^aG_a + 32RR^{\dagger}.
\label{B.89bis}
\end{equation}
Interestingly enough the curvature scalar is necessarily accompanied by
the D-term superfield ${\cd}^{\al}X_{\al} = -2 {\bf D}$ of the $U(1)$
gauge sector, described in terms of the gaugino superfields
$X_\al$ and ${\bX}^{\da}$ subject to
the usual chirality and reality conditions
\begin{equation}
{\cd}_\al {\bX}^{\da} \,=\,0, \cem {\cd}^{\da} X_\al \,=\,0,
\label{GRA.102} \end{equation}
\begin{equation}
{\cd}^\al X_\al - {\cd}_{\da} {\bX}^{\da}  \,=\, 0.
 \label{GRA.103}
\end{equation}
This shows very clearly that generic $U(1)$ superspace provides
the natural framework for the description of gauged
R-transformations \cite{Fre77}, \cite{BFNS82}, \cite{Ste82},
\cite{CD96}, \cite{CFM96}. Relation (\ref{B.89bis}) shows that
supersymmetric completion of the (canonically normalized)
curvature scalar action induces a Fayet-Iliopoulos term for
gauged R-transformations.

\indent

{\em At this point we wish to make a digression to indicate how
the superspace geometry described above can be related to that of
\cite{Mul86} and restricted to the superspace geometry relevant to
the minimal supergravity multiplet. To this end, call $A_0$ the
$U(1)$ gauge potential of the superspace geometry described here
and $A_1$ the $U(1)$ gauge potential of \cite{Mul86}. The two
(equivalent) descriptions are related through
\begin{equation}
A_1 \,=\,A_0 - \f{3i}{2} E^a G_a. \label{GRA.105} \end{equation}

On the other hand, the superspace geometry of \cite{WB83} is
recovered by
\begin{equation}
A_1 \,=\,0, \cem X_\al \,=\,0, \cem {\bX}^{\da} \,=\,0,
\label{GRA.106} \end{equation}
 giving rise (among other things) to
\begin{eqnarray}
 {{T^{o.m}}_{\gamma b}}^\al & = & + \f{3i}{2}
\dtdu{\gamma}{\al} G_b + \f{i}{2} G^c {(\si_c \sib_b)_\gamma}^\al,
\label{GRA.107} \\ {T^{o.m {\dg}}}_{b \da} & = & - \f{3i}{2}
\dtud{\dg}{\da} G_b - \f{i}{2} G^c {(\sib_c \sib_b)^{\dg}}_{\da}
\label{GRA.108}
\end{eqnarray}
 and}
\begin{equation}
{\cd}_\al R \,=\,{\cd}^{\da} G_{\al \da} \cem
{\cd}^{\da} R^{\dagger} \,=\, - {\cd}_\al G^{\al \da}.
\label{GRA.109}
\end{equation}
In this sense $U(1)$ superspace is the underlying framework for
both minimal supergravity and its coupling to matter. Note, en
passant, that in \cite{Mul86} the other two supergravity
multiplets, non-minimal and new minimal, have been derived from
generic $U(1)$ superspace as well.

\subsec{ Formulation in K\"ahler Superspace}
{\label{GRA3}

As pointed out earlier, the description of supersymmetric
nonlinear sigma models \cite{Zum79b} as well as the construction
of supergravity/matter couplings
\cite{CJS+78b,CJS+79,CFGvP82,CFGvP83,Bag83,Bag85,BGGM87b,BGGM87a}
is based on an intriguing analogy between K\"ahler geometry and
supersymmetric gauge theory, which are both defined by means of
differential constraints. In K\"ahler geometry the fundamental
2-form of complex geometry is required to be closed whereas
supersymmetric gauge theory is characterized by covariant
constraints as explained in section \ref{RS3}. The constraints
imply that the K\"ahler metric is expressed in terms of
derivatives of the K\"ahler potential whereas, on the other hand,
the superspace gauge potential is expressed in terms of a
pre-potential. Pre-potential transformations, which are chiral
superfields should then be compared to K\"ahler transformations
which are holomorphic functions of the complex coordinates.

Matter superfields, on the other hand, are given by chiral
superfields. It remains to promote the complex coordinates of the
K\"ahler manifold to chiral superfields: holomorphic functions of
chiral superfields are still chiral superfields. Correspondingly,
the K\"ahler potential becomes a function of the chiral and
antichiral superfield coordinates. The geometry of the
supersymmetry coupling is then obtained by replacing the gauge
potential in $U(1)$ superspace by the superfield K\"ahler
potential \cite{BGGM87b,BGGM87a,Gri90a}.

In section \ref{GRA31} we present the basic features of this
geometric structure in a self-contained manner. In section
\ref{GRA32} we include Yang-Mills interactions (-cf. appendix
\ref{appB} for their formulation in $U(1)$ superspace). Gauged
superfield isometries of the K\"ahler metric are treated in
appendix \ref{appC}. We also study carefully the supergravity
transformations of the whole system. Finally in section
\ref{GRA33} invariant superfield actions and the corresponding
superfield equations of motion will be discussed.

\subsubsection{ Definition and properties  of K\"ahler superspace}
{\label{GRA31}

K\"ahler superspace geometry is defined as $U(1)$ superspace
geometry, presented in section \ref{GRA23}, with suitable
identification of the $U(1)$ pre-potential and pre-gauge
transformations with the K\"ahler potential and K\"ahler
transformations. The relevant version of $U(1)$ superspace
geometry is the one where the $U(1)$ structure group
transformations are realized in terms of chiral and antichiral
superfields as described in (\ref{RS22}) for the
$(\dem,\dem)$ basis, where most of the work has already been
done. As a matter of fact, the structures developed there in the
framework of rigid superspace are very easily generalized to the
present case of curved $U(1)$ superspace geometry. To begin with,
the solution of (\ref{GRA.92}) is given as
 \begin{eqnarray} A_\al & = & - T^{-1}
{E_\al}^M \prt_M T,  \label{GRA.110} \\ A^{\da} & = & -U^{-1}
E^{\da M} \prt_M U, \label{GRA.111} \end{eqnarray}
 with ${E_A}^M$
now the full (inverse) frame of $U(1)$ superspace geometry.
 As
anticipated in section (\ref{RS22}) the geometric structure
relevant to the superspace formulation of supergravity/matter
coupling is the basis $(a,b)=(\dem,\dem)$. In this basis one
has
  \begin{eqnarray}
   A_\al (\dem) & = & - \f {1}{2} W^{-1} {E_\al}^M (\dem)
\prt_M W,
\label{GRA.129}
\\
A^{\da} (\dem) & = & + \f {1}{2} W^{-1} E^{\da M}
(\dem) \prt_M W,
\label{GRA.130}
\end{eqnarray}
where
$W=T\,U^{-1}$ transforms as given in (\ref{RSWtf}). For the
vielbein we have
\begin{equation}
E^A(\dem) \, \mapsto \, \left[ \overline{\bf P}\, {\bf Q}
\right]^{-\frac{w(A)}{2}}\, E^A(\dem) \label{GRA.132}
\end{equation}
 and
\begin{eqnarray}
   A_\al (\dem) &\mapsto& (\overline{\bf P}\,{\bf Q} )^{1/2}
    \left[\, A_\al(\dem) + \f {1}{2} {E_\al}^M (\dem)
\prt_M\log {\bf Q} \right], \label{GRA.133} \\ A^{\da} (\dem)
&\mapsto& (\overline{\bf P}\,{\bf Q} )^{-1/2} \left[\,
A^{\da}(\dem) + \f {1}{2}  E^{\da M} (\dem) \prt_M
\log\overline{\bf P} \right]. \label{GRA.134}
\end{eqnarray}

\indent
 In order to make contact with the
superspace structures obtained in section \ref{RS2}, we relate $W$
to the K\"ahler-potential $K(\phi, \phib)$ and $\overline{{\bf
P}}$ and  ${\bf Q}$ to the K\"ahler transformations $F(\phi)$ and
$F(\phib)$. It is very easy to convince oneself that the
identifications
\begin{eqnarray}
 W & = & \exp(- K(\phi, \phib)\, /\,2),
\label{GRA.135}
\\
\overline{{\bf P}} & = & \exp\, (- {\bF} (\phib)\, /\,2),
 \label{GRA.136}
\\
{\bf Q} & = & \exp\,(+ F(\phi)\, /\,2),
\label{GRA.137}
\end{eqnarray}
reproduce exactly the geometric structures obtained
at the end of section \ref{GRA2} after superfield rescalings. The
primed quantities defined there are identical with the $U(1)$
superspace geometry in the $(\dem,\dem)$ basis after the
identifications (\ref{GRA.135}) - (\ref{GRA.137}), \ie
\begin{eqnarray}
 {E'}^A &=& E^A(\dem),
\label{GRA.138} \\ {A'} &=& A(\dem). \label{GRA.139}
\end{eqnarray}
 In particular, from (\ref{RSWtf}) we recover the K\"ahler
transformations
\begin{equation}
K(\phi, \phib) \, \mapsto \, K(\phi, \phib) + F(\phi) +
{\bF}(\phib).
\label{GRA.140}
\end{equation}
 Moreover,
(\ref{GRA.129}), (\ref{GRA.130}) reproduce (\ref{GRA.42}), (\ref{GRA.43}),
and (\ref{GRA.133}), (\ref{GRA.134}) correspond
exactly to (\ref{GRA.56}), (\ref{GRA.57}).

\indent {\em We have thus constructed the superspace geometry
relevant for the description of supergravity/matter couplings and
at the same time established the equivalence with the more
traditional formulation.}

 In this new kind of superspace geometry, called
{\em K\"ahler superspace geometry}, or {\em $U_K(1)$ superspace geometry},
the complete action for the kinetic terms of both supergravity and matter
fields is given by the superdeterminant of the frame in superspace.
 Expression of
this superfield action in terms of component fields leads to the
correctly normalized component field actions without any need for
rescalings. Invariance under superfield K\"ahler transformations
is achieved {\em 'ab initio}} without any need for compensating
transformations.

The local frame
$E^A$ is subject to both Lorentz and K\"ahler transformations in a
well defined way. Covariance of the torsion 2-form is
achieved with the help of gauge potentials ${\phi_B}^A$ and $A$
for Lorentz and K\"ahler transformations respectively,
\begin{equation}
T^A \,=\,dE^A + E^B {\phi_B}^A + w(E^A) E^A A. \label{GRA.141}
\end{equation}
 The complete expression is the same as in
$U(1)$ superspace geometry, except that the chiral gauge potential
is no longer an independent field but rather expressed in terms of
the K\"ahler potential $K(\phi, \phib)$. Hence this superspace
torsion contains at the same time supergravity and matter fields!
The K\"ahler transformations of $A$ are induced from those of the
K\"ahler potential, \ie
\begin{equation}
K(\phi, \phib) \ \mapsto \ K(\phi, \phib) + F(\phi) +
{\bF}(\phib), \label{GRA.142} \end{equation}
 to be
\begin{equation}
A \ \mapsto \ A + \f {i}{2}d \ima F. \label{GRA.143}
\end{equation}
 At the same time the frame is required to undergo
the chiral rotation
\begin{equation}
E^A \, \mapsto \, E^A \, e^{-\frac{i}{2} \, w(E^A) \, \ima F},
\label{GRA.144}
\end{equation}
ensuring a covariant transformation
law of the superspace torsion,
\begin{equation}
T^A \, \mapsto \, T^A \, e^{-\frac{i}{2} \, w(E^A) \, \ima F}.
\label{GRA.145}
\end{equation}
 Its coefficients are subject to the
same constraints as those of $U(1)$ superspace and therefore the
tensor decompositions as obtained from the analysis of superspace
Bianchi identities remain valid. For details we refer to appendix
\ref{appB}.

We shall, however, present in  detail the structure of the $U(1)$
gauge sector, in particular the special properties which arise
from the parametrization of $A$ in terms of the K\"ahler potential
$K(\phi, \phib)$, namely
\begin{equation}
 A_\al \, = \, \frac{1}{4} {E_\al}^M \prt_M K(\phi, \phib), \cem
 A^\da \, = \, - \frac{1}{4} E^\da{}^M \prt_M K(\phi, \phib),
\label{GRA.146a}
\end{equation}
\begin{equation}
 A_{\al \da}- \frac{3i}{2}G_{\al \da} \, = \, \frac{i}{2}
  \lp \cd_\al A_\da + \cd_\da A_\al \rp.
\label{GRA.146}
\end{equation}
It follows that its field strength 2-form, $F = dA$, has the spinor
coefficients
\begin{equation}
F_{\bt\al} \,=\,0,\cem F^{\db {\da}} \,=\,0,\cem {F_\bt}^{\da} \ =
\ 3{({\si}^a \eps)_{\bt}}^{\da} G_a. \label{GRA.147}
\end{equation}
 Of course, this reproduces the
structure of the constraints already encountered in $U(1)$
superspace which implies also
\begin{eqnarray}
{F_{\bt a}}-\f {3i}{2}{\cd}_{\bt}G_a
 & = & + \f {i}{2} \si_{a \bt \db} \bX^{\db}, \\[1.5mm]
{F^\db}_a
- \f {3i}{2}{\cd}^\db G_a & = & - \f
{i}{2}{\sib_a}^{\db \bt} X_{\bt},
\label{GRA.148}
\end{eqnarray}
with
 \begin{eqnarray}
  X_\al  \; & = &{\cd}_{\al}R
-{\cd}^{\da} G_{\al {\da}},
\label{GRA.149} \\[1.5mm]
 {\bX}^{\da} & =
&{\cd}^{\da} R^{\dagger} + {\cd}_\al G^{\al {\da}}.
\label{GRA.150} \end{eqnarray}
 In the absence of matter, the
superfields $X_\al, \bX^{\da}$ vanish and we are left with
standard superspace supergravity. In the presence of matter they
are given in terms of the K\"ahler potential as
\begin{eqnarray}
X_\al \; & = & -\f {1}{8}({\cdb}^2 -8R)
 {\cd}_{\al}K(\phi,\phib),
\label{GRA.151} \\[1.5mm]
{\bX}^{\da} & = & -\f {1}{8}({\cd}^2
-8R^{\dagger}) {\cd}^{\da} K(\phi,\phib). \label{GRA.152}
\end{eqnarray}
 These expressions are simply a
consequence of the explicit definitions given so far.

In an alternative, slightly more illuminating way, we may
write $A$
as\footnote{Note that the term containing $G_a$
originates from our particular
choice of constraint (\ref{GRA.93}) \ie
$F_{\bt}{}^\da  \,=\, 3 {{(\si^a \eps)_\beta}^{\da}} \, G_a$.}
\begin{equation}
A \,=\,\f {1}{4} (K_k d{\phi}^k -K_{\bar{k}}d{\phib}^{\bar{k}})
+\f {i}{8}E^a \lp 12G_a + {\sib}_a^{{\da} \al}g_{k{\bar{k}}} {\cd}_\al
{\phi}^k {\cd}_{\da}{\phib}^{\bar{k}} \rp,
\label{GRA.153}
\end{equation}
 where $K_k$ and $K_{\bk}$ stand for the derivatives
of the K\"ahler potential with respect to ${\phi}^k$ and
${\phib}^{\bk}$, this way of writing $A$ is more in line with
K\"ahler geometry. The exterior derivative of $A$,
\begin{equation}
F \,=\,dA \,=\,\f {1}{2} g_{k{\bar{k}}}d\phi^k d{\phib}^{\bar{k}}
+\f {i}{8} d \left[ E^a \lp 12G_a+{\sib}_a^{{\da}
\al}g_{k{\bar{k}}}{\cd}_\al {\phi}^k {\cd}_{\da}{\phib}^{\bar{k}}
\rp \right], \label{GRA.154}
\end{equation}
 yields the superspace analogue of
the fundamental form in ordinary K\"ahler geometry, with complex
coordinates replaced by chiral superfields (the additional term is
not essential and could have been absorbed in a redefinition of
the vector component of $A$).

This form of $F$ is also very convenient to derive directly the
explicit expression of $X_\al$ and of $\bX^{\da}$  in terms of the
matter superfields,  avoiding
 explicit evaluation of the spinor derivatives in (\ref{GRA.152}),
(\ref{GRA.153}). A straightforward identification in
$F_{\bt a}$ resp. $F^\db{}_a$
shows that
\begin{eqnarray}
 X_\al \, & = & -\f {i}{2}g_{k{\bar{k}}}
{\si}^a_{\al{\da}} {\cd}_a{\phi}^k {\cd}^{\da} {\phib}^{\bar{k}}
+\f {1}{2}g_{k{\bar{k}}}{\cd}_\al {\phi}^k \, {\bF}^{\bar{k}},
\label{GRA.155} \\[1.5mm]
{\bX}^{\da} & = &- \f {i}{2}g_{k
{\bar{k}}}{\sib}^{a{\da} \al}
{\cd}_a{\phib}^{\bar{k}}{\cd}_{\al}{\phi}^k + \f {1}{2}g_{k
{\bar{k}}}{\cd}^{\da}{\phib}^{\bar{k}}F^k.
\label{Gra.156}
\end{eqnarray}
 Here we have used the definitions,
\begin{equation}
F^k \,=\,- \f {1}{4}{\cd}^2{\phi}^k, \cem {\bF}^{\ \bk} \,=\,- \f
{1}{4}{\cdb}^2{\phib}^{\bk}. \label{GRA.157}
\end{equation}
The covariant derivatives are defined as
\begin{equation}
{\cd}_\al \phi^k \,=\,{E_\al}^M \partial_M \phi^k, \cem
{\cd}^{\da} {\phib}^{\bk} \,=\,E^{\da M} \partial_M {\phib}^{\bk},
\label{GRA.158}
\end{equation}
\begin{eqnarray}
{\cd}_B {\cd}_\al \phi^k &=& {E_B}^M \partial_M {\cd}_\al \phi^k -
{\phi_{B\al}}{}^\varphi {\cd}_\varphi \phi^k - A_B {\cd}_\al
\phi^k + {\Gamma^k}_{ij} {\cd}_B \phi^i{\cd}_\al \phi^j,
\label{GRA.159}\\[2mm]
{\cd}_B {\cd}^{\da} \phib^{\bk} &=& {E_B}^M \partial_M {\cd}^{\da}
\phib^{\bk} - {{\phi_B}^{\da}}{}_{\dv} {\cd}^{\dv}\phib^{\bk} +
A_B {\cd}^{\da} \phib^{\bk} + {\Gamma^{\bk}}_{\bi\bj} {\cd}_B
\phib^{\bi}{\cd}^{\da}\phib^{\bj},
\label{GRA.160}
\end{eqnarray}
 assuring covariance with respect to Lorentz and K\"ahler
transformations and (ungauged) isometries of the K\"ahler metric.
The Levi-Civita symbols
\begin{equation}
{\Gamma^k}_{ij} \,=\,g^{k\bl}g_{i\bl,j} , \cem
{\Gamma^{\bk}}_{{\bi}{\bj}} \,=\,g^{l{\bk}}g_{l{\bi},{\bj}} ,
\label{GRA.161} \end{equation}
 are now, of course, functions of
the matter superfields. Do not forget that, due to their
geometric origin, the superfields $X_{\al},  {\bX}^{\da}$ have the
properties
\begin{equation}
{\cd}^{\da}X_{\al} \,=\,0 ,\cem {\cd}_{\al} {\bX}^{\da} \,=\,0,
\label{GRA.162} \end{equation}
\begin{equation}
{\cd}^{\al} \! X_{\al} \,=\,{\cd}_{\da} {\bX}^{\da}. \label{GRA.163}
\end{equation}

As we shall see later on, the lowest components of the superfields
$X_{\al},  {\bX}^{\da}$, as well as that of ${\cd}^{\al} \!
X_{\al}$,
 appear in the construction of the component field
action. In order to prepare the ground for this construction we
display here the superfield expression of the K\"ahler D-term. It
is
 \begin{eqnarray}
-\f {1}{2}{\cd}^\al \! X_\al
&=&-g_{k{\bk}}{\eta}^{ab} {\cd}_b {\phi}^k {\cd}_a
{\phib}^{\bk} -\f
{i}{4}g_{k{\bk}}{\si}^a_{\al{\da}}{\cd}^{\al}{\phi}^k {\cd}_a
{\cd}^{\da} {\phib}^{\bk} \nonumber \\ & &-\f
{i}{4}g_{k{\bk}}{\si}^a_{\al{\da}}{\cd}^{\da}{\phib}^{\bk}
 {\cd}_a {\cd}^{\al}{\phi}^k
+g_{k{\bk}}F^k {\bF}^{\bk} \nonumber \\[1mm]
& &+\f {1}{16} R_{k{\bk} j {\bj}}{\cd}^{\al}{\phi}^k
 {\cd}_{\al}{\phi}^j {\cd}_{\da}{\phib}^{\bk}
{\cd}^{\da}{\phib}^{\bj},
\label{GRA.164} \end{eqnarray}
with covariant derivatives as defined above in (\ref{GRA.159}),
(\ref{GRA.160}).
The Riemann tensor is given as
\begin{equation}
R_{k{\bk} j{\bj}} \,=\,g_{k{\bk},j{\bj}}- g^{l{\bar{l}}}
g_{k{\bar{l}},j} g_{l{\bk},{\bj}}\; . \label{GRA.165}
\end{equation}
 The terminology employed here concerning the notion
of a D-term may appear unusual but it is perfectly adapted to the
construction in curved superspace, where explicit superfield
expansions are replaced by successively taking covariant spinor
derivatives and projecting to lowest superfield components. In
this sense the lowest component of the superfield ${\cd}^\al \! X_\al$
indeed provides the complete and invariant geometric
definition of the component field D-term.

In our geometric formulation, this K\"ahler D-term appears very
naturally in the superfield expansions of the superfields $R,
R^{\dagger}$ of the supergravity sector. To see this in more
detail,
 recall first of all the chirality properties,
\begin{equation}
{\cd}_{\al}R^{\dagger} \,=\,0,\cem {\cd}^{\da} \! R \,=\,0,
\label{GRA.166} \end{equation}
 with $R, R^{\dagger}$ having chiral
weights $w(R)=2$ and $w(R^{\dagger})=-2$, respectively. For the
spinor derivatives of the opposite chirality the Bianchi
identities imply
\begin{eqnarray}
{\cd}_\al R
&=&
-\f {1}{3}X_\al
-\f {2}{3}{({\si}^{cb} \eps)}_{\al\varphi}{T_{cb}}^\varphi,
\label{GRA.167} \\
{\cd}^{\da} \! R^{\dagger}
&=&
-\f {1}{3} {\bX}^{\da}
-\f {2}{3} {({\sib}^{cb}\eps)}^{{\da}{\dot{\varphi}}}
{T_{cb}}_{\dot{\varphi}}.
\label{GRA.168}
\end{eqnarray}
Applying once more suitable spinor
derivatives and making use of the Bianchi identities yields
\begin{equation}
{\cd}^2 R+{\cdb}^2 R^{\dagger}
 \,=\,- \f {2}{3}{R_{ba}}^{ba}-\f {2}{3}{\cd}^{\al}X_{\al}
+4G^a G_a +32 RR^{\dagger}. \label{GRA.169} \end{equation}
 This
relation will turn out to be crucial for the construction of the
component field action.

\subsubsection{ The supergravity/matter/Yang-Mills system}
\label{GRA32}

Having established \ka superspace geometry as a general framework
for the coupling of supergravity to matter, it is quite natural
to include couplings to supersymmetric Yang-Mills theory as well.
In terms of superspace the basic geometric objects for this
construction are
\begin{itemize}
\item $E^A=dz^M {E_M}^A \hs{1.3cm}$        the frame of superspace,
\item ${\phi}^k, {\phib}^{\bk}\cem \hs{2.05cm}$  the chiral matter superfields,
\item ${\ca}^{(r)}=dz^M {{\ca}_M}^{(r)}\cem$  the Yang-Mills potential.
\end{itemize}As we have already
pointed out in section \ref{RS3}, Yang-Mills couplings of
supersymmetric matter are described in terms of covariantly chiral
superfields. It remains to couple the matter/Yang-Mills system as
described in section \ref{RS3} to supergravity, in combination
with the structure of \ka superspace. This is very easy. All we
have to do is to write all the equations of section \ref{RS3} in
the background of \ka superspace. This will define the underlying
geometric structure of the {\em supergravity/matter/Yang-Mills
system}\footnote{More generally, the complex manifold of chiral
matter superfields, in the sense of \ka geometry, could be
endowed with gauged isometries, compatible with supersymmetry. We
have deferred the description of the corresponding geometric
structure in superspace to appendix \ref{appC}, see also
\cite{BGG90}.}.

As to the geometry of the supergravity/matter sector, the \ka
potential is now understood to be given in terms of covariantly
chiral superfields. As a consequence, the composite $U(1)$ \ka
connection $A$, given before in (\ref{GRA.153}), becomes now
\begin{equation}
A \, = \,
\f{1}{4}K_k \, {\cd}{\phi}^k
-\f {1}{4}K_{\bk} \, {\cd}{\phib}^{\bk}
+ \f{i}{8}E^a \lp 12G_a
   +{\sib}_a^{\da\al}g_{k{\bk}}{\cd}_{\al}
    {\phi}^k{\cd}_{\da}{\phib}^{\bk} \rp,
\label{kahlcon}
\end{equation}
simply as a consequence of covariant chirality conditions, the expressions
(\ref{GRA.146a}), (\ref{GRA.146}) for the components $A_A$ being still
valid. The covariant exterior derivatives
\begin{equation}
{\cd}{\phi}^k \, = \, d {\phi}^k
- \ca^{(r)} \lp {\bf T}_{(r)} \phi\rp^k, \cem
{\cd}{\phib}^{\bk} \, = \, d {\phib}^{\bk}
+   \ca^{(r)} \lp \phib {\bf T}_{(r)} \rp^\bk,
\end{equation}
appearing here are now defined in the background of \ka superspace. The
superfields $X_\al$, $\bX^\da$, previously given in
(\ref{GRA.155}), (\ref{Gra.156}), are still identified as the field
strength components $F_{\bt a}$ resp. $F^\db{}_a$. They take now the form
\begin{eqnarray}
X_\al  &=& - \f{i}{2} \, g_{k \bk} \,
 \cd_a \phi^k \, \si^a_{\al \da} \cd^{\da} \! \phib^{\bk}
+ \f{1}{2} \, g_{k \bk} \, {\bF}^{\bk} \, \cd_\al \phi^k
-\f{1}{2} \, \cw_\al^{(r)} \, {\cal K}_{(r)}
\label{XYM}
\\[2mm]
\bX^{\da} &=& -\f{i}{2} \, g_{k \bk} \,
 \cd_a \phib^{\bk} \, \sib^{a \, \da \al} \cd_\al \phi^k
+ \f {1}{2} \, g_{k \bk} \, F^k \, \cd^\da \! \phib^{\bk}
-\f{1}{2} {\cw}^{(r) {\da}} \, {\cal K}_{(r)}.
\label{CPN.46}
\end{eqnarray}
The derivatives are covariant with respect to the Yang-Mills gauge
structure and we have defined
\begin{equation}
{\cal K}_{(r)} \, = \, K_k \lp{{\bf
T}_{(r)}}{\phi} \rp^k  + K_\bk \lp \phib{{{\bf T}}_{(r)}}\rp^{\bk}.
\end{equation}
Likewise, the \ka D-term superfield -cf. (\ref{GRA.164}),
\begin{eqnarray}
- \f {1}{2} \, \cd^\al X_\al
&=&
- g_{k \bk} \, \eta^{ab} \, \cd_a \phi^k \, \cd_b \phib^{\bk}
- \f{i}{4} \, g_{k \bk} \, \si^a_{\al \da} \,
\cd^\al \! \phi^k \, {\cd}_a \cd^{\da} \! {\phib}^{\bk}
\nonumber \\[1mm]
&&
-\f{i}{4} \, g_{k \bk} \, \si^a_{\al \da} \,
\cd^\da \! \phib^{\bk} \, {\cd}_a \cd^{\al} \! \phi^k
+ g_{k \bk} \, F^k {\bF}^{\bk}
\nonumber \\[2mm]
&&
+ \f{1}{16} \, R_{j \bj k \bk}
\, \cd^\al \phi^k \, \cd_\al \phi^j
\, \cd_{\da} \phib^{\bk} \, \cd^{\da} \phib^{\bj}
\nonumber \\[2mm]
&&
- g_{k \bk}\lp {\phib {\bf T }_{(r)}} \rp ^{\bk} \cw_\al^{(r)}
\, \cd^\al \phi^k
+ g_{k \bk} \lp {{{\bf T}_{(r)}}\phi}\rp ^k \cw_\da^{(r)}
\, \cd^{\da} \! \phib^{\bk}
\nonumber \\[2mm]
&&
+ \f{1}{4} \, \cd^\al \cw_\al^{(r)} \kil,
\label{CPN.48}
\end{eqnarray}
receives additional terms due to the Yang-Mills couplings.
Observe that
covariant derivatives refer to all symmetries, the definitions
(\ref{GRA.159}), (\ref{GRA.160}) are replaced by
\begin{eqnarray}
{\cd}_B {\cd}_\al \phi^k
&=&
{E_B}^M \partial_M {\cd}_\al \phi^k
- {\phi_{B\al}}{}^\varphi {\cd}_\varphi \phi^k
- {\ca}^{(r)}_B \lp{\bf T}_{(r)} {\cd}_\al \phi \rp^k
\nonumber \\[2mm]
& &
- A_B {\cd}_\al \phi^k +
{\Gamma^k}_{ij} {\cd}_B \phi^i{\cd}_\al \phi^j ,
\label{GRA.170}\\[2mm]
{\cd}_B {\cd}^{\da} \phib^{\bk}
&=&
{E_B}^M \partial_M {\cd}^{\da} \phib^{\bk}
- {{\phi_B}^{\da}}{}_{\dv} {\cd}^{\dv}\phib^{\bk}
+  {\ca}^{(r)}_B \lp {\cd}^{\da} {\phib}{\bf T}_{(r)} \rp^{\bk}
\nonumber \\[2mm]
& &
+ A_B {\cd}^{\da} \phib^{\bk}
+ {\Gamma^{\bk}}_{\bi\bj} {\cd}_B \phib^{\bi}{\cd}^{\da}\phib^{\bj},
\label{GRA.171}
\end{eqnarray}
with $A_B$ identified in (\ref{kahlcon}). In terms of these covariant
derivatives the superfields $F^k$
and ${\bF}^{\bk}$ are still defined as in (\ref{GRA.157}).

Based on this geometric formulation, we can now proceed to derive
supersymmetry transformations in terms of superfields, as in
appendix \ref{appC2}, and in component fields, as in section
\ref{CPN3}. Invariant actions in superspace and superfield
equations of motion are discussed below, section \ref{GRA33}, and
in appendix \ref{appD}, whereas component field actions, derived
from superspace, are given in sections \ref{CPN4} and \ref{CPN5}.

\subsubsection{ Superfield actions and equations of motion}
{\label{GRA33}

Invariant actions in superspace supergravity are obtained upon
integrating superspace densities over the commuting and
anticommuting directions of superspace. Densities, in this case,
are constructed with the help of $E$, the superdeterminant of
${E_M}^A$. As we have already alluded to above, the supergravity action
in standard superspace geometry is just the
volume of superspace.
In our present situation where both
supergravity and matter occur together in a generalized superspace
geometry, the volume element
corresponding to this superspace geometry yields the complete
kinetic actions for the supergravity/matter
system.
To be more precise, the kinetic terms for the
supergravity/matter system in our geometry are obtained from
\begin{equation}
\ASM \,=\,-3\int_\ast \! \! E,
\label{GRA.240}
\end{equation}
where the asterisk denotes integration over space-time {\em and}
superspace. The action of the kinetic terms of the Yang-Mills multiplet,
coupled to supergravity and matter, is given as
\begin{equation}
\AYM \, = \,
\f{1}{8} \int_\ast \! \frac{E}{R} f_{(r)(s)}(\phi)
\, {\cw}^{(r){\al}}{\cw}_{\al}^{(s)}   +
\f{1}{8} \int_\ast \! \frac{E}{R^{\dagger}}
{{\bar{f}}}_{(r)(s)}({\phib}) \,
{{\cw}}_{\da}^{(r)}{{\cw}}^{(s){\da}},
\label{GRA.241}
\end{equation}
whereas the superpotential coupled to supergravity is obtained from
\begin{equation}
\ASPOT \, = \, {\f{1}{2}} \int_\ast \! \frac{E}{R}\
e^{K/2} W(\phi) +  {\f{1}{2}} \int_\ast \! \frac{E}
{R^{\dagger}}e^{K/2}{\overline{W}}({\phib}).
 \label{GRA.242}
\end{equation}

\indent
Clearly, these actions are invariant under superspace coordinate
transformations, what about invariance under K\"ahler
transformations?

First of all, the superfields $R$ and
$R^\dagger$ have chiral weights $w(R)=2$ and $w(R^\dagger)=-2$,
respectively, so their K\"ahler transformations are
\begin{equation}
R \, \mapsto \, R \, e^{-2i \, \ima F}, \cem
R^{\dagger} \, \mapsto \, R^{\dagger} \, e^{+2i \, \ima F}.
\label{GRA.244}
\end{equation}
 The
Yang-Mills action is invariant provided the symmetric functions
$f_{(r)(s)}({\phi})=f_{(s)(r)}({\phi})$ and
${{\bar{f}}}_{(r)(s)}({\phib}) ={{\bar{f}}}_{(s)(r)}({\phib})$ are
inert under K\"ahler transformations. The superpotential terms are
invariant, provided the superpotential transforms as
\begin{equation}
 W(\phi) \, \mapsto \, e^{-F} \, W(\phi), \cem
{\overline{W}}({\phib})  \, \mapsto \, e^{-\bF} \,
{\overline{W}}({\phib}).
\label{GRA.245}
\end{equation}
 In this case, although neither the K\"ahler potential nor the
superpotential are tensors with respect to K\"ahler
transformations, the combinations
\begin{equation}
e^{K/2} \, W ,\cem  e^{K/2} \, {\overline{W}}, \label{GRA.247}
\end{equation}
 have perfectly well defined chiral weights, namely:
\begin{equation}
w(e^{K/2} \, W) \,=\,2,\cem  w(e^{K/2} \, {\overline{W}}) \,=\,-2.
\label{GRA.248} \end{equation}
 As to Yang-Mills symmetries, the kinetic term
of the supergravity/matter system is obviously invariant, so is
the superpotential term, by construction. The Yang-Mills term
itself is invariant provided
 \begin{eqnarray}
  i \lp {\bf
T}_{(p)} \phi \rp^k \frac{\prt}{\prt \phi^k} f_{(r)(s)}(\phi) &=&
{c_{(p)(r)}}^{(t)} f_{(t)(s)}(\phi)
                     + {c_{(p)(s)}}^{(t)} f_{(t)(r)}(\phi), \label{GRA.249} \\
- i \lp \phib {\bf T}_{(p)} \rp^\bk \frac{\prt}{\prt \phib^\bk}
  \bar{f}_{(r)(s)}(\phib) &=& {c_{(p)(r)}}^{(t)}
 \bar{f}_{(t)(s)}(\phib)
                     + {c_{(p)(s)}}^{(t)} \bar{f}_{(t)(r)}(\phib),
\label{GRA.250}
\end{eqnarray}
 that is, provided $f_{(r)(s)}(\phi)$ and
$\bar{f}_{(r)(s)}(\phib)$ transform as the symmetric product of
two adjoint representations of the Yang-Mills structure group.
\indent

We still have to justify that the superfield actions presented
above indeed correctly describe the dynamics of the
supergravity/matter system. One way to do so is to simply work out
the corresponding component field actions -  this will be done in
the next chapter. Another possibility is to derive the superfield
equations of motion - this will be done here. To begin with, the
variation of the action $\ca = \int d^4 x \cl(x)$ for the
supergravity/matter kinetic terms can be written as
\begin{equation}
\delta {\ca}_{\mathrm{Supergravity+Matter}}
\,=\,-3\int_\ast \! \! E{H_A}^A (-)^a,
\label{GRA.253}
\end{equation}
where we have defined
\begin{equation}
{H_B}^A \,=\,{E_B}^M \delta {E_M}^A. \label{GRA.254}
\end{equation}
 This is not the end of the story, however. The
vielbein variations by themselves are not suitable, because of the
presence of the torsion constraints. Solving the variational
equations of the torsion constraints allows to express the
vielbein variations in terms of unconstrained superfields and to
derive the correct superfield equations of motion \cite{WZ78a}. In
our case the matter fields must be taken into account as well.
Again, their variations themselves are not good - we have to solve
first the variational equations for the chirality constraints to
identify the unconstrained variations. Similar remarks hold for
the Yang-Mills sector. In appendix \ref{appD} a detailed
derivation of the equations of motion is presented; here we
content ourselves to state the results:

\indent

The complete action is given as
\begin{equation}
{\ca} \,=\, {\ca}_{\mathrm{Supergravity+Matter}}
+ \ca_{\mathrm{Yang-Mills}} + \ca_{\mathrm{Superpotential}}.
\label{GRA.255}
\end{equation}
The superfield equations of motion are then:

\begin{itemize}
\item  Supergravity sector
 \begin{eqnarray}
  R-\f{1}{2} e^{K/2}\
W({\phi}) &=& 0, \label{GRA.256} \\ R^{\dagger} -\f{1}{2} e^{K/2}\
{\overline{W}}({\phib})&=&0, \label{GRA.257} \end{eqnarray}
\begin{equation}
G_b +\f{1}{8}{\sib}_b^{\da \al} g_{k {\bk}}{\cd}_{\al} {\phi}^k
{\cd}_{\da}{\phib}^{\bk} -\f{1}{8} {\sib}_b^{\da \al} \left(
f+{\bar{f}} \right)_{(r)(s)} {\cw}_{\al}^{(r)}\ {\cw}_{\da}^{(s)}
\,=\,0. \label{GRA.258} \end{equation}
\item Yang-Mills sector
 \begin{eqnarray}
  &&\f{1}{2} f_{(r)(s)}({\phi}) {\cd}^{\al}
{\cw}_{\al}^{(s)} -\f{1}{2} \frac{\prt f_{(r)(s)}}{\prt {\phi}^k}
{\cd}_{\al} {\phi}^k {\cw}^{(s) \al } \nn \\  &+& \f{1}{2} \kil+
h.c. \,=\,0, \label{GRA.259} \end{eqnarray}
\item Matter sector
\begin{equation}
g_{k{\bk}}{\bF}^{\bk}+\frac{1}{4} \frac{\prt f_{(r)(s)}}
{\prt{\phi}^k} {\cw}^{{(r)} \al} {\cw}^{ (s)}_{\al} +e^{K/2}W
\frac{\prt}{\prt{\phi}^k} log \left(e^K W \right) \,=\,0,
\label{GRA.260} \end{equation}
\begin{equation}
g_{k {\bk}}{F}^k+\frac{1}{4}\frac{\prt {\bar{f}}_{(r)(s)}}{\prt
{\phib}^{\bk}} {\cw}^{(r)}_{\da}  {\cw}^{(s) \da}
+e^{K/2}{\overline{W}}\frac{\prt}{\prt {\phib}^{\bk}} log
\left(e^K{\bar{W}} \right) \,=\,0. \label{GRA.261} \end{equation}
\end{itemize}
\indent

The lowest components in the
superfield expansion provide the algebraic equations for the auxiliary
fields. The equations of motion of all the other component fields
of the supergravity/matter system are contained at higher orders
in the superfield expansion. They are most easily obtained by
suitably applying spinor derivatives and projecting afterwards to
lowest superfield components.

\newpage
\sect{ COMPONENT FIELD FORMALISM} \label{CPN}

The superspace approach presented in the previous section provides
a concise and coherent framework for the component field
construction of the general supergravity/matter/\-Yang-Mills
system. Supersymmetry and \ka transformations of the component
fields derive directly from the geometric structure, the
corresponding invariant component field action has a canonically
normalized curvature scalar term, without any need of component field
Weyl rescalings. This should be contrasted with the original
component field approach \cite{CJS+78b,CJS+79,CFGvP82,CFGvP83},
where normalization of the action and invariance under \ka phase
transformations appeared only after a Weyl rescaling of the
component fields or, equivalently, a conformal gauge fixing
\cite{KU83a,KU83b}.

Anticipating on our results, we will see that the supergravity/matter
Lagrangian (\ref{GRA.240}), when projected to component fields, exhibits
the kinetic Lagrangian density of the matter sector as a
Fayet-Iliopoulos D-term, \ie it has the decomposition
\begin{equation}
{\cl}_{\mathrm{supergravity+matter}} \, = \,
{\cl}_{\mathrm{supergravity}} + e \, \DM. \nn \label{CPN.1}
\end{equation}
Here $e$ denotes the usual vierbein determinant $e = {\rm det}({e_m}^a)$
and $\DM$ is the D-term pertaining to the Abelian
\ka gauge structure of the previous section. More precisely, the
component field D-term derived from \ka superspace has the form
\begin{equation}
\DM \,=\, -  \f {1}{2} {\cal D}^\al X_\al {\loco} + \f{i}{2}
{\psi_m}^\al \si^m_{\al \da} \bX^{\da} {\loco} + \f{i}{2}
\psib_{m \,\da} \sib^{m \, \da \al} X_\al {\loco}, \label{CPN.2}
\nn \end{equation} where the vertical bars denote projections to
lowest superfield components of the superfields given
respectively in
(\ref{CPN.48}), (\ref{XYM}), (\ref{CPN.46}).
Recall that a D-term in global supersymmetry may be understood as the
lowest component of the superfield $D^\al X_\al$ with
\begin{equation}
X_\al \, = \,  - \f{1}{8} \bar{D}^2 D_\al K(\phi, \phib).
\label{CPN.3} \nn
\end{equation}
In this sense the \ka superspace construction is
the natural generalization of Zumino's construction \cite{Zum79b}
of supersymmetric sigma models.

In subsection \ref{CPN1} we identify component fields and provide a
method to derive super-covariant component field strength and space-time
derivatives. In  subsection \ref{CPN2} we discuss some more of the basic
building blocks useful for the component field formulation, in particular
for the geometric derivation of supersymmetry transformations of all the
component fields, which are given explicitly in subsection \ref{CPN3}, and
the component field actions, constructed in subsections \ref{CPN4} and
\ref{CPN5}.

\subsec{ Definition of Component Fields} \label{CPN1}

As explained already in section \ref{RS}, component fields are obtained as
projections to lowest components of superfields. A supermultiplet is
defined through successive application of covariant spinor derivatives and
subsequent projection to lowest components, as for instance for the chiral
multiplet in section \ref{RS23}. Defined in this manner the component fields
are related in a natural way by Wess-Zumino transformations. The structure
of a supersymmetric theory, in particular the construction of invariant
actions, as in section \ref{RS24}, is then completely determined by the
algebra of covariant derivatives. This approach avoids cumbersome
expansions in the anticommuting variables and provides a geometric
realization of the Wess-Zumino gauge. It is of particular importance in
the case of the component field formalism for supergravity, as will be
pointed out here.

In a first step we are going to identify the vierbein and the
Rarita-Schwinger fields. They appear as the $dx^m$ coefficients of the
differential form $E^A = dz^M E_M{}^A$.
It is therefore convenient to define systematically an operation which
projects at the same time on the $dx^m$ coefficients and on lowest
superfield components, called the
{\em double bar projection} \cite{BBG87}. To be more
precise, we define
\begin{equation}
E^a \doubar \, = \, e^a \, = \, dx^m e_m{}^a(x),
\label{CPN.5a}
\end{equation}
\begin{equation}
E^\alpha \doubar \, = \, e^\alpha \, = \, \f {1}{2} dx^m
\psi_m{}^\al(x), \cem E_{\dot \alpha} \doubar \, = \, e_{\dot
\alpha} \, = \, \f {1}{2} dx^m \psib_m{}_{\da}(x). \label{CPN.5b}
\end{equation}
This identifies the vierbein field $e_m{}^a(x)$ and thereby the usual
metric tensor
\begin{equation}
 g_{mn} \,=\,{e_m}^a {e_n}^b {\eta}_{ab},
\label{CPN.4}
\end{equation}
as well as the gravitino field ${\psi_m}^\al$, $\psib_{m \da}$,
which is at the same time a vector and a Majorana spinor. The factors
$1/2$ are included for later convenience in the construction of the
Rarita-Schwinger action.

The definition of component fields as lowest superfield components defines
unambiguously their chiral $U_K(1)$ weights due to the geometric
construction of the previous section. As a consequence, the vierbein has
vanishing weight whereas the Rarita-Schwinger field is assigned
chiral weights
\begin{equation}
w(\psi_m{}^\al) \, = \, +1, \cem
w(\psib_m{}_{\da}) \, = \, -1.
\label{RSweights}
\end{equation}
The remaining component fields are defined as
\begin{equation}
R{\loco} \,=\,- \f{1}{6} M, \cem
R^{\dagger}{\loco} \,=\,- \f{1}{6} \ovM, \cem
G_a{\loco} \,=\,- \f{1}{3} b_a,
\label{CPN.7}
\end{equation}
with chiral $U_K(1)$ weights
\begin{equation}
w(M) \, = \, +2, \cem w(\ovM) \, = \, -2, \cem w(b_a) \, = \, 0.
\end{equation}

The vierbein and Rarita-Schwinger fields together with $M$, $\ovM$
and $b_a$ are the components of the supergravity
sector, $M$, $\ovM$ and $b_a$ will turn out to describe non-propagating, or
auxiliary fields.

Supergravity in terms of component fields is quite complex. However, when
derived from superspace geometry a number of elementary building blocks
arise in a natural way, allowing to gather complicated expressions
involving the basic component fields and their derivatives in a compact
and concise way.

As a first example we consider the spin connection. In ordinary
gravity with vanishing torsion, the spin connection is given in
terms of the vierbein and its derivatives. In the supergravity
case it acquires additional contributions, as we explain now.  To
begin with, consider the torsion component $T^a = dE^a + E^b
{\phi_b}^a$, which is a superspace 2-form. The component field
spin connection is identified upon applying the double bar
projection to ${\phi_b}^a$,
\begin{equation}
{\phi_b}^a {\doubar} \, = \, \om_b{}^a \, = \, dx^m {\om_{m\, b}}^a(x).
\label{CPN.8}
\end{equation}
Defining
\begin{equation}
{\phi_\beta}^\al \doubar \, = \, {\om_{\beta}}^\al
     \, = \, dx^m {\om_{m \, \beta}}^\al(x), \cem
{{\phi}^{\db}}_{\da} \doubar \, = \, {{\om}^{\db}}_{\da}
\, = \, dx^m {{\om_m}^{\db}}_{\da}(x),
\end{equation}
for the spinor components, (\ref{spincom}) gives rise to the usual relations
\begin{equation}
{\om_{m \beta}}^\al \, = \, -
\f {1}{2} {(\si^{ba})_\beta}^\al \, \om_{m\, ba}, \cem
 {{\om_m}^\db}_{\da}  \, = \, - \f {1}{2}
{({\sib}^{ba})^\db}_{\da} \, \om_{m\, ba}. \label{CPN.20}
\end{equation}
Then, applying the double bar projection to the full torsion yields
\begin{equation}
T^a \doubar \, = \, \f{1}{2} dx^m dx^n \, {T_{nm}}^a {\loco}
\, = \, de^a + e^b \om_b{}^a \, = \, D e^a.
\end{equation}
In this expression the exterior derivative is purely space-time. Using
moreover
\begin{equation}
{T_{nm}}^a {\loco} \,=\,\cd_n {e_m}^a - \cd_m {e_n}^a,
\label{CPN.11}
\end{equation}
the component field covariant derivative of the vierbein is identified
as
\begin{equation}
\cd_n {e_m}^a \,=\,\prt_n {e_m}^a + {e_m}^b {\om_{n\, b}}^a .
\label{CPN.12}
\end{equation}
Seemingly this is the same expression as in ordinary gravity, so
how does supersymmetry modify it? To this end, we note that the
double bar projection can be employed in an alternative way, in
terms of the covariant component field differentials $e^A$
defined above. Taking into account the torsion constraints, in
particular ${T_{cb}}^a = 0$, this reads simply
\begin{equation}
T^a \doubar \, = \, e_\db \, e^\gm \, T_\gm{}^\db {}^a {\loco},
\end{equation}
where only the constant torsion coefficient
${T_\gm}^{\db a} = -2i {(\si^a \eps)_\gm}^{\db}$
survives.
Combining the two alternative expressions for $T^a \doubar$
gives rise to
\begin{equation}
\cd_n {e_m}^a - \cd_m {e_n}^a \,=\,\f{i}{2} (\psi_n \si^a \psib_m - \psi_m
\si^a \psib_n). \label{CPN.13} \end{equation}
In view of the explicit form of the covariant derivatives, it is a
matter of straightforward algebraic manipulations to arrive at
$(\si_m = {e_m}^a \si_a)$
\begin{eqnarray}
\om_{mnp}  \,=\, e_{pa} {e_n}^b {\om_{mb}}^a \ &=& \ \f {1}{2}
({e_m}^a \prt_n e_{pa} - {e_p}^a \prt_m e_{na} - {e_n}^a \prt_p \,
e_{ma})
\nonumber \\ &&
- \f {1}{2} ({e_m}^a  \prt_p e_{na} - {e_n}^a \prt_m {e_p}^a
- {e_p}^a \prt_n e_{ma})
\nonumber \\ &&
+ \f{i}{4} (\psi_p \, \si_m \psib_n - \psi_m \si_n \psib_p -
\psi_n\si_p \psib_m)
\nonumber \\ &&
- \f{i}{4} (\psi_n \si_m \psib_p - \psi_m \si_p \psib_n - \psi_p
\, \si_n \psib_m). \label{CPN.15}
\end{eqnarray}
This shows how $\om_{m \, b}{}^a$ is expressed in terms of the
vierbein, its derivatives and, in the supersymmetric case, with
additional terms quadratic in the gravitino (Rarita-Schwinger)
field.

The Rarita-Schwinger component field strength is given terms of the
covariant derivative of the gravitino field. As a consequence
of the non-vanishing chiral $U_K(1)$ weights (\ref{RSweights}),
contributions from the matter sector arise due to the presence
of the component
\begin{equation}
A \doubar \, = \, dx^m \, A_m(x),
\label{CPN.21}
\end{equation}
of the $U_K(1)$ gauge potential. In order to work out the explicit form
of $A_m(x)$, the double bar projection must be applied to the
superspace 1-form
\begin{equation}
A \, = \,
\f{1}{4}K_k \, {\cd}{\phi}^k
-\f {1}{4}K_{\bk} \, {\cd}{\phib}^{\bk}
+ \f{i}{8}E^a \lp 12G_a
   +{\sib}_a^{\da\al}g_{k{\bk}}{\cd}_{\al}
    {\phi}^k{\cd}_{\da}{\phib}^{\bk} \rp,
\end{equation}
as given in (\ref{kahlcon}). This in turn means that we
need to define first matter and Yang-Mills component fields and their
covariant derivatives. Recall that
the exterior Yang-Mills covariant derivatives are defined as
\begin{equation}
{\cd}{\phi}^k \, = \, d {\phi}^k
- \ca^{(r)} \lp {\bf T}_{(r)} \phi\rp^k, \cem
{\cd}{\phib}^{\bk} \, = \, d {\phib}^{\bk}
+   \ca^{(r)} \lp \phib {\bf T}_{(r)} \rp^\bk.
\end{equation}
This shows that, for the definition of the component field \ka
connection $A_m$, we need at the same time the component fields
for the matter and Yang-Mills sectors. The components of chiral,
resp. antichiral superfields $\phi^k$, resp. $\phib^{\bk}$ are
defined as
 \begin{eqnarray}
\phi^k {\loco} & = & A^k, \cem
\cd_\al \phi^k{\loco} \,=\,\sqrt{2} \chi_\al^k, \cem
\cd^\al \cd_\al \phi^k {\loco} \,=\,-4F^k,
\label{CPN.28}
\\[3mm]
\phib^{\bk} {\loco} & = & \bA^{\bk}, \cem
\cd_\da
\phib^{\bk}{\loco} \,=\,\sqrt{2} {\chib_\da}^{\bk}, \cem
\cd_\da \cd^\da \phib^{\bk} {\loco} \,=\, -4{\bF}^{\bk},
\label{CPN.29}
\end{eqnarray}
with indices $k$, $\bk$ referring to the K\"ahler manifold (not to be
confused with space-time indices).
As to the Yang-Mills potential we define
\begin{equation}
\ca \doubar \, = \, i {\aym} \, = \, i dx^m {\aym}_m,
\end{equation}
whereas the remaining covariant components of the Yang-Mills multiplet are
defined as
\begin{equation}
{\cw}^{\db} {\loco}\,=\,i \lab^{\db}, \cem
{\cw}_\beta {\loco}\,=\,- i\la_\beta, \cem
\cd^\al {\cw}_\al {\loco}\,=\,-2 {\bf D}.
\label{CPN.30}
\end{equation}
Recall that all the components of this multiplet are Lie algebra
valued, corresponding to their identification in $\ca=\ca^{(r)}
{\bf T}_{(r)}$ and $\cf=\cf^{(r)} {\bf T}_{(r)}$. We can now apply
the double bar projection to $A$ and identify
$A \doubar \, = \, dx^m \, A_m(x)$,
where, for reasons of
notational economy, the same symbol $A_m$ for the
superfield and its lowest component, \ie $A_m(x) = A_m {\loco}$,
is used. We obtain the explicit component
field form by the double bar projection of the covariant exterior
derivatives of the matter superfields, \ie
\[
D \phi^k \doubar \, = \,
dx^m \lp \prt_m A^k - i {\aym}_m^{(r)} \lp {\bf T}_{(r)} A \rp^k
\rp, \cem
D \phib^\bk \doubar \, = \,
dx^m \lp \prt_m {\bar A}^\bk
+ i {\aym}_m^{(r)} \lp \bA {\bf T}_{(r)} \rp^\bk \rp,
\]
suggesting the definitions
\begin{equation}
\cd_m A^k \, = \, \prt_m A^k
- i {\aym}_m^{(r)} \lp {\bf T}_{(r)} A \rp^k, \cem
\cd_m {\bar A}^\bk \, = \,
      \prt_m {\bar A}^\bk
+ i {\aym}_m^{(r)} \lp \bA {\bf T}_{(r) }\rp^\bk.
\label{CPN.31}
\end{equation}
It is then straightforward to read off the explicit component
field expression
\begin{equation}
{A}_m + \f{i}{2} {e_m}^a b_a
\,=\,\f{1}{4} K_k \, \cd_m A^k -
\f{1}{4} K_{\bk} \, \cd_m A^{\bk}
+ \f{i}{4} g_{k {\bk}} \, \chi^k \si_m \chib^{\bk},
\label{CPN.40}
\end{equation}
this field dependent \ka connection will show up in any covariant
derivative acting on components with non-vanishing $U_K(1)$
weights. The spinor components of the \ka connection are field
dependent as well, they are given as -cf. (\ref{GRA.146})-
\begin{equation}
{A}_\al {\loco} \,=\,\frac{1}{2 \sqrt{2}} \, K_k \, \chi^k_\al ,
\cem
{A}_ \da {\loco} \,=\,
      - \frac{1}{2 \sqrt{2}} \, K_{\bk} \, \chib^{\bk}_{\da}.
\label{CPN.44}
\end{equation}
These terms will appear explicitly in various places of component field
expressions later on as well.

We can now turn to the construction of the
super-covariant component field strength
${T_{cb}}^{\undal}{\loco}$ for the gravitino.
The relevant superspace
2-forms are $T^\al = dE^\al + E^\beta {\phi_\beta}^\al + E^\al \, {A}$
and its conjugate $T_\da$.
The double bar projection of the field strength itself is then
$(\undal = \al, \da)$
\begin{equation}
T^\undal \doubar \, = \, \f{1}{2} dx^m dx^n \, {T_{nm}}^\undal {\loco},
\end{equation}
where
\begin{equation}
{T_{nm}}^{\undal}{\loco} \,=\,\f {1}{2} (\cd_n {\psi_m}^{\undal} -
\cd_m {\psi_n}^{\undal}), \cem  \label{CPN.16}
\end{equation}
contains the covariant derivatives
\begin{eqnarray}
\cd_n {\psi_m}^\al & = &
\prt_n {\psi_m}^\al + {\psi_m}^\beta {\om_{n \beta}}^\al
+ {\psi_m}^\al \, {A}_n,
\label{CPN.17} \\
\cd_n {\psib_{m \da}} & = &
\prt_n {\psib_{m \da}} + {\psib_{m \db}} {{\om_{n}}^{\db}}_{\da}
- \psib_{m \da} \, {A}_n.
\label{CPN.18}
\end{eqnarray}
On the other hand, we employ the double bar projection in terms of
the covariant differentials,
\begin{equation}
T^\al \doubar \, = \, \f{1}{2} e^b e^c \, T_{cb}{}^\al \loco
                        + e^b e^\gm \, T_{\gm b}{}^\al \loco
                        + e^b e_\dg \, T^\dg{}_b{}^\al \loco,
\end{equation}
and similarly for $T_\da$. Using the explicit form of the torsion
coefficients appearing here, and comparing the two alternative forms of
$T^\undal \doubar$ gives rise to the component field expressions
\begin{eqnarray}
{T_{cb}}^\al {\loco}
& = &
\f {1}{2} {e_b}^m {e_c}^n ( {\cd}_n {\psi_m}^\al
- \cd_m {\psi_n}^\al)
\nonumber \\ &&
+ \f{i}{12} ({e_c}^m \psi_m \si_a \sib_b
- {e_b}^m \psi_m \si_a \sib_c)^\al \, b^a
\nonumber \\ &&
- \f{i}{12} ({e_c}^m \psib_m \sib_b
- {e_b}^m \psib_m \sib_c)^\al M,
\label{CPN.23}
\end{eqnarray}
and
\begin{eqnarray}
T_{cb \da} {\loco}
& = &
\f {1}{2} {e_b}^m {e_c}^n ( {\cd}_n {\psib_{m \da}}
- \cd_m {\psib_{n \da}})
\nonumber \\ &&
- \f{i}{12} ({e_c}^m \psib_m \sib_a \si_b -
{e_b}^m \psib_m \sib_a \si_c)_{\da} \, b^a
\nonumber \\ &&
- \f{i}{12} ({e_c}^m \psi_m \si_b - {e_b}^m \psi_m \si_c)_{\da} \, \ovM,
\label{CPN.24}
\end{eqnarray}
for the super-covariant gravitino field strength. The contributions of
the matter and Yang-Mills sector are hidden in the covariant derivatives
through the definitions given above.

Yet another important object in the component field formulation is
the super-covariant version of the curvature scalar, identified as
${R_{ab}}^{ab}{\loco}$. We use the same method as before for its
evaluation; the relevant superspace quantity is the curvature
2-form
\begin{equation}
R_b{}^a \, = \, d \phi_b{}^a + \phi_b{}^c \phi_c{}^a.
\end{equation}
The double bar projection yields
\begin{equation}
R_b{}^a \doubar \, = \, \f{1}{2} dx^m dx^n \, R_{nm \, b}{}^a \loco,
\end{equation}
where $R_{nm \, b}{}^a \loco$ is given in terms of ${\om_{m \, b}}^a$.
Note that, in distinction to ordinary gravity, the explicit form of
${\om_{m \, b}}^a$, given above in (\ref{CPN.15}) contains quadratic
gravitino terms, which will give rise to complicated additional
contributions in $R_{nm \, b}{}^a \loco$. Fortunately enough, in the
present formulation, the projection technique takes care of these
complications automatically in a concise way. As to the curvature scalar,
we use the notation
\begin{equation}
{\cal R}(x) \, = \, {e_a}^n {e_b}^m {R_{nm}}^{ab}{\loco}.
\label{CPN.26}
\end{equation}
The relation between ${R_{ab}}^{ab}{\loco}$ and ${\cal R}(x)$ is once
more obtained after employing the double bar projection in terms of
covariant differentials, \ie
\begin{equation}
R_b{}^a \doubar \, = \, \f{1}{2} e^c e^d R_{dc \, b}{}^a \loco
                        + e^c e^{\unddt} R_{\unddt c \, b}{}^a \loco
        + \f{1}{2} e^\undgm e^{\unddt} R_{\unddt \undgm \, b}{}^a \loco,
\end{equation}
Although our formalism is quite compact it requires still some
algebra (the values of the curvature tensor components present on
the right-hand side can be found in appendix \ref{appB3}.) to
arrive at the result
\begin{eqnarray}
{R_{ab}}^{ab}{\loco} &=& {\cal R} + 2i {e_b}^m (\psi_m \si_a
\eps)^{\dv} {T^{ab}}_{\dv}{\loco} +2i {e_b}^m (\si_a
\psib_m)_\varphi T^{ab\varphi}{\loco}
              \nonumber \\[2mm]
&& - \f{1}{3} \ovM \psi_m \si^{mn} \psi_n - \f{1}{3} M \psib_m
\sib^{mn} \psib_n - \f{i}{3} \vep^{klmn}b_k \psi_l\si_m\psib_n.
\label{CPN.27}
\end{eqnarray}
Observe that this simple looking expression hides quite a number
of complicated terms, in particular Rarita-Schwinger fields up to
fourth order as well as contributions from the matter and
Yang-Mills sectors.

\indent

Fully covariant derivatives for the components of the chiral superfields
(to make things clear we write
the spin term, the $U_K(1)$ term, the Yang-Mills term and the one
with K\"ahler Levi-Civita symbol - in this order) are defined as
\begin{eqnarray}
\cd_m \chi_\al^k &  =  &
\prt_m \chi_\al^k - {\om_{m \al}}^\varphi {\chi}_{\varphi}^k
-{A}_m \chi_\al^k
\nn \\[2mm] && \cem
-i \aym_m^{(r)} \lp {\bf T}_{(r)} \chi_\al \rp^k
+\chi^i_\al\, {\Gamma^k}_{ij} \, {\cal D}_m A^j,
\label{CPN.33}
\\[2mm]
\cd_m \chib^{\da \bk} & = &
\prt_m \chib^{\da \bk} - {{\om_{m}}^\da}_{\dv} \chib^{\dv \bk}
+ {A}_m \chib^{\da \bk}
\nn \\[2mm] && \cem
+i  {\aym}_m^{(r)} \lp \chib^\da {\bf T}_{(r)}\rp^\bk
+  \chib^{\da\bi} \ {\Gamma^\bk}_{\bi \bj} \, {\cal D}_m
{\bA}^{\bj}.
\label{CPN.34}
\end{eqnarray}
In the Yang-Mills sector we apply the double bar projection to
the field strength
\begin{equation}
\cf \,=\,d{\ca} + {\ca}{\ca} \,=\,\f {1}{2} E^A E^B \cf_{BA}.
\label{CPN.35} \end{equation} Taking into account coefficients
\begin{equation}
\cf_{\beta a} {\loco} \, = \, - (\si_a \lab)_{\beta} , \cem
{\cf^\db}_a {\loco} = - (\sib_a \la)^{\db}, \label{CPN.36}
\end{equation}
given in terms of the gaugino field, we establish the expression
\begin{eqnarray}
\cf_{b a} {\loco} & = & i {e_b}^n {e_a}^m (\prt_n {\aym}_m -
\prt_m {\aym}_n - i[{\aym}_n, {\aym}_m])
\nonumber \\[2mm] &&
+ \f{1}{2} {e_b}^n (\psi_n \si_a \lab)
- \f {1}{2} {e_a}^n (\psi_n \si_b \lab)
\nonumber \\[2mm] &&
+ \f {1}{2} {e_b}^n (\psib_n \sib_a \la)
- \f {1}{2} {e_a}^n (\psib_n \sib_b \la),
\label{CPN.37}
\end{eqnarray}
for the super-covariant field strength.
The covariant derivatives of the gaugino field read
\begin{eqnarray}
\cd_m \la_\al \
&=&
\partial_m \la_\al
-{\om_{m\al}}^\varphi \la_\varphi
+i [{\aym}_m , \la_\al] + {A}_m \, \la_\al ,
\label{CPN.38} \\[2mm]
\cd_m \lab^\da
&=&
\partial_m \lab^\da -{\om_m}^{\da}{}_{\dv} \lab^\dv
+i [{\aym}_m , \lab^\da] - {A}_m \, \lab^\da.
\label{CPN.39}
\end{eqnarray}


\subsec{ Some Basic Building Blocks \label{CPN2}}

We indicated above that one of the necessary tasks to obtain the
Lagrangian is to derive the components of the chiral superfields
${X}_\al,\bX^{ \da}$. Their superfield explicit form was already
derived -cf. (\ref{XYM}), (\ref{CPN.46})- but for the
 sake of simplicity, we give them here again,
\begin{eqnarray}
 {{X}}_\al  &=& - \f{i}{2} g_{i \bj}
\cd_{\al \dv} \phi^i \cd^{\dv} \phib^{\bj} + \f{1}{2}
{\bF}^{\bj}g_{i \bj} \cd_\al \phi^i -\f{1}{2} \cw_\al^{(r)}\kil
\nn \\[2mm]
\bX^{\da} &=&
-\f{i}{2} g_{i \bj}
\cd^{\varphi \da} \phib^{\bj} \cd_{\varphi} \phi^i
+ \f {1}{2} F^i g_{i \bj} \cd^\da \phib^{\bj}
-\f{1}{2} {\cw}^{(r) {\da}} \kil.
\nn
\end{eqnarray}
 One infers -cf. (\ref{CPN.48})-
  \begin{eqnarray}
 - \f {1}{2} \, \cd^\al {X}_\al
&=&
- g_{i \bj} \, \eta^{ab} \, \cd_a \phi^i \, \cd_b \phib^{\bj}
- \f{i}{4} g_{i \bj} \, \si^a_{\al \da} \,
\cd^\al \! \phi^i \, {\cd}_a \cd^{\da} \! {\phib}^{\bj}
\nonumber \\[1mm]
&&
-\f{i}{4} g_{i \bj} \, \si^a_{\al \da} \,
\cd^\da \! \phib^{\bj} \, {\cd}_a \cd^{\al} \! \phi^i
+ g_{i \bj} \, F^i {\bF}^{\bj}
\nonumber \\[2mm]
&&
+ \f{1}{16} R_{j \bj k \bk}
\, \cd^\al \phi^k \, \cd_\al \phi^j
\, \cd_{\da} \phib^{\bk} \, \cd^{\da} \phib^{\bj}
\nonumber \\[2mm]
&&
- g_{i \bj}\lp { \phib{\bf T }_{(r)}} \rp ^{\bj} {\cw_\al}^{(r)}
\, \cd^\al \phi^i
+ g_{i \bj} \lp {{{\Bbb T}_{(r)}}\phi}\rp ^i {\cw_{\da}}^{(r)}
\, \cd^{\da} \! \phib^{\bj}
\nonumber \\[2mm]
&&
+ \f {1}{4} \cd^\al {\cw_\al}^{(r)} \kil,
\label{CPN.48double}
\end{eqnarray}
where
\begin{equation}
R_{i \bi j \bj} \,=\,\prt_i \prt_{\bi} g_{j \bj} - g^{k \bk} g_{i
\bk, j} \  g_{k \bi, \bj} \ =\ \prt_i \prt_{\bi} g_{j \bj} -
{\Gamma^k}_{ij} \  g_{k \bi, \bj}. \label{CPN.49} \end{equation}
We see that the main effort is to obtain the component field
expressions of super-covariant derivatives. Special
attention should be paid to the super-covariant derivatives with
respect to Lorentz indices. As an example, we detail the
computation of $\cd_a \phi^i {\loco}$. The starting point is the
superspace exterior derivative $\cd \phi^i$, whose double bar
projection reads
\begin{equation}
\cd \phi^i \doubar \, = \, dx^m \cd_m A^i(x).
\end{equation}
On the other hand, in terms of covariant differentials and due to
the chirality of $\phi^i$, we have
\begin{equation}
\cd \phi^i \doubar \, = \, e^a \cd_a \phi^i \loco
     + \sqrt{2}e^\al \chi_\al.
\end{equation}
Combination of these two equations gives immediately
\begin{equation}
\cd_a \phi^i {\loco} \, = \,
     {e_a}^m (\cd_m A^i - \frac{1}{\sqrt{2}} {\psi_m}^\al \chi^i_\al ).
 \label{CPN.50}
\end{equation}
Similarly,
\begin{equation}
\cd_a \phib^{\bj} {\loco} \,=\,{e_a}^m  (\cd_m \bA^{\bj} -
\frac{1}{\sqrt{2}} \psib_{m \da} \chib^{\bj \da}). \label{CPN.51}
\end{equation}
The lowest components of the superfields ${{X}}_\al,
\bX^{\da}$ are then obtained as
\begin{eqnarray}
{{X}}_\al {\loco} &=& -\frac{i}{\sqrt{2}}\
g_{k\bk}\si^m_{\al\da} {\chib}^{\bk\da}
(\cd_mA^k-\frac{1}{\sqrt{2}}\psi_m^\beta\chi_\beta^k)
               \nonumber \\
&& +\frac{1}{\sqrt{2}}g_{k\bk}\chi^k_\al{\bF}^{\bk}
+\f{i}{2}\la^{(r)}_\al \kilc, \label{CPN.52} \\
\bX^{\da} {\loco} &=& -\frac{i}{\sqrt{2}}\
g_{k\bk}\sib^{m\da\al} \chi^k_\al
(\cd_m\bA^{\bk}-\frac{1}{\sqrt{2}} \psib_{m\db}\chib^{\bk\db})
\nonumber \\ && +\frac{1}{\sqrt{2}}g_{k\bk}\chib^{\bk\da}F^k
-\f{i}{2}\lab^{(r)\da}\kilc. \label{CPN.53}
\end{eqnarray}

As to $-\f {1}{2} \cd^\al {X}_\al{\loco}$, we infer that the
first term in (\ref{CPN.48double}) reads
\begin{eqnarray}
 - g_{i \bj}
\eta^{ab} \cd_a \phi^i \cd_b \phib^{\bj} {\loco} &=& - g_{i \bj}
g^{mn} \cd_m A^i \cd_n \bA^{\bj} \nonumber \\ && +
\frac{1}{\sqrt{2}} g_{i \bj} g^{mn} \cd_m A^i \psib_{n\da}
\chib^{\bj \da}+ \frac{1}{\sqrt{2}} g_{i \bj} g^{mn} \cd_m
\bA^{\bj} {\psi_n}^\al {\chi^i}_\al \nonumber  \\ && -  \f {1}{2}
g_{i \bj} g^{mn} {\psi_m}^\al {\chi^i}_\al \psib_{n \da}\chib^{\bj
\da} . \label{CPN.54} \end{eqnarray}

We see that this term provides the kinetic term for the scalar
components of the (anti)chiral matter supermultiplets (as
promised, $\DM$ contains all the derivative
interactions of such fields). Likewise,
\begin{equation}
{\cd}_a \cd^{\da} \phib^{\bj} {\loco} \, = \,
{e_a}^m  \left[
\sqrt{2} {\cd}_m \chib^{\da \bj} - {\psib_m}^{\da} {\bF}^{\bj}
- i (\psi_m \si^n \eps)^{\da} (\cd_n \bA^{\bj}
- \frac{1}{\sqrt{2}} \psib_{n \dv} \chib^{\dv \bj})
\right],
\label{CPN.56a}
\end{equation}
\begin{equation}
{\cd}_a \cd^{\al} \phi^i {\loco} \, = \,
{e_a}^m \left[
\sqrt{2} {\cd}_m {\chi}^{\al i}
- {\psi_m}^{\al} F^i + i (\psib_m \sib^n)^{\al} (\cd_n A^i
- \frac{1}{\sqrt{2}} {\psi_n}^{\varphi} \chi_{\varphi}^i)
\right].
\label{CPN.56}
\end{equation}
Hence the second term in (\ref{CPN.48double}) yields
\begin{eqnarray}
& - & \f{i}{4} g_{i \bj} \si^a_{\al \da} \cd^\al \phi^i {\cd}_a
\cd^{\da} \phib^{\bj} {\loco}  - \f{i}{4} g_{i \bj} \si^a_{\al
\da} \cd^{\da} \phib^{\bj} {\cd}_a \cd^\al \phi^i {\loco}
\nonumber
\\ & = & - \f{i}{2} \chi^{\al i} g_{i \bj} \si^m_{\al
\da} {\cd}_m \chi^{\da \bj} + \f{i}{2} ({\cd}_m \chi^{\al i})
\si^m_ {\al \da} g_{i \bj} \chib^{\da \bj} \nonumber \\ && +
\frac{i}{2 \sqrt{2}} (\chi^i \si^m \psib_m) g_{i \bj} {\bF}^{\bj}
- \frac{i}{ 2 \sqrt{2}} (\psi_m \si^m \chib^{\bj}) g_{i \bj} F^i
\nonumber \\ && - \frac{1}{2 \sqrt{2}} (\psi_m \si^n \sib^m
\chi^i) g_{i \bj} (\cd_n \bA^{\bj} - \frac{1}{ \sqrt{2}} \psib_{n
\dv} \chib^{\dv \bj}) \nonumber \\ && - \frac{1}{ 2 \sqrt{2}}
(\psib_m \sib^n \si^m \chib^{\bj}) g_{i \bj} (\cd_n A^i -
\frac{1}{ \sqrt{2}} {\psi_{n}}^{\varphi} \chi^i_\varphi).
\label{CPN.57}
\end{eqnarray}
 We stress the
presence of the kinetic term for the fermionic component of the
matter supermultiplet.

Altogether we obtain from (\ref{CPN.48double})
 \begin{eqnarray}
 - \f
{1}{2} \cd^\al {X}_\al {\loco} & = & - g_{i \bj} g^{mn} \cd_m
A^i \cd_n \bA^{\bj} - \f{i}{2} \chi^{\al i} g_{i \bj} \si^m_{\al
\da} {\cd}_m \chib^{\bj \da} \nonumber
\\ && + \f{i}{2}({\cd}_m \chi^{\al i}) \si^m_{\al \da}
g_{i \bj} \chib^{\da \bj}  + g_{i \bj} {F}^i {{\bF}}^{\bj} + \f
{1}{2} g_{i \bj} g^{mn} (\psi_m \chi^i)(\psib_n \chib^{\bj})
\nonumber \\ && + \f{1}{4} R_{i \bi j \bj} (\chi^i \chi^j)
(\chib^{\bi} \chib^{\bj}) -i \sqrt{2} (\chi^i \la^{(r)}) g_{i
\bj}\lp { \bA{\bf T }_{(r)}} \rp ^{\bj} +i \sqrt{2} (\chib^{\bj}
{\lab}^{(r)})g_{i \bj}\lp {{{\bf T}_{(r)}} A}\rp ^i \nonumber \\
&& -\f{1}{2} {\bf D}^{(r)}\kilc \nn \\ & & - \frac{i}{ 2\sqrt{2}}
(\psib_m \sib^m \chi^i) g_{i \bj} {{\bF}}^{\bj} - \frac{i}{
2\sqrt{2}} ({\psi_{m}} \si^m \chib^{\bj}) g_{i \bj} {F}^i
\nonumber \\ && - \frac{1}{ 2 \sqrt{2}} (\psib_m \sib^n \si^m
\chib^{\bj} - 2 \psib_m \chib^{\bj} g^{nm}) g_{i \bj} (\cd_n A^i -
\frac{1}{ \sqrt{2}} {\psi_{n}} \chi^i) \nonumber \\ && - \frac{1}{
2 \sqrt{2}} (\psi_m \si^n \sib^m \chi^i - 2 \psi_m \chi^i g^{nm})
g_{i \bj} (\cd_n \bA^{\bj} - \frac{1}{ \sqrt{2}} {\psib_{n}}
\chib^{\bj}). \label{CPN.58} \end{eqnarray}

\indent

It is straightforward to obtain the other terms in $\DM$, the
final result reads
\begin{eqnarray}
 \DM & = & -
g_{i \bj} \, g^{mn} \cd_m A^i \cd_n \bA^{\bj} - \f{i}{2}g_{i
\bj}\,\lp \chi^{i}  \si^m {\nabla}_m \chib^{\bj} \rp
 + \f{i}{2}g_{i \bj}\, \lp{\nabla}_m \chi^{i}
\si^m \chib^{\bj}\rp \nonumber \\ && + g_{i \bj} F^i {\bF}^{\bj}
 + \f{1}{4} R_{i \bi j \bj} (\chi^i \chi^j) (\chib^{\bi} \chib^{\bj})
- \f {1}{2} g_{i \bj}  (\chi^i \si^a \chib^{\bj}) b_a \nonumber
\\ && - \frac{1}{ \sqrt{2}} (\psib_m \sib^n \si^m \chib^{\bj})
g_{i \bj} \cd_n A^i - \frac{1}{ \sqrt{2}} (\psi_m \si^n \sib^m
\chi^i) g_{i \bj} \cd_n \bA^{\bj} \nonumber \\ && -\frac{i}{2}
g_{i \bj}\,\vep^{klmn}\lp \chi^i \si_k \chib^{\bj}\rp  \lp \psi_l
\si_m \psib_n \rp - \frac{1}{ 2} g_{i \bj} g^{mn} (\psi_m
\chi^i)(\psib_n \chib^{\bj})
\nonumber \\ &&
-i \sqrt{2} (\chi^i \la^{(r)}) g_{i \bj}\lp {
\bA{\bf T }_{(r)}} \rp ^{\bj} +i \sqrt{2} (\chib^{\bj} \lab^{(r)})
g_{i \bj}\lp {{{\bf T}_{(r)}} A}\rp ^i
\\ &&
 -\f{1}{2}
\left[ {\bf D}^{(r)} + \f {1}{2} (\psib_m \sib^m \la^{(r)} -
\psi_m \si^m \lab^{(r)} )\right] \kilc.
\nn
\label{CPN.59}
\end{eqnarray}
In this expression, the covariant derivatives $\cd_m A^i, \cd_m
\bA^{\bj}$ are defined in (\ref{CPN.31}) and (\ref{CPN.31}). The
derivatives ${\nabla}_m \chi_\al^i, {\nabla}_m \chib^{\bj \da}$
differ from $\cd_m \chi_\al^i, \cd_m \chib^{\bj \da}$ already
introduced in (\ref{CPN.33}) and (\ref{CPN.34}) by the
contribution of $\frac{i}{2}{e_m}^a b_a$ to ${ A}_m$ -cf.
(\ref{CPN.40}). This allows to keep track of
 the complete dependence in the auxiliary field $b_a$ in order to
solve its equation of motion later. Explicitly,
 \begin{eqnarray}
  {\nabla}_m \chi_\al^i & = & \prt_m
\chi_\al^i - {\om_{m \al}}^\varphi \chi_{\varphi}^i
 - i {\aym}^{(r)}_m \lp{{\bf T}_{(r)}} \chi_\al \rp^i
 -\f{1}{4} ( K_j \cd_m A^j -  K_{\bj} \cd_m \bA^{\bj})
\chi_\al^i
  \nonumber \\ &&
 - \f{i}{4} g_{j\bk} (\chi^j \si_m\chib^\bk)\ \chi^i_\al
 +\chi^j_\al \ {\Gamma^i}_{jk} {\cal D}_m A^k,
\label{CPN.60} \\ {\nabla}_m \chib^{\bj \da} & = & \prt_m
\chib^{\bj \da} - {{\om_{m}}^{\da}}_{\dv} \chib^{\bj \dv} + i
{\aym}^{(r)}_m \lp  \chib^ \da{{\bf T}_{(r)}} \rp^\bj + \f{1}{4} (
K_k \cd_m A^k -  K_{\bk} \cd_m \bA^{\bk}) \chib^{\bj \da}
 \nonumber \\ &&
 +\f{i}{4} g_{j\bk} (\chi^j \si_m\chib^\bk)\ \chib^{\bj\da}
+ \chib^{\da\bi}\ {\Gamma^\bj}_{\bi \bk} {\cal D}_m {\bA}^{\bk}.
\label{CPN.61} \end{eqnarray}
Finally, using the set of equations
 \begin{eqnarray}
  \cd_\beta
\cw_\al + \cd_\al \cw_\beta & = & -2 (\si^{ba} \eps)_{\beta \al}
\cf_{ba} \nonumber \\ \cd_\beta \cw_\al - \cd_\al \cw_\beta & = &
+ \eps_{\beta \al} \cd^{\varphi} \cw_\varphi \label{CPN.62} \\
\cd_{\db} {\cw}_{\da} + \cd_{\da} {\cw}_{\db}  & = & -2 ( \eps
\sib^{ba})_{\db \da} \cf_{ba} \nonumber \\ \cd_{\db} {\cw}_{\da} -
\cd_{\da} {\cw}_{\db}  & = & - \eps_{\db \da} \cd_{\dv}
{\cw}^{\dv}, \label{CPN.63}
\end{eqnarray}
 we obtain,
along the same lines as before, the lowest components of the
super-covariant derivative of the Yang-Mills superfields
($\cf_{ba}{\loco}$ has been given in (\ref{CPN.37})),
 \begin{eqnarray}
\cd_a \cw^\al{\loco} &=& {e_a}^m \left[ -i\cd_m\la^\al - \f {1}{2}
(i {\fym}_{pq} + \psi_p\si_q \lab + \psib_p \sib_q \la)\,(\psi_m
\si^{pq})^\al - \f {1}{2} {\bf D}\, \psi_m{}^\al \right],
\label{CPN.64} \nn \\
\cd_a \cw_\da{\loco} &=& {e_a}^m \left[
 +i\cd_m\lab_\da + \f {1}{2} ( i {\fym}_{pq} + \psi_p\si_q \lab +
\psib_p \sib_q \la)\,(\psib_m \sib^{pq})_\da - \f {1}{2}  {\bf
D}\,\psib_{m\da} \right],
\nn \\ &&
\label{CPN.65}
\end{eqnarray}
where
\begin{equation}
{\fym}_{mn} \,=\,\prt_m {\aym}_n - \prt_n {\aym}_m - i[{\aym}_m,
{\aym}_n], \label{CPN.66} \end{equation}
 and the covariant
derivatives $\cd_m \la^\al, \cd_m\lab_\da$ are defined in
(\ref{CPN.38}), (\ref{CPN.39}).

 \subsec{ Supersymmetry Transformations \label{CPN3}}

In the superspace formalism, supersymmetry transformations are
identified as special cases of superspace diffeomorphisms. The
general form of these diffeomorphisms is given in appendix
\ref{appC2} and we will use the results obtained there.

Before writing these transformations at the component field level,
we would like to stress a point of some importance in the process
of generalizing supergravity transformations to the K\"ahler
superspace. For this we need the transformation law of
the vielbein and of a generic (spinless) superfield $\Phi$
 under diffeomorphisms
($\xi^C$), Lorentz $({\Lambda_B}^A)$ and K\"ahler $(\Lambda)$
transformations
\begin{eqnarray}
 \delta {E_M}^A  &=&  \cd_M \xi^A + {E_M}^B \xi^C
{T_{CB}}^A + {E_M}^B ({\Lambda_B}^A - \xi^C {\phi_{CB}}^A)
\nonumber \\ && \;\;\;\;\;\; +  w(E^A)(\Lambda - \xi^C A_C)
{E_M}^A ,
\label{CPN.67}\\[2mm]
\delta \Phi \ &=& \ \xi^B \cd_B \Phi +
w(\Phi)  (\Lambda - \xi^C A_C)\Phi. \label{CPN.68}
\end{eqnarray}
Supergravity transformations are defined \cite{WB83} by
compensating the term $\xi^C {\phi_{CB}}^A$ with a field-dependent
Lorentz transformation
\begin{equation}
{\Lambda_B}^A \,=\,\xi^C {\phi_{CB}}^A. \label{CPN.69}
\end{equation}
The point is that the same procedure cannot be followed for the
K\"ahler transformation since $\Lambda$ is fixed to be of the form
\begin{equation}
\Lambda \,=\,- \frac{F(\phi) - {\bF}(\phib)}{ 4} \label{CPN.70}
\end{equation}
 and generic terms proportional to the K\"ahler connection
appear in the supergravity transformations, weighted by the
K\"ahler weight of the field considered.

Supergravity transformations, denoted by the symbol ${\delta}_{\WZ}$,
are discussed in detail in appendix \ref{appC}. As in the remainder of this
section we will be exclusively concerned with supergravity transformations,
we will drop from now on the subscript in ${\delta}_{\WZ}$, supergravity
variations will be denoted $\dt$.

\begin{itemize}

\item
{\bf Supergravity sector}

The transformations of vierbein and gravitino
are derived from (\ref{C.134}), which reads
 \begin{eqnarray}
{\delta}{E_M}^A &=& {\cd}_M {\xi}^A +{E_M}^B {\xi}^C
{T_{CB}}^A \nonumber \\ & &- \f{1}{4}w(E^A){E_M}^A
{\xi}^B(K_k{\cd}_B {\phi}^k -K_{\bk} {\cd}_B{\phib}^{\bk})
\nonumber \\ & &-\f{i}{8}w(E^A){E_M}^A{\xi}^b (12G_b +
{\sib}_b^{{\da} \al}g_{k{\bk}} {\cd}_\al {\phi}^k
{\cd}_{\da}{\phib}^{\bk}), \label{CPN.71}
\end{eqnarray}
Projecting to lowest components and using
(\ref{CPN.5a}), (\ref{CPN.5b}), (\ref{CPN.7}),
together with the
torsions summarized in appendix \ref{appB}, and
\begin{equation}
\xi^a {\loco} \,=\,0, \cem \xi^\al {\loco} \,=\,\xi^\al, \cem
\xi_{\da}{\loco} \ =\ \xi_{\da},
\label{CPN.73}
\end{equation}
gives rise to
\begin{eqnarray}
 \delta {e_m}^a & = & i
\xi \si^a \psib_m + i \xib \sib^a \psi_m,
\label{CPN.74} \\[2mm]
\delta
{\psi_m}^\al & = & 2 \cd_m \xi^\al - \f{i}{3} (\xi \si^a
\sib_m)^\al b_a + \f{i}{3} (\xib \sib_m)^\al M
\nonumber \\ &&
- \frac{1}{ 2 \sqrt{2}} {\psi_m}^\al (K_i \xi \chi^i -
K_{\bj} \xib \chib^{\bj}),
\label{CPN.75} \\[2mm]
\delta \psib_{m \da}
& = & 2 \cd_m \xib_{\da} + \f{i}{3} (\xib {\sib^a} \si_m)_{\da}
b_a + \f{i}{3} (\xi \si_m)_{\da} \ovM
                        \nonumber \\
&& + \frac{1}{ 2 \sqrt{2}} \psib_{m \da} (K_i \xi \chi^i - K_{\bj}
\xib \chib^{\bj}),
\label{CPN.76}
\end{eqnarray}
 with
\begin{eqnarray}
 \cd_m \xi^\al & = & \prt_m \xi^\al +
\xi^\beta {\om_{m \beta}}^\al + \xi^\al {A}_m, \label{CPN.77}
\\ \cd_m \xib_{\da} & = & \prt_m \xib_{\da} + \xib_{\db}
{{\om_m}^{\db}}_{\da} - \xib_{\da} {A}_m, \label{CPN.78}
\end{eqnarray}
 and ${A}_m$ given in (\ref{CPN.40}).
For future use, note that the determinant of the
vielbein transforms as
\begin{equation}
\delta e = e {e_a}^m \delta {e_m}^a
        \,=\,e ( i\xi \si^m \psib_m + i \xib \sib^m \psi_m),
\label{CPN.79} \end{equation}
 and the $\si_m, \sib^m$ matrices
as
\begin{eqnarray}
 \delta \si_{m\al\da} &=& \delta ({e_m}^a
\si_{a\al\da} )
   \,=\,+i\si_{n\al\da}(\xi \si^n \psib_m + \xib \sib^n \psi_m),
      \label{CPN.80} \\
\delta \sib^{m\da\al} &=& \delta ( \sib^{a\da\al} {e_a}^m )
   \,=\,-i\sib^{n\da\al}(\xi \si^m \psib_n + \xib \sib^m \psi_n).
      \label{CPN.81}
\end{eqnarray}

The supersymmetry transformations of the components
$M,\ovM$ and $b_a$ are derived from the supergravity
transformations (\ref{C.137})
\begin{eqnarray}
{\delta}{\Phi}&=& {\xi}^A{\cd}_A {\Phi}
-\f{1}{4}w({\Phi}){\xi}^A (K_k{\cd}_A{\phi}^k
-K_{\bk}{\cd}_A{\phib}^{\bk}){\Phi}  \nonumber \\ &
&-\f{i}{8}w({\Phi}){\xi}^b(12G_b + {\sib}_b^{{\da} \al}
g_{k{\bk}}{\cd}_\al {\phi}^k {\cd}_{\da}{\phib}^{\bk}){\Phi},
\label{CPN.72}
\end{eqnarray}
of the generic superfield $\Phi$ after suitable
specification. In a first step, projection to lowest components yields
\begin{equation}
\delta\Phi{\loco} \,=\,\xi^\al \cd_\al \Phi{\loco}
+\xib_\da\cd^\da \Phi{\loco} -\frac{1}{ 2\sqrt{2}}\, w(\Phi)
\left( K_k\xi\chi^k-K_\bk \xib\chib^\bk \right) \Phi{\loco}.
\label{CPN.82}
\end{equation}
Substituting $R, R^\dagger$ and $G_a$ for $\Phi$
and  using the information given in appendix \ref{appB}, in
particular (\ref{B.77}) - (\ref{B.80}), it is straightforward to
arrive at the transformation laws
 \begin{eqnarray}
  \delta M &=&  -i \sqrt{2} \
g_{i \bj} (\xi \si^m \chib^{\bj})(\cd_m A^i - \frac{1}{\sqrt{2}}
\psi_m \chi^i) + \sqrt{2} \ g_{i \bj} \ (\xi \chi^i) {\bF}^{\bj}
\nonumber \\
         & & + i \ (\xi \la^{(r)})\kilc
- \frac{1}{\sqrt{2}} M (K_i \ \xi \chi^i - K_{\bj} \ \xib \chib^{\bj})
\nonumber \\[1mm]
& &
          + 4 (\xi \si^{nm} \cd_n \psi_m) -
i (\xi \si^m \sib_a \psi_m) b^a - i(\xi \si^m \psib_m)M  ,
  \label{CPN.83} \\[3mm]
\delta \ovM &=& -i\sqrt{2} \ g_{i \bj} (\xib \sib^m \chi^i)(\cd_m
\bA^{\bj} - \frac{1}{\sqrt{2}} \psib_m \chib^{\bj}) + \sqrt{2} \
g_{i \bj} \ {F}^i (\xib \chib^{\bj})
\nonumber \\
         & & -i (\xib \lab^{(r)})\kilc
+ \frac{1}{\sqrt{2}} \ \ovM (K_i \ \xi \chi^i - K_{\bj} \ \xib
\chib^{\bj})
\nonumber \\[1mm]
         & &
+ 4 (\xib \sib^{nm} \cd_n
\psib_m) + i (\xib \sib^m \si_a \psi_m) b^a - i (\xib \sib^m
\psi_m) \ovM ,   \label{CPN.84}
   \\[3mm]
\delta b_a &=&
\f {1}{2}\lp \xi\si_a\sib^{nm}
            -3 \  \xi\si^{nm}\si_a \rp \cd_n \psib_m
             - \f {1}{2}\lp \xib\sib_a\si^{nm}
            -3 \  \xib\sib^{nm}\sib_a\ \rp \cd_n\psi_m
\nn \\
& &
-\f{i}{2}{e_a}^m (\xi\si^d\psib_m +\xib\sib^d \psi_m)b_d
     -\f{i}{2} {e_a}^m (\psib_m \xib) M
     +\f{i}{2} {e_a}^m (\xi\psi_m) \ovM
\nn \\
& &
-\frac{i}{ \sqrt{2}} g_{k\bk}(\xi\si_a\sib^m\chi^k)
(\cd_m\bA^{\bk}-\frac{1}{ \sqrt{2}}\psib_m\chib^{\bk}) +\frac{1}{
\sqrt{2}}g_{k\bk}(\xi\si_a\chib^\bk)F^k
\nn \\
& &
-\frac{i}{
\sqrt{2}} g_{k\bk}(\xib\sib_a\si^m\chib^\bk) (\cd_m A^k-\frac{1}{
\sqrt{2}}\psi_m\chi^k) +\frac{1}{
\sqrt{2}}g_{k\bk}(\xib\sib_a\chi^k){\bF}^\bk
\nn \\[1mm]
& &
- \f{i}{2} \lp\xi\si_a\lab^{(r)}
   - \xib\sib_a\la^{(r)}\rp \kilc.
\label{CPN.85}
\end{eqnarray}

\item
{\bf Matter sector}

Let us first discuss the chiral superfield $\phi^i$. The supersymmetry
transformation of the component field $A^i$ is derived from
(\ref{C.135})
\begin{equation}
{\delta}{\phi}^i \, = \, {\xi}^A{\cd}_A{\phi}^i,
\label{CPN.86}
\end{equation}
upon straightforward projection to lowest components. As to the
components $\chi_\al^i$ and $F^i$ the situation is slightly more
involved. They are identified in the lowest components of the
superfields $\cd_\al \phi^i$ and $\cd^\al \cd_\al \phi^i$ of
respective chiral weights $-1$ and $-2$. They are particular
cases of a generic superfield of the type ${\bf U}^i$, with some
chiral weight. The relevant equations in appendix \ref{appC} are
(\ref{C.128}) - (\ref{C.133}) and (\ref{C.137}), (\ref{C.138}).
We have to consider a superfield ${\bf U}^i$ (which is actually a
mixture of the superfields $\Phi$ and ${\bf U}^i$ of appendix
\ref{appC}) with supergravity transformation
\begin{eqnarray}
{\delta}{\bf U}^i &=& {\xi}^A{\cd}_A {\bf
U}^i+{{\Gamma}^i}_{jk}{\xi}^A {\cd}_A {\phi}^j {\bf U}^k
\nonumber \\
& &
-\f{1}{4}w({\bf U}^i) {\bf U}^i {\xi}^A
(K_k{\cd}_A{\phi}^k -K_{\bk}{\cd}_A{\phib}^{\bk})
\nonumber \\
& &
-\f{i}{8}w({\bf U}^i) {\bf U}^i {\xi}^b(12G_b +
{\sib}_b^{{\da} \al} g_{k{\bk}}{\cd}_\al {\phi}^k
{\cd}_{\da}{\phib}^{\bk}).
 \label{CPN.88}
\end{eqnarray}
This provides the supergravity transformations for $\chi_\al^i$ and $F^i$,
once ${\bf U}^i$ is replaced by  $\cd_\al \phi^i$ and $\cd^\al \cd_\al
\phi^i$, and the result projected to lowest components.
Intermediate steps in the computation
involve the covariant derivative relations
\begin{eqnarray}
\cd^{\db} \cd_{\al} {\phi}^i & = & 2i {(\si^a \eps)_\al}^{\db}
\cd_a \phi^i,
\label{CPN.89}
\\[1mm]
\cd_\beta \cd^\al \cd_\al \phi^i & = & \frac{2}{ 3} \{
\cd_\beta , \cd^\al \} \cd_\al \phi^i = 8 R^{\dagger} \cd_\beta
\phi^i,
\label{CPN.90} \\[1mm]
\cd^{\db} \cd^\al \cd_\al \phi^i & = &
-4i {(\si^a \eps)_\al} ^{\db} \cd_a \cd^\al \phi^i + {4(\si^a
\eps)_\al}^{\db} G_a \cd^{\al} {\phi}^i \nonumber
\\[1mm]
 &&+ {R^i}_{j k \bk} \cd^{\db} {\phib}^{\bk} \cd^{\al}  {\phi}^k
 {\cd}_{\al} {\phi}^j
 - 8{\cw}^{(r) \db}\lp {{{\bf T}_{(r)}}\phi}\rp ^i. \label{CPN.91}
\end{eqnarray}
As a final result we obtain the component field transformations
\begin{eqnarray}
 \delta A^i & = & \sqrt{2} \, \xi \chi^i ,
\label{CPN.92} \\[2mm]
\delta \chi^i_\al & = & i
\sqrt{2} (\xib \sib^m \eps)_\al (\cd_m A^i - \frac{1}{ \sqrt{2}}
\psi_m \chi^i) + \sqrt{2} \xi_\al F^i
\nonumber
\\ &&
+ \frac{1}{ \sqrt{2}} \xi_\al {\Gamma^i}_{jk}
(\chi^{j}\chi^k) + \frac{1}{ 2 \sqrt{2}} \chi_\al^i (K_k \xi
\chi^k - K_{\bk} \xib \chib^{\bk}), \label{CPN.93}
\\[3mm]
\delta F^i & = &
i \sqrt{2} \lp \xib \sib^m  {\nabla}_m \chi^i \rp - i(\xib
\sib^m \psi_m ) F^i
\nonumber
\\ &&
+ (\xib \sib^m \si^n \psib_m)(\cd_n A^i - \frac{1}{
\sqrt{2}} \psi_n \chi^i) \nonumber \\ && + \frac{\sqrt{2}}{3} \ovM
\xi \chi^i + \frac{\sqrt{2}}{ 3 }
 (\xib \sib^a \chi^i) b_a - 2i \xib \lab^{(r)}\lp {{{\bf T}_{(r)}} A}\rp ^i \nonumber \\
&& + \sqrt{2} {\Gamma^i}_{jk} ( \xi \chi^j) F^k - \frac{1}{
\sqrt{2}} {R^i}_{jk\bk} (\chi^{j} \chi^k) (\xib \chib^{\bk})
\nonumber \\ && + \frac{1}{ \sqrt{2}} F^i (K_k \xi \chi^k -
K_{\bk} \xib \chib^{\bk}), \label{CPN.94}
\end{eqnarray}
 where the relevant covariant
derivatives are given in (\ref{CPN.31}) and (\ref{CPN.60}). The
supersymmetry transformations for a general chiral superfield of
non-zero weight $w$ will be given in the next subsection
-cf. (\ref{CPN.111}) - (\ref{CPN.113}).

\indent

Similarly, for an antichiral superfield $\phib^{\bj}$ of supergravity
transformation
\begin{equation}
{\delta}_{\xi}{\phib}^{\bi} \, = \, {\xi}^A {\cd}_A {\phib}^{\bi},
\label{CPN.87}
\end{equation}
we use the relations
\begin{eqnarray}
 \cd^{\db} \cd_\al \phib^{\bj} & = & 2i
{(\si^a \eps)_\al}^{\db} \cd_a \phib^{\bj},
\label{CPN.95}
\\[2mm]
\cd^{\db} \cd_{\da} \cd^{\da} \phib^{\bj} & = & 8R \ \cd^{\db}
\phib^{\bj},
\label{CPN.96} \\[2mm]
\cd_\beta \cd_{\da} \cd^{\da}
\phib^{\bj} & = & -4i \si^a_{\beta \db} \cd_a \cd^{\db}
\phib^{\bj} - 4 {\si^a}_{\beta \db} G_a \cd^{\db} \phib^{\bj}
\nonumber \\ &&+{R^\bj}_{\bi \bk k} \cd_{\beta}{\phi}^{k} \cd_\da
{\phib}^\bk \cd^{\da} {\phib}^\bi
 + 8 {\cw_\beta}^{(r)}\lp {
\phib {\bf T }_{(r)}} \rp ^{\bj},
\label{CPN.97}
\end{eqnarray}
to arrive at the component field transformations
\begin{eqnarray}
 \delta \bA^{\bj} & = &  \sqrt{2} \, \xib \chib^{\bj},
\label{CPN.98} \\[1mm]
\delta \chib^{\da \bj} & = & i \sqrt{2} (\xi
\si^m \eps)^{\da} (\cd_m \bA^{\bj} - \frac{1}{\sqrt{2}} \psib_n
\chib^{\bj}) + \sqrt{2} \xib^{\da} {\bF}^{\bj} \nonumber
\\ && + \frac{1}{\sqrt{2}} \xib^\da {\Gamma^\bi}_{\bj\bk}
(\chib^{\bj} \chib^{\bk}) - \frac{1}{ 2 \sqrt{2}} \chib^{\da \bj}
(K_k \xi \chi^k - K_{\bk} \xib \chib^{\bk}),
\label{CPN.99} \\[3mm]
\delta {\bF}^{\bj} & = & i \sqrt{2} \lp\xi \si^m {\nabla}_m
\chib^{\bj}\rp - i(\xi \si^m \psib_m ) {\bF}^{\bj} \nonumber \\ &&
+ (\xi \si^m \sib^n \psi_m)(\cd_n \bA^{\bj} - \frac{1}{ \sqrt{2}}
\psib_n \chib^{\bj}) \nonumber
\\ &&  + \frac{\sqrt{2}}{ 3 } M \xib \chib^{\bj} +
\frac{\sqrt{2}}{ 3 } (\xi \si^a \chib^{\bj})b_a + 2i \xi
\la^{(r)}\lp { \bA{\bf T }_{(r)}} \rp ^{\bj} \nonumber \\ && +
\sqrt{2} {\Gamma^\bj}_{\bi\bk} (\xib \chib^{\bi})\bF^\bk -
\frac{1}{ \sqrt{2}} {R^\bj}_{\bi\bk k} (\chib^{\bi} \chib^{\bk})
(\xi \chi^k) \nonumber \\ && - \frac{1}{ \sqrt{2}} {\bF}^{\bj}
(K_k \xi \chi^k - K_{\bk} \xib \chib^{\bk}), \label{CPN.100}
\end{eqnarray}
after suitable projection to lowest components.

\item
{\bf Yang-Mills sector}\footnote{All the fields below belong to
the adjoint representation of the Yang-Mills group,
$({\aym}_m,\la, \lab, {\bf D}) = ({{\aym}_m}^{(r)},\la^{(r)},
\lab^{(r)}, \bf D^{(r)})\cdot{\bf T }_{(r)}$.}

 As to the supergravity transformation of the gauge
potential ${\aym}_m =-i\ca_m{\loco}$, we project (\ref{C.125})
\begin{equation}
\delta \ca_M \,=\,{E_M}^B \xi^C {\cf_{CB}}^A
\label{CPN.101}
\end{equation}
to lowest components and use (\ref{CPN.37}) to obtain
\begin{equation}
\delta {\aym}_m \,=\,i (\xi \si_m \lab) + i (\xib \sib_m \la).
\label{CPN.102} \end{equation} Concerning the fermionic
components $\la^\al, {\lab_{\da}}$ defined in (\ref{CPN.30}), the
supersymmetry transformations are obtained after identification
of $\Phi$  in
 (\ref{CPN.72}) with $\cw^\al$ resp. $\bar{\cw}_{\da}$
and subsequent projection to lowest components. Using
(\ref{CPN.62}), (\ref{CPN.63}) and the explicit form of
$\cf_{ba}{\loco}$ in (\ref{CPN.37}), we obtain
 \begin{eqnarray}
\delta \la^\al & = & (\xi \si^{mn})^\al  ( - {\fym}_{mn} + i
\psi_m \si_n \lab + i  \psib_m \sib_n \la) + i \xi^\al {\bf D}
\nonumber \\ && - \frac{1}{2\sqrt{2}} \la^\al (K_k \xi \chi^k -
K_{\bk} \xib \chib^{\bk}),
\label{CPN.103}\\[2mm]
\delta \lab_{\da} & =
& (\xib \sib^{mn})_{\da} ( - {\fym}_{mn} + i \psi_m \si_n \lab + i
\psib_m \sib_n \la) - i \xib_{\da} {\bf D} \nonumber
\\ && + \frac{1}{2\sqrt{2}} \lab_{\da} (K_k \xi \chi^k - K_{\bk}
\xib \chib^{\bk}), \label{CPN.104}
\end{eqnarray}
with ${\fym}_{mn}$ defined in (\ref{CPN.66}). Finally, the
transformation
\end{itemize}
\begin{eqnarray}
 \delta {\bf D} &=& -\xi
\si^m\cd_m\lab +\xib \sib^m\cd_m\la + \f{i}{2}
(\psib_m\sib^m\xi +\psi_m\si^m\xib) \ {\bf D} \nonumber \\
&& +\f{1}{2} (\psib_m\sib^{kl}\sib^m\xi-\psi_m\si^{kl}\si^m\xib)
({\fym}_{kl} - i \psi_k\si_l\lab - i \psib_k\sib_l\la),
\label{CPN.105}
\end{eqnarray}
\hspace{10mm}
of the auxiliary field is obtained along the same lines.

\subsec{ Generic Component Field Action \label{CPN4}}

Although superfield actions, as discussed in section \ref{GRA33}, are quite
compact, and invariance under supersymmetry transformations is rather
transparent, their component field expansions are notoriously complicated.
In section \ref{GRA33} we have seen that the chiral volume element
provides the generalization of the F-term construction to the case of local
supersymmetry. The superfield actions for the supergravity/matter system,
the Yang-Mills kinetic terms and the superpotential in (\ref{GRA.240}),
(\ref{GRA.241}) and (\ref{GRA.242}) are all of the generic form
\begin{equation}
\ca ({\bf r}, {\bf \bar r})\, = \, \int_\ast \! \frac{E}{R} \, \rgen
      \ \ + \ \ \mbox{h. c.} \ ,
 \label{GRAgen.242}
\end{equation}
with $\rgen$ a chiral superfield of $U(1)$ weight $w(\rgen) =2$. The
various superfield actions are then obtained from identifying $\rgen$
respectively with
\begin{equation}
\rSM \, = \, - 3R,
\end{equation}
\begin{equation}
\rYM \, = \, \f{1}{8} f_{(r)(s)}(\phi)
\, {\cw}^{(r){\al}}{\cw}_{\al}^{(s)},
\end{equation}
and
\begin{equation}
\rSPOT \, = \, {\f{1}{2}} e^{K/2} W(\phi).
\end{equation}
We will proceed, in a first step, with the construction of a
locally supersymmetric component field action a generic
chiral superfield $\rgen$, starting from the definition
\begin{equation}
 \ca ({\bf r}, {\bf \bar r})\loco \, = \, \int_\ast \! \frac{E}{R} \,
\rgen \loco
     \ \ + \ \ \mbox{h. c.} \
              \, = \, \int \! \! d^4x \ \cl ({\bf r}, {\bf \bar r}).
\end{equation}
In the following we will determine $\cl ({\bf r}, {\bf \bar r})$ as a
suitably modified F-term for the superfield $\rgen$.
Defining the components of ${\bf r}$ as usual,
\begin{equation}
{\bf r} \,=\, \rgen \loco,
\cem
{\bf s}_\al \,=\,\frac{1}{\sqrt{2}} \cd_\al \rgen \loco,
\cem
{\bf f} \,=\, -\f{1}{4} \cd^\al \cd_\al \rgen \loco,
\label{CPN.106}
\end{equation}
it should be clear that the F-term space-time density, \ie the
component field $e \, {\bf f}$ alone is not invariant under
supergravity transformations. Calling
\begin{equation}
  l_1 \, = \, e \, {\bf f},
\label{CPN.107}
\end{equation}
we allow for additional terms
\begin{eqnarray}
l_2 & = & \la_2^\al \, {\bf s}_\al,
\label{CPN.108}
\\
l_3 & = & \la_3 \, {\bf r},
\label{CPN.109}
\end{eqnarray}
with field dependent coefficients $\la_2^\al$, $\la_3$ of respective
$U(1)$ weights $-1, -2$.
The strategy is then to use the supersymmetry transformations of the gravity
sector, which are already known, and those of the generic multiplet
to determine $l_2$
and $l_3$, \ie $\la_2^\al$ and $\la_3$, such that $l_1 + l_2 + l_3$
is invariant under supersymmetry, up to a total space-time derivative. The
reader not interested in the details of the computation can go
directly to (\ref{CPN.122}), (\ref{CPN.123}) which summarize the results.

\indent

The supersymmetry transformation laws for the components of a
superfield $\rgen$ of K\"ahler weight $w \equiv w(\rgen)$ are
be obtained
from the general procedure exposed in section \ref{CPN3},
they read
\begin{eqnarray}
 \delta {\bf r} & = & \sqrt{2} \,\xi {\bf s} - \frac{w}{2 \sqrt{2}}
  (K_k \ \xi \chi^k - K_{\bk} \ \xib \chib^{\bk})\, {\bf r},
\label{CPN.111} \\
 \delta {\bf s}_\al & = &
\sqrt{2} \ \xi_\al {\bf f}
+ i \sqrt{2} \ (\si^m \xib)_\al \,
      (\cd_m {\bf r}
      - \frac{1}{\sqrt{2}} \ \psi_m{} {\bf s}
      + \frac{i w}{2} {e_m}^a \ b_a {\bf r})
\nonumber \\ &&
- \frac{w-1}{2 \sqrt{2}} \, ( K_k \ \xi \chi^k - K_{\bk} \ \xib
\chib^{\bk}) \, {\bf s}_\al ,
\label{CPN.112} \\[2mm]
\delta {\bf f} & = &
i \sqrt{2} \, (\xib \sib^m \cd_m {\bf s})
- i (\xib \sib^m \psi_m)\, {\bf f}
\nonumber \\[2mm] &&
+ (\xib \ \sib^m \si^n \psib_m)
(\cd_n {\bf r}
- \frac{1}{\sqrt{2}} \psi_n {\bf s}
+ \frac{iw}{2} {e_n}^a b_a  {\bf r})
\nonumber \\ && +
\frac{\sqrt{2}}{3} \ovM \xi {\bf s}
- \frac{\sqrt{2}}{6} (3w-2)(\xib \sib^a {\bf s}) \,  b_a
+ wr \, \xib_{\da} {\bX}^{\da}
\nonumber \\ &&
- \frac{w-2}{2\sqrt{2}}  (K_k \ \xi \chi^k - K_{\bk} \
\xib \chib^{\bk}) \, {\bf f}. \label{CPN.113}
\end{eqnarray}
Thus, specifying to the case $w = 2$ and using (\ref{CPN.79}) and
(\ref{CPN.113}), gives rise to
\begin{eqnarray}
\frac{1}{e} \ \delta l_1 & = &
i \lp \xi \si^m \psib_m \rp {\bf f}
- \frac{2\sqrt{2}}{3} (\xib \sib^a {\bf s}) \,  b_a
+ i \sqrt{2} \ (\xib \sib^m \cd_m {\bf s})
 \\ &&
+ (\xib \sib^m \si^n \psib_m)
(\cd_n {\bf r} - \frac{1}{\sqrt{2}} \psi_n {\bf s}
+ i {e_n}^a b_a {\bf r})
+ \frac{\sqrt{2}}{3} \ovM \xi {\bf s}
+2r \, \xib_\da \bX^\da.
\nonumber
\label{CPN.114} \end{eqnarray}
A glance at the transformation law (\ref{CPN.112}) shows
that the first term can be cancelled in choosing
\begin{equation}
l_2 \,=\,\frac{ie}{\sqrt{2}} (\psib_m \sib^m)^\al {\bf s}_\al.
\label{CPN.115}
\end{equation}
In the next step we work out the supersymmetry transformation of
the sum $l_1 + l_2$. Using (\ref{CPN.76}) and (\ref{CPN.81}) we
obtain
\begin{eqnarray}
\frac{1}{e} \ \delta(l_1 + l_2) & = &
\sqrt{2} \, \xi {\bf s} \, \ovM
+ 2 {\bf r} \, \xib_\da \, \bX^\da
\nonumber \\ &&
+i \sqrt{2} \, (\xib \sib^m \, \cd_m {\bf s})
+ i \sqrt{2} \, (\cd_m \xib \, \sib^m {\bf s})
\nonumber \\ && +
4(\xib \sib^{mn} \psib_m)
(\cd_n {\bf r}
- \frac{1}{\sqrt{2}} \psi_n {\bf s}
+ i {e_n}^a b_a {\bf r})
\nonumber \\ &&
+ \frac{1}{\sqrt{2}} (\psib_m \sib^n {\bf s}) \,
(\xi \si^m \psib_n + \xib \sib^m \psi_n)
\nonumber \\ &&
- \frac{1}{\sqrt{2}} (\psib_m \sib^m {\bf s}) \,
(\xi \si^n \psib_n + \xib \sib^n \psi_n).
\label{CPN.116} \end{eqnarray}
Again, requiring cancellation of the first term suggests to choose
\begin{equation}
l_3 \,=\,- e \, \ovM \, {\bf r}. \label{CPN.117}
\end{equation}
Taking into account the supergravity transformation law
(\ref{CPN.84}), we now obtain
\begin{eqnarray}
\frac{1}{e} \delta (l_1 + l_2 + l_3) & = &
4 (\cd_n \xib \, \sib^{nm} \psib_m) \, {\bf r}
-i(\xib \sib^a \si^m \psib_m) \, b_a {\bf r}
- i (\xi \si^m \psib_m) \ovM {\bf r}
\nonumber \\ &&
+ i \sqrt{2} \, \cd_m (\xib \sib^{m} {\bf s})
- 4 \, \cd_n ( \xib \sib^{nm} {\psib_m}\, {\bf r} )
\nonumber \\ &&
+ \frac{4}{\sqrt{2}} (\xib \sib^{nm} \psib_m)(\psi_n {\bf s})
\nonumber \\ &&
+ \frac{1}{\sqrt{2}} (\psib_m \sib^n {\bf s})
(\xi \si^m \psib_n + \xib \sib^m \psi_n)
\nonumber \\ &&
- \frac{1}{\sqrt{2}} (\psib_m \sib^m {\bf s})
(\xi \si^n \psib_n + \xib \sib^n \psi_n).
\label{CPN.118}
\end{eqnarray}
Here, the first term can be cancelled with the help of another term of the
type
$l_3$. Indeed, the transformation law (\ref{CPN.76}) suggests to take
\begin{equation}
l'_3 \,=\,-e \, {\bf r} \, \psib_m \sib^{mn} \psib_n.
\label{CPN.119}
\end{equation}
Using (\ref{CPN.76}) and (\ref{CPN.80}), (\ref{CPN.81}), we find
\begin{eqnarray}
 \frac{1}{e} \ \delta l'_3 & = &
-4(\cd_n \xib \, \sib^{nm} \psib_m) {\bf r}
+ i(\xib \sib^a \si^m \psib_m)b_a {\bf r}
+ i (\xi \si^m \psib_m) \ovM {\bf r}
\nonumber \\ &&
- \sqrt{2} (\psib_m \sib^{mn} \psib_n)(\xi {\bf s})
+ 2i {\bf r} (\psib_m \sib^{kn} \psib_n) (\xi \si^m \psib_k
+ \xib \sib^m \psi_k)
\nonumber \\ &&
-i {\bf r} (\psib_m \sib^{mn} \psib_n)(\xi \si^k \psib_k +
\xib \sib^k \psi_k).
\label{CPN.120}
\end{eqnarray}
Using the relation
\begin{equation}
e \, {e_a}^m \, \cd_m  v^a \,=\,\prt_m (e v^a {e_a}^m) + \f{ie}{2}
{(\si^b \eps)_\beta}^{\db} v^a ({e_b}^n {e_a}^m - {e_b}^m {e_a}^n)
{\psi_n}^\beta \psib_{m \db}, \label{CPN.121}
\end{equation}
for integration by parts at the component field level and after
some algebra together with (\ref{cycsig}), we finally obtain
\begin{equation}
\delta (l_1 + l_2 + l_3 + l'_3) \,=\,\prt_m \left[i \sqrt{2} \ e
(\xib \sib^m {\bf s}) - 4e (\xib \sib^{mn} \psib_m) {\bf r} \right].
\label{CPN.122} \end{equation}
This shows that the Lagrangian density
\begin{eqnarray}
\cl ({\bf r}, {\bf \bar r}) &=&
e \lp {\bf f} + {\bf \bar f} \rp
+ \frac{ie}{\sqrt{2}}
\lp \psi_m \si^m {\bf \bar s} +\psib_m \sib^m {\bf s} \rp
\nn \\ &&
- e \, {\bf \bar r} \lp M + \psi_m \, \si^{mn} \psi_n \rp
- e \, {\bf r} \lp \ovM + \psib_m \, \sib^{mn} \psib_n \rp,
\label{CPN.123}
\end{eqnarray}
constructed with the components
(\ref{CPN.106}) of a generic chiral superfield of chiral weight $w=2$
provides a supersymmetric action.

\subsec{ Invariant Actions \label{CPN5}}

The generic construction can now be applied
 to derive the component field versions of
the superfield actions discussed in section \ref{GRA33}, namely
$\ASM$, $\ASPOT$ and $\AYM$ given respectively in eqs.
(\ref{GRA.240}), (\ref{GRA.242}) and (\ref{GRA.241}).

 \subsubsection{ Supergravity and matter \label{CPN51}}
Identifying the generic superfield such that
\begin{equation}
\rSM \,=\,-3R,
\label{CPN.124}
\end{equation}
determines component fields correspondingly. The lowest
component is given as
\begin{equation}
{\bf r} \,=\,\frac{M}{2}.
\label{CPN.125}
\end{equation}
As a consequence of (\ref{GRA.167}) the spinor component takes
the form
\begin{equation}
{\bf s}_\al \,=\,\frac{1}{\sqrt{2}} { X}_\al {\loco} + \sqrt{2}
(\si^{cb} \eps)_{\al \varphi} {T_{cb}}^{\varphi}{\loco}.
\label{CPN.126}
\end{equation}
In the construction of the component field Lagrangian this appears
in the combination
\begin{eqnarray}
\frac{i}{\sqrt{2}} (\psib_m \sib^m {\bf s}) &=& \f{i}{2} \psib_m \sib^m
{ X}{\loco} + i{e_b}^m (\psib_m \sib_a \eps)_\varphi T^{ab
\varphi}{\loco} \nonumber
\\ && + \f {1}{2} \vep^{mnpq} \psib_m \sib_n \cd_p \psi_q +
\f{i}{6} \vep^{mnpq} \psib_m \sib_n \psi_q b_p \nonumber \\ && +
\f{i}{6} \left( \psi_n \si^m \psib_m - \psi_m \si^m \psib_n
\right)b^n + \f{1}{3} \psib_m \sib^{mn} \psib_n M,
\label{CPN.127}
\end{eqnarray}
where we have used (\ref{A.49}) and (\ref{CPN.23}) as
well as other formulas given in appendix \ref{appA}.
Finally, from (\ref{GRA.169}) and (\ref{CPN.27}), we infer
\begin{eqnarray}
{\bf f} + \bar {\bf f} & = & - \f{1}{4} {\cal R} - i{e_b}^m
(\psib_m \sib_a \eps )_\varphi T^{ab \varphi} {\loco} - \f{1}{4}
\cd^\al { X}_\al {\loco} +\f{1}{6} b^a b_a \nn + \f{1}{3} M
\ovM\\ & & + \f{i}{12} \vep^{mnpq} \psib_m \sib_n \psi_q b_p  +
\f{1}{6} M \psib_m \sib^{mn} \psib_n
\ \ + \ \ \mbox{h. c.} \  , \label{CPN.128}
\end{eqnarray}
with the curvature scalar ${\cal R}$
defined in (\ref{CPN.26}).

Recapitulating, the Lagrangian (\ref{CPN.123}) becomes
\begin{eqnarray}
 e^{-1} \LSM & = & - \f{1}{4}{\cal R} + \f
{1}{2} \vep^{mnpq} \psib_m \sib_n (\cd_p \psi_q + \f{i}{2} b_p
\psi_q ) \nonumber \\ && - \f{1}{6} M \ovM + \f{1}{6} b^a b_a -
\f{1}{4} \cd^\al { X}_\al {\loco} + \f{i}{2} \psib_m \sib^m
{ X}{\loco} \ \ + \ \ \mbox{h. c.} \
\label{CPN.129}  \nn \\[2mm]
& = & - \f {1}{2}
{\cal R} + \f {1}{2} \vep^{mnpq} (\psib_m \sib_n {\nabla}_p \psi_q
- \psi_m \si_n {\nabla}_p \psib_q ) \nonumber \\ && - \f{1}{3} M
\ovM + \f{1}{3} b^a b_a + \DM.
\label{CPN.130}
\end{eqnarray}
The cancellation of the $\psi_m b_n \psib_p$
terms with those coming from (\ref{CPN.40}) is manifest in terms of
the new covariant derivatives
  \begin{eqnarray}
   {\nabla}_n
\psi^\al_m & = & \prt_n \psi^\al _m
+ \psi_m^\beta \, {\om_{n \beta}}^\al
\nonumber
\\[1mm] &&
+ \f{1}{4} {\psi_m}^\al \left( K_k \cd_n A^k
-  K_{\bk} \cd_n A^{\bk}
+ i g_{k \bk} \chi^k \si_n \chib^{\bk} \right),
\label{CPN.131} \\[2mm]
{\nabla}_n \psib_{m \da} & = & \prt_n \psib_{m \da}
+ \psib_{m \db} \, {{\om_n}^{\db}}_{\da}
\nonumber
\\[1mm] &&
 - \f{1}{4} \psib_{m \da} \left( K_k \cd_n A^k
-  K_{\bk} \cd_n A^{\bk}
+ i g_{k \bk} \chi^k \si_m \chib^{\bk} \right),
\label{CPN.132}
\end{eqnarray}
which are fully
Lorentz, K\"ahler and gauge covariant derivatives.
 Finally, the
expression of
$\DM$, defined in (\ref{CPN.2}), in terms of
the component fields has been given explicitly in (\ref{CPN.59}).

We now see explicitly what was stressed in the introduction to
this section: the explicit dependence in the matter fields appears
only through the D-term induced by the K\"ahler structure $e \DM$;
the rest of the Lagrangian has the form of
the standard supergravity Lagrangian. It should be kept in mind, however,
that all the covariant derivatives in
$\LSM$ are now covariant also with respect to the K\"ahler and
Yang-Mills transformations.

\subsubsection{ Superpotential \label{CPN52}}
We now turn to the potential term in the Lagrangian and consider
\begin{equation}
\rSPOT \,=\, e^{K/2} W.
\label{CPN.133}
\end{equation}
In order to identify the corresponding component fields we have to apply
covariant spinor derivatives. Since neither $K$ nor $W$ are tensors with
respect to the \ka phase transformations we make use of
$\cd_\al {\bf r} \,= \,{E_\al}^M \prt_M {\bf r} + 2 {A}_\al {\bf r}$,
before applying the product rule.
Recall that in (\ref{kelkon}), the explicit form of
${A}_\al$  is given as
\begin{equation}
{A}_\al \,= \,\f{1}{4} K_k \cd_\al \phi^k , \label{CPN.134}
\end{equation}
in terms of the usual Yang-Mills covariant derivative.
Using furthermore the requirement that $W$ as well as $K$ are Yang-Mills
invariant, we obtain
\begin{equation}
{E_\al}^M \prt_M W \,=\, W_k \cd_\al \phi^k, \cem {E_\al}^M \prt_M
K \,=\, K_k \cd_\al \phi^k . \label{CPN.138} \end{equation}
 Adding
these three contributions yields
\begin{equation}
\cd_\al r \,=\, e^{K/2} ( K_k W + W_k ) \cd_\al \phi^k.
\label{CPN.139} \end{equation} Let us note that the combination $
( K_k W + W_k )$ behaves as $W$ under \ka transformations, \ie
\begin{equation}
W \,\mapsto \, e^{-F} W \cem \mbox{then}\cem  ( K_k W + W_k )
\,\mapsto e^{-F}  ( K_k W + W_k ). \nn
\end{equation}
This suggests to denote
\begin{equation}
 ( K_k W + W_k )= D_k W,
\end{equation}
and we obtain
\begin{equation}
{\bf s}_\al
         \,=\, e^{K/2} \, \chi^k_\al D_k W . \label{CPN.140}
\end{equation}
The evaluation of $\cd^\al \cd_\al \rSPOT$ proceeds along the same lines.
Taking carefully into account the \ka structure leads to
 \begin{eqnarray}
  \cd^\al \cd_\al {\bf r} \,&=&\, + e^{K/2} (K_k W
+ W_k) \cd^\al \cd_\al \phi^k
+ e^{K/2} \left[
(K_{ij} - K_k {\Gamma^k}_{ij} + K_i K_j) W \right.
    \nonumber \\[1mm]
& &
\left. + ( W_{ij} - W_k {\Gamma^k}_{ij}+ W_j K_i +
W_iK_j)\right] \cd^\al \phi^i \cd_\al \phi^j.
\label{CPN.141}
\end{eqnarray}
Observe that the expression inside brackets is just equal to
$(\prt_i+K_i)D_j W-{\Gamma^k}_{ij} D_k W$ and transforms as $W$
and $D_i W$ under \ka (the presence of the Levi-Civita
symbol ensures the covariance of the derivatives with
respect to K\"ahler manifold indices). Again, this suggests the definition
\begin{equation}
D_i D_j W \, = \,  (\prt_i+K_i)D_j W-{\Gamma^k}_{ij} D_k W,
\end{equation}
giving rise to the compact expression
  \begin{equation}
{\bf f}
\, = \, e^{K/2}\left[ F^k D_k W
        - \f{1}{2} \, \chi^i\chi^j \, D_i D_j W \right],
\label{CPN.142}
\end{equation}
for the F-term component field.
Substituting in the generic formula
(\ref{CPN.123}), yields the Lagrangian
\begin{eqnarray}
 e^{-1}\LSPOT &=& e^{K/2}
\left[ F^k D_k W + \bF^\bk  D_\bk \ovW  -M\ovW -\ovM W
 \right]
\nonumber \\
& &
- \frac{e^{K/2}}{2}\left[
\chi^i\chi^j \, D_i D_j W  +
\chib^\bi \chib^\bj \, D_\bi D_\bj \ovW
\right]
\nonumber
\\ & &
+  \frac{e^{K/2}}{\sqrt{2}}\left[
i (\psib_m \sib^m \chi^k) D_k W
+ i (\psi_m \si^m \chib^\bk) D_\bk\ovW\right.
\nonumber
\\ & &\cem \ \  \left.
- \sqrt{2} (\psib_m \sib^{mn} \psib_n) W
- \sqrt{2} ( \psi_m \si^{mn} \psi_n) \ovW \,\right] .
\label{CPN.143}
\end{eqnarray}

\subsubsection{ Yang-Mills}
\label{CPN53}
 Finally, to obtain the Yang-Mills Lagrangian, we start from the superfield
\begin{equation}
\rYM \,=\,\f{1}{4} f_{(r)(s)} \cw^{(r)\al} \cw^{(s)}_\al,
\label{CPN.144}
\end{equation}
with lowest component
\begin{equation}
{\bf r} \,=\,- \f{1}{4} f_{(r)(s)} (\la^{(r)}\la^{(s)}).
\label{CPN.145}
\end{equation}
Applying a covariant spinor
derivative to $\rYM$ and using the
transformation properties of $f_{(r)(s)}$ and $ {\bar f}_{(r)(s)}$ as given
in (\ref{GRA.249}), (\ref{GRA.250}), together with
(\ref{CPN.62}) yields
\begin{eqnarray}
 \cd_{\! \al} \, \rYM &=& -\f{1}{4}
f_{(r)(s)} \cw^{(r)}_\al \cd^\vp \cw^{(s)}_\vp + \f{1}{2}
f_{(r)(s)} ( \si^{ba} \cw^{(r)})_\al \cf^{(s)}_{ba} \nonumber
\\[2mm] & &
+ \f{1}{4} \frac{\prt f_{(r)(s)}}{\prt \phi^i } \cd_\al
\phi^i \cw^{(r)\varphi} \cw^{(s)}_\varphi.
\label{CPN.148}
\end{eqnarray}
It remains to project to the lowest superfield components
-cf. (\ref{CPN.30}) (\ref{CPN.37}), (\ref{CPN.106}), giving rise to
 \begin{eqnarray}
 {\bf s}_\al &=& \frac{-i}{2\sqrt{2}} f_{(r)(s)}
\left[
\la^{(r)}_\al
{\bf D}^{(s)} + ( \si^{mn} \la^{(r)})_\al
(i {\fym}^{(s)}_{mn} +
\psi_m\si_n \lab^{(s)} + \psib_m\sib_n\la^{(s)}) \right]
\nonumber
\\ & &
- \f{1}{4} \frac{\prt f_{(r)(s)}}{\prt A^i } \chi^i_\al
(\la^{(r)} \la^{(s)}), \label{CPN.149}
\end{eqnarray}
with ${\fym}^{(s)}_{pq}$ defined in (\ref{CPN.66}).
Similarly, using (\ref{CPN.62}) and (\ref{B.129}), we obtain
\begin{eqnarray}
 \cd^\al \cd_\al {\bf r} &=& -\f{1}{2} f_{(r)(s)} \left(
\f{1}{2} (\cd^\al \cw_\al^{(r)})(\cd^\beta \cw_\beta^{(s)}) +
\cf^{(r)ba} \cf^{(s)}_{ba} + \f{i}{2} \eps^{dcba} \cf^{(r)}_{dc}
\cf^{(s)}_{ba} \right)   \nonumber \\ && +\f{1}{2} f_{(r)(s)}
\cw^{(r)\al} \left( 12 R^\dagger \ \cw^{(s)}_\al + 4i
\si^a_{\al\da} \cd_a \cw^{(s)\da} \right) \nonumber \\ &&
-\frac{\prt f_{(r)(s)}}{\prt\phi^i} \left( \f{1}{2} \cd^\al\phi^i
\ \cw_\al^{(r)} \cd^\beta \cw_\beta^{(s)} - \cd^\al\phi^i
{(\si^{ba})_\al}^\beta \cw_\beta^{(r)} \cf_{ba}^{(s)} \right)
    \nonumber \\
&& + \f{1}{4} \frac{\prt f_{(r)(s)}}{\prt\phi^i} \ \cd^\al\cd_\al
\phi^i \ \cw^{(r)\varphi} \cw_\varphi^{(s)} \nonumber \\ && +
\f{1}{4} \left( \frac{\prt^2 f_{(r)(s)}}{\prt\phi^i\prt\phi^j} -
{\Gamma^l}_{ij} \frac{\prt f_{(r)(s)}}{\prt\phi^l} \right) \cd^\al
\phi^i \cd_\al \phi^j \ \cw^{(r)\varphi} \cw^{(s)}_\varphi.
\label{CPN.150}
\end{eqnarray}

One recognizes in the last line the covariant derivative of
$f_{(r)(s)}$ with respect to K\"ahler manifold indices. The
corresponding component field expression is
\begin{eqnarray}
 {\bf f} &=& - \f{1}{4} f_{(r)(s)} \left[\  \f{1}{2}
{\fym}^{(r)mn} {\fym}^{(s)}_{mn} +  \f{i}{4} \vep^{mnpq} \
{\fym}^{(r)}_{mn} {\fym}^{(s)}_{pq} + 2i \la^{(r)} \si^m \cd_m
\lab^{(s)} \right. \nonumber \\ &&\hs{19mm} - {\bf D}^{(r)}{\bf
D}^{(s)} + \ovM \la^{(r)} \la^{(s)} - (\la^{(r)} \si^m \psib_m)
{\Bbb D}^{(s)} \nonumber \\ &&\hs{19mm}\left. -i( \psi_m \si^{pq}
\si^m \lab^{(r)} + \psib_m \sib^{pq} \sib^m \la^{(r)} - \psib_m
\sib^m \si^{pq} \la^{(r)}) \ {\fym}^{(s)}_{pq} \right]
   \nonumber \\
&& + \f{1}{4} f_{(r)(s)} \lp \f{1}{2} \psi_m \si^{pq} \si^m
\lab^{(r)} + \psib_m \sib^{pq} \sib^m \la^{(r)} - \f{1}{2} \psib_m
\sib^m \si^{pq} \la^{(r)}\rp \lp\psi_p \si_q \lab^{(s)} + \psib_p
\sib_q \la^{(s)}\rp \nonumber
\\
&& + \f{1}{4}\frac{\prt f_{(r)(s)}}{\prt A^i} \left[ - \sqrt{2}
(\chi^i \si^{mn} \la^{(r)}) {\fym}^{(s)}_{mn} + i\sqrt{2} (\chi^i
\la^{(r)}) {\bf D}^{(s)} - F^i (\la^{(r)} \la^{(s)})\right.
\nonumber \\ && \left. \hs{22mm} + i\sqrt{2} (\chi^i \si^{mn}
\la^{(r)}) (\psi_m\si_n\lab^{(s)} + \psib_m \sib_n \la^{(s)})
\right] \nonumber
\\
&&+\f{1}{8} \left(\frac{\prt^2 f_{(r)(s)}}{\prt A^i \prt A^j} -
{\Gamma^l}_{ij} \frac{\prt f_{(r)(s)}}{\prt A^l} \right)
(\chi^i\chi^j) (\la^{(r)} \la^{(s)}), \label{CPN.151}
\end{eqnarray}
 where the covariant derivative $\cd_m\lab^{(s)}$ is
defined in (\ref{CPN.39}).
Making heavy use of the
relations (\ref{A.45}) - (\ref{A.54}), we finally obtain
 \begin{eqnarray}
\lefteqn{  e^{-1} \LYM = - \f{1}{4} f_{(r)(s)}\, \times} \nn \\
[0.5ex]&& \left[ \ \f{1}{2} {\fym}^{(r)mn} {\fym}^{(s)}_{mn} +
\f{i}{4} \vep^{mnpq} \ {\fym}^{(r)}_{mn} {\fym}^{(s)}_{pq} + 2i
\la^{(r)} \si^m {\nabla}_m \lab^{(s)} - {\bf D}^{(r)}{\bf D}^{(s)}
- ( \la^{(r)} \si^a \lab^{(s)})b_a   \right. \nonumber
\\ &&\  -i {\fym}^{(r)mn} (\psi_m \si_n \lab^{(s)} + \psib_m \sib_n
\la^{(s)}) + \f{1}{2} \vep^{mnpq} {\fym}^{(r)}_{mn} (\psi_p \si_q
\lab^{(s)} - \psib_p \sib_q \la^{(s)}) \nonumber
\\ &&\ + \f{1}{8} (\la^{(r)}\la^{(s)}) (4\psib_m\psib^m +
\psib_m \sib^{m} \si^{n} \psib_n)+ \f{1}{8} (\lab^{(r)}\lab^{(s)})
(4\psi_m\psi^m + \psi_m \si^{m} \sib^{n} \psi_n) \nonumber \\ && \
\left. - \f{1}{2} ( g^{mp}g^{nq} - g^{mq}g^{np} -i \vep^{mnpq})
(\psib_m \sib_n \la^{(r)})(\psi_p \si_q \lab^{(s)}) \right]
\nonumber \\
&& - \f{1}{4} \frac{\prt f_{(r)(s)}}{\prt A^i} \left[ \sqrt{2}
(\chi^i \si^{mn} \la^{(r)}) {\fym}^{(s)}_{mn} - i\sqrt{2} (\chi^i
\la^{(r)}) {\bf D}^{(s)} + F^i (\la^{(r)} \la^{(s)})\right.
\nonumber \\ &&\hs{21mm} \left. -i \f{\sqrt{2}}{4} (\la^{(r)}
\la^{(s)}) (\psib_m \sib^m \chi^i) - i\sqrt{2}
(\psi_m\si_n\lab^{(r)}) (\chi^i \si^{mn} \la^{(s)})
  \right] \nonumber \\
&&+\f{1}{8} \left( \frac{\prt^2 f_{(r)(s)}}{\prt A^i \prt A^j} -
{\Gamma^l}_{ij} \frac{\prt f_{(r)(s)}}{\prt A^l} \right)
(\chi^i\chi^j) (\la^{(r)} \la^{(s)})
\ \ + \ \ \mbox{h. c.} \  .
\label{CPN.152}
\end{eqnarray}
The Yang-Mills field strength ${\fym}^{(r)}_{mn}$ is defined in
(\ref{CPN.66}). The covariant derivatives
\begin{eqnarray}
 {\nabla}_m \la_\al^{(r)}   & = &
\prt_m \la_\al^{(r)} - {\om_{m \al}}^\varphi \la_{\varphi}^{(r)}-
 {\aym}^{(t)}_m   {c_{(s)(t)}}^{(r)} \la^{(s)}_\al
 \nonumber \\ && + \f{1}{4} (K_j \cd_m A^j -  K_{\bj} \cd_m
\bA^{\bj}) \la_\al^{(r)} + \f{i}{4} g_{j\bk} (\chi^j
\si_m\chib^\bk)\ \la^{(r)}_\al ,
                \label{CPN.153} \\[2mm]
{\nabla}_m \lab^{{(r)}\da} & = & \prt_m \lab^{{(r)} \da} -
{{\om_{m}}^{\da}}_{\dv} \lab^{{(r)} \dv} -
{\aym}_m^{(t)}{c_{(s)(t)}}^{(r)} \lab^{{(s)} \da} \nonumber \\ &&
- \f{1}{4} ( K_k \prt_m A^k -  K_{\bk} \prt_m \bA^{\bk})
\lab^{{(r)} \da}
 -\f{i}{4} g_{j\bk} (\chi^j \si_m\chib^\bk)\
\lab^{{(r)}\da} . \label{CPN.154} \end{eqnarray}
differ from the covariant derivatives $\cd_m \la^{(r)}_\al$ and $\cd_m
\lab^{(r)\da}$ introduced in (\ref{CPN.38}), (\ref{CPN.39}) by the
covariant $b_a$ dependent term appearing in the definition of
${A}_m$, in analogy with previous definitions
-cf. (\ref{CPN.60}), (\ref{CPN.61}) and (\ref{CPN.131}), (\ref{CPN.132}).

\subsubsection{ Recapitulation}
\label{CPN54}

The complete Lagrangian describing the
interaction of Yang-Mills and chiral supermultiplets with supergravity
is given by the sum of (\ref{CPN.130}), (\ref{CPN.143}), and
(\ref{CPN.152}), with the matter D-term given in (\ref{CPN.59}). In taking
the sum, we diagonalize in the auxiliary field sector, with the result
\begin{eqnarray}
\lefteqn{ e^{-1} \cl =}\nn \\[0.8ex] && - \f {1}{2} {\cal R} + \f
{1}{2} \vep^{mnpq} (\psib_m \sib_n {\nabla}_p \psi_q - \psi_m
\si_n {\nabla}_p \psib_q )  \nonumber
\\[0.5ex] &&
  - g_{i \bj} \, \cd^m A^i
 \cd_m \bA^{\bj} - \f{i}{2}g_{i \bj}\,\lp \chi^{i}  \si^m
{\nabla}_m \chib^{\bj}
 +\chib^{\bj}\sib^m\,{\nabla}_m \chi^{i}\rp \nn \\[0.5ex] &&
  - \f{1}{4}\rea f_{(r)(s)}\,{\fym}^{(r)mn} {\fym}^{(s)}_{mn}
  +\f{1}{8}\ima f_{(r)(s)}\,\vep_{mnpq}\,{\fym}^{(r)mn} {\fym}^{(s)pq}
 \nn \\[0.5ex] && -\f{i}{2}\left[ f_{(r)(s)}
   \la^{(r)} \si^m{\nabla}_m \lab^{(s)} +\bar{f} _{(r)(s)}\lab^{(r)} \sib^m{\nabla}_m
   \la^{(s)}\right] \nn \\[0.5ex] &&+ 3 e^{K} |W|^2 - g^{i \bj}\, e^K
 D_iW \, D_\bj \ovW - \frac{e^{K/2}}{2}
 \left[\,D_i D_j W \, (\chi^i\chi^j) + D_\bi D_\bj \ovW \,
 (\chib^\bi \chib^\bj)\right] \nonumber \\[0.5ex] &&
 +\f{1}{4}\lp R_{i \bj k \bl}+ \f{3}{2} g_{i \bj}\, g_{k \bl}\rp
 \, (\chi^i \chi^k) (\chib^{\bj} \chib^{\bl})- \f{3}{4} g_{i \bj}\,\rea f_{(r)
(s)}\,(\chi^i \la^{(r)} )(\chib^\bj \lab^{(s)})
 \nonumber \\[0.5ex] && -i \sqrt{2} (\chi^i \la^{(r)}) g_{i \bj}\lp {
\bA{\bf T }_{(r)}} \rp ^{\bj} +i \sqrt{2} (\chib^{\bj} \lab^{(r)})
g_{i \bj}\lp {{{\bf T}_{(r)}} A}\rp ^i \nonumber \\[0.5ex] && -
\f{1}{2 \sqrt{2}}\left[ \frac{\prt f_{(r)(s)}}{\prt A^i}
 \,(\chi^i \si^{mn} \la^{(r)})+
 \frac{\prt {\bar{f}}_{(r)(s)}}{\prt \bA^\bi}
 \, (\chib^\bi \sib^{mn} \lab^{(r)})\right] {\fym}^{(s)}_{mn}\nn
 \\[0.5ex]
&& +\f{1}{8}\left[ \left( \frac{\prt^2 f_{(r)(s)}}{\prt A^i \prt
A^j} - {\Gamma^l}_{ij} \frac{\prt f_{(r)(s)}}{\prt A^l} \right)
(\chi^i\chi^j)+2 g^{i \bj} e^{K/2}\,D _\bj \ovW\,\frac{\prt
f_{(r)(s)}}{\prt A^i}\, \right](\la^{(r)} \la^{(s)})  \nonumber
\\[0.5ex] && +\f{1}{8}\left[ \left( \frac{\prt^2 \bar{f}_{(r)(s)}}{\prt \bA^\bi
\prt \bA^\bj} - {\Gamma^\bl}_{\bi \bj} \frac{\prt
\bar{f}_{(r)(s)}}{\prt \bA^\bl} \right) (\chib^\bi \chib^\bj)+2
g^{i \bj}\, e^{K/2}\,D_i W\,
 \frac{\prt {\bar{f}}_{(r)(s)}}{\prt \bA^\bj}\,
\right] (\lab^{(r)} \lab^{(s)}) \nn \\[0.5ex] &&+\lp \f{3}{8} \rea
f_{(r) (t)}\, \rea f_{(s) (u)}\,  -\f{1}{16} g^{i \bj}\,
\frac{\prt f_{(r)(s)}}{\prt A^i}\, \frac{\prt
{\bar{f}}_{(t)(u)}}{\prt \bA^\bi}\,\rp(\la^{(r)} \la^{(s)})
(\lab^{(t)} \lab^{(u)})\nn \\[0.5ex] &&
 +\f{1}{2} \lp \rea \,f_{(r) (s)} \rp^{-1} \left[ K_i \lp {{{\bf T}_{(r)}} A}\rp ^i
 -\f{i}{\sqrt{2}}\frac{\prt f_{(r)(t)}}{\prt A^i}\, (\chi^i
 \la^{(t)}) \right] \nn \\[0.5ex] && \hs{27.5mm} \times \left[K_\bj \lp {
\bA{\bf T }_{(s)}} \rp ^{\bj}  +\f{i}{ \sqrt{2}} \frac{\prt
{\bar{f}}_{(s)(u)}}{\prt
 \bA^\bj}\,(\chib^\bj \lab^{(u)}) \right]
\nonumber \\[0.5ex]
 && - \frac{1}{ \sqrt{2}} (\psib_m \sib^n \si^m \chib^{\bj})
g_{i \bj} \cd_n A^i - \frac{1}{ \sqrt{2}} (\psi_m \si^n \sib^m
\chi^i) g_{i \bj} \cd_n \bA^{\bj}\nn \\[0.5ex] &&-\f{1}{4}(\psib_m
\sib^m \la^{(r)} - \psi_m \si^m \lab^{(r)} ) \kilc \nn
\\[0.5ex]
 && +i \f{1}{ 8\sqrt{2}}\left[\frac{\prt f_{(r)(s)}}{\prt A^i}
(\la^{(r)} \la^{(s)}) (\psib_m \sib^m\chi^i)+ \frac{\prt
{\bar{f}}_{(r)(s)}}{\prt \bA^\bi}(\lab{(r)} \lab^{(s)}) (\psi_m
\sib^m\chib^\bi) \right] \nn \\[0.5ex] &&
 + i\f{1}{2\sqrt{2}} \left[\frac{\prt f_{(r)(s)}}{\prt A^i}
  (\psi_m\si_n\lab^{(r)}) (\chi^i \si^{mn}
\la^{(s)})+\frac{\prt \bar{f}_{(r)(s)}}{\prt \bA^\bi}
  (\psib_m\sib_n \la^{(r)}) (\chib^\bi \sib^{mn}
\lab^{(s)})\right]
  \nonumber \\[0.5ex] && +\f{i}{2} \rea
f_{(r) (s)} {\fym}^{(r)mn}\left[ (\psi_m \si_n \lab^{(s)} +
\psib_m \sib_n \la^{(s)}) - \f{i}{2}\vep_{mnpq}
 (\psi^p \si^q \lab^{(s)} - \psib^p \sib^q
\la^{(s)})\right] \nonumber
\\[0.5ex] && +\, e^{K/2}\,
\left[\frac{i}{\sqrt{2}} (\psib_m \sib^m \chi^k)\, D_k W
+\frac{i}{\sqrt{2}} (\psi_m \si^m \chib^\bk)\, D_\bk \ovW -\psib_m
\sib^{mn} \psib_n W + \psi_m \si^{mn} \psi_n \ovW \, \right]
\nonumber \\[0.5ex] &&
-\frac{i}{2} g_{i \bj}\,\vep^{klmn}\lp \chi^i \si_k \chib^{\bj}\rp
\lp \psi_l \si_m \psib_n \rp - \frac{1}{ 2} g_{i \bj}\, g^{mn}
(\psi_m \chi^i)(\psib_n \chib^{\bj})
 \nonumber \\[0.5ex] && + \f{1}{16} \rea f_{(r) (s)}\,\left[(\la^{(r)}\la^{(s)})\, \psib_m (3
g^{mn}+ 2 \sib^{mn}) \psib_n+  (\lab^{(r)}\lab^{(s)})\, \psi_m (3
g^{mn}+ 2 \si^{mn}) \psi_n \right] \nonumber \\[0.5ex] && +
\f{1}{4}\left[\rea f_{(r)(s)}\,  ( g^{mp}g^{nq} -
g^{mq}g^{np})+\ima f_{(r) (s)} \, \vep^{mnpq}\right] (\psib_m
\sib_n \la^{(r)})(\psi_p \si_q \lab^{(s)})
\nn \\[0.5ex] && + \, \,
-\f{1}{3} {\bf M} {\bf \ovM}
+\f{1}{3} {\bf b}^a {\bf b}_a
+ g_{ i \bj} {\bf F}^i {\bf \bF}^\bj
+\f{1}{2}\rea f_{(r)(s)} {\bf \hat D}^{(r)} {\bf \hat D}^{(s)}.
\end{eqnarray}
The diagonalized auxiliary fields, defined as
\begin{eqnarray}
{\bf M} &=& M +3e^{K/2} W,
\\[1mm]
{\bf \ovM} &=& \ovM +3e^{K/2}\ovW,
\\[1mm]
{\bf b}^a &=& b^a - \f{3}{4} g_{i\bj} (\chi^i \si^a \chib^\bj)
+ \f{3}{4}\rea f_{(r)(s)} (\la^{(r)} \si^a \lab^{(s)}),
\\[1mm]
{\bf F}^i &=& F^i + e^{K/2} g^{i\bk}\, D_\bk \ovW  + \f{1}{4}
g^{i\bk}\,  \frac{\prt \bar{f}_{(r)(s)}}{\prt
{\bA}^\bk}(\lab^{(r)} \lab^{(s)}),
\\[1mm]
{\bf \bF}^\bj &=& \bF^\bj + e^{K/2} g^{l\bj}\, D_l W +
\f{1}{4} g^{l\bj} \, \, \frac{\prt {f}_{(t)(u)}}{\prt
{A}^l}(\la^{(t)} \la^{(u)}),
\\[1mm]
{\bf \hat D}^{(r)} &=& {\bf D}^{(r)}
-\lp\rea f_{(r)(t)} \rp^{-1}\,\left(
K_k \lp {{{\bf T}_{(t)}} A}\rp ^k -\frac{i}{\sqrt{2}}  \frac{\prt
f_{(t)(v)}}{\prt A^k}\, (\la^{(v)}\chi^k)\,\right),
\end{eqnarray}
have trivial equations of motion
which coincide with the lowest components of those
found in (\ref{GRA.256}) - (\ref{GRA.261}) in superfield language.

\indent

We would like to end this section with one comment: it was first
realized in \cite{CJS+78b,CJS+79,CFGvP82,CFGvP83} that the
Lagrangian depends only on the combination
\begin{equation}
{\cal G}\,=\,K + \ln |W|^2, \end{equation}
 and not independently on
the K\"ahler potential $K$ and the superpotential $W$. This can be
made clear in a straightforward manner in the K\"ahler superspace
formalism. Indeed, performing a K\"ahler transformation
-cf. (\ref{GRA.7})- with $F = \ln W$ yields
\begin{equation}
e^{-1} \LSPOT\,=\,\f {1}{2} \int \frac{E}{R} e^{{\cal G}/2}
\ \ + \ \ \mbox{h. c.} \ .
\end{equation}
This field dependent redefinition, which has the form of a \ka
transformation, must of course be performed in the whole geometric structure,
leading to a new superspace geometry which is completely inert under \ka
transformations. The component field expressions in this new basis, with \ka
inert components, have the same form as the previous ones, with
 $K$ replaced by ${\cal G}$
in all the implicit dependence on the K\"ahler potential and $W$
and $\ovW$ set to one. It was actually given in this basis
in \cite{BGGM87b}.

\newpage
\sect{ LINEAR MULTIPLET AND SUPERGRAVITY} \label{2F}

The antisymmetric tensor gauge potential, $b_{mn} \,=\,- b_{nm}$,
first discussed in \cite{OP67}, appears naturally in the context
of string theory \cite{KR74}.
At the dynamical level it is related to a real massless scalar field
through a duality transformation.

In supersymmetry, the antisymmetric tensor is part of the linear
multiplet \cite{FZW74,Sie79}, together with a real scalar and a Majorana
spinor. The duality with a massless scalar multiplet is most easily
established in superfield language \cite{LR83}.

Postponing the discussion of the relevance of the linear multiplet and
its couplings in low energy effective superstring theory to
the closing section \ref{conclusions}, we concentrate here on the general
description of linear multiplets in superspace and couplings to the full
supergravity/matter/Yang-Mills system, including Chern-Simons forms.

The basic idea of the {\em linear superfield formalism} is to describe
a 2-form gauge potential in the background of $U_K(1)$ superspace and to
promote the \ka potential to a more general superfield function, which
not only depends on the chiral matter superfields but also on linear
superfields.

In order to prepare the ground, section \ref{RS5} provides an elementary
and quite detailed introduction to the antisymmetric tensor gauge potential
and to linear superfields without supergravity. Whereas the superspace
geometry of the 2-form in $U_K(1)$ superspace is presented in \ref{2F2},
component fields are identified in section \ref{compfields}. In section
\ref{2F3} we explain the coupling of the linear superfield to the
supergravity/matter/Yang-Mills system. Duality transformations in this
general context, including Chern-Simons forms are discussed in section
\ref{2F4}, relating the {\em linear superfield formalism} to the
{\em chiral superfield formalism}. In section \ref{NHGC} we show that the
linear superfield formalism provides a natural explanation of
non-holomorphic gauge coupling constants. Finally, in section \ref{sevlin}
we extend our analysis to the case of several linear multiplets.

\subsec{ The Linear Multiplet In Rigid Superspace \label{RS5}}

\subsubsection{ The antisymmetric tensor gauge field
\label{RS51}}

Consider first the simple case of the antisymmetric tensor gauge
potential $b_{mn}$ in four dimensions with gauge transformations
parametrized by a four vector $\beta_m$ such that
\begin{equation}
b_{mn} \ \mapsto \ b_{mn} + \prt_m \beta_n - \prt_n \beta_m,
\end{equation}
 and with invariant field strength given as
\begin{equation}
h_{0 \, lmn} \,=\,\prt_l b_{mn} + \prt_m b_{nl} + \prt_n b_{lm}.
\end{equation}
 The subscript $0$ denotes here the absence of Chern-Simons
forms. As a consequence of its definition the field strength
satisfies the Bianchi identity
\begin{equation}
\vep^{klmn} \prt_k h_{0 \, lmn}\,=\,0. \end{equation}
 The invariant kinetic action is given as
\begin{equation}
\cl \,=\,\f12 \, { {}^* h}_0^m \,{ {}^* h}_{0 \, m}, \label{hh}
\end{equation}
 with ${}^* h_0^k = \f{1}{3!} \vep^{klmn} h_{0 \, lmn}$
denoting the Hodge dual of the field strength tensor.

\indent

Consider next the case where a Chern-Simons term for a Yang-Mills
potential ${\aym}_m$, such as
\begin{equation}
Q_{lmn} \,=\,- \tr \lp {\aym}_{[l} \prt_m {\aym}_{n]} - \f{2i}{3}
{\aym}_{[l} {\aym}_m {\aym}_{n]} \rp, \label{cs}
 \end{equation}
with $[lmn]=lmn + mnl + nlm - mln - lnm - nml$, is added to the
field strength,
\begin{equation}
h_{lmn} \,=\,h_{0 \, lmn}+ k \, Q_{lmn}. \label{modh}
\end{equation}
 Here $k$
is a constant which helps keeping track of the terms induced by
the inclusion of the Chern-Simons combination. The Chern-Simons
term is introduced to compensate the Yang-Mills gauge
transformations to the antisymmetric tensor, thus rendering the
modified field strength invariant. The Bianchi identity is
modified as well; it now reads
\begin{equation}
\vep^{klmn} \prt_k h_{lmn} \,=\,-\f{3}{2} k \, \vep^{klmn}
\tr({\fym}_{kl}{\fym}_{mn}). \end{equation}

A dynamical theory may then be obtained from the invariant action
\begin{equation}
\cl \,=\,\f12 \, { {}^* h}^m \,{{}^* h}_m -\f{1}{4} \,
\tr({\fym}^{mn}{\fym}_{mn}), \label{hf} \end{equation}
 with ${
{}^* h}^k $ the dual of $h_{lmn}$. This action describes the
dynamics of Yang-Mills potentials ${\aym}_m(x)$ and an
antisymmetric tensor gauge potential $b_{mn}$ with effective
$k$-dependent couplings induced through the Chern-Simons form.

\indent

This theory is dual to another one where the antisymmetric tensor
is replaced by a real pseudoscalar $a(x)$ in the following sense:
one starts from a first order action describing a vector $X^m(x)$,
a scalar $a(x)$ and the Yang-Mills gauge potential ${\aym}_m(x)$,
\begin{equation}
\cl \,=\,( X^m - k \, { {}^* Q}^m ) \prt_m a
         + \f12 \, X^m X_m -\f{1}{4} \, \tr({\fym}^{mn}{\fym}_{mn}),
\label{ford} \end{equation}
 where the gauge Chern-Simons form is
included as
\begin{equation}
{ {}^* Q}^k \,=\,\f{1}{3!} \vep^{klmn} Q_{lmn}
      \,=\,- \vep^{klmn} \, \tr \lp {\aym}_l
      \prt_m {\aym}_n -\f{2i}{3} {\aym}_l {\aym}_m {\aym}_n \rp.
      \label{Qstar}
\end{equation}
 Variation of the first order action with respect to the field $a$
  gives rise to an equation of motion which is solved in
terms of an antisymmetric tensor
\begin{equation}
\prt_m (X^m -k\, { {}^* Q}^m) \,=\, 0, \hspace{.7cm} \Rightarrow
\hspace{.7cm} X^k - k \, { {}^* Q}^k \,=\,\f12 \vep^{klmn} \prt_l
b_{mn}.
 \end{equation}
Substituting back shows that the first term in (\ref{ford})
becomes a total derivative and we end up with the previous action
(\ref{hf}) where ${ {}^* h}^m=X^m$, describing an antisymmetric
tensor gauge field coupled to a gauge Chern-Simons form.

On the other hand, varying the first order action with respect to
$X^m$ yields
\begin{equation}
X_m \,=\,- \prt_m a. \end{equation}
 In this case, substitution of
the equation of motion, together with the divergence equation for
the Chern-Simons form, \ie
\begin{equation}
\prt_k { {}^* Q}^k \,=\,- \f{1}{4} \vep^{klmn} \tr \lp {\fym}_{kl}
{\fym}_{mn} \rp . \end{equation}
 gives rise to a
theory describing a real scalar field with an axion coupling term
\begin{equation}
\cl \,=\,- \frac{1}{2} \prt^m a(x) \, \prt_m a(x)
          - \frac{1}{4} \tr({\fym}^{mn}{\fym}_{mn})
          - \frac{k}{4} \, a(x) \, \vep^{klmn} \tr({\fym}_{kl} {\fym}_{mn}).
\label{af} \end{equation} It is in this sense that the two actions
(\ref{hf}) and (\ref{af}) derived here from the first order one
(\ref{ford}) are dual to each other. They describe alternatively
the dynamics of an antisymmetric tensor gauge field or of a real
pseudoscalar, respectively, with special types of Yang-Mills
couplings. Indeed, the pseudoscalar field is often referred to as
an axion because of its couplings (\ref{af}) to Yang-Mills fields
(although it is not necessarily the QCD axion). Note that the
kinetic term of the Yang-Mills sector is not modified in this
procedure.

\subsubsection{ The linear superfield \label{RS52}}

As already mentioned, the linear supermultiplet consists of an
antisymmetric tensor, a real scalar and a Majorana spinor. In
string theories, the real scalar is the dilaton found among the
massless modes of the gravity supermultiplet. As $b_{mn}$ is the
coefficient of a 2-form, we can describe its supersymmetric
version by a 2-form in superspace with appropriate constrains and
build the corresponding supermultiplet by solving the Bianchi
identities. We shall proceed this way in section \ref{2F2}.
 In superfield
language it is described by a superfield $L_0$, subject to the
constrains
\begin{equation}
D^2 L_0 \,=\,0, \cem \bD^2 L_0 \,=\,0. \label{lincon}
\end{equation}
 Again, the subscript $0$ means that we do not
include, for the moment, the coupling to Chern-Simons forms. The
linear superfield $L_0$ contains the antisymmetric tensor only
through its field strength $h_{0 \, lmn}$. Indeed, the superfield
$L_0$ is the supersymmetric analogue of $h_{0 \, lmn}$ (it
describes the multiplet of field strengths) and the constrains
(\ref{lincon}) are the supersymmetric version of the Bianchi
identities. The particular form of these constraints implies that
terms quadratic in $\th$ and $\thb$ are not independent component
fields; it is for this reason that $L_0$ has been called a {\em
linear superfield} \cite{FZW74}.

As before, component fields are identified as projections to
lowest superfield components. To begin with, we identify the real
scalar $L_0(x)$ of the linear multiplet as the lowest component
\begin{equation}
L_0{\loco} \,=\,L_0(x). \end{equation}
 The spinor
derivatives of superfields are again superfields and we define the
Weyl components $(\La_\al (x), \Lab^\da (x))$ of the Majorana
spinor of the linear multiplet as
\begin{equation}
D_\al L_0{\loco} \,=\,\La_\al (x), \cem D^\da L_0{\loco}
\,=\,\Lab^\da (x). \end{equation}
 The
antisymmetric tensor appears in $L_0$ via its field strength
identified as
\begin{equation}
\l[ D_\al, D_\da \r] L_0{\loco} \, = \, - \frac{1}{3} \si_{k \,
\al \da} \, \vep^{klmn} h_{0 \, lmn} \,=\, -2\si_{k \, \al \da}
{}^*{h_{0}}^k,
\end{equation}
 thus completing the identification of the independent
component fields contained in $L_0$.

 The canonical supersymmetric
kinetic action for the linear multiplet is then given by the
square of the linear superfield integrated over superspace, \ie in
the language of projections to lowest superfield components,
 \begin{eqnarray}
\cl \ &=& \ -\f{1}{32} \lp D^2 \bD^2 + \bD^2 D^2 \rp
(L_0)^2{\loco} \nn \\ &=& \ \f12 { {}^* h}_0^m \, { {}^* h}_{0 \,
m}  -  \f12 \prt^m L_0 \, \prt_m L_0
                   -\f{i}{2} \si^m_{\al \da} ( \La^\al \prt_m \Lab^\da
                                        +\Lab^\da \prt_m \La^\al ),
\end{eqnarray}
 generalizing the purely bosonic action (\ref{hh}) given above
and showing that there is no auxiliary field in the linear
multiplet.

\indent

In order to construct the supersymmetric version of (\ref{hf}), we
come now to the supersymmetric description of the corresponding
Chern-Simons forms. They are described in terms of the
Chern-Simons superfield $\Om$, which has the properties
\begin{equation}
 \tr(\cw^\al \cw_\al) \ =\  \f{1}{2} \bD^2 \Om,\cem
  \tr(\cw_\da \cw^\da) \ =\ \f{1}{2} D^2 \Om. \label{om2}
\end{equation}
The appearance of the differential operators $D^2$ and $\bD^2$ is
due to the chirality constraint (\ref{RS.120}) on the gaugino
superfields $\cw^\al, \cw_\da$, whereas the additional constraint
(\ref{RS.121}) is responsible for the fact that one and the same
real superfield $\Om$ appears in both equations. The component
field Chern-Simons form (\ref{cs}) is then identified in the
lowest superfield component
\begin{equation}
[D_\al , D_\da] \, \Om{\loco} \,=\,- 2 \si_{k \al \da} \,
{{}^*Q}^k
         - 4 \, \tr(\la_\al \lab_\da),
\end{equation}
 with ${{}^*Q}^k$ given in (\ref{Qstar}).

Since the terms on the left-hand side in (\ref{om2}) are gauge
invariant, it is clear that a gauge transformation adds a linear
superfield to $\Om$. The explicit construction given in appendix
\ref{appF2}, in the full supergravity context, shows that, up to a
linear superfield, we may identify
\begin{equation}
 L \,=\,L_0 + k \, \Om,     \end{equation}
such that $L$ is gauge invariant. However, this superfield $L$
satisfies now the modified linearity conditions
 \begin{eqnarray}
  \bD^2 L &=& 2k \,
\tr(\cw^\al \cw_\al), \label{modlin1}
\\
D^2 L &=& 2k \, \tr(\cw_\da \cw^\da), \label{modlin2}
\end{eqnarray}
 Again, these equations together with
\begin{equation}
[D_\al , D_\da] L \,=\,- \f{1}{3} \si_{d \al \da} \, \eps^{dcba}
H_{cba}
                               - 4k \, \tr(\cw_\al \cw_\da),
\end{equation}
 have an interpretation as Bianchi identities in superspace
geometry. The last one shows how the usual field strength of the
antisymmetric tensor together with the Chern-Simons component
field appears in the superfield expansion of $L$,
\begin{equation}
[D_\al , D_\da] L{\loco} \, = \,- \si_{k \al \da} \, \vep^{klmn}
\lp \prt_l b_{mn} + \f{k}{3} Q_{lmn} \rp
 - 4k \, \tr(\la_\al \lab_\da).
\end{equation}
The invariant action for this supersymmetric system is given as
the lowest component of the superfield
\begin{equation}
\cl \,=\,-\f{1}{32} \lp D^2 \bD^2 + \bD^2 D^2 \rp L^2
         -\f{1}{16} D^2 \tr \cw^2
         -\f{1}{16} \bD^2 \tr \cwb^2.
\label{sushf} \end{equation}
 This action describes the
supersymmetric version of the purely bosonic action (\ref{hf}).
Its explicit component field gestalt will be displayed and
commented on in a short while.

\indent

The notion of duality  can be extended to supersymmetric theories
as well \cite{LR83}; this is most conveniently done in the
language of superfields. The supersymmetric version of the first
order action (\ref{ford}) is given as
\begin{equation}
\cl \,=\,-\f{1}{32} \lp D^2 \bD^2 + \bD^2 D^2 \rp
                   \lp X^2 + \s2 (X - k \Om)(S + \bS) \rp
         -\f{1}{16} D^2 \tr \cw^2 -\f{1}{16} \bD^2 \tr \cwb^2.
\label{susford} \end{equation} Here, $X$ is a real but otherwise
unconstrained superfield, whereas $S$ and $\bS$ are chiral,
\begin{equation}
D_\al \bS \,=\,0, \cem \bD^\da S \,=\,0. \end{equation}
 Of
course, the chiral multiplets are going to play the part of the
scalar field $a(x)$ in the previous non-supersymmetric discussion.

Varying the first order action with respect to the superfield $S$
or, more correctly, with respect to its unconstrained
pre-potential $\Si$, defined as $S=\bD^2 \Si$, the solution of
the chirality constraint, shows immediately (upon integration by
parts using spinor derivatives) that the superfield $X$ must
satisfy the modified linearity condition. It is therefore
identified with $L$ and we recover the action (\ref{sushf}) above.

On the other hand, varying the first order action (\ref{susford})
with respect to $X$ yields the superfield equation of motion
\begin{equation}
 X \,=\,-\f{1}{\s2} (S + \bS).
\end{equation}
 Substituting for $X$ in (\ref{susford}) and
observing that the terms $S^2$ and $\bS^2$ yield total derivatives
which are trivial upon superspace integration, we arrive at
\begin{equation}
\cl \,=\,\f{1}{32} \lp D^2 \bD^2 + \bD^2 D^2 \rp
                   \lp \bS S + k \s2 \, \Om \, (S + \bS) \rp
         -\f{1}{16} D^2 \tr \cw^2 -\f{1}{16} \bD^2 \tr \cwb^2.
\end{equation}
 One recognizes the usual superfield
kinetic term for the chiral multiplet and the Yang-Mills kinetic
terms. It remains to have a closer look at the terms containing
the Chern-Simons superfield. Taking into account the chirality
properties for $S$ and $\bS$ and the derivative relations
(\ref{om2}) for the Chern-Simons superfields we obtain, up to a
total derivative,
\begin{eqnarray}
 \cl &=& \f{1}{32} \lp D^2 \bD^2 + \bD^2
D^2 \rp \bS S
             -\f{1}{16} D^2 \tr \cw^2 -\f{1}{16} \bD^2 \tr \cwb^2 \nn \\[2mm]
       & &   +\f{k \s2}{8} D^2 \lp S \, \tr \cw^2 \rp
             +\f{k \s2}{8} \bD^2 \lp \bS \, \tr \cwb^2 \rp.
\label{susaf} \end{eqnarray}
 This action is the supersymmetric version
of the action (\ref{af}).

\indent

The component field expressions for the two dual versions
(\ref{sushf}) and (\ref{susaf}) of the supersymmetric construction
are then easily derived. In the antisymmetric tensor version, the
complete invariant component field action deriving from
(\ref{sushf}) is given as
 \begin{eqnarray}
  \cl &=& \f12 { {}^*
h}^m { {}^* h}_m - \f12 \prt^m L \ \prt_m L
                  -\f{i}{2} \si^m_{\al \da} \lp \La^\al \prt_m \Lab^\da
                           + \Lab^\da \prt_m \La^\al \rp \nn \\[2mm]
& & + (1+2kL) \, \tr \left[ - \f{1}{4} {\fym}^{mn} {\fym}_{mn}
                   - \f{i}{2} \si^m_{\al \da} \lp \la^\al \cd_m \lab^\da
            + \lab^\da \cd_m \la^\al \rp
  + \f{1}{2} {\bf{\hat D}} \, {\bf{\hat D}} \right] \nn \\[2mm]
& & -k \, { {}^* h}^m \, \tr(\la \si_m \lab)
    -k \, \La \, \si^{mn} \, \tr(\la {\fym}_{mn})
    -k \, \Lab \, \sib^{mn} \, \tr(\lab {\fym}_{mn}) \nn \\[2mm]
& & - \f{k^2}{4}(1+2kL)^{-1} \left[ \La^2 \, \tr \la^2 + \Lab^2 \,
\tr \lab^2
             - 2 \La \si^m \Lab \ \tr(\la \si_m \lab) \right] \nn \\[2mm]
& & - \f{k^2}{2} \left[ \tr \la^2 \ \tr \lab^2
                  - \tr (\la \si^m \lab) \ \tr (\la \si_m \lab) \right].
\label{comphf} \end{eqnarray}
 This is the supersymmetric version
of (\ref{hf}). The redefined auxiliary field
\begin{equation}
{\bf{\hat D}} \,=\,{\bf D} + {ik}\lp{1+2kL}\rp^{-1} (\La \la -
\Lab \lab),
\end{equation}
 has trivial equation of motion.

On the other hand, in order to display the component field
Lagrangian in the chiral superfield version, we recall the
definition of the component field content of the chiral
superfields
\begin{equation}
S{\loco} \,=\,S(x), \cem D_\al S{\loco} \,=\,\s2 \, \chi_\al(x),
\cem D^2 S{\loco} \,=\,-4F(x),
\end{equation}
 and
\begin{equation}
\bS{\loco} \,=\,\bS(x), \cem D^\da \bS{\loco} \ = \ \s2 \,
\chib^\da (x), \cem \bD^2 \bS{\loco} \,=\, -4\bF(x).
\end{equation} The component field action in the dual
formulation, derived from the superfield action (\ref{susaf})
takes then the form
\begin{eqnarray}
 \cl &=& - \prt^m \bS \, \prt_m S
        - \f{i}{2} \si^m_{\al \da} \lp \chi^\al \prt_m \chib^\da
                                       +\chib^\da \prt_m \chi^\al \rp
+ \widehat{F} \widehat{\bF} \nn \\[2mm] & & + \lp 1-k\s2 \,
(S+\bS) \rp \tr \left[ -\f{1}{4} {\fym}^{mn}{\fym}_{mn}
            -\f{i}{2} \si^m_{\al \da} \lp \la^\al \cd_m \lab^\da
                     + \lab^\da \cd_m \la^\al \rp
  +\f12 {\bf{\hat D}} \, {\bf{\hat D}} \right] \nn \\[2mm]
& & -\f{k}{4i\s2}(S-\bS) \left[ \vep^{klmn} \tr({\fym}_{kl}
{\fym}_{mn})
                        + 4 \prt_m \tr (\la \si^m \lab) \right] \nn \\[2mm]
& & +k \, \chi \si^{mn} \, \tr (\la {\fym}_{mn})
    +k \, \chib \sib^{mn} \, \tr (\lab {\fym}_{mn})
        -\f{k^2}{8} \, \tr \la^2 \, \tr \lab^2 \nn \\[2mm]
& & -\f{k^2}{4} \lp 1-k\s2 \, (S+\bS) \rp^{-1}
 \left[ \chi^2 \, \tr \la^2 + \chib^2 \, \tr \lab^2
             -2 (\chi \si^m \chib) \, \tr (\la \si_m \lab) \right].
\label{compaf} \end{eqnarray}
 This is the supersymmetric version of
(\ref{af}). Again, we have introduced the diagonalized
combinations for the auxiliary fields
\begin{equation}
\widehat{F} \,=\,F + \frac{k\s2}{4} \, \tr \lab^2, \cem
\widehat{\bF} \,=\,\bF + \frac{k\s2}{4} \, \tr \la^2,
\end{equation}
 and
\begin{equation}
{\bf{\hat D}} \,=\,{\bf D} -{ik} \left[{1-k\s2 \,
(S+\bS)}\right]^{-1} \, (\chi \la - \chib \lab). \end{equation}

\indent

The two supersymmetric actions (\ref{comphf}) and (\ref{compaf})
are dual to each other, in the precise sense of the construction
performed above. In both cases the presence of the Chern-Simons
form induces $k$-dependent effective couplings, in particular
quadri-linear spinor couplings. Also, we easily recognize in the
second version the axion term already encountered in the purely
bosonic case discussed before.

A striking difference with the non-supersymmetric case, however,
is the appearance of a $k$-dependent gauge coupling function,
multiplying the Yang-Mills kinetic terms. This shows that
supersymmetrization of (\ref{hf}) and (\ref{af}) results not only
in supplementary fermionic terms, but induces also genuinely new
purely bosonic terms.

In this line of construction, one can imagine an extension
of Zumino's construction of the nonlinear sigma model
\cite{Zum79b}, \cite{FG80}, \cite{AGF83}, where we replace the
K\"ahler potential $ K(\phi , \phib) $ by a more general function
$ K(\phi , \phib , L) $ which not only depends on {\em{complex}}
chiral and antichiral superfields,
but also on a number of {\em{real}} linear superfields.

\subsec{ The Geometry of the 2-Form}
{\label{2F2}

The linear multiplet has a geometric interpretation as a 2-form
gauge potential in superspace geometry. Since we wish to construct
theories where the linear multiplet is coupled to the
supergravity-matter system, we will formulate this 2-form geometry
in the background of $U_K(1)$ superspace. The basic object is the
2-form gauge potential defined as
\begin{equation}
 B \,=\,
\frac{1}{2} dz^M dz^N \, B_{NM}, \end{equation}
 subject to
gauge transformations of parameters $\beta =dz^M \beta_M$ which
are themselves 1-forms in superspace
\begin{equation}
 B \, \mapsto \, B + d \beta,
\end{equation}
\ie,
\begin{equation}
  B_{NM} \, \mapsto \, B_{NM} + \prt_N \beta_M - (-)^{nm} \prt_M \beta_N.
\end{equation}
 The invariant field strength is a 3-form,
\begin{equation}
 H \,=\,dB \,=\,\frac{1}{3!} E^A E^B E^C H_{CBA}, \end{equation}
 with $E^A$ the
frame of $U_K(1)$ superspace. As a consequence of $dd=0$ one
obtains the Bianchi identity, $dH = 0$, which fully developed
read
\begin{equation}
\frac{1}{4!} E^A E^B E^C E^D \lp 4 \, \cd_D H_{CBA}
                + 6 \, T_{DC}{}^F H_{FBA} \rp \,=\,0.
\end{equation}
The linear superfield is recovered from this general structure in
imposing  covariant
constraints on the field strength coefficients $H_{CBA}$ such that
$(\undal = \al,\da)$
\begin{equation}
H_{\undgm \undbt \undal} \,=\,0, \cem H_{\gm \bt a} \,=\,0, \cem
H_{\dg \db a} \,=\,0. \label{conH} \end{equation}
As consequences of these constraints we find (by explicitly solving them
in terms of {\em{unconstrained}} pre-potentials or else working through
the covariant Bianchi identities) that all the field strength
components of the 2-form are expressed in terms of one real superfield.
In the absence of Chern-Simons forms -cf. also section \ref{RS52},
it will be denoted by $L_0$. It is identified in
\begin{equation}
H_{\gm}{}^\db{}_a \,=\,-2i \, (\sigma_a \eps)_\gm{}^\db L_0.
\label{l0def}
\end{equation}
Explicitly we obtain
\begin{equation}
  H_{\gm ba}
\, = \, 2 (\sigma_{ba}{)_\gm}^\vp \, \cd_\vp L_0, \cem H^\dg_{\ \ ba}
\, = \, 2 (\sib_{ba}{)^\dg}_\dv \, \cd^\dv \! L_0,
\end{equation}
and
\begin{equation}
 -\frac{1}{3} \sigma_{d \, \al \da} \, \eps^{dcba} H_{cba}
 \,=\, \lp [\cd_\al, \cd_\da] - 4 \, \sigma^a_{\al \da} \, G_a \rp L_0.
\label{Hcba}
\end{equation}
This equation identifies the super-covariant field strength $H_{cba}$
in the superfield expansion of $L_0$.
Compatibility of the constraints imposed above with the structure
of the Bia\-nchi iden\-ti\-ties then implies the linearity
conditions
\begin{equation}
 \proki L_0 \, = \,  0, \cem \prokib L_0 \, = \,  0,
\end{equation}
for a linear superfield in interaction with the supergravity-matter system.

\indent

In general, when the linear multiplet is coupled
to the supergravity/matter/Yang-Mills
system, we will have to allow for Chern-Simons couplings as well. As
gravitational Chern-Simons forms are beyond the scope of this report, we
will restrict ourselves to the Yang-Mills case. Recall that the
Chern-Simons 3-form of a Yang-Mills potential $\ca$ in superspace is
defined as \cite{GG87a}
\begin{equation}
\cq^{\ym} \,=\,\tr \lp \ca d \ca + \f{2}{3} \ca \ca \ca \rp.
\label{csdef}
\end{equation}
Its exterior derivative yields a field strength squared term
\begin{equation}
d \cq^{\ym} \,=\, \tr \lp \cf \cf \rp.
\label{qff}
\end{equation}

The coupling to the antisymmetric tensor multiplet is
obtained by incorporating this Chern-Simons form into the field
strength of the 2-form gauge potential
\begin{equation}
H^{\ym} \,=\,dB + k\, \cq^{\ym}.
\end{equation}
 The superscript
${}^{\ym}$ indicates the presence of the Yang-Mills Chern-Simons form in
the definition of the field strength. Note that if
$\aym_m$, the Yang-Mills potential and
$b_{mn}$, the antisymmetric tensor gauge potential have the conventional
dimension of a mass (the corresponding kinetic actions are then
dimensionless) the constant
$k$ has dimension of an inverse mass.

Since $\cq^{\ym}$ changes under
gauge transformations of the Yang-Mills connection $\ca$ with the
exterior derivative of a 2-form $\Dt({\ca, \gym})$,
\begin{equation}
\cq^{\ym} \ \mapsto \ {}^{\gym} \cq^{\ym} \,=\,\cq^{\ym} + d \Dt({\ca, \gym}),
\label{cstrf}
\end{equation}
 covariance of $H^{\ym}$ can be
achieved in assigning an inhomogeneous compensating gauge
transformation
\begin{equation}
B \ \mapsto \ {}^{\gym} B \,=\,B - k \Dt({\ca, \gym}),\label{btf}
\end{equation}
 to the 2-form gauge potential. Finally, the addition of the
Chern-Simons forms gives rise to the modified Bianchi identities
\begin{equation}
d H^{\ym} \,=\,k \, \tr \lp \cf \cf \rp. \end{equation}

 A question of compatibility arises when the two superspace
structures are combined in the way we propose here, since the
linear multiplet corresponds to a 2-form geometry with constraints
on the 3-form field strength and the Yang-Mills field strength
$\cf$ is constrained as well.
As it turns out \cite{Gri85, GG87a}, assuming the usual constraints for
$\cf$, the modified field strength $H^{\ym}$ may be constrained in the
same way as $H$, without any contradiction.
The most
immediate way to see this is to investigate explicitly the
structure of the modified Bianchi identities
\begin{equation}
 \frac{1}{4!} E^A E^B E^C E^D \lp 4
\, \cd_D H_{CBA} + 6 \, T_{DC}{}^F H_{FBA}-6k \tr(\cf_{DC}
\cf_{BA}) \rp \,=\,0. \label{mbi}
\end{equation} Assuming for $H^{\ym}$ the same
constraints as for $H$ -cf. (\ref{conH}), (\ref{l0def})- and
replacing $L_0$ by $L^{\ym}$
on the one hand and taking into account the special properties of the $\cf
\cf$ terms arising from the Yang-Mills constraints on the other hand, one
can show that the linearity conditions (\ref{mbi}) are replaced by the
{\em modified linearity conditions}
 \begin{eqnarray}
\prokib L^{\ym} &=& 2k \, \tr \lp \cw_\da \cw^\da \rp, \label{mlc1}
\\ \proki L^{\ym} &=& 2k \, \tr \lp \cw^\al \cw_\al \rp. \label{mlc2}
\end{eqnarray}
Likewise, (\ref{Hcba}) acquires an additional term,
\begin{equation}
\lp \l[ \cd_\al, \cd_\da \r] - 4 \sigma^a_{\al \da} G_a \rp L^{\ym}
\, = \,
         -\f{1}{3} \sigma_{d \al \da} \eps^{dcba} H^{\ym}_{cba}
         -4k \, \tr \lp \cw_\al \cw_\da \rp.
\label{commrel}
\end{equation}
The special properties of $\cw_{\undal}$ allow to express the
quadratic gaugino contributions in (\ref{mlc1}), (\ref{mlc2}) in terms of
a single {\em Chern-Simons superfield} $\Om^{\ym}$,
\begin{equation}
 \tr \lp \cw_\da \cw^\da \rp \, = \, \f{1}{2} \prokib \Om^{\ym}, \cem
 \tr \lp \cw^\al \cw_\al\rp \, = \, \f{1}{2} \proki \Om^{\ym}. \label{CSsfi}
 \end{equation}
The existence of $\Om^{\ym}$ and its explicit construction -cf.
appendix F- rely on the similarity of Chern-Simons forms with a
generic 3-form gauge potential $C$. The Chern-Simons form
(\ref{csdef}) plays the role of a 3-form gauge potential
(\ref{cstrf}) and $\tr \lp \cf \cf \rp$ corresponds to its field
strength (\ref{qff}). Given the identification
\begin{equation}
\Si \,=\,\f{1}{4!} E^A E^B E^C E^D {\Si}_{DCBA} \,=\,
          \f{1}{4!} E^A E^B E^C E^D \, 6 \,
           \tr \lp \cf_{DC} \, \cf_{BA} \rp,
\end{equation}
 and the constraints on $\cf$ it is immediate to deduce
that indeed
\begin{equation}
{\Si}_{\unddt \, \undgm \, \undal \, A} \,=\,0,
\end{equation}
which are just the constraints of the 4-form field strength in the
generic case. Anticipating part of the discussion of the next
section, we observe that, as a consequence of the constraints, all
the components of the generic 4-form field strength are expressible in
terms of chiral superfields $Y$ and $\ovY$
($\cd_\al \ovY=0,\,\cd^\da Y = 0 $) identified in
\begin{equation}
\Si_{\dt \gm \, ba} \, = \, \f12 (\sigma_{ba}\eps)_{\dt \gm} \, \ovY,
\cem
\Si^{\dd \dg}{\,}_{ba} \, = \, \f12 (\sib_{ba}\eps)^{\dd \dg} \, Y.
\end{equation}
For the remaining coefficients, \ie $\Si_{\unddt \, cba}$
and $\Si_{dcba}$, respectively, we obtain then
\begin{equation}
\Si_{\dt \, cba} \, = \,
    -\frac{1}{16} \, \sigma^d_{\dt \dd} \, \eps_{dcba} \cd^\dd \ovY,
\cem
\Si^\dd {}_{cba}\, = \,
    +\frac{1}{16} \, \sib^{d \, \dd \dt} \, \eps_{dcba} \cd_\dt Y,
\end{equation}
 and
\begin{equation}
 \Si_{dcba}\,=\,\frac{i}{16}\ \eps_{dcba}
 \left[ \ \lp \cd^2 - 24 \rd \rp Y
- \lp \cdb^2 - 24 R \rp \ovY\, \right].
\end{equation}

This last equation should be understood as a further constraint between
the chiral superfields $Y$ and $\ovY$, thus describing the
supermultiplet of a 3-form gauge potential in $U_K(1)$ superspace.

From the explicit solution of the constraints,
one finds that $Y$ and $\ovY$ are given as the chiral projections
of $U_K(1)$ superspace geometry acting on one and the same
pre-potential $\Om$,
\begin{equation}
Y \, = \, -4\proki \Om ,\cem \ovY \, = \, -4 \prokib \Om.
\end{equation}
Due to the same constraint structure of $\Si$ and $\tr(\cf \cf)$,
this analysis applies to the case of Chern-Simons forms as well.
We identify
\begin{equation}
 Y^{\ym} \ = \ -8 \, \tr \lp \cw^\al \cw_\al \rp, \cem \ovY^{\ym}
\,=\,-8 \, \tr \lp \cw_\da \cw^\da \rp.
\end{equation}
Correspondingly, $\Om$ the generic pre-potential, is identified as
$\Om^{\ym}$, the Chern-Simons superfield, expressed in terms of the
unconstrained Yang-Mills pre-potential. A detailed account of this
analysis is given in appendix F.

It is instructive to investigate the
relation between the superfields $L^{\ym}$ and $L_0$ in this context.
As we have seen,
$L_0$ and $L^{\ym}-k \Om^{\ym}$ satisfy the same linearity
conditions. As a consequence, they can be identified up
to some linear superfield, \ie
\begin{equation}
L^{\ym} \,=\, L_0 + k \, \beta^{\ym} + k \, \Om^{\ym}.
\end{equation}
Here $\beta^{\ym}$ is a pre-potential dependent linear superfield whose
explicit form, irrelevant for the present discussion, may be read
off from the equations in appendix F.
Note that $\Om^{\ym}$
changes under Yang-Mills gauge transformations by a linear
superfield (hence (\ref{CSsfi}) are unchanged), whereas the
combination $\Om^{\ym} + \beta^{\ym}$ is gauge invariant, in
accordance with the gauge invariance of $L_0$ and $L^{\ym}$.

We have tried to make clear in this section that the superspace geometry
of the 3-form gauge potential provides a generic framework for the
discussion of Chern-Simons forms in superspace. Established in
full detail for the Yang-Mills case, this property can be
advantageously exploited \cite{GG99} in the much more involved
gravitational case, relevant in the Green-Schwarz mechanism in
superstrings.

As we will consider the Yang-Mills case only,
we drop the ${}^{\ym}$
superscript from now on, a superfield $L$ being
supposed to satisfy the modified linearity conditions.

\subsec{ Component Fields \label{compfields}}

When coupled to the supergravity/matter/Yang-Mills system, the components
\begin{equation}
b_{mn}(x) , \hspace{4mm} L(x), \hspace{4mm} \La_\al(x),
\hspace{4mm} \Lab_\da(x),
\end{equation}
of the linear multiplet are still defined as lowest superfield
components, but now in the framework of $U_K(1)$ superspace geometry.
For the covariant components $L(x)$, $\La_\al(x)$, $\Lab_\da(x)$, we
define
\begin{equation}
L{\loco} \,=\,L(x),\cem \cd_\al L{\loco} \,=\,\La_\al (x), \cem
\cd^\da L{\loco} \, = \, \Lab^\da (x),
\end{equation}
whereas the antisymmetric tensor gauge field is identified as
\begin{equation}
B {\doubar} \,=\,b \,=\,\f{1}{2} dx^m dx^n b_{nm}(x).
\end{equation}
The double bar projection, as defined in \ref{CPN}, is particularly useful
for the determination of the lowest component of $H_{cba}$, the
super-covariant field strength of the antisymmetric tensor.
Recall that the component field expression of the Chern-Simons form,
in terms of $\ca \doubar = i dx^m {\aym}_m(x)$, is given as
\begin{equation}
Q \doubar \,=\, \f{1}{3!} dx^n dx^m dx^l Q_{lmn} \, = \,
- \f{1}{3!} dx^n dx^m dx^l \tr \lp {\aym}_{l} \prt_m {\aym}_{n} -
\f{2i}{3} {\aym}_{l} {\aym}_m {\aym}_{n} \rp.
\end{equation}
The double bar projection is then
applied in two ways. On the one hand, we have
\begin{equation}
H {\doubar} \,=\,\f{1}{3!} dx^l dx^m dx^n h_{nml},
\label{Hbb}
\end{equation}
with
$h_{nml} \,=\,\prt_n b_{ml} + \prt_m b_{ln} + \prt_l b_{nm}
+ k \, Q_{nml}$.
The super-covariant field strength $H_{cba} \loco $, on the other hand,
is identified in employing the double bar projection in terms of the
covariant component field differentials $e^A$, defined in
(\ref{CPN.5a}), (\ref{CPN.5b}), and taking into account the constraints on
$H_{CBA}$. As a result, we find
\begin{equation}
H {\doubar} \,=\,\f{1}{3!} e^a e^b e^c H_{cba}{\loco}
        + \f{1}{2} e^a e^b e^\gm H_{\gm ba}{\loco}
        + \f{1}{2} e^a e^b e_\dg {H^\dg}_{ba}{\loco}
        + e^a e_\db e^\gm H_\gm{}^{\db}{}_a{\loco}.
\end{equation}
Inserting the explicit expressions for $H_{\gm ba}$, ${H^\dg}_{ba}$ and
$H_\gm{}^{\db}{}_a$ yields then in a straightforward way
\begin{equation}
\f{1}{3!} \eps^{dcba} H_{cba}{\loco} \,=\,
\f{1}{3!} e_n{}^d \vep^{nmlk} \lp h_{mlk}
+ 3i L  \psi_m \sigma_l \psib_k \rp
+ i \, e_n{}^d\lp \psi_m \sigma^{nm} \La  -\psib_m \sib^{nm} \Lab \rp,
\end{equation}
Note that the super-covariant field strength $H_{cba} \loco$, one of the
basic building blocks in the construction of component field actions,
exhibits terms linear and quadratic in the Rarita-Schwinger field.
Details on the geometric derivation of supersymmetry transformation laws
and the construction of invariant component field actions are presented
in appendix \ref{appE}.

\subsec{ Linear Multiplet Coupling}
{\label{2F3}

For the coupling of the linear multiplet to the general
supergravity/matter/Yang-Mills system we may imagine to follow the
same steps as before, but with the {\em{\ka potential}} replaced
by an $L$ dependent superfield $K(\phi, \phib, L)$
\cite{BGGM87a,BGG91,ABGG93}, which we shall call the {\em{kinetic
potential}}.
Let us note that $L$ being real, the interpretation of $K$ as a
potential of \ka geometry is partly lost.

As we now explain, such a
construction does not yield a canonically normalized Einstein term.
To begin with, we note that the curvature scalar still appears in the
combination
\begin{equation}
{\cd}^2R + {\cdb}^2 R^{\dagger}
 \,=\,
- \f {2}{3}{R_{ba}}^{ba}-\f {1}{3}\lp {\cd}^{\al} \! X_{\al}
+\cd_\da \bX^\da \rp +4G^a G_a +32 RR^{\dagger},
\label{2F.30}
\end{equation}
where the combination ${\cd}^{\al} \! X_{\al}+ \cd_\da
\bX^\da$ should now be evaluated using $K\,=\,K(\phi,\phib, L)$ as
a starting point.
This generates extra $\cd^2 R + \cdb^2 \rd$
terms. Indeed, recall the $U_K(1)$ relations
(\ref{B.81}), (\ref{B.82})
\begin{equation}
 -3\,{\cd}_{\al}R \, = \,  X_{\al} + 4 S_{\al}, \cem
-3\,{\cd}^{\da}R^{\dagger} \, = \,  X^{\da} -4 S^{\da},
\label{relDR}
\end{equation}
and the definitions
\begin{eqnarray}
X_\al  & = &
-\f {1}{8}({\cdb}^2 -8R) \, {\cd}_{\al}K(\phi,\phib, L),
\label{2F.31} \\
{\bX}^{\da} & = & -\f {1}{8}({\cd}^2
-8R^{\dagger}) {\cd}^{\da} K(\phi,\phib, L). \label{2F.32}
\end{eqnarray}
In the $L$ independent case these relations serve to identify
$\cd_\al R$ and $\cd^\da \! \rd$ as superfields, roughly
speaking, depending through $X_\al, \bX^\da$ on the matter sector and
through $S_\al, \bS^\da$ on the gravity sector. In the $L$
dependent case,
due to the presence of $R$, $\rd$ in the linearity conditions,
successive
spinor derivatives generate extra $\cd_\al R$, resp. $\cd^\da \! \rd$
  terms in the expressions of $X_\al$, resp. $\bX^\da$. We can make
  explicit such contributions and write ($K_L \equiv {\prt
  K}/{\prt L}$)
\begin{equation}
 X_\al \,=\, -L\,K_L \, \cd_\al R + Y_\al,\cem
\bX^\da\,=\,-L\, K_L \, \cd^\da \! \rd + \bY^\da,
\label{2FXmod}
\end{equation}
where $Y_\al$ and $\bY^\da$ contain all remaining contributions
including those stemming from the Chern-Simons forms.
 Hence, in
this case $\cd_\al R$ and $\cd^\da \! \rd$ are still identified as
dependent superfields, but the relations (\ref{relDR}) take a modified
form,
\begin{eqnarray}
\lp L K_L-3 \rp {\cd}_{\al}R &=& Y_\al + 4 S_{\al},
\\
\lp L K_L-3 \rp {\cd}^{\da} \! \rd &=& \bY^\da +4 S^{\da}.
\end{eqnarray}

This, in turn, implies that the basic geometric relation
(\ref{2F.30}) takes a modified form as well
\begin{eqnarray}
&&\lp 1-\f{1}{3}LK_L \rp \lp{\cd}^2 R +{\cdb}^2 \rd \rp
 \, = \, - \f{2}{3}{R_{ba}}^{ba}  +4G^a G_a +32 R\,\rd
\nn \\ &&\ - \f{1}{3}\lp {\cd}^{\al}Y_{\al}+ \cd_\da \bY^\da \rp
+\f{1}{3} \cd^\al \lp LK_L\rp\cd_\al R
 +\f{1}{3} \cd_\da \lp LK_L\rp \cd^\da \rd.  \label{2F.33}
\end{eqnarray}
Evaluating the component field action, following the
procedure of section \ref{CPN5}, we obtain an Einstein term with a
field dependent normalization $\lp 1-\frac{1}{3}L K_L \rp^{-1}$.
In other terms, in the linear superfield formalism, a superfield action
which is just the integral over the superdeterminant of the frame, leads
to a non-canonical normalization of the Einstein term.

In order to have more flexibility for the normalization function we
consider from now on a general superfield action,
\begin{equation}
{\cal{L}} \,=\,-3\, \int \! \! E \, F(\phi, \phib , L),
\label{3.13}
\end{equation}
where the {\em subsidiary function} $F$ depends in a yet unspecified
manner on the chiral and linear superfields.
Observe that the kinetic potential $K(\phi, \phib , L)$ is implicit in
$E$ through the $U_K(1)$ construction. The component field version of
this generalized superfield action is evaluated using the chiral
superfield,

\begin{equation}
{\bf r} \, = \, - \f{1}{8} \proki F(\phi, \phib , L),
\end{equation}
and its complex conjugate in the generic construction of section
\ref{CPN5}. A straightforward calculation shows that in this case
the Einstein term is multiplied by the {\em normalization function}
\begin{equation}
N(\phi, \phib , L) \, = \, \frac{F-L F_L}{1-\frac{1}{3} L K_L}.
\label{normf}
\end{equation}
Requiring $N = 1$, or
\begin{equation}
F - L F_L \,=\,1 - \frac{1}{3} L K_L,
\label{3.13l}
\end{equation}
ensures that we get a canonically normalized Einstein term.

Note that in the case of $L$-independent functions $F$ and $K$,
this equation implies simply $F = 1$. In the general case,
the solution of (\ref{3.13l})
reads
\begin{equation}
F(\phi,\phib,L)\,=\,1 + L V (\phi,\phib)
  +\frac{L}{3} \int \frac{d \la}{\la} K_\la (\phi,\phib,\la).
\label{3.14}
\end{equation}
We see that the only term in $F(\phi,\phib,L)$ which is not fixed
by the choice of the K\"ahler potential is the term
$L V (\phi,\phib)$, the ``integration constant'' of the
differential equation (\ref{3.13l}). Indeed, one can check that,
in the Lagrangian (\ref{3.13}), only a term linear in $L$, viz.
\begin{equation}
{\cal{L}}_{\rm{lin}}\,=\,- 3 \int \! \! E \, L V(\phi,\phib),
\label{3.15}
\end{equation}
cannot be set to 1 by a superfield rescaling since the Weyl
weights of $E$ and $L$ sum up to zero ($\sigma (E) =-2,\sigma (L)=2$).

As we discuss now, the real function $V(\phi, \phib)$ plays an important
role in the discussion of certain anomaly cancellation mechanisms.
From now on we refer to it as {\em linear potential}.
To be more definite, consider  the {\em effective
transformation}
\begin{equation}
V(\phi,\phib) \mapsto V(\phi,\phib)
           + H(\phi) + \bar{H}(\phib),
\label{3.16}
\end{equation}
with $H$ a chiral superfield which is a holomorphic
function of the chiral matter fields. How does the Lagrangian
${\cal{L}}_{\rm{lin}}$ change under such a transformation?
To see this more explicitly, use integration by parts and
apply the modified linearity conditions,
\begin{equation}
\int \! \! E \, L H \, = \,
-\frac{1}{8} \int \! \! \frac{E}{R} H \proki L \, = \,
\int \! \! \frac{E}{R} H \tr (\cw^{\al} \cw_{\al}).
\label{efftrf}
\end{equation}
Note the appearance of the chiral volume
element in superspace. Therefore, (\ref{3.16}) gives rise to the
effective transformation
\begin{equation}
{\cal{L}}_{\rm{lin}} \mapsto {\cal{L}}_{\rm{lin}} +
          \frac{3k}{4} \int \! \! \frac{E}{R}
           H(\phi)  \tr (\cw^{\al} \cw_{\al})
         + \frac{3k}{4}  \int \! \! \frac{E}{R^{\dagger}}
           \bar{H}(\phib)  \tr (\cw_{\da} \cw ^{\da} ).
\label{3.21}
\end{equation}
This shows that in the absence of Chern-Simons forms, $k=0$, the
transformation (\ref{3.16}) is a symmetry of the theory. In the presence
of Chern-Simons forms it creates an Abelian anomaly term, multiplied by
$H- \bar H$ and gives rise, at the same time, to a Yang-Mills kinetic term
multiplied by $H +\bar H$. We will come back to this issue later on.

\subsec{ Duality Transformations}
{\label{2F4}

As is well known and has been stressed in section \ref{RS5}, the
antisymmetric tensor/real scalar duality extends to the supersymmetric
case, where it becomes a linear/chiral multiplet duality.
This duality should now be explored for the case of a linear
multiplet coupled to the general supergravity/matter/Yang-Mills system,
the so-called {\em linear superfield formalism}, in relation to the {\em
chiral superfield formalism}, where only chiral multiplets occur.

It is not surprising that the subsidiary function $F(\phi,\phib,L)$,
introduced in the previous subsection, be of some importance. As a
matter of fact, the normalization condition (\ref{3.13l}), justified
previously at the component field level will reappear in an intriguing
way in the superfield duality transformation mechanism in curved
superspace.
Let us consider the {\em first order formalism} Lagrangian
\begin{equation}
{\cal{L}}_{\! \mbox{\tiny{FOF}}}\,=\,-3 \int \! \! E \left[
F(\phi,\phib,X)
    + X(S + {\bar{S}}) \right],
\label{3.2}
\end{equation}
where $S$ is a chiral superfield, ${\cal{D}}^{\da} S = 0$, and $X$ is an
unconstrained superfield. The kinetic potential $K(\phi,\phib,X)$ and the
normalization function $F(\phi,\phib,X)$ are supposed to be given in
terms of this unconstrained superfield.

Variation of (\ref{3.2}) with respect to $X$ gives rise to
\begin{equation}
(S+\bar{S}) \left( 1 - \frac{1}{3} X K_X \right)
\, = \,
\frac{1}{3} F K_X - F_X,
\label{3.8}
\end{equation}
where we have used
\begin{equation}
\delta_{X} E\,=\,- \frac{1}{3} E K_X \delta X,
\label{3.9}
\end{equation}
as derived from (\ref{D.130}) and (\ref{D.90}). For given $F$ and $K$
functions, (\ref{3.8}) should allow to express $X$ as a function
of $\phi, \phib$ and of $S+\bar{S}$, such that the resulting
Lagrangian in the {\em chiral superfield formalism} is given as
\begin{equation}
{\cal{L}}_{\! \mbox{\tiny{CSF}}}
\, = \,
-3 \int \! \! E
\left[
F \lp \phi, \phib, X(\phi, \phib, S+\bar{S}) \rp
+ (S + \bar{S}) \, X(\phi, \phib, S+\bar{S})
\right].
\label{3.10}
\end{equation}
Clearly, this Lagrangian will not necessarily
yield the canonical normalization of the curvature scalar term.
On the other hand, we have
shown in section \ref{GRA2} that the Lagrangian, built with
(anti)chiral superfields, which gives a correct Einstein term is
simply
\begin{equation}
{\cal{L}}\,=\,- 3 \int \! \! E. \label{3.11}
\end{equation}
This form of (\ref{3.10}) can be obtained in requiring
\begin{equation}
F(\phi,\phib,X) + X (S+\bar{S})\,=\,1, \label{3.12}
\end{equation}
where $X$ is the solution of (\ref{3.8}). Formally, these two equations
combine into
\begin{equation}
F - X F_X \,=\,1 - \frac{1}{3} X K_X.
\label{3.13a}
\end{equation}
This means that for a theory with canonical Einstein term, $F$ and
$K$ cannot be chosen independently, they should satisfy (\ref{3.13a}),
which has the same form as (\ref{3.13l}), but with $L$ replaced by
$X(\phi, \phib, S+\bar{S})$. Likewise,
$F( \phi, \phib, X(\phi, \phib, S+\bar{S}))$ should have the same
functional dependence on $X$ as it had before on $L$. These relations are
of fundamental importance if we want to make meaningful
comparisons between different theories (or compare for example the
tree level and one-loop effective actions).

\indent

Alternatively, we can vary (\ref{3.2}) with respect to $S$ or
$\bS$. Due to chirality, they can be written as
\begin{equation}
S\,=\, \proki \Sigma, \cem \bar S\,=\, \prokib \bar \Sigma,\label{3.5}
\end{equation}
where $\Sigma$, $\bar \Sigma$ are unconstrained superfields.

Variation of (\ref{3.2}) with respect to $\Sigma$, $\bar \Sigma$ yields
 after integration by parts:
\begin{equation}
(\cdb^2 - 8 R) X \, = \, 0, \cem
 (\cd^2 - 8 R^{\dagger}) X \, = \, 0. \label{3.7}
\end{equation}
We conclude that $X$ is a linear superfield, which we identify
with $L_0$. An integration by parts (linear $\times$ chiral
integrates to zero) then shows that (\ref{3.2}) reproduces
(\ref{3.13}) and we are back with the {\em linear superfield formalism}
discussed in the previous subsection.

There, however, the linear multiplet was coupled to Chern-Simons forms.
How does this coupling affect the duality structure? It is clear that in
the linear superfield formalism we should reproduce the modified
linearity conditions. Therefore, the {\em first order formalism} should
include the Chern-Simons superfield $\Om$, such that
\begin{equation}
{\cal{L}}_{\! \mbox{\tiny{FOF}}}
\,=\,-3 \int \! \! E \, [F(\phi,\phib,X)+(X -\,k\,\Omega)
(S+\bar{S})].
\label{3.17P}
\end{equation}
Varying with respect to $\Sigma$, $\bar \Sigma$ establishes then the
modified linearity conditions.
On the other hand, varying\footnote{Due to the variation law
$\delta_{X} \Omega = \frac{1}{3} \Omega K_X \delta X$,
the terms proportional to
the Chern-Simons form cancel out in this equation, as expected
from gauge invariance considerations.}
(\ref{3.17P}) with respect to $X$ gives rise to the same equation
(\ref{3.8}) as before. Imposing moreover a canonical Einstein term,
using  (\ref{3.12}), the Lagrangian in
the {\em chiral superfield formalism} then reads
\begin{equation}
{\cal{L}}_{\! \mbox{\tiny{CSF}}}
\,=\,-3 \int \! \! E \, [1 - \,k\,\Omega (S+\bar{S})],
\label{3.17}
\end{equation}
To put the new terms, arising from the Chern-Simons couplings, in a more
familiar form, we write them as
\begin{equation}
{\cal{L}}_{\! \mbox{\tiny{CSF}}}
\, = \, - 3 \int \! \! E - \frac{3}{8} \int \! \!
\frac{E}{R}
 S  (\cdb^2 - 8 R) \Omega - \frac{3}{8} \int \! \!
\frac{E}{R^{\dagger}}
 \bar{S} (\cd^2 - 8 R^{\dagger}) \Omega,
\label{3.18}
\end{equation}
where the derivative terms vanish upon integration by parts ($S$ and
$R$ are chiral superfields), and use (\ref{CSsfi}) to obtain
\begin{equation}
{\cal{L}}_{\! \mbox{\tiny{CSF}}}
\, = \, - 3 \int \! \! E  - \frac{3}{4} k \int \! \! \frac{E}{R}
 S \, \tr (\cw^{\al} \cw_{\al}) -
 \frac{3}{4} k  \int \! \! \frac{E}{R^{\dagger}}
 \bar{S} \, \tr (\cw_{\da} \cw^{\da} ).
\label{3.19}
\end{equation}
We therefore recover the standard formulation of matter coupled to
supergravity with a holomorphic gauge coupling function
\begin{equation}
f(S)\,=\,-\, 6 k S. \label{3.20}
\end{equation}
Comparing this to (\ref{3.21}) suggests that the effective
transformations
(\ref{3.16}) should be realized in the chiral superfield formalism as
field dependent shifts of the chiral superfield $S$, \ie $S \mapsto S
-H(\phi)$ and
$\bar S \mapsto \bar S - \bar H(\phib)$

Let us stress that the duality between the linear superfield formulation
and the chiral superfield formulation, discussed here for the case of one
single linear superfield, extends quite obviously to the case of several
linear superfields and suitable Chern-Simons couplings.
We will come back to this after the next subsection.

 We close this subsection on an example \cite{BGGM87b,BGG90}
which plays an important role in
 superstring models. We take for the \ka potential:
\begin{equation}
K\,=\,K_0 (\phi, \phib) + \al \log L
\label{3.25}
\end{equation}
where it was already stressed that $L$ plays the r\^ole of the string coupling.
The corresponding solution of (\ref{3.13}) is:
\begin{equation}
F\,=\,1 - \al/3 + LV(\phi,\phib).
\label{3.26}
\end{equation}
The solution of (\ref{3.12}) reads
\begin{equation}
\frac{\al}{3L}\,=\,S + \bS + V(\phi, \phib) \label{3.27}
\end{equation}
and
\begin{equation}
K(\phi,\phib,S+\bS)\,=\,K_0(\phi,\phib) + \al \log \frac{\al}{3}
    -\al \log (S+\bS + V(\phi,\phib)).
\label{3.28}
\end{equation}
It is interesting to discuss equation (\ref{3.27}) in the context of the
one-loop renormalization of the gauge coupling performed by Dixon,
Kaplunovsky and Louis \cite{DKL91}: $S+\bS$ is interpreted as the
tree level gauge coupling and $V(\phi,\phib)$ is a generic
(non-holomorphic) threshold correction. We thus see that, up to a
normalization factor, it is $L^{-1}$ which must be interpreted as
the renormalized gauge coupling. Thus,
 {\em{ the natural framework to
perform the renormalization of the gauge coupling functions is the
linear multiplet formulation}}.

 We note also that the \ka potential in (\ref{3.28}) is invariant under
 the effective transformations  (\ref{3.16}) together with
 $S \mapsto S -H(\phi)$ and $\bar S \mapsto \bar S - \bar H(\phib)$.

Adding terms of order $L^n$  ($n
\geq 2$) in (\ref{3.25}) would include higher order corrections, if any, but we
can note here the special status played by one-loop corrections.
The explicit computation of Ref.\cite{DKL91} indicates that, in
this context, $V(\phi,\phib)$ contains a piece which is nothing
else but $K_0(\phi,\phib)$. This fact has been stressed by
Derendinger et al. \cite{DFKZ92} and is in agreement with the
K\"ahler properties of $V(\phi,\phib)$ -cf. ({\ref{3.16}).

\subsec{ Non Holomorphic Gauge Couplings}
\label{NHGC}

In general, as explained in section \ref{GRA33}, supersymmetric
Yang-Mills theory allows for arbitrary holomorphic gauge coupling
functions in terms of the complex matter scalar fields. The corresponding
invariant supergravity action (\ref{GRA.241}) is given as a F-term in
$U_K(1)$ superspace.

Superstring theory, in its effective low energy limit, seems to suggest
non-holomorphic gauge coupling functions \cite{SV91}, \cite{Lou91} as
well. From the formal point of view, such non-canonical structures arise
naturally in the linear superfield formalism \cite{BGG91}, \cite{DFKZ92}.

Independently of the relation to string theory, it is instructive in
itself to elucidate the origin of non-holomorphic gauge couplings in the
linear superfield formalism. The crucial ingredient is the coupling of
Chern-Simons forms to linear multiplets, as described in sections
\ref{2F2} and \ref{2F3}. In this context, the modified linearity
conditions (\ref{mlc1}), (\ref{mlc2}) are of utmost importance. In
the following we will point out schematically how non-holomorphic gauge
couplings appear in the component field theory, starting from the
geometric superspace description.

Recall that the basic object for the construction of the component field
action are the chiral superfields ${\bf r}$ and ${\bf {\bar r}}$ given as
\begin{equation}
{\bf r} \, = \, -\frac{1}{8} \proki F(\phi, \phib, L), \cem
{\bf {\bar r}} \, = \, -\frac{1}{8} \prokib F(\phi, \phib, L).
\label{genact}
\end{equation}
Working through the generic construction of section \ref{CPN4} allows to
determine unambiguously the complete component field action.
As we are interested only in the gauge coupling function, it is not
necessary to go through all these steps in full detail.

For the sake of a schematical discussion recall first of all that the
gauge kinetic terms arise from the lowest component of the superfield
\begin{equation}
\cd^2 \tr \cw^2 + \bar \cd^2 \tr \bar \cw^2.
\label{FF}
\end{equation}
On the other hand, the complete set of kinetic terms of all the component
fields is identified in
\begin{equation}
\cd^2 {\bf r} + \bar \cd^2 {\bf \bar r}.
\label{kint}
\end{equation}
The procedure consists in evaluating the spinor derivatives in
(\ref{kint}) and in isolating terms proportional to (\ref{FF}).
In a first step we identify relevant terms in
\begin{equation}
\cd^2 {\bf r} + \bar \cd^2 {\bf \bar r} \, \stackrel{RT}{=} \,
    F ( \cd^2 R + \bar \cd^2 \rd)
   -\frac{1}{8}  ( \cd^2 \bar \cd^2 + \bar \cd^2 \cd^2) F.
\end{equation}
The symbol $ \, \stackrel{RT}{=} \, $ indicates that we only
retain the terms relevant for our discussion, making the
arguments more transparent. The first term on the right contains
the contribution originating from the $L$ dependence of $K$.
Using (\ref{2F.30}), we obtain
\begin{equation}
\cd^2 {\bf r} + \bar \cd^2 {\bf \bar r} \, \stackrel{RT}{=} \,
    -\frac{1}{3} F ( \cd^\al \! X_\al + \cd_\da \bar X^\da)
   -\frac{1}{8}  F_L ( \cd^2 \bar \cd^2 + \bar \cd^2 \cd^2) L.
\end{equation}
Next we insert the explicit expression for $X^\al$ in terms of
$K(\phi, \phib, L)$, \ie
\begin{equation}
\cd^\al \! X_\al + \cd_\da \bar X^\da  \, \stackrel{RT}{=} \,
 -\frac{1}{8} K_L ( \cd^2 \bar \cd^2 + \bar \cd^2 \cd^2) L,
\label{xk}
\end{equation}
to arrive at the intermediate result
\begin{equation}
\cd^2 {\bf r} + \bar \cd^2 {\bf \bar r} \, \stackrel{RT}{=} \,
    -\frac{1}{8} (F_L -\frac{1}{3} F K_L) \,
    ( \cd^2 \bar \cd^2 + \bar \cd^2 \cd^2) L.
\end{equation}
In the next step we are going to exploit the modified linearity conditions
\begin{equation}
\bar \cd^2 L \, = \, 8 R L + 2k \tr \cw^2, \cem
\cd^2 L \, = \, 8 \rd L + 2k \tr \bar \cw^2.
\end{equation}
As a consequence we find
\begin{equation}
( \cd^2 \bar \cd^2 + \bar \cd^2 \cd^2) L \, \stackrel{RT}{=} \,
8 L ( \cd^2 R + \bar \cd^2 \rd)
+2k \, (\cd^2 \tr \cw^2 + \bar \cd^2 \tr \bar \cw^2).
\end{equation}
Using once more (\ref{xk}), \ie
\begin{equation}
\cd^2 R + \bar \cd^2 \rd \, \stackrel{RT}{=} \,
\frac{1}{24} K_L ( \cd^2 \bar \cd^2 + \bar \cd^2 \cd^2) L,
\end{equation}
yields
\begin{equation}
( \cd^2 \bar \cd^2 + \bar \cd^2 \cd^2) L \, \stackrel{RT}{=} \,
\frac{2k}{1-\frac{1}{3} L K_L} \ (\cd^2 \tr \cw^2 + \bar \cd^2 \tr \bar
\cw^2).
\end{equation}
The final result is then
\begin{equation}
\cd^2 {\bf r} + \bar \cd^2 {\bf \bar r} \, \stackrel{RT}{=} \,
-\frac{k}{4} \,
\frac{F_L -\frac{1}{3} F K_L}{1-\frac{1}{3} L K_L}
\ (\cd^2 \tr \cw^2 + \bar \cd^2 \tr \bar \cw^2),
\end{equation}
which allows to identify the {\em gauge coupling function}
\begin{equation}
\Gamma(\phi, \phib, L)
\, = \, \frac{F_L -\frac{1}{3} F K_L}{1-\frac{1}{3} L K_L}.
\label{nhgcf}
\end{equation}
Recall that in the standard case the gauge coupling is the sum of a
holomorphic and an antiholomorphic function. In the  more
general formulation given here, non-holomorphic coupling
functions are allowed.

At this point it is important to note that
so far we did not make any reference to possible normalizations of the
Einstein term, appearing in the same action. In section \ref{2F3} we have
identified the normalization function
\begin{equation}
N(\phi, \phib , L) \, = \, \frac{F-L F_L}{1-\frac{1}{3} L K_L}.
\end{equation}
A glance at the explicit form of $\Gamma$ and $N$ shows that they are
related to $F$ through the simple relation
\begin{equation}
L \Gamma + N \, = \, F.
\label{GFN}
\end{equation}
Finally, the same Lagrangian contains also a kinetic term for $L$,
\begin{equation}
\frac{1}{4L} \left[ 3 N_L + K_L ( L N_L - N) \right]
g^{mn} \partial_m L \, \partial_n L,
\end{equation}
whose normalization function is expressed in terms of previously defined
quantities.
Note that, in view of the normalization of the curvature scalar,
\ie
\begin{equation}
-\frac{N}{2} \car,
\end{equation}
it should be clear that the conformally trivial combination is obtained
from the choice $N=L$; remember that $L$ has Weyl weight
$\sigma(L) = - 2$.

\indent

Let us now turn to a discussion of the duality transformation in
this general case, \ie in the presence of non trivial
normalization function $N$, gauge coupling $\Gamma$, and
subsidiary function $F$. The relevant first order action is still
(\ref{3.17P}). The linear superfield formalism discussed above is
obtained in the usual way, varying with respect to the
unconstrained pre-potentials of the chiral superfield $S$. The
chiral superfield formalism, on the other hand, is obtained from
variation with respect to $X$. As before, the corresponding
equation of motion (\ref{3.8}) should be understood as an
expression which determines $X$ in terms of $\phi$, $\phib$ and
$S+ \bar S$. The chiral superfield formalism is then obtained
from (\ref{3.17P}), but with $X$ now a function $X(\phi, \phib,
S+\bar S)$. As to the gauge coupling function we are back to the
holomorphic case.

From what we have learned before, it should be clear that the superfields
underlying the component field construction of the action are now
\begin{equation}
{\bf r} \, = \, -\frac{1}{8} \proki F + \frac{k}{4} \tr \cw^2, \cem
{\bf {\bar r}} \, = \, -\frac{1}{8} \prokib F + \frac{k}{4} \tr \cwb^2.
\label{genactym}
\end{equation}
It is instructive to identify the normalization function of the curvature
scalar and the gauge coupling function, using a similar reasoning as
before in the linear superfield formalism. Working through the
successive application of spinor derivatives in
$\cd^2 {\bf r} + \cdb^2 {\bf {\bar r}}$ and keeping track only of terms
relevant for our purpose we find
\begin{equation}
\cd^2 {\bf r} + \cdb^2 {\bf {\bar r}} \, \stackrel{RT}{=} \,
-\frac{2}{3} (F + X (S + \bar S) ) \, R_{ba}{}^{ba}
+\frac{k}{4} (S + \bar S)
\, (\cd^2 \tr \cw^2 + \bar \cd^2 \tr \bar \cw^2 \, ).
\end{equation}
The gauge coupling function is simply proportional to $S+\bar S$, in
accordance with (\ref{3.8}) and the definition (\ref{nhgcf}). As to the
normalization function of the Einstein term we observe that, using
formally (\ref{3.8}) together with (\ref{GFN}), means simply that
\begin{equation}
F + X (S + \bar S) \, = \, N,
\end{equation}
with the $X$ dependent function $N$ written in terms of
$X(\phi, \phib, S+\bar S)$. The determination of the normalization of the
kinetic terms of $S$, $\bar S$ is left as an exercise.

\subsec{ Several Linear Multiplets}
\label{sevlin}

The linear superfield formalism can be easily generalized to accommodate
several linear multiplets. Noting
$L^\uti$, with I $ =0,1,...,n$, the $n+1$ copies of linear superfields
we will have a set of $n+1$ modified linearity conditions
\begin{eqnarray}
\prokib L^\uti &=& 2k^\uti_{\mbox{\tiny{{\bf G}}}} \, \tr
\cw_{\mbox{\tiny{{\bf G}}}}^2 , \label{gmlc1}
\\ \proki L^\uti &=& 2k^\uti_{\mbox{\tiny{{\bf G}}}}
\, \tr  \cwb_{\mbox{\tiny{{\bf G}}}}^2 . \label{gmlc2}
\end{eqnarray}
Here the subscript ${\mbox{\tiny{{\bf G}}}}$ indicates that different
linear combinations of Chern-Simons  forms (Yang-Mills potentials for
different gauge groups) may couple to different antisymmetric tensors.

In this general scenario the kinetic potential $K$ and the
subsidiary function $F$ will be functions of the $n+1$ superfields $L^\uti$.
The superfield action
\begin{equation}
{\cal{L}} \,=\,-3\,\int \! \! E F(\phi, \phib , L^\uti),
\label{g3.13}
\end{equation}
depends implicitly on $K(\phi, \phib , L^\uti)$ through $E$ due to the
geometric construction.

The presence of several linear superfields implies that different gauge
sectors may have different gauge coupling functions. The determination
of the explicit form of the gauge coupling and normalization functions
follows exactly the same steps as in the case of a single linear
superfield, taking now the chiral  superfields ${\bf r}$ and
${\bf {\bar r}}$ to be
\begin{equation}
{\bf r} \, = \, -\frac{1}{8} \proki F(\phi, \phib, L^\uti), \cem
{\bf {\bar r}} \, = \, -\frac{1}{8} \prokib F(\phi, \phib, L^\uti).
\label{ggenact}
\end{equation}
As a result, the normalization function takes the form
\begin{equation}
N(\phi, \phib , L^\uti) \, =
\, \frac{F-L \cdot F_L}{1-\frac{1}{3} L \cdot K_L},
\label{gnormf}
\end{equation}
whereas the gauge coupling functions are given as
\begin{equation}
\Gamma_{\mbox{\tiny{{\bf G}}}}(\phi, \phib, L^\uti)
\, = \,( F_\dti-\frac{N}{3}  K_\dti ) \, k^\uti_{\mbox{\tiny{{\bf G}}}}.
\label{gnhgcf}
\end{equation}
We use here the notation
$L \cdot F_L = L^\uti F_\dti$, with $F_\dti$ denoting
the derivative of $F$ with respect to $L^\uti$, and the same for $K$.
The gauge coupling and normalization functions satisfy the sum rule
\begin{equation}
L^\uti \Gamma_{\bti} + N \, = \, F,
\label{gsumrule}
\end{equation}
with $\Gamma_{\bti}$ identified as
$\Gamma_{\mbox{\tiny{{\bf G}}}}
= \Gamma_{\bti} \, k^\uti_{\mbox{\tiny{{\bf G}}}}$. The brackets
indicate that the enclosed subscript does not refer to a derivative. It
is also interesting to note that the kinetic term
$g^{mn}
\prt_m L^\utj
\prt_n L^\uti$ is multiplied by a function
\begin{equation}
G_{\mbox{\tiny{(IJ)}}} \, = \, F_{\dti \dtj}
       - \frac{1}{3} ( N K_{\dti \dtj} + N_\dti K_\dtj + N_\dtj K_\dti ).
\label{GIJ}
\end{equation}
The effective transformations in the case of a single linear multiplet
generalize as well. To this end we observe first of all that a
replacement
\begin{equation}
F(\phi, \phib , L^\uti) \, \mapsto \,
     F(\phi, \phib , L^\uti) + L^\uti V_{\bti}(\phi, \phib),
\end{equation}
leaves the normalization function (\ref{gnormf}) as well as the sum rule
(\ref{gsumrule}) invariant, whereas the gauge coupling function changes
as
\begin{equation}
\Gamma_{\bti}(\phi, \phib , L^\uti) \, \mapsto \,
     \Gamma_{\bti}(\phi, \phib , L^\uti) + V_{\bti}(\phi, \phib).
\end{equation}
The counterpart of the effective action (\ref{3.15}) in the presence
of several multiplets becomes
\begin{equation}
{\cal{L}}_{\rm{lin}}\,=\,- 3 \int \! \! E \,
L^\uti V_{\bti}(\phi,\phib),
\label{g3.15}
\end{equation}
with effective transformations
\begin{equation}
V_{\bti}(\phi,\phib) \, \mapsto \, V_{\bti}(\phi,\phib)
           + H_{\bti}(\phi) + \bar{H}_{\bti}(\phib),
\label{g3.16}
\end{equation}
giving rise to
\begin{equation}
{\cal{L}}_{\rm{lin}} \, \mapsto \, {\cal{L}}_{\rm{lin}} +
          \frac{3k}{4} \int \! \! \frac{E}{R}
     H_{\bti}(\phi) \, k^\uti_{\mbox{\tiny{{\bf G}}}}
     \tr \cw_{\mbox{\tiny{{\bf G}}}}^2
         + \frac{3k}{4}  \int \! \! \frac{E}{R^{\dagger}}
     \bar{H}_{\bti}(\phib) \, k^\uti_{\mbox{\tiny{{\bf G}}}}
      \tr \cwb_{\mbox{\tiny{{\bf G}}}}^2 .
\label{g3.21}
\end{equation}
This shows that the case of several linear multiplets is more flexible in
view of possible applications to anomaly cancellation mechanisms.

\indent

As to the duality transformations between the linear and the chiral
superfield formalism we will make use of $n+1$ unconstrained real
superfields $X^\uti$ together with the real combination $S_\dti+\bar{S}_\dti$ of
chiral superfields. The first order action (\ref{3.17P})
generalizes then to
\begin{equation}
{\cal{L}}_{\! \mbox{\tiny{FOF}}}
\,=\,-3 \int \! \! E \left[ F(\phi,\phib,X^\uti)
   +(X^\uti -\,k^\uti_{\mbox{\tiny{{\bf G}}}}\,\Omega_{\mbox{\tiny{{\bf G}}}})
(S_\dti+\bar{S}_\dti) \right],
\label{g3.17P}
\end{equation}
with $\Omega_{\mbox{\tiny{{\bf G}}}}$ the Chern-Simons superfield
pertaining to the gauge sector specified by the subscript
${\mbox{\tiny{{\bf G}}}}$. Variation with respect to $S_\dti$, resp.
$\bar{S}_\dti$ gives back the theory in the linear superfield formalism,
whereas variation with respect to $X^\uti$ gives rise to the equation
\begin{equation}
\left( \, S_\dti + \bar{S}_\dti \, \right)
\left( 1 - \frac{1}{3} \, X \cdot K_X\right) \, = \,
 \frac{F}{3} \, K_\dti - F_\dti.
\end{equation}
Again, this should be understood as an equation which expresses, for
given kinetic potential $K$ and subsidiary function $F$, the previously
unconstrained real superfields $X^\uti$ in terms of $\phi$, $\phib$ and
$S_\dti + \bar{S}_\dti$.

\indent

Coming back to the linear superfield formalism, we note that
the particular form (\ref{gnormf}) of the normalization function $N$
suggests to introduce projective variables for the set of linear
superfields. Choosing a particular linear superfield of reference, say
$L^0$, we define
\begin{equation}
L_0 \, = \, L, \cem \cem
\xi^\uti \, = \, \frac{L^\uti}{L^0} \, ,
\end{equation}
with I ranging from $1$ to $n$ whenever attached to a projective
variable $\xi$. The kinetic potential
$K$ and the subsidiary function $F$ are now supposed to be given
in terms of $L$ and $\xi^\uti$. In this parametrization the
normalization function $N$ takes the form
\begin{equation}
N(\phi, \phib , L, \xi^\uti)
  \, = \, \frac{F-L F_L}{1-\frac{1}{3} L K_L}.
\label{normfpro}
\end{equation}
Here only derivatives with respect to the particular superfield
$L$ occur. This closely resembles (\ref{normf}), except for the
additional dependence on the projective variables $\xi^\uti$.
Likewise, in the effective Lagrangian density one may parametrize
\begin{equation}
 L^\uti V_{\bti}(\phi, \phib) \, = \,
  L \, {\cv}(\phi, \phib, \xi^\uti),
\end{equation}
with (identifying $V_{(0)} = V$)
\begin{equation}
{\cv}(\phi, \phib, \xi^\uti) \, = \, V(\phi, \phib)
+\xi^\uti V_{\bti}(\phi, \phib).
\end{equation}
Observe that we could have chosen, instead of $L^0$, another superfield
of reference, without changing the reasoning. Different choices are
related in terms of reparametrizations in an obvious way.

\indent

As a last remark consider the linear superfield formalism for the case
of a trivial coupling function $N=1$. From the previous discussion, it
should be clear that we recover the same type of differential
equation (\ref{3.13l}) as in the case of a single linear multiplet
\begin{equation}
F - L F_L \,=\, 1 - \frac{1}{3} L K_L,
\label{3.23}
\end{equation}
which is solved in the same way, \ie
\begin{equation}
F\,=\,1 + L \, \cv (\phi,\phib,\xi^\uti)
  +\frac{L}{3} \int \frac{d \la}{\la} K_\la
     (\phi,\phib,\la,\xi^\uti).
\label{3.24}
\end{equation}
In conclusion, the linear superfield formalism for the case of several
linear multiplets exhibits a quite intriguing structure which clearly
should be further investigated. It would be interesting to pursue this
approach in the context of duality transformations and the construction
of the respective component field actions.

\newpage
\sect{ 3-FORM COUPLING TO SUPERGRAVITY \label{3F}} \subsec{
General remarks \label{3F1}}

The 3-form supermultiplet is, besides the chiral and linear
multiplet, yet another supermultiplet describing helicity
$(0,1/2)$. It consists of a three-index antisymmetric gauge
potential $C_{lmn}(x)$, a complex scalar $Y(x)$, a Majorana
spinor with Weyl components $\eta_\al(x)$, $\eta^\da(x)$ and a
real scalar auxiliary field $H(x)$.

In superfield language
\cite{Gat81}, \cite{BPGG96} it is described by a chiral
superfield
\begin{equation}
D^\da Y \,=\,0, \cem D_\al \ovY \,=\,0,
\end{equation}
which is subject to the additional constraint
\begin{equation}
D^2 \ovY - \bD^2 Y \,=\,\f{8i}{3} \vep^{klmn} \Si_{klmn},
\label{con3f}
\end{equation}
with the field strength of the 3-form gauge potential defined as
\begin{equation}
\Si_{klmn} \,=\,\prt_k C_{lmn} - \prt_l C_{mnk}
               + \prt_m C_{nkl} - \prt_n C_{klm}.
\end{equation}
It is invariant under the transformation
\begin{equation}
C_{lmn} \ \mapsto \ C_{lmn} + \prt_l \La_{mn}
                    + \prt_m \La_{nl} + \prt_n \La_{lm},
\end{equation}
where the gauge parameters $\La_{mn} = - \La_{nm}$ have an
interpretation as a 2-form coefficients.

The component fields of the 3-form multiplet are propagating:
supersymmetry couples the rank-3 antisymmetric tensor gauge
potential with {\em dynamical} degrees of freedom. This should be
compared to the non-supersymmetric case, discussed in the context
of the cosmological constant problem
\cite{Haw84,Col88,DvN80,Duf89}, where the 3-form does not imply
dynamical degrees of freedom.

In section \ref{3F2} the superspace formulation of \cite{Gat81} will be
adapted to the background of $U_K(1)$ superspace, providing the geometric
structure underlying the coupling of the 3-form multiplet to the general
supergravity/matter/Yang-Mills system (and to linear multiplets, if
desired). We discuss in particular the 3-form Bianchi identities in the
presence of appropriate constraints and define supergravity
transformations on the superfield and component field levels.

As constraint chiral superfields, subject to the additional
constraint (\ref{con3f}),
$Y$ and $\ovY$ derive from one and the same real pre-potential $\Om$
superfield such that
\begin{equation}
Y \,=\,-4 {\bar D}^2 \Om, \cem \ovY \,=\,-4 D^2 \Om.
\label{eq:prepotential}
\end{equation}
In appendix \ref{appF} we present a detailed derivation of the
explicit solution of the 3-form constraints in the background of
$U(1)$ superspace and identify the unconstrained pre-potential
$\Om$ in this general geometric context.

The 3-form superfields $Y$ and $\ovY$ differ from usual chiral
superfields, employed for the description of matter multiplets in yet
another respect: they have non-vanishing chiral weights. This property
modifies considerably the possible supergravity couplings, compared to
the case of vanishing chiral weights. In section \ref{3F3} we give a very
detailed account of the couplings of the 3-form multiplet to supergravity
and matter.

Although the study of the 3-form multiplet is interesting in its
own right, it has an interesting application in the description of
gaugino condensation. There, as a consequence of the chirality of
the gaugino superfields, the composite superfields $\tr(\cw^2)$
and $\tr(\cwb^2)$ obey chirality conditions
\begin{equation}
D^\da \, \tr(\cw^2) \,=\,0, \cem D_\al \, \tr(\cwb^2) \,=\,0,
\end{equation}
as well. On the other hand, the gaugino superfields are subject
to the additional constraint (\ref{con3f}), which translates into
an additional equation for the composites, corresponding to
\ref{con3f}. At the component field level this implies the
identification
\begin{equation}
D^2 \tr(\cw^2) \loco - \bD^2 \tr(\cwb^2) \loco \,=\,
i \vep^{klmn} \tr({\fym}_{kl} {\fym}_{mn}),
\end{equation}
 where the topological density
\begin{equation}
\vep^{klmn} \tr({\fym}_{kl}{\fym}_{mn}) \,=\,-\f{2}{3} \vep^{klmn}
\prt_k Q_{lmn},
\end{equation}
plays now the role of
the field-strength and the Chern-Simons form (which, under
Yang-Mills transformations changes indeed by the derivative of a
2-form) the role of the 3-form gauge potential. The analogy between
the Chern-Simons forms in superspace and the 3-form geometry is
discussed in detail in appendices \ref{appF2}, \ref{appF3}, and has
already been exploited in section \ref{2F2}.

\subsec{ The 3-Form Multiplet Geometry \label{3F2}}
The superspace geometry of the 3-form multiplet has been known for
some time \cite{Gat81}. Its coupling to the general
supergravity/matter/Yang-Mills system is most conveniently described in the
framework of $U_K(1)$ superspace -cf. section \ref{GRA3}.
This approach is particularly useful in view of the non trivial
\ka transformations of the 3-form superfield $Y$. Moreover, it provides a
concise way to derive supergravity transformations of the component fields.

\subsubsection{ Constraints and Bianchi identities \label{3F21}}
The basic geometric object is the 3-form gauge potential
\begin{equation}
 C \,=\,\frac{1}{3!} dz^L dz^M dz^N C_{NML},
\end{equation}
subject to 2-form gauge transformations of parameter
$\La = \frac{1}{2} dz^M dz^N \La_{NM}$ such that
\begin{equation}
C \, \mapsto \, C + d \La. \label{GFB3}
\end{equation}
 The invariant field strength
 \begin{equation}
  \Si \,=\,d C \  = \ \frac{1}{4!}E^A E^B E^C E^D \, \Si_{DCBA},
\end{equation}
is a 4-form in superspace with coefficients
\begin{equation}
\frac{1}{4!} E^A E^B E^C E^D \, \Si_{DCBA} \, =\, \frac{1}{4!} E^A
E^B E^C E^D \lp 4 \; \cd_D C_{CBA} + 6 \; T_{DC}{}^F C_{FBA}
\rp.  \label{3FBI}\end{equation}
 Here, the full
$U_K(1)$ superspace covariant derivatives and torsions appear.
Likewise, the Bianchi identity, $d \Si \,=\,0$, is a 5-form with
coefficients
\begin{equation}
\frac{1}{5!} E^A E^B E^C E^D E^E \left( 5 \; \cd_E \Si_{DCBA} + 10
\; T_{ED}{}^F \Si_{FCBA} \right) \,=\,0.
\label{4FBI}\end{equation}
 In these
formulas we have kept the covariant differentials in order to keep
track of the graded tensor structure of the coefficients.

The multiplet containing the  3-form gauge potential is obtained
after imposing constraints on the covariant field-strength
coefficients. Following \cite{Gat81} we require
\begin{equation}
\Si_{\unddt \, \undgm \, \undbt \ A} \,=\,0, \label{sugraconst}
\end{equation}
where
$ \ \undal \sim \al, \da \ $ and $ \ A \sim a, \al, \da$.
The consequences of these constraints can be studied by
analyzing consecutively the Bianchi identities, from lower to
higher canonical dimensions.
The tensor structures of the coefficients of $\Si$ at higher
canonical dimensions are then subject to restrictions due to the
constraints. In addition, covariant superfield conditions
involving spinor derivatives will emerge. The constraints serve to
reduce the number of independent component fields to those of the 3-form
multiplet, but do not imply any dynamical equations.

As a result of this analysis (alternatively, appendix \ref{appF1} provides
the explicit solution of the constraints in terms of an unconstrained
pre-potential), all the coefficients of the 4-form field strength
$\Si$ can be expressed in terms of the two superfields $\ovY$ and $Y$, which
are identified  in the tensor decompositions
 \begin{equation}
  \Si_{\dt \gm \ ba}=  \frac{1}{2} (\si_{ba} \eps)_{\dt \gm} \ovY, \cem
   {\Si^{\dd \dg}}_{\ ba} =  \frac{1}{2} ({\sib}_{ba} \eps)^{\dd {\dg}} Y.
\label{7.12} \end{equation}
As a consequence, the $U_K(1)$ weights of $Y$ and $\ovY$ are
\begin{equation}
 w(Y) \,=\,+2, \cem w(\ovY) \,=\,-2.
\end{equation}
This implies that the covariant exterior derivatives
\begin{equation}
 \cd Y \,=\,d Y +2\, {A}\, Y, \cem \cd \ovY \,=\,d \ovY -2\, {A}\, \ovY,
\end{equation}
contain ${A} = E^M {A}_M$, the $U_K(1)$ gauge potential. On
the other hand, the {\em {Weyl}} weights are determined to be
\begin{equation}
 \om(Y) \,=\,\om (\ovY) \,=\,+3.
\end{equation}
 By a special choice of conventional constraints,
\ie a covariant redefinition of $C_{cba}$, it is possible to
impose
\begin{equation}
{{\Si_\dt}^{\dg}}_{\ ba} \,=\,0.
\end{equation}
The one spinor-three vector components of $\Si$ are given as
\begin{equation}
 \Si{}_{\, \dt \ cba} \, = \,
- \frac{1}{16} \si^d_{\dt \dd} \, \eps_{dcba} \cd^\dd \ovY, \cem
\Si^\dd{}_{\, cba} \, = \,
+\frac{1}{16} {\sib}^{d \dd \dt } \, \eps_{dcba} \cd_\dt Y.
\end{equation}
 At the same time, the superfields $\ovY$ and
$Y$ are subject to the chirality conditions
\begin{equation}
\cd_\al \ovY \,=\,0, \cem \cd^\da Y \,=\,0 \label{con1}
\end{equation}
and are further constrained by the relation
\begin{equation}
\frac{8i}{3} \eps^{dcba} \Si_{dcba} \,=\,
      \left( \cd^2 - 24 \rd \right) Y
    - \left( \cdb^2 - 24 R \right) \ovY,
\label{con2}
\end{equation}
indicating that the imaginary part of the F-term of the 3-form superfield is
given as the curl of the 3-form gauge potential, with a number of
additional nonlinear terms due to the coupling to supergravity.

In conclusion, we have seen that all the coefficients of the
superspace 4-form $\Si$, subject to the constraints, are given in
terms of the superfields  $\ovY$ and $Y$ and their spinor
derivatives. It is a matter of straightforward computation to show
that all the remaining Bianchi identities do not contain any new
information.

\subsubsection{ Component fields and supergravity transformations
\label{3F22}}
As usual, we define component fields as lowest components of superfields.
First of all, the 3-form gauge potential is identified as
\begin{equation}
 C_{klm}{\loco} \,=\,C_{klm}(x).
\end{equation}
 As to the components of $Y$ and $\ovY$ we define
\begin{equation}
Y{\loco} \,=\,Y (x), \cem \cd_\al Y{\loco} \,=\,\s2 \,
\eta_\al(x),
\end{equation}
 and
\begin{equation}
\ovY{\loco} \,=\,\ovY(x), \cem \cdb^\da \ovY{\loco} \,=\,\s2 \,
\etab^\da (x).
\end{equation}
 At the level of two
covariant spinor derivatives we define the component $H(x)$ as
\begin{equation}
\cd^2 Y{\loco} + \cdb^2 \ovY{\loco} \,=\,-8 \, H(x).
\label{defH}\end{equation}
The orthogonal combination
however is not an independent component field. Projection to lowest
components of (\ref{con2}) shows that it is given as
\begin{eqnarray}
\cd^2 Y{\loco} - \cdb^2 \ovY{\loco} \ &=& \
    - \frac{32i}{3} \vep^{klmn} \prt_k C_{lmn}
 + 2 \s2 i \, (\psib_m \sib^m)^\al \eta_\al
   - 2 \s2 i \, (\psi_m \si^m)_\da \etab^\da \nn \\[2mm]
&& - 4 (\ovM + \psib_m \sib^{mn} \psib_n) \, Y
     + 4 (M + \psi_m \si^{mn} \psi_n) \, \ovY.
\label{dC}
\end{eqnarray}
This expression provides the supercovariant component field
strength of the 3-form gauge potential, displaying  the
modifications which arise from the coupling to supergravity: here
the appearance of the Rarita-Schwinger field and the supergravity
auxiliary field, in the particular combination $M \ovY - \ovM Y$ .

The component fields in the supergravity,
matter and Yang-Mills sectors are defined as usual -cf. section
\ref{CPN1}. Some new aspects arise in the treatment of the field
dependent $U_K(1)$ pre-potential due to the presence of the fields
$Y$ and $\ovY$, carrying non-vanishing $U_K(1)$ weights. It is for
this reason that we refrain from calling $K$ a \ka potential, we
rather shall refer to the field dependent  $U_K(1)$ pre-potential
as {\em kinetic potential}.

Before turning to the derivation of the supergravity
transformations we shortly digress on the properties of the
composite $U_K(1)$ connection arising from the kinetic
pre-potential
\[ \ K(\phi,Y,\phib,\ovY), \]
subject to \ka transformations
\[ \ K(\phi,Y,\phib,\ovY) \ \mapsto \ K(\phi,Y,\phib,\ovY) + F(\phi) +
\bF(\phib). \]
Requiring invariance of the kinetic potential under $U_K(1)$ transformations
of the superfields $Y$ and $\ovY$, implies the relation
\begin{equation}
Y K_{Y} = \ovY K_{\ovY},
\label{Kinv}
\end{equation}
which we shall use systematically\footnote{The special kinetic potential
\[K(\phi,\phib,Y,\ovY) = \log \left[ X(\phi,\phib) + Z(\phi,\phib) \,
\ovY Y \right], \] where $X$ and $Z$ are functions of the matter
fields, is a non-trivial example which satisfies this condition.
\label{foot}}. The composite $U_K(1)$ connection derives from the
commutator term $ [\cd_\al,\cd_\da] K$, which, in the presence of
the 3-form superfields is given as
\begin{eqnarray}
 [\cd_\al,\cd_\da] K &=& 2i K_k \cd_{\al \da}
\phi^k-2i K_\bk \cd_{\al \da} \phib^\bk
 +2i K_Y \cd_{\al \da} Y - 2i K_{\ovY}
\cd_{\al \da} \ovY \nn \\ &&+2 K_{\ca \cab} \cd_\al \Psi^\ca
\cd_\da \Psib^\cab
   +6 \, (Y K_Y + \ovY K_{\ovY}) \,  G_{\al \da}, \end{eqnarray}
where we use the shorthand notation $\Psi^\ca= (\phi^k, Y)$, and
$\Psib^\cab  = (\phib^\bk, \ovY)$, with obvious meaning for
$K_{\ca \cab}$. The important point is that on the right hand the
$U_K(1)$ connection, ${A}$, appears in the covariant derivatives
of $Y$ and $\ovY$ due to their non-vanishing $U_K(1)$ weights.
Explicitly one has
\[ \cd_{\al \da} \ovY{\loco} \,=\,
\si^m_{\al \da}\left( \prt_m \ovY -2 {A}_m \ovY - \f{1}{\s2}
\psib_{m \, \dv} \etab^\dv \right),  \]
\[ \cd_{\al \da} Y{\loco} \,=\,
\si^m_{\al \da}\left( \prt_m Y +2 {A}_m Y - \f{1}{\s2}
\psib_m{}^\vp \eta_\vp \right). \]
 Substituting in the defining
equation for ${A}_m$ (\ref{GRA.146}) and factorizing gives then
rise to
\begin{eqnarray}
{A}_m(x) + \f{i}{2} \, e_m{}^a b_a \ &= &\ \frac{1}{4} \
\frac{1}{1-Y K_Y} \lp K_k \, \cd_m A^k - K_\bk \, \cd_m \bA^\bk
\right.  \nn \\[1mm] && \left. + K_Y \, \prt_m Y - K_{\ovY} \,
\prt_m \ovY + i \, \sib_m^{\da \al} K_{\ca \cab} \, \Psi_\al^\ca
\, \Psib_\da^\cab \ \rp . \label{Am}
\end{eqnarray}
As above, we use the shorthand notation
$\Psi_\al^\ca=(\chi_\al^k, \eta_\al)$ and
$\Psib_\da^\cab=(\chib_\da^\bk, \etab_\da)$. As is easily verified
by an explicit calculation, ${A}_m$ defined this way transforms
as it should under the $U_K(1)$ transformations given above, \ie
\[ {A}_m \ \mapsto \ {A}_m + \f{i}{2} \prt_m \ima F.\]
Observe that the factor $({1-Y K_Y})^{-1}$ accounts for the
non-trivial $U_K(1)$ phase transformations
\[Y \ \mapsto \ Y e^{-i \ima F}, \cem
        \ovY \ \mapsto \ \ovY e^{+i \ima F}, \]
of the 3-form superfields.

We turn now to the derivation of supergravity transformations. In section
\ref{GRA32} they were defined as combinations of superspace diffeomorphisms
and field dependent gauge transformations.
In the case of the 3-form one has
\begin{equation}
\delta C \,=\,(\imath_\xi d + d \imath_\xi) C + d \La
           \,=\,\imath_\xi \Si + d \left( \La + \imath_\xi C \right),
 \end{equation}
the corresponding supergravity transformation is defined as a
diffeomorphism of parameter $\xi^A=\imath_\xi E^A$ together with a
compensating infinitesimal 2-form gauge transformation of
parameter $\La \,=\,- \imath_\xi C$, giving rise to
\begin{equation}
 \delta_{\WZ} C \,=\,\imath_\xi \Si \, =\, \f{1}{3!} E^A E^B E^C \xi^D
  \Si_{DCBA}. \end{equation}
   The supergravity transformation of the component 3-form gauge field
$C_{klm}$ is then simply obtained from the double bar projection
\cite{BBG87} (simultaneously to lowest superfield components and
to space-time differential forms) as
\begin{equation}
 \delta_{\WZ} C  {\doubar} \
= \ \f{1}{3!} dx^k dx^l dx^m \delta_{\WZ} C_{mlk} \,=\,\f{1}{3!} e^A e^B
e^C \xi^{\unddt} \ \Si_{\unddt \ CBA}{\loco}. \end{equation}
Taking into account the definition $e^A = E^A {\doubar}$
(\ref{CPN.5a}, \ref{CPN.5b}) and the particular form of the
coefficients of $\Si$ we obtain
\begin{equation}
 \delta_{\WZ} C_{mlk} \,=\,\frac{\s2}{16} \left(
\xib \sib^n \eta - \xi \si^n \etab \right) \vep_{nmlk} +
\frac{1}{2} \oint_{mlk} \left[ (\psi_m \si_{lk} \xi) \, \ovY
                          +(\psib_m \sib_{lk} \xib) \, Y \right]. \end{equation}

Let us turn now to the transformations of the remaining
components. To start, note that at the superfield level, one has
\begin{eqnarray}
 \dt_{\WZ} Y &=& \imath_\xi d Y \,=\,\imath_\xi \cd Y -2
\imath_\xi {A} \, Y, \\ \dt_{\WZ} \ovY &=& \imath_\xi d \ovY \,=\,
\imath_\xi \cd \ovY +2 \imath_\xi {A} \, \ovY.
\end{eqnarray}
 Taking
into account the explicit form of the field-dependent factor
$\imath_\xi {A}{\loco} = \xi^\undal {A}_\undal{\loco}$ -compare
to (\ref{CPN.44})- one finds
 \begin{eqnarray}
\dt_{\WZ} Y &=& \s2 \xi^\al
     \left\{ (1-\f12 Y K_Y)\eta_\al - \f12 Y K_k \chi^k_\al
      \right\} +\f{1}{\s2} \xib_\da Y
     \left\{ K_{\ovY} \etab^\da + K_\bk \chib^{\da \bk}\right\},\nn \\[2mm]
 \dt_{\WZ} \ovY &=& \s2 \xib_\da
     \left\{ (1-\f12 \ovY K_{\ovY})\etab^\da -\f12 \ovY K_\bk \chib^{\da
\bk}\right\}  + \f{1}{\s2} \xi^\al \ovY
     \left\{K_Y \eta_\al + K_k \chi^k_\al \right\}. \nn \\ &&
\end{eqnarray}
 It is more convenient to use a notation where one keeps the
combination
\begin{equation}
 \Xi \,=\, \xi^\undal {A}_\undal{\loco} \,=\,
\f{1}{2\s2} \xi^\al \left( K_k \chi^k_\al + K_Y \eta_\al \right)
-\f{1}{2\s2} \xib_\da \left( K_\bk \chib^{\da \bk} + K_{\ovY}
\etab^\da \right), \end{equation}
 giving rise to a  compact form
of the supersymmetry transformations
\begin{equation}
 \dt_{\WZ} Y \, = \, \s2 \, \xi^\al \eta_\al -2 \, \Xi \, Y, \cem
    \dt_{\WZ} \ovY \, = \, \s2 \, \xib_\da \etab^\da +2 \, \Xi \, \ovY.
\end{equation}
The transformation law for the 3-"forminos" comes out as
\begin{eqnarray}
 \dt_{\WZ} \eta_\al &=& \s2 \xi_\al H
                   +\f{4i\s2}{3} \xi_\al \vep^{klmn} \prt_k C_{lmn}
                   +i\s2 (\xib\sib^m \eps)_\al \nabla_m Y
                   - \Xi \, \eta_\al \nn \\
&&-\f{i}{2} \xi_\al \left( \psib_m \sib^m \eta - \psi_m \si^m
\etab \right)
  -i (\xib\sib^m \eps)_\al \psi_m{}^\vp \eta_\vp \nn \\
&& + \f{1}{\s2} \xi_\al \left\{ (\ovM + \psib_m \sib^{mn} \psib_n)
\, Y
     - (M + \psi_m \si^{mn} \psi_n) \, \ovY \right\},
\end{eqnarray}
 and
  \begin{eqnarray}
   \dt_{\WZ} \etab^\da &=& \s2 \xib^\da H +i \s2
(\xi \si^m \eps)^\da \nabla_m \ovY
                 -\f{4i\s2}{3} \xib^\da \vep^{klmn} \prt_k C_{lmn} + \Xi \,
\etab^\da \nn \\ && +\f{i}{2} \xib^\da \left( \psib_m \sib^m \eta
- \psi_m \si^m \etab \right)
    -i (\xib\sib^m \eps)_\al \psib_{m \dv} \etab^\dv \nn \\
&& - \f{1}{\s2} \xib^\da \left\{ (\ovM + \psib_m \sib^{mn}
\psib_n) \, Y
     - (M + \psi_m \si^{mn} \psi_n) \, \ovY \right\}.
\end{eqnarray}
 Finally, the supergravity transformation of $H$ is given as
\begin{eqnarray}
 \dt_{\WZ} H &=& \f{1}{\s2} (\xib\sib^m)^\al \nabla_m \eta_\al
          + \f12 (\xib\sib^m \si^n \psib_m)
                   (\nabla_n Y - \f{1}{\s2} \psi_m{}^\vp \eta_\vp) \nn
\\[1mm]
      & & \f{1}{\s2} (\xi \si^m)_\da \nabla_m \etab^\da
          + \f12 (\xi \si^m \sib^n \psi_m)
                   (\nabla_n \ovY - \f{1}{\s2} \psib_{m \dv} \etab^\dv) \nn
\\[1mm] && + \f{1}{3 \s2} \ovM \xi^\al \eta_\al
   + \f{1}{3 \s2} M \xib_\da \etab^\da
   + \f{1}{3 \s2} (\xib \sib^a \eta + \xi \si^a \etab) b_a \nn \\[1mm]
&& + Y \, \xib_\da {\bar{X}}^\da{\loco}  + \ovY \, \xi^\al {
X}_\al{\loco}
   - \f{i}{\s2} (\xib \sib^m \psi_m + \xi \si^m \psib_m) H \nn \\[1mm]
&& + \f{2}{3}(\xib \sib^p \psi_p - \xi \si^p \psib_p)
                   \vep^{klmn} \prt_k C_{lmn}
 -\f{1}{4 \s2}(\xib \sib^n \psi_n - \xi \si^n \psib_n)
 (\psib_m \sib^m \eta - \psi_m \si^m \etab) \nn \\[1mm]
&&  -\f{i}{4}(\xib \sib^l \psi_l - \xi \si^l \psib_l)
           \left\{ (\ovM + \psib_m \sib^{mn} \psib_n) \, Y
         - (M + \psi_m \si^{mn} \psi_n) \, \ovY \right\}. \label{delH}
 \end{eqnarray}

Note that in the above equations we changed $\cd$-derivatives into
$\nabla$-derivatives as in section \ref{CPN2} -cf. (\ref{CPN.60}),
(\ref{CPN.61})- using a redefined ${U_k}(1)$ connection $ v_m(x) =
{A}_m(x) + \f{i}{2} \, e_m{}^a b_a$. This allows to keep track of
the auxiliary field $b_a$, otherwise concealed in
the numerous covariant derivatives occurring in the Lagrangian.
We still have to work out the component field expressions for
${X}_\al \loco$ and ${\bar{X}}^\da \loco$ from the superfields
\begin{equation}
 {X}_\al \,=\,-\f{1}{8} \lp \cdb^2 - 8 R \rp \cd_\al K, \cem
{\bar{X}}^\da \,=\,-\f{1}{8} \lp \cd^2 - 8 \rd \rp \cdb^\da K,
\end{equation}
given in terms of the matter and 3-form superfield dependent kinetic
potential $K$. This can be achieved in
successively applying the spinor derivatives to $K$.
Alternatively, one may use the expression
\begin{eqnarray}
 {A} &=& \f{1}{4} K_\ca \cd \Psi^\ca - \f{1}{4} K_\cab
\cd \Psib^\cab
            + \f{i}{8} E^a \sib_a^{\da \al}
                  K_{\ca \cab} \cd_\al \Psi^\ca \cd_\da \Psib^\cab \nn \\
&&            + \f{3i}{2} E^a G_a \lp 1 - \f12 (Y K_Y + \ovY
K_{\ovY})\rp, \end{eqnarray}
 for the composite $U_K(1)$ connection,
take the exterior derivative $d{A} = {F}$ and identify
${\bar{X}}^\da$ and ${X}_\al$ in the 2-form coefficients
\begin{equation}
{F}_{\bt a} \,=\,+ \f{i}{2} \si_{a \, \bt \db} {\bar{X}}^\db
                    + \f{3i}{2} \cd_\bt G_a, \cem
   {F}^\db {}_a \,=\,- \f{i}{2} \sib_a^{\bt \db} {X}_\bt
                    + \f{3i}{2} \cd^\db G_a. \end{equation}
A straightforward calculation then yields the component field
expression\footnote{ We make use, in the Yang-Mills sector, of the
suggestive notations
\[
K_\bk \, (\lab^\da \! \! \cdot \! \bA)^\bk \,= \,
\lab^{{(r)} \, \da} K_\bk \left({\bf T}_{(r)} \bA \right)^\bk,
\cem
K_k \, (\la_\al \! \cdot \! A)^k \, = \,
       \la_\al^{(r)}  K_k \lp {\bf T}_{(r)} A\rp^k .
\]}
\begin{eqnarray}
&& {\bar{X}}^\da (1-\ovY K_{\ovY}){\loco} \, = \,
     -\f{i}{\s2} K_{\ca \cab} \Psi_\al^\ca \sib^{m \, \da \al}
            \lp \nabla_m \Psib^\cab -\f{1}{\s2} \psib_{m \, \dv} \Psib^{\dv
\cab} \rp  \nn \\[1mm] &&   - \f{\s2}{8} \cd^2 \phi^k{\loco} \,
K_{k \cab} \Psib^{\da \cab}
   + \f{1}{\s2} H K_{Y \cab} \Psib^{\da \cab}
   + \f{4i}{3 \s2} \,\vep^{klmn} \prt_k C_{lmn}\,\Psib^{\da \cab} K_{\cab Y}
\nn \\[1mm] && -\f{1}{2 \s2} K_{\cab \cb \cc} \, \Psi^{\al \cc} \,
\Psi_\al^\cb \, \Psib^{\da \cab}
   - i K_\bk \, (\lab^\da \! \! \cdot \! \bA)^\bk   - \f{i}{4} \Psib^{\da \cab} K_{\cab Y}
           (\psib_m \sib^m \eta - \psi_m \si^m \etab) \nn \\[1mm]
&&+ \f{1}{2 \s2} \Psib^{\da \cab} K_{\cab Y}
         \left\{ (\ovM + \psib_m \sib^{mn} \psib_n) \, Y
         - (M + \psi_m \si^{mn} \psi_n) \, \ovY \right\}
\end{eqnarray}
 and
  \begin{eqnarray}
&&  {X}_\al (1-Y K_Y){\loco} \,=\,
         -\f{i}{\s2} K_{\ca \cab} \Psi^{\da \ca} \, \sib^m_{\al \da}
            \lp \nabla_m \Psi^\ca -\f{1}{\s2} \psi_m{}^\vp \Psi_\vp^\ca \rp
\nn
\\[1mm]
&&   - \f{\s2}{8} \cdb^2 \phib^\bk \, K_{\ca \bk} \Psi_\al^\ca
   + \f{1}{\s2} H K_{\ca \ovY  } \Psi_\al^\ca
   - \f{4i}{3 \s2} \, \vep^{klmn} \prt_k C_{lmn} \,\Psi_\al^\ca K_{\ca \ovY  }
\nn \\[1mm] && -\f{1}{2 \s2} K_{\ca \cbb \ccb} \, \Psib_\da^\ccb
\,
                   \Psib^{\da \cbb} \, \Psi_\al^\ca
 + i K_k \, (\la_\al \! \! \cdot \! A)^k + \f{i}{4} \Psi_\al^\ca K_{\ca \ovY  }
           (\psib_m \sib^m \eta - \psi_m \si^m \etab) \nn \\[1mm]
&&- \f{1}{2 \s2} \Psi_\al^\ca K_{\ca \ovY  }
         \left\{ (\ovM + \psib_m \sib^{mn} \psib_n) \, Y
         - (M + \psi_m \si^{mn} \psi_n) \, \ovY \right\}.
\end{eqnarray}
 These are the component field expressions which are to be used in the
transformation law of $H$ (\ref{delH}). The same expressions will
be needed later on in the construction of the invariant action.

\subsec{ General Action Terms  \label{3F3}}
In section \ref{CPN5} we have explained in detail the construction of
supersymmetric and $U_K(1)$ invariant component field Lagrangians starting
from a generic chiral superfield ${\bf r}$ of $U_K(1)$ weight
$w({\bf r}) = +2$ and its complex conjugate ${\bf \bar r}$ of weight
$w({\bf \bar r}) = -2$. We will apply this construction
to the case of 3-form superfields coupled to the
supergravity/matter/Yang-Mills system.
The generic Lagrangian -cf. (\ref{CPN.123})- is given as
\begin{eqnarray}
\cl ({\bf r}, {\bf \bar r}) &=&
e \lp {\bf f} + {\bf \bar f} \rp
+ \frac{ie}{\sqrt{2}}
\lp \psi_m \si^m {\bf \bar s} +\psib_m \sib^m {\bf s} \rp
\nn \\ &&
- e \, {\bf \bar r} \lp M + \psi_m \, \si^{mn} \psi_n \rp
- e \, {\bf r} \lp \ovM + \psib_m \, \sib^{mn} \psib_n \rp.
\label{genact3F}
\end{eqnarray}
 Particular component field
actions are then obtained by choosing ${\bf r}$ and ${\bf \br}$
appropriately. The complete action we are going to consider here will
consist of three separately supersymmetric pieces,
\begin{equation}
\cl  \,=\, \LSM  + \LSPOT + \LYM  \, .
\end{equation}

In the following we shall discuss one by one the three individual
contributions to the total Lagrangian.

 \subsubsection{ Supergravity and matter \label{CPN51bis}}
The starting point is the same as in section \ref{CPN51}, we replace the
generic superfield ${\bf r}$ with
\begin{equation}
\rSM \,=\,-3R.
\label{CPN.124bis}
\end{equation}
The difference with section \ref{CPN51} is that now the component
field Lagrangian must be evaluated in the presence of the 3-form
gauge field. As in  \ref{CPN51} we decompose the
supergravity/matter action such that
\begin{equation}
\LSM \,=\,\cl_{\mathrm{supergravity}} + e \, {\bf D}_{\mathrm{matter}}  ,
\end{equation}
where the pure supergravity part is given by the usual expression,
\ie
\[ \cl_{\mathrm{supergravity}} \,=\,
- \f{e}{2} \, \car
    + \f{e}{2} \, \vep^{klmn} \lp \psib_k \sib_l \cd_m \psi_n
                           - \psi_k \si_l \cd_m \psib_n \rp
      - \f{e}{3} \, \ovM M + \f{e}{3} \, b^a b_a,
\]
 except that the $U_K(1)$ covariant derivatives of the
Rarita-Schwinger field contain now the new composite $U_K(1)$
connection as defined above. For the matter part, the D-term
matter component field ${\bf D}_{\mathrm{matter}}$ is defined in
(\ref{CPN.2}) in terms of the $U_K(1)$
gaugino superfield ${X}_\al$. We therefore have to evaluate the
superfield  $\cd^\al {X}_\al$ in the presence of the 3-form
multiplet \ie apply the spinor derivative to the superfield expression
 \begin{eqnarray}
2i \, {X}_\al \lp 1-Y K_Y \rp \, &=& \, K_{\ca \cab} \, \cdb^\da
\Psib^\cab \, \nabla_{\al \da} \Psi^\ca
   - \f{i}{4} K_{\ca \cab} \, \cd_\al \Psi^\ca \, \cdb^2 \Psib^\cab
\nn \\[1mm]
&-& \f{i}{4} K_{\ca \cbb \ccb} \,
        \cdb_\da \Psib^\ccb \, \cdb^\da \Psib^\cbb \, \cd_\al \Psi^\ca
    -2i \, K_k \lp  \cw_\al \! \cdot \phi \rp^k.
\label{Xal} \end{eqnarray}
 Remember here, that we are using the
space-time covariant derivative $\nabla_{\al \da}$, which by definition
does not depend on the superfield $G_{\al \da}$. In full detail
\begin{equation}
 \cd_{\al \da} \ovY \,=\, \nabla_{\al \da} \ovY -3i \, G_{\al \da}
\ovY, \cem
   \cd_{\al \da} Y \,=\,\nabla_{\al \da} Y +3i \, G_{\al \da} Y,
\end{equation}
 \begin{equation}
  \cd_{\al \da} \, \cd_\db \ovY = \nabla_{\al \da}
\, \cd_\db \ovY
   -\f{3i}{2} \, G_{\al \da} \, \cd_\db \ovY, \cem
\cd_{\al \da} \, \cd_\bt Y = \nabla_{\al \da} \, \cd_\bt Y
   +\f{3i}{2} \, G_{\al \da} \, \cd_\bt Y.
\end{equation}
 In deriving the explicit expression for $\cd^\al {X}_\al$, we
  make systematic use of this derivative, which somewhat
simplifies the calculations and is useful when passing to the
component field expression later on. In applying the spinor
derivative to (\ref{Xal}) it is convenient to make use of the
following relations
 \begin{eqnarray}
  \cd_\al \cdb_\da \ovY
&=& -2i \, \nabla_{\al \da} \ovY, \\ \cd_\al \cdb^2 \ovY &=& -4i \,
\nabla_{\al \da} \cd^\da \ovY
                     +2 \, G_{\al \da} \, \cd^\da \ovY -8 {X}_\al \ovY, \\
\cd_\al \cdb^2 \phib^\bk &=& -4i \, \nabla_{\al \da} \cd^\da \phib^\bk
      +2 \, G_{\al \da} \, \cd^\da \phib^\bk + 8 \lp  \cw_\al \! \cdot \phib
\rp^\bk. \end{eqnarray}
 In order to obtain a  compact form for
$\cd^\al \! {X}_\al$, we introduce $K^{\cab \ca}$ as the inverse
of $K_{\ca \cab}$ and we define
 \begin{eqnarray}
  -4 \, F^\ca &=&
\cd^2 \Psi^\ca
                  + \Gm^\ca{}_{\cb \cc} \cd^\al \Psi^\cb \cd_\al \Psi^\cc,\label{4F} \\
-4 \, \bF^\cab &=& \cdb^2 \Psib^\cab
                  + \Gmb^\cab{}_{\cbb \ccb} \cdb_\da \Psib^\cbb \cdb^\da
\Psib^\ccb,\label{4bF} \end{eqnarray}
 with
\begin{equation}
\Gm^\ca{}_{\cb \cc} \,=\,K^{\cab \ca} K_{\cab \cb \cc}, \cem
\Gmb^\cab{}_{\cbb \ccb} \,=\,K^{\cab \ca} K_{\ca \cbb \ccb}.
\end{equation}
 Moreover we define the new covariant derivatives
\begin{eqnarray}
\hat \nabla_{\al \da} \cd^\al \Psi^\ca &=& \nabla_{\al \da}
\cd^\al \Psi^\ca
         + \Gm^\ca{}_{\cb \cc} \, \nabla_{\al \da} \Psi^\cb \, \cd^\al
\Psi^\cc, \\
\hat \nabla_{\al \da} \cdb^\da \Psib^\cab &=& \nabla_{\al \da} \cdb^\da
\Psib^\cab
         + \Gmb^\cab{}_{\cbb \ccb} \, \nabla_{\al \da} \Psib^\cbb \,
\cdb^\da
\Psib^\ccb. \end{eqnarray}
 Then, the superfield
expression of $\cd^\al \! {X}_\al$ becomes simply
 \begin{eqnarray}
\lefteqn{2i \, \cd^\al \! {X}_\al \lp 1-\ovY K_{\ovY} \rp \, = \,
       4i \, \ovY K_{\ca \ovY  } \, {X}^\al \cd_\al \Psi^\ca
      +4i \, Y K_{Y \cab} \, {\bar{X}}_\da \cd^\da \Psib^\cab} \nn \\[1mm]
&& -2i \, K_{\ca \cab} \, \nabla^{\al \da} \Psib^\cab \, \nabla_{\al \da}
\Psi^\ca
   - 4i \, K_{\ca \cab} \, F^\ca \bF^\cab \nn \\[1mm]
&&   -K_{\ca \cab} \,  \cdb^\da \Psib^\cab \,
           \hat \nabla_{\al \da} \cd^\al \Psi^\ca
     -K_{\ca \cab} \, \cd^\al \Psi^\ca \,
           \hat \nabla_{\al \da} \, \cdb^\da \Psib^\cab \nn \\[1mm]
&&  -\f{i}{4} \car_{\ca \cb \cab \cbb} \, \cd^\al \Psi^\ca \,
\cd_\al \Psi^\cb \,
       \cdb_\da \Psib^\cab \, \cdb^\da \Psib^\cbb \nn \\[1mm]
&& -3i \, K_{\ca \cab} \, \cd^\al \Psi^\ca \, \cdb^\da \Psib^\cab
\,
            G_{\al \da}\nn
   +2i \, K_\bk \lp \cd^\al\cw_\al \! \cdot \phib \rp^\bk\\[1mm]
&& -4i \, K_{k \cab} \, \cdb_\da \Psib^\cab \, (\cw^\da \! \cdot
\phi)^k
   -4i \, K_{\ca \bk} \, \cd^\al \Psi^\ca \, (\cw_\al \! \cdot \phib)^\bk.
\end{eqnarray}
 This looks indeed very similar to the usual case (\ref{CPN.58}).
  One of the
differences however is that the F-terms and their complex
conjugates for the superfields $Y$ and $\ovY$ have special forms.
So we obtain for the matter part
\begin{eqnarray}
 \lefteqn{(1 - Y K_Y) \, {\bf D}_{\mathrm{Matter}} \,=\,
     - \s2 \, {X}^\al{\loco} \, \Psi^\ca_\al \ \ovY K_{\ca \ovY}
     - \s2 \ {\bar{X}}_\da{\loco} \, \Psib^{\da \cab} \, Y K_{Y \cab} } \nn \\[1mm]
&&-g^{mn} K_{\ca \cab} \ \nabla_m \Psi^\ca \ \nabla_n \Psi^\cab + K_{\ca
\cab} \ F^\ca \ \bF^\cab \nn \\[1mm] && - \f{i}{2} \, K_{\ca
\cab} \, \Psib^{\da \cab} \ \si^m_{\al \da}
       \hat \nabla_m \Psi^{\al \ca}
- \f{i}{2} \, K_{\ca \cab} \, \Psi^{\al \ca} \ \si^m_{\al \da}
       \hat \nabla_m \Psib^{\da \cab} \nn \\[1mm]
&& + \f{1}{4} \, \car_{\ca \cb \cab \cbb} \, \Psi^{\al \ca} \,
\Psi^\cb_\al
          \, \Psib_\da^\cab \, \Psib^{\da \cbb}
   - \f12 \, K_{\ca \cab} \, \Psi^{\al \ca} \, \Psib^{\da \cab}
                   b_{\al \da} \nn \\[1mm]
&& - \f{1}{\s2} \, (\psib_m \sib^n \si^m \Psib^\cab) \
        K_{\ca \cab} \ \nabla_n \Psi^\ca
- \f{1}{\s2} \, (\psi_m \si^n \sib^m \Psi^\ca)
        \ K_{\ca \cab} \ \nabla_n \Psib^\cab \nn \\[1mm]
&& - (\psi_m \si^{mn}\Psi^\ca) \, K_{\ca \cab} \, (\psib_n
\Psib^\cab)
   - (\psib_m \sib^{mn}\Psib^\cab) \, K_{\ca \cab} \, (\psi_n \Psi^\ca)
\nn \\[1mm] && - \f12 \, K_{\ca \cab} \, g^{mn} \, (\psi_m
\Psi^\ca) (\psib_n \Psib^\cab) - \f{1}{2} {\bf D}^{(r)} \kilc \nn
\\[1mm] &&
   + i \s2 \, K_{k \cab} \, \Psib_\da^\cab \, (\lab^\da \cdot A)^k
   - i \s2 \, K_{\ca \bk} \, \Psi^{\al \ca} \, (\la_\al \cdot \bA)^\bk
\nn \\[1mm] && - \f12 (\psib_m \sib^m)^\al \, K_k \, (\la_\al
\cdot A)^k
   + \f12 (\psi_m \si^m)_\da \, K_\bk \, (\lab^\da \cdot \bA)^\bk,
\end{eqnarray}
 with the terms in the first line given as
  \begin{eqnarray}
\lefteqn{- \s2 \, {X}^\al{\loco} \, \Psi^\ca_\al \ \ovY K_{\ca
\ovY}
     - \s2 {\bar{X}}_\da{\loco} \, \Psib^{\da \cab} \, Y K_{Y \cab} \,=\,} \\[2mm]
 \frac{1}{1- Y K_Y}  \hs{-6mm} && \left[
  + i \, \ovY K_{\cb \ovY} \, K_{\ca \cab} \, \Psi^{\al \cb} \, \Psib^{\da \cbb}
          \si^m_{\al \da}
      \, \lp \nabla_m \Psi^\ca - \f{1}{\s2} \, \psi_m{}^\vp \Psi^\ca_\vp \rp
     \right. \nn \\[1mm]
&& \left. \ + i \, Y K_{Y \cbb} \, K_{\ca \cab} \, \Psib^{\da
\cbb}
          \, \Psi^{\al \ca} \si^m_{\al \da}
      \, \lp \nabla_m \Psib^\cab - \f{1}{\s2} \, \psib_{m \, \dv}
                     \, \Psib^{\da \cab} \rp
         \right. \nn \\[1mm]
&& \left.
 - Y K_{Y \cbb} \, K_{\ca \cab} \, \Psib_\da^\cbb
        \, \Psib^{\da \cab} F^\ca
 - \ovY K_{\cb \ovY} \, K_{\ca \cab} \, \Psi^{\al \cb}
       \, \Psi^\ca_\al \, \bF^\cab \right. \nn \\[1mm]
&& \left.\ - i \, \s2 \, \ovY K_{\ca \ovY} \, \Psi^{\al \ca} \,
             K_k \, (\la_\al \cdot A)^k
            + i \, \s2 \, Y K_{Y \cab} \, \Psib_\da^\cab \,
              K_\bk \, (\lab^\da \cdot \bA)^\bk \ \right] \, . \nn
\end{eqnarray}

\subsubsection{ Superpotential \label{CPN52bis}}

In the usual case where we consider only $U_K(1)$ inert
superfields like $\phi^k$ and $\phib^\bk$, the Lagrangian is
obtained from identifying the generic superfield {\bf r} with
\begin{equation}
\rSPOT \,=\, e^{K/2} W,
\label{CPN.133bis}
\end{equation}
as in (\ref{CPN.133}) of section \ref{CPN52}. In the present case the
superfield $W$ is allowed to depend on the 3-form superfield as well.
As we wish to maintain the transformation $W(\phi) \mapsto e^{-F} W(\phi)$
for the more general superpotential $W(\phi, Y)$,
we must proceed with care due to the non zero weight of $Y$.
In order to distinguish this more general situation from the usual case,
we use the symbol ${\cal P}$ for the chiral superfield of weight
$w({\cal P})=2$, defined as
\begin{equation}
{\cal P} \, =\, e^{K/2} \ W(\phi, Y)\, =\, \sum e^{{\al}_n K/2}
W_n (\phi){Y}^n,
\end{equation}
where we have allowed for a parameter $\al_n$. What happens under a
\ka transformations ? Assigning a holomorphic transformation law
$W_n \mapsto e^{{-\beta}_n F } W_n$ to the coefficient superfields, we find
\begin{eqnarray}
 {\cal P} &\mapsto&  e^{-i\, \ima F} {\cal P} \nn \\
  e^{{ \al}_n K/2} W_n(\phi){Y}^n &\mapsto& e^{(\, {\al}_n \rea F-
  {\beta}_n F -in \ima F)}\  e^{{\al}_n K/2} W_n(\phi){Y}^n.
\end{eqnarray}
Consistency with the transformations of $W$ and $Y$ then requires
${\al}_n = {\beta}_n =1-n$, hence
\begin{equation}
{\cal P} \, =\, e^{K/2} \sum  W_n (\phi)\left[e^{-K/2}{Y}\right]^n.
\end{equation}
This suggest to define the superfields
 \begin{equation}
y \,=\,e^{-K/2} \, Y, \cem \by \,=\,e^{-K/2} \, \ovY,
\label{superpotential}
\end{equation}
as the basic variables in the construction of the superpotential term,
\ie
\begin{equation}
{\cal P} \, =\, e^{K/2}\ W(\phi,\ {y}), \cem {\bar{\cal P}} \
=e^{K/2}\ {\ovW}(\phib,\ {\by}).
\end{equation}
Note that, by construction, $y$ transforms as a holomorphic
section. We can now proceed with the construction of $\LSPOT$,
taking ${\cal P}$ as starting point in the canonical procedure.

We parametrize the covariant spinor derivatives of ${\cal  P}$
such that
\begin{equation}
  \cd_\al {\cal  P} \,=\,\Si_\ca \, \cd_\al \Psi^\ca
\end{equation}
 and
 \begin{equation}
  \cd^2 {\cal  P} \,=\,-4 \, \Si_\ca \, F^\ca
                  + \Si_{\ca \cb} \, \cd^\al \Psi^\ca \, \cd_\al \Psi^\cb. \end{equation}
The various components of the coefficients $\Si_\ca$ and $\Si_{\ca
\cb}$ are given as
 \begin{eqnarray}
  \Si_{k} &=& e^{K/2} (W_k + K_k
W) - Y W_{y} K_k \label{Sik}, \\ \Si_{Y} &=& e^{K/2} W K_{Y} +
W_{y} (1 - Y K_{Y}) \label{SiX} \end{eqnarray} and
\begin{eqnarray}
 \Si_{kl} &=& (e^{K/2} W - Y W_y)(K_{kl} +
K_k K_l) \nn
\\[1mm] & & - Y (W_{k y} K_l + W_{l y} K_k)
    + e^{K/2} (W_{kl} + W_k K_l + W_l K_k) \nn \\[1mm]
& & + e^{-K/2} \, Y^2 \, K_k K_l W_{y y}
    -\Si_{\ca} {\Gamma^{\ca}}_{kl}, \label{Sikl} \\[2mm]
\Si_{k Y} &=&  (e^{K/2} W - Y W_y)(K_{k Y}
         + K_k K_Y) \nn \\[1mm]
& & + W_{k y} \, (1-Y K_Y) + e^{K/2} W_k K_Y \nn \\[1mm] & & -
e^{-K/2} \, Y K_k W_{y y} \, (1-Y K_Y)
    -\Si_{\ca} {\Gamma^{\ca}}_{k Y},  \label{SikX} \\[2mm]
\Si_{Y Y} &=&  (e^{K/2} W
       - Y W_y)(K_{Y Y} + K_Y K_Y) \nn \\[1mm]
& & + e^{-K/2} W_{y y} \, (1-Y K_Y)^2
    -\Si_{\ca} {\Gamma^{\ca}}_{Y Y}. \label{SiXX}
\end{eqnarray}
 Complex conjugate expressions are obtained from
\begin{equation}
{\bar {\cal  P}} = e^{K/2} {\overline W}(\phib,\by),
\end{equation}
 with
$\by = e^{-K/2} \ovY$.
Making use of the
superpotential superfield and the corresponding definitions given
above one derives easily the component field expression
\begin{eqnarray}
\f{1}{e}\LSPOT &=& \, \Si_\ca F^\ca
        - \frac{1}{2} \, \Si_{\ca \cb} \Psi^{\al \ca} \Psi_\al^\cb
        + \frac{i}{\s2} \, \Si_\ca \lp \psib_m \sib^m \Psi^\ca \rp \nn \\
&& -  \, e^{K/2} W \lp \bar{M} + \psib_m \sib^{m n} \psib_n \rp
           \ \ + \ \ \mbox{h.c.} \ . \label{supercomp}
\end{eqnarray}

\subsubsection{ Yang-Mills}
\label{CPN53bis}

Finally the Yang-Mills action is obtained in replacing the generic
superfield ${\bf r}$ with
\begin{equation}
\rYM \,=\, \f{1}{4} f_{(r)(s)} \cw^{(r)\al} \cw^{(s)}_\al,
\label{CPN.144bis}
\end{equation}
in the same way as in (\ref{CPN.144}) of section \ref{CPN53}.
Assuming the gauge coupling functions to be independent of the 3-form
superfields, the resulting component field expression has the same form as
in (\ref{CPN.66}), which we display here in the form
\begin{eqnarray}
 \f{1}{e} \LYM &=& -\f{1}{4} f_{{(r) (s)}}
\left[ \frac{}{}
      {\fym}^{{(r)} \, mn} \, {\fym}^{{(s)}}_{mn}
     + 2i \, \la^{(r)} \si^m \nabla_m \lab^{(s)}
     + 2i \, \lab^{(s)}\sib^m \nabla_m \la^{(r)} \right. \nn \\[1mm]
&& \hs{17mm} \left.
     - 2 \, {\bf D}^{(r)} {\bf D}^{(s)}+ \f{i}{2} \vep^{klmn} \,
             {\fym}^{{(r)}}_{kl} \, {\fym}^{{(s)}}_{mn}
   - 2 \, (\la^{(r)} \si^a \lab^{(s)}) \, b_a \frac{}{} \right] \nn \\[1mm]
&& -\f{1}{4} \frac{\prt f_{{(r) (s)}}}{\prt A^i} \left[
  \s2 \, (\chi^i \si^{mn} \la^{(r)})
{\fym}^{{(s)}}_{mn}
  - \s2 \, (\chi^i \la^{(r)}) \, {\bf D}^{(s)}
  + (\la^{(r)} \la^{(s)}) \, F^i \right] \nn \\[1mm]
&& -\f{1}{4} \frac{\prt \bar{f}_{{(r) (s)}}}{\prt \bA^\bi} \left[
  \s2 \, (\chib^\bi \sib^{mn} \lab^{(r)})
{\fym}^{{(s)}}_{mn}
  - \s2 \, (\chib^\bi \lab^{(r)}) \, {\bf D}^{(s)}
  + (\lab^{(r)} \lab^{(s)}) \, \bF^\bi \right] \nn \\[1mm]
&& + \f18 \lp \frac{\prt^2 f_{{(r) (s)}}}{\prt A^k \prt A^l}
               -  \frac{\prt f_{{(r) (s)}}}{\prt A^i} \ \Gm^i{}_{kl} \rp
      (\chi^k \chi^l) (\la^{(r)} \la^{(s)}) \nn \\[1mm]
&& + \f18 \lp \frac{\prt^2 \bar{f}_{{(r) (s)}}}{\prt \bA^\bk \prt
\bA^\bl}
          -  \frac{\prt \bar{f}_{{(r) (s)}}}{\prt \bA^\bi}
             \ \oGm^\bi{}_{\bk \bl} \rp
      (\chib^\bk \chib^\bl) (\lab^{(r)} \lab^{(s)}) \nn \\[1mm]
&& \mbox{ plus $\psi_m, \, \psib_m$ dependent terms} .
\end{eqnarray}
In the covariant derivatives of the gauginos
 \begin{eqnarray}
  \nabla_m \la^{(r)}_\al &=& \prt_m \la^{(r)}_\al
            - \om_{m \, \al}{}^\vp \la^{(r)}_\vp
            + v_m \la^{(r)}_\al
            - \aym^{(t)}_m \la^{(s)}_\al c_{{(s) (t)}}{}^{(r)}, \\[1mm]
\nabla_m \lab^{{(r)} \, \da} &=& \prt_m \lab^{{(r)} \, \da}
            -\om_m{}^\da{}_\dv \lab^{{(r)} \, \dv}
            -v_m \lab^{{(r)} \, \da}
            -\aym^{(t)}_m \lab^{{(s)}\, \da} c_{{(s) (t)}}{}^{(r)},
\end{eqnarray}
defined as in (\ref{CPN.60}), (\ref{CPN.61}) the composite \ka connection
is now given in terms of (\ref{Am}), displaying the dependence on the
3-form multiplet.
The Yang-Mills field strength tensor is given as usual
 \begin{equation}
{\fym}^{(r)}_{mn} \,=\,
    \prt_m \aym^{(r)}_n - \prt_n \aym^{(r)}_m
        + \aym^{(s)}_m \, \aym^{(t)}_n c_{{(s) (t)}}{}^{(r)}.
\end{equation}

\subsubsection{ Solving for the auxiliary fields  \label{3F31}}
Although this is standard stuff, we detail the calculations to
make clear some subtleties related to the inclusion of the 3-form.
In the different pieces of the whole Lagrangian, we isolate the
contributions containing auxiliary fields and proceed sector by
sector as much as possible.

 Diagonalization in $b_a$ makes use of the terms
\begin{equation}
\La_b \,=\,\f13 b^a b_a
        -\f12 M_{\ca \cab} \lp \Psi^\ca \si^a \Psi^\cab \rp b_a
        +\f12 f_{{(r) (s)}} \lp \la^{(r)} \si^a \lab^{(s)}\rp b_a,
\end{equation}
 with
 \begin{equation}
  M_{\ca \cab} \,=\, \frac{1}{1- Y K_Y} \ K_{\ca
\cab}, \end{equation}
 whereas the relevant terms for the
Yang-Mills auxiliary sector are
\begin{eqnarray}
\La_{\bf D} \ &=& \ \f12 f_{{(r) (s)}} {\bf D}^{(r)} {\bf D}^{(s)}
        + \frac{1}{1- Y K_Y} \,
           {\bf D}^{(s)} K_\bl \,\lp  \bA {\bf T}_{(s)} \rp^{\bl}
\nn \\[1mm] && \hs{2cm}+ \f{\s2}{4} \, {\bf D}^{(s)}
     \left( \frac{\prt f_{{(r) (s)}}}{\prt A^k} (\chi^k \la^{(r)})
            +\frac{\prt \bar{f}_{{(r) (s)}}}{\prt \bA^\bk}
                   (\chib^\bk \lab^{(r)}) \right) .
\end{eqnarray}
 The F-terms of chiral matter and the 3-form appear in the
general form
\begin{equation}
\La_{F,\bF} \,=\,F^\ca M_{\ca \cab} \bF^\cab
         +F^\ca P_\ca + \ovP_\cab \bF^\cab,
\end{equation}
 with the definitions
 \begin{eqnarray}
  P_k &=& \Si_k
   -\frac{1}{4} \frac{\prt f_{{(r) (s)}}}{\prt A^k} (\la^{(r)} \la^{(s)})
   - Y M_{Y \cbb} \, M_{k \cab}
               \, \Psib_\da^\cbb \, \Psib^{\da \cab}, \\[1mm]
P_Y &=& \Si_Y - Y M_{Y \cbb} \, M_{Y \cab}
               \, {\Psib}_\da^\cbb \, \Psib^{\,\da \cab}.
\end{eqnarray}
 We write this expression as
\begin{equation}
\La_{F,\bF} \,=\,\cf^k M_{k \bk} \, \ovcf^{\, \bk}
         - \ovP_{\! \cab} \, M^{\cab \ca} \, P_\ca
         + \cf^Y \! \frac{1}{M^{\ovY Y}} \, \ovcf^{\, \ovY},
\label{quad} \end{equation}
 where $M^{\cab \ca}$ is the inverse of
$M_{\ca \cab}$ and in particular
\begin{equation}
 \frac{1}{M^{\ovY Y}}\,=\,M_{Y \ovY}-M_{Y \bk} \,
 \gom^{\,\bk k} \,  M_{k \ovY},
\end{equation}
 with $\gom^{\,\bk k}$ the inverse of the submatrix $M_{k \bk}$,
 related to the usual \ka metric. Moreover
\begin{eqnarray}
 \cf^k &=& F^k + \lp \ovP_\bk + F^Y M_{Y \bk} \rp
\gom^{\,\bk k}, \\ \ovcf^{\, \bk} &=& \bF^\bk + \gom^{\,\bk k} \lp
P_k + M_{k \ovY} \bF^{\ovY} \rp, \end{eqnarray}
 and
\begin{equation}
\cf^Y \,=\,F^Y + \ovP_\cab M^{\cab Y}, \cem \ovcf^{\, \ovY} \, =
\, \bF^{\ovY} + M^{\ovY \ca} P_\ca. \end{equation}
 We use now the
particular structure of the 3-form multiplet to further specify
these F-terms. Using (\ref{defH}), (\ref{dC}), (\ref{4F}) and
(\ref{4bF}) we parametrize
\begin{eqnarray}
 \cf^Y \ &=& \ H + i\lp \Dt + \frac{\ovM Y - M \ovY}{2i}
\rp + f^Y,
 \\
\ovcf^{\, \ovY} \ &=& \ H - i\lp \Dt +\frac{\ovM Y - M \ovY}{2i
}\rp
 + \bar{f}^{\, \ovY},
\end{eqnarray}
 with
 \begin{eqnarray}
  f^Y &=& -\frac{1}{4}\Gm^Y{}_{\cb
\cc} \ \cd^\al \Psi^\cb \cd_\al \Psi^\cc
           + \ovP_\cab M^{\cab Y}, \\
\bar{f}^{\, \ovY} &=& -\frac{1}{4} \oGm^{\ovY}{}_{\cbb \ccb}
                    \ \cd_\da \Psib^\cbb \cd^\da \Psib^\ccb
           + M^{\ovY \ca} P_\ca,
\end{eqnarray}
 as well as
 \begin{eqnarray}
  \Dt &=& \f{4}{3} \vep^{klmn}
\prt_k C_{lmn} - \f{1}{2 \s2} \lp \psib_m \sib^m \eta
   - \psi_m \si^m \etab \rp \nn \\
&& + \frac{1}{2i} \left[ (\psib_m \sib^{mn} \psib_n) \, Y
     - (\psi_m \si^{mn} \psi_n) \, \ovY \right].
\end{eqnarray}

In terms of these notations the last term in (\ref{quad}) takes
then the form
\begin{eqnarray}
 \cf^Y \! \frac{1}{M^{\ovY Y}} \,
\ovcf^{\, \ovY} \ &=& \
      \frac{1}{M^{\ovY Y}} \ \lp H+ \frac{ f^Y + \bar{f}^{\, \ovY}}{2} \rp ^2
 \nn \\
    &+&   \ \frac{1}{M^{\ovY Y}}
      \left(\Dt+ \frac{\ovM\, Y - M\, \ovY}{2i}
 +\frac{ f^Y - \bar{f}^{\, \ovY}}{2i}\right)^2 .
\end{eqnarray}
 In this equation the last term makes a contribution to the
sector $M ,\ovM$ and the 3-form we consider next.
 Except for this term, the sum of $ \La_b, \La_{\bf D}, \La_{F,\bF} $ will
give rise to the diagonalized expression
 \begin{eqnarray}
\frac{1}{e} \cl (F^k, \bF^\bk, b_a, {\bf D}^{(r)}, H) &=&
 \frac{1}{3} \hat{b}_a \hat{b}^a + \frac{1}{2} \widehat{\bf D}^{(r)} f_{{(r) (s)}}
\widehat{\bf D}^{(s)}+\cf^k M_{k \bk} \, \ovcf^{\, \bk} \nn \\
 &+&       \frac{1}{M^{\ovY Y}} \ \lp H+ \frac{ f^Y + \bar{f}^{\, \ovY}}{2}
\rp ^2 - \frac{3}{16} \Bbb{B}_a \Bbb{B}^a \nn \\ &-&\frac{1}{2}
\Bbb{D}_{(r)} (f^{-1})^{{(r) (s)}} \Bbb{D}_{(s)}
 - \ovP_{\! \cab} \, M^{\cab \ca} \, P_\ca ,
\end{eqnarray}
 where $\hat{b}_a = b_a +\Bbb{B}_a $ with
\begin{equation}
 {\Bbb{B}}_a =  - M_{\ca \cab} \lp \Psi^\ca \si_a \Psib^\cab \rp
        + f_{{(r) (s)}} \lp \la^{(r)} \si_a \lab^{(s)}\rp ,
\end{equation}
 and ${\widehat{\bf D}}^{(r)} = {\bf D}^{(r)} +(f^{-1})^{(r)
(s)} \Bbb{D}_{(s)}$ with
\begin{eqnarray}
 {\Bbb{D}}_{{(r)}} &=&
- \frac{1}{1- Y K_Y}   \lp K_k \, {{\bf T}}_{{(r)}} .A \rp^k \nn
\\
 &+& \frac{\s2}{4} \,
     \left( \frac{\prt f_{{(r) (s)}}}{\prt A^k} \; (\chi^k \la^{(s)})
            +\frac{\prt \bar{f}_{{(r) (s)}}}{\prt \bA^\bk} \;
                   (\chib^\bk \lab^{(s)}) \right) .
\end{eqnarray}
 Use of the equations of motion simply sets to zero the first
four terms, leaving for the Lagrangian
 \begin{eqnarray}
\frac{1}{e} \cl  &=& - \frac{3}{16} \Bbb{B}_a \Bbb{B}^a
-\frac{1}{2} \Bbb{D}_{(r)} (f^{-1})^{{(r) (s)}}
\Bbb{D}_{(s)}-{\ovP}_{\ovY} \frac{1}{M_{Y \ovY}} P_{Y} \nn \\
 &-& \lp \ovP_{\! \bk}  -  \ovP_{\!\ovY} {\frac{M_{Y \bk}}{M_{Y \ovY}}}
 \rp
 \, M^{\bk k} \, \lp P_k -{\frac{M_{k \ovY}}{M_{Y
\ovY}}} P_{Y} \rp, \end{eqnarray}
 where we have block
diagonalized $ M^{\cab \ca}$.

As to the $M,\ovM$ dependent
terms of the full action we observe that they are intricately
entangled with the field strength tensor of the 3-form, a novel
structure compared to the usual supergravity-matter couplings. The
relevant terms for this sector are identified to be
\begin{eqnarray}
 \La_{M,\ovM} &=& 3 e^K \,|W|^2 -\frac{1}{3}
\, |M+3e^{K/2} W|^2 \nn \\ &+& \frac{1}{M^{\ovY Y}}\left[ \Dt
-\frac{1}{2i} \lp
 M\,\ovY-\ovM\, Y \rp + \frac{1}{2i}\lp
f^Y - \bar{f}^{\, \ovY} \rp  \right]^2 .\label{LaMM}
\end{eqnarray}
\\
 One recognizes in the first
two terms the usual superpotential contributions whereas the last
term is new. This expression contains all the terms of the full
action which depend on $M$, $\ovM$ or the 3-form $C_{klm}$. The
question we have to answer is how far the $M$, $\ovM$ sector and
the 3-form sector  can be disentangled, if at all. Clearly, the
dynamical consequences of this structure deserve careful
investigation. \indent

The 3-form contribution is not algebraic, so we cannot use the
solution of its equation of motion
 in the Lagrangian
\cite{Duf89}. One way out is to derive the equations of motion
and look for an equivalent Lagrangian giving rise to the same equations of
motion. Explicitly  we obtain for the 3-form
\begin{equation}
 \prt_k  \left\{    \frac{1}{M^{\ovY Y}}\left[ \Dt -\frac{1}{2i} \lp
 M\ovY-\ovM Y \rp + \frac{1}{2i}\lp
f^Y - \bar{f}^{\, \ovY} \rp  \right] \right\} =0 , \end{equation}
solved by setting
\begin{equation}
 \frac{1}{M^{\ovY Y}}\left[ \Dt -\frac{1}{2i} \lp
 M\ovY-\ovM Y \rp + \frac{1}{2i}\lp
f^Y - \bar{f}^{\, \ovY} \rp  \right] =c \, , \end{equation}
 where c
is a real constant. Then the {\em e.o.m.}'s for $M$ and $\ovM$
read
\begin{equation}
M+3e^{K/2} W = -3ic Y, \cem \cem \ovM+3e^{K/2} \ovW =3ic \ovY .
\end{equation}
 At last, we consider the {\em e.o.m.} for \eg $\ovY$, in
which we denote by $\cl(\ovY)$ the many contributions of $\ovY$ to
the Lagrangian, except for $ \La_{M,\ovM}$,
\begin{equation}
\prt_m{\frac{\dt{\cl(\ovY)}}{\dt{\prt}_m \ovY}} -\frac{\dt{\cl
(\ovY)}}{\dt \ovY} -\frac{\dt  \La_{M,\ovM}}{\dt \ovY} =0 .
\end{equation}
 Using (3.42) and (3.43) the last term takes
the form
\begin{eqnarray}
 \frac{\dt  \La_{M,\ovM}}{\dt \ovY}  &=& \frac{\dt }{\dt \ovY}
 \left\{ 3 e^K \, |W+icy|^2  -c^2 M^{\ovY Y} -ic( f^Y -f^\ovY)  \right. \nn \\
   && \cem   - ic \left[ (\psib_m \sib^{mn} \psib_n) \, Y
     - (\psi_m \si^{mn} \psi_n) \, \ovY \right] \nn \\
  && \cem \left. - \f{c}{\s2} \lp \psib_m \sib^m \eta
   - \psi_m \si^m \etab \rp \right\} .
\end{eqnarray}
 This suggests that the equations of motion can be derived
from an equivalent Lagrangian obtained by dropping the 3-form
contribution and shifting the superpotential $W$ to
 $W+~icy$. This can be seen more clearly by restricting our attention
  to the scalar degrees of freedom as in the next section.

\subsubsection{ The scalar potential  \label{3F32}}
The analysis presented above allows to obtain the scalar potential
of the theory as
 \begin{eqnarray}
  V &=&   \lp \oSi_{\! \bk}  -
(\oSi_{\!\ovY} -ic) {\frac{M_{Y \bk}}{M_{Y \ovY}}}
 \rp
 \, M^{\bk k} \, \lp \Si_k \, -{\frac{M_{k \ovY}}{M_{Y
\ovY}}} (\Si_{Y} +ic) \rp \nn \\
 &+&({\oSi}_{\ovY} -ic) \frac{1}{M_{Y \ovY}} (\Si_{Y}+ic)
- 3 e^{K}\, |W+icy|^2 \nn \\ &+&\f12  \frac{1}{1- Y K_Y} K_{\bk}
\, \lp {{\bf T}}_{(r)} .\bA \rp^{\bk}  (f^{-1})^{{(r) (s)}}
\frac{1}{1- Y K_Y}
   K_k \,
\lp{{\bf T}}_{(s)}.A \rp^k . \label{V1} \end{eqnarray}
 We note that
the shift $W \mapsto W+icy $ induces  $\Si_k \mapsto \Si_k $ and
$\Si_Y \mapsto \Si_Y +ic $, which are precisely the combinations
which appear in (\ref{V1}).

In fact (\ref{V1}) is nothing but the scalar potential of some
matter fields $\phi^k$ of K\"ahler weight $0$ plus a field $Y = y
e^{K/2}$ of K\"ahler weight $2$ with a superpotential $W+icy$ in
the usual formulation of supergravity. In order to show this, let
us consider $y$ and $\by$ as our new field variables and define
\begin{equation}
K(Y, \phi, \ovY, \phib) =\ck (y, \phi, \by, \phib),
\end{equation}
 Taking as an example the \ka potential in footnote (\ref{foot}) with $Z=1$, we
find
\begin{equation}
y \, = \, {Y  (X + Y \ovY)^{-1/2}}, \cem
\by \, = \, {\ovY  (X + Y \ovY)^{-1/2}},
\end{equation}
and therefore
\begin{equation}
\ck(y,\by) \, = \, \log X(\phi, \phib) - \log (1- y\by).
\label{ex1'}
\end{equation}
which is a typical \ka potential with $SU(1,1)$ noncompact
symmetry.

We can express the matrix $M_{\ca \cab}$ and its inverse $M^{\cab
\ca}$ in terms of the derivatives of $\ck$, namely $\ck_{\ca
\cab}$ and of its inverse $\ck^{\cab\ca}$ ( $\ca $ denotes $k,y$
as well as $k, Y$ depending on the context).
 Then it appears that the expression of the scalar
potential becomes very simple as we use the relevant relations.
 Indeed, using the following definitions
\begin{equation}
{\hhW} =W+icy, \cem \cem D_{\ca}{\hhW}= {\hhW}_{\ca}+
{\ck}_{\ca}{\hhW}, \end{equation}
we obtain
  \begin{eqnarray}
   V &=&e^{\ck} \lp D_{\cab}{\hhW} \, {\ck}^{\cab \ca} \,
D_{\ca}{\hhW}-3 \, |\hhW|^2 \rp \nn \\ &+& \f12 {\ck}_{\bk} \,\lp
{{\bf T}}_{(r)} .\bA \rp^{\bk}
  (f^{-1})^{{(r) (s)}}   {\ck}_k \,\lp{{\bf T}}_{(s)}.A \rp^k,
\end{eqnarray}
 which is the familiar expression of the scalar potential of
the scalar fields $\phi^k$ and $y$ in the standard formulation of
supergravity.

\newpage
\renewcommand{\theequation}{\thesection.\arabic{equation}}

\sect{     CONCLUSIONS \label{conclusions}}

Since the upsurge of supersymmetry, a number of formalisms have
been developed in order to cope with the notorious complexity of
this Fermi-Bose symmetry, in particular in the context of
supergravity, for a sample of review articles see for instance
\cite{deWit:1981fp,VanProeyen:1983tx,VanProeyen:1983wk,
Ferrara:1985bb,Ferrara:1986ra,
Fer87a,Fer87b,Jac86,
CDF90,D'Auria:1989qm}
. Among these formalisms are tensor calculus, the
superconformal compensator method and the group manifold approach.
It would be an interesting undertaking
to establish explicitly the relation among these different
approaches and to superspace geometry, which is however certainly
beyond the scope of this report.

Methods of superspace geometry are convenient in the discussion of the
conceptual aspects of supersymmetric theories and useful in the derivation
of component field expressions and have a wide range of applications.

In this report we have focused on the \ka superspace approach to
the construction of the general couplings of matter and
Yang-Mills theory to supergravity. As a solid understanding of
this subject is central for further applications and
developments, we have made an effort to present the conceptual
foundations and the technical ramifications in full detail. In
order to demonstrate the way the geometrical formulation works, we
included a detailed description of the couplings of linear and
3-form multiplets to supergravity.

There are other topics, which have been discussed in this
geometric context, but which are not included in this report.
Among them are the algebraic description of anomalies in
supersymmetric theories \cite{GGS85} and the construction of the
geometric BRS transformations \cite{BBG87}.

We also refrained from a discussion of conformal supergravity and the
construction of curvature-squared terms and supersymmetric topological
invariants. Gravitational Chern-Simons forms, which are closely related to
the 3-form geometry presented here, and their coupling to linear
multiplets have a rather transparent formulation in the geometric context.

Let us also mention the systematic description of the alternative
incarnations of supergravity, new minimal and non-minimal, in the
framework of superspace geometry in relation with the identification of
the reducible multiplet.

Finally, we have restricted ourselves to $N=1, D=4$ supersymmetry.
Superspace geometry has been widely employed in the investigations of
extended and higher and lower dimensional supersymmetry.


The methods discussed in this report have a potential interest for discussing
effective superstring field theories and have been extensively used in this
respect. We discuss in what follows some of these potential applications.

As stressed in section \ref{RS5}, the linear multiplet plays a central role
in the field theory limit of superstring theories.
Its bosonic component consists of a
scalar field associated with dilatation symmetry, the dilaton, and
of a pseudoscalar field which has many properties in common with
an axion field. Its fermionic component, sometimes called the
dilatino, may be a component of the goldstino field whose
presence in a supersymmetric theory is the sign of the spontaneous
breakdown of supersymmetry.

The close connections of dilatation symmetry with the
vanishing of the cosmological constant, of axionic couplings with
the cancellation of chiral anomalies and of the goldstino with the
super-Higgs mechanism certainly make the dilaton-axion-dilatino
set of fields a system  worthy of detailed studies. Supergravity
theories provide the natural setting for such studies, given the
intimate connections noted above with gravity and supersymmetry
(the dilaton as a Brans-Dicke scalar, the dilatino associated with
the possible breaking of local supersymmetry).

In the effective 4-dimensional supergravity theory of weakly
coupled 10-dimensional string theories, the axion field does not
appear as such in the spectrum.
 Indeed, the massless string modes include a dilaton and
an antisymmetric tensor which, together with a dilatino spinor
field, form a linear multiplet which plays an important role in
the effective field theory. As we have seen in section \ref{RS5}, a
supersymmetric duality transformation  relates this linear
supermultiplet to a chiral supermultiplet \cite{LR83} whose
content includes the original scalar field as well as the
pseudoscalar ( with axion-like couplings) dual to the
antisymmetric tensor\footnote{Such a duality transformation may be
related to a string duality in the case of some moduli fields.}.
However such a
transformation only establishes a relationship on shell and some
relevant properties or some transparence might be lost or hidden in the chiral
supermultiplet formulation. Moreover, in the context of
superstring theories it appears that it is the linear multiplet,
$L$, which plays the role of string loop expansion parameter.
Therefore stringy corrections (perturbative and non-perturbative)
are naturally parametrized by $L$, which then allows to
disentangle purely stringy effects from field theoretical ones.
This is very clear in the study of gauge coupling renormalization
and gaugino condensation in superstring effective
theories (see below).

A classical example is the way modular invariance is realized at the quantum
level in these theories. This invariance involves transformations of the
moduli fields, which are described by chiral superfields in the case of
a weakly coupled string theory. The corresponding invariance is realized
through some \ka transformation.
The simplest example is the case of a single superfield $T$ with \ka potential
$K(T, \bar T) =- 3 \ln (T + \bar T)$. The modular transformation is then
simply a $SL(2,Z)$ symmetry~:
\begin{equation}
T \mapsto {aT-ib \over icT+d}, \ \ ab-cd=1, \ \ a,b,c,d \in Z,
\end{equation}
which amounts to the \ka transformation
\begin{equation}
K \mapsto K + F + \bar F, \ \  \ {\rm with}\ \ F = 3 \ln \left(
icT+d \right)
\end{equation}
This invariance is violated by radiative corrections generated by
quantum loops of massless particles \cite{Lou91,CO92a,DFKZ92}.
These anomalies are cancelled by two types of counterterms. The
first one is model independent and is a 4-dimensional version
\cite{CO92a,DFKZ92} of the Green-Schwarz \cite{GS84} anomaly
cancellation mechanism. As is well-known, this mechanism makes
use of the presence of the antisymmetric tensor and thus, in four
dimensions, it involves the linear multiplet $L$. The other part
\cite{DKL91} which is model-dependent involves string threshold
corrections depending on the  moduli fields.

These terms play an important role when one discusses issues such as
supersymmetry breaking. For example, in the classical scenario of gaugino
condensation, it proves to be very useful, in order to take into account these
important one-loop effects, to make a supersymmetric description of the
dynamics in terms of the dilaton linear multiplet. It turns out
\cite{BDQQ95,BGT95} that, in the
effective theories below the scale of condensation, a single vector superfield
$V$ incorporates the degrees of freedom of the original linear multiplet $L$
as well as the gaugino and gauge field condensates. The one-loop terms
discussed above, {\em i.e.} Green-Schwarz counterterm and moduli-dependent
string threshold corrections, play an important dynamical role
\cite{BGW96,BGW97a,BGW97b} in this mechanism.

As we see, one-loop terms play a crucial role in all these
applications. Since supergravity is not a renormalisable theory,
great care must be used in the regularization procedure. In a
major effort, Gaillard and collaborators
\cite{GJ94,GJK96,GJK97,Gail98,Gail00} have used Pauli-Villars
regulators (carefully chosen not to break supersymmetry nor the
symmetries of the theory) to compute the full one-loop
corrections to the  supergravity effective superstring theories
theory in the \ka superspace formalism.


Similar to the duality between a rank-2 antisymmetric tensor and a
pseudoscalar, a rank-3 antisymmetric tensor is dual to a constant
scalar field. Indeed,  such a relation was considered some time
ago in connection with the cosmological constant problem
\cite{Haw84,Col88,DvN80,Duf89}. As we have seen in section
\ref{3F}, the role of supersymmetry is striking when one
considers the rank-3 antisymmetric tensor. Whereas in the
non-supersymmetric case such a field does not correspond to any
physical degree of freedom (through its equation of motion, its
field strength is a constant 4-form), supersymmetry couples it
with propagating fields. Indeed, the 3-form supermultiplet
\cite{Gat81} can be described by a chiral superfield $Y$ and an
antichiral field $\ovY$ subject to a further constraint
(\ref{con2})
\begin{equation}
\frac{8i}{3} \eps^{dcba} \Si_{dcba} \,=\,
      \left( \cd^2 - 24 \rd \right) Y
    - \left( \cdb^2 - 24 R \right) \ovY,
\label{constraint}\nn
\end{equation}
where $\Sigma$ is the gauge-invariant field strength of the rank-3
gauge potential superfield, $C_{klm}$ \ie $\Sigma=dC$.
Its superpartners, identified as
component fields of the (anti)chiral superfield $Y$ and $\ovY$,
are propagating. Supersymmetry couples the rank-3
antisymmetric tensor with {\em dynamical} degrees of freedom,
while respecting the gauge invariance associated with the 3-form.
Let us emphasize (see appendix \ref{appF}) that $Y$ is not a
general chiral superfield since it must obey the constraint above
(\ref{constraint}), which is possible only if $Y$ derives from a
pre-potential $\Om$ which is real:
\begin{equation}
 \ovY = -4 \prokib \Om , \ \ Y = -4 \proki \Om.
\label{eq:pretential}
\end{equation}

Rank-3 antisymmetric tensors might play an important role in
several problems of interest, connected with string theories. One
of them is the breaking of supersymmetry through gaugino
condensation. Indeed, as we have noted above, the composite
degrees of freedom are described, in the effective theory below
the scale of condensation, by a vector superfield $V$ which
incorporates also the components of the fundamental linear
multiplet $L$. The chiral superfield
\begin{equation}
U = - (\cdb^2 - 8 R) V, \end{equation}
 has the same
quantum numbers (in particular the same K\"ahler weight) as the
superfield $W^\al W_\al$. Its scalar component, for instance, is
interpreted as the gaugino condensate.

Alternatively, the vector superfield is interpreted as a
``fossil'' Chern-Simons field \cite{BGT95,BG96} which includes
the fundamental degrees of freedom of the dilaton supermultiplet.
It can be considered as a pre-potential for the chiral superfield
$U$ as in (\ref{eq:pretential}).

 Another interesting appearance of the 3-form supermultiplet
occurs in the context of strong-weak coupling duality. More
precisely, the dual formulation of 10-dimensional supergravity
\cite{Cha81,GN86,GN87,NG87,GV87} appears as an effective field
theory of some dual formulation of string models, such as
5-branes \cite{Duf88,Str90,DL91a,DL91b,CHS91a,CHS91b,DDP92}. The
Yang-Mills field strength which is a 7-form in 10 dimensions may
precisely yield in 4 dimensions a 4-form field strength.

\vskip 2cm
{\bf Acknowledgments}
\vskip 1cm
 We would like to stress our debt to Martin M\"uller who has been a key
member of our collaborative effort to set up the framework of K\"ahler
superspace.
We wish to thank Mary K. Gaillard for her interest in this approach, and
for urging us, especially one of us,
for years to write such a review.
And, all the
more now that the work is completed, we are grateful to Jon Bagger who was
the one to finally convince us, as editor of Physics Reports C.

\newpage
\appendix
\renewcommand{\theequation}{\thesection.\arabic{equation}}
\renewcommand{\theequation}{\thesubsection.\arabic{equation}}
\addtocontents{toc}{\protect\vspace{2ex}}
\addcontentsline{toc}{section}{\hs{8mm}  \bf APPENDICES}
 \sect{ Technicalities \label{appA}}
 We collect here some definitions, conventions
and identities involving quantities which are frequently used in
superspace calculations. We do not aim at any rigorous
presentation but try to provide a compendium of formulae and
relations which appear useful when performing explicit
computations. We use essentially the same conventions as
\cite{WB83}, {\em{except}} for $\eps_{0123}$ and $\si^0$ defined
with opposite signs.

 \subsec{Superforms Toolkit \label{appA0}}

 Coordinates of curved superspace are denoted
$z^M\,=\,(x^m,\th^\mu,\thb_\dmu)$ and differential elements
$dz^M\,=\,(dx^m,d\th^\mu,d\thb_\dmu)$, with their wedge product (
$\wedge$ is understood)
\begin{equation}
  dz^M \,dz^N\,=\,-(-)^{mn}dz^N\, dz^M,\end{equation}
  $m,n$ are the gradings of the indices $M,N$: $0$ for the vector
  ones, $1$ for the spinors.
We define p-superforms with the following ordering convention
\begin{equation}
  \Omega_p \, =\, \frac{1}{p!} dz^{M_1}...dz^{M_p}
  \,\Omega_{M_{p}...M_{1}}. \label{Formdef}
\end{equation}
The coefficients $\Omega_{M_{p}...M_{1}}$ are superfields and
graded antisymmetric tensors in their indices, \ie
\begin{equation}\Omega_{{M_1}...{M_i}...{M_j}...{M_p}}\,=\,
  -(-)^{{m_i}m_{j}}(-)^{(m_i+m_j)(m_{j-1}+...+m_i)}
\,\Omega_{{M_1}...{M_j}...{M_i}...{M_p}}.\end{equation} In
agreement with (\ref{Formdef}), we define the wedge product of two
(super)forms as follows,
\begin{eqnarray}
\Omega_p\, \Omega_q \,&=&\, \frac{1}{p! q!}dz^{M_1}...dz^{M_p}
  \Omega_{M_{p}...M_{1}}dz^{N_1}...dz^{N_q}
  \Omega_{N_{q}...N_{1}} \nonumber \\
  &=&\, \frac{1}{p! q!}dz^{M_1}...dz^{M_p}dz^{N_1}...dz^{N_q}\,
  \Omega_{N_{q}...N_{1}}\, \Omega_{M_{p}...M_{1}}.
  \end{eqnarray}
 The exterior
derivative, $d=dz^M \prt_M \ \mbox{such that}\ d^2=0$, transforms
a p-superform into a $(p+1)$-superform
\begin{equation}
d \Omega_p =\frac{1}{p!} dz^{M_1}...dz^{M_p}dz^L
\prt_L\,\Omega_{M_{p}...M_{1}}
\end{equation}
and obeys the Leibniz rule
\begin{equation}\label{Leib}
  d\,(\Omega_p \Omega_q)\,=\,\Omega_p\,d\, \Omega_q
  +(-)^q\,d\Omega_p\, \Omega_q.
\end{equation}
The interior product, denoted $\iota_\xi$, transforms a
$p$-superform into a $(p-1)$-superform, it depends on a vector
field, \eg $\xi$, with which one operates the contraction
\begin{equation}\label{Ixi}
  \iota_\xi\,dz^M\,=\,\xi^M
\ \ \ \Rightarrow \ \ \
\iota_\xi\, \Omega_p\,=\,
  \frac{1}{(p-1)!} dz^{M_1}...dz^{M_{p-1}}\xi^{M_p}
  \,\Omega_{M_{p}...M_{1}}.
\end{equation}
 Using the analogue of Cartan's local
  frame we can define quantities in the local flat tangent
  superspace (flat indices are traditionally $A,B...,H$; $A=a,\al,\da)$
\begin{equation} E^A\,=\, dz^M\, {E_M}^A(z), \cem
  dz^M\,=\,E^A \,{E_A}^M(z).
\end{equation}
 ${E_M}^A(z)$ is called the (super)vielbein and ${E_A}^M(z)$ its
 inverse, they fulfill
\begin{equation}
{E_M}^A(z)\,{E_A}^N(z)\,=\,{\delta_M}^N, \cem
{E_A}^M(z)\,{E_M}^B(z)\,=\, {\delta_A}^B.
\end{equation}
The $E^A$'s are the basis 1-forms in the tangent superspace. As we
defined superforms on the $dz^M$ basis, we can equally well define
them on the $E^A$ basis
\begin{equation}
  \Omega_p = \frac{1}{p!} E^{A_1}...E^{A_p}\, \Omega_{{A_p}...{A_1}}
\end{equation}
and $d=E^A\,D_A$. As above
\begin{equation}\label{IxiEa}
  \iota_\xi\,E^A\,=\,\xi^A \ \Rightarrow\ \iota_\xi\, \Omega_p\,=\,
  \frac{1}{(p-1)!} E^{A_1}...E^{A_{p-1}}\xi^{A_p}
  \,\Omega_{A_{p}...A_{1}}.
\end{equation}
Relating the coefficients in
one basis to the ones in the other implies the occurrence of many
vielbeins or their inverses, \eg for a 2-form
\begin{equation}
B\, =\,\frac{1}{2} dz^M dz^N B_{NM} =\frac{1}{2}E^A
{E_A}^M\,E^B\,{E_B}^N\,B_{NM}\,=\, (-)^{b(m+a)} \frac{1}{2} E^A
E^B \, {E_A}^M {E_B}^N B_{NM},
\end{equation}
so that
\begin{eqnarray}
  B_{BA}\,&=&\,(-)^{b(m+a)}{E_A}^M {E_B}^N B_{NM}\nonumber \\
  B_{NM}\,&=&\,(-)^{n(m+a)}{E_M}^A {E_N}^B B_{BA}.
\end{eqnarray}

\subsec{Basic Quantities in $SO(1,3)$ and $SL(2,C)$.
\label{appA1}}

In our notations, the metric tensor ${\eta}_{ab}$  with
$a,b=0,1,2,3$ is defined as
\begin{equation}
\left[\eta_{ab}\right]  \,=\,diag(-1,+1,+1,+1), \label{A.1}
\end{equation}
 with inverse
\begin{equation}
\eta_{ac}\eta^{cb} \,=\,\dtdu{a}{b}. \label{A.2} \end{equation}
The totally antisymmetric symbols $\eps_{abcd}$ is normalized such
that
\begin{equation}
\eps_{0123} \,=\,+1, \cem \eps^{0123} \,=\,-1. \label{A.3}
\end{equation}
 The product of two $\eps$-symbols is the given as
\begin{equation}
\eps^{abcd}\eps_{efgh} \,=\,-\delta^{abcd}_{efgh}, \label{A.4}
\end{equation}
 where the multi index Kronecker delta is defined as
\begin{equation}
{\delta}^{abcd}_{efgh} \,\equiv \, \det\left[{{\delta}^i}_j
\right] , \label{A.5}
\end{equation}
 with $i=a,b,c,d$ and $j=e,f,g,h$. In
somewhat more explicit notation this can be written as
\begin{eqnarray}
 \delta^{abcd}_{efgh} &=& \dtdu{e}{a}
\delta^{bcd}_{fgh} -\dtdu{f}{a} \delta^{bcd}_{ghe} +\dtdu{g}{a}
\delta^{bcd}_{hef} -\dtdu{h}{a} \delta^{bcd}_{efg}, \label{A.6} \\
\delta^{bcd}_{fgh} &=& \dtdu{f}{b} \delta^{cd}_{gh}
                        +\dtdu{g}{b} \delta^{cd}_{hf}
                        +\dtdu{h}{b} \delta^{cd}_{fg}, \label{A.7} \\
\delta^{cd}_{gh} &=& \dtdu{g}{c} \dtdu{h}{d} - \dtdu{h}{c}
\dtdu{g}{d}. \label{A.8} \end{eqnarray}
 Accordingly, the respective
contractions of indices yield
 \begin{eqnarray}
{\eps}^{abcd}{\eps}_{efgd} &=& -{\delta}^{abc}_{efg}, \label{A.9}
\\ {\eps}^{abcd}{\eps}_{efcd} &=& -2\,{\delta}^{ab}_{ef}, \\
{\eps}^{abcd}{\eps}_{ebcd} &=& -6 \,\dtud{a}{e}, \label{A.10}
\\ {\eps}^{abcd}{\eps}_{abcd} &=& -24. \label{A.11} \end{eqnarray}
In curved space we use the totally antisymmetric tensor
$\vep_{klmn}$, defined by
\begin{equation} \vep_{klmn}\,=\, {e_k}^a {e_l}^b {e_m}^c {e_n}^d
\, \eps_{abcd}, \end{equation} with ${e_k}^a $ the moving frame.
$SL(2,C)$ spinors carry undotted and dotted indices, ${\al}=1,2$ and
${\da}=\dot{1},\dot{2}$. For the case of undotted indices, the
symbol ${\eps}_{\al \bt}=-{\eps}_{\bt \al}$ is defined by
\begin{equation}
{\eps}_{21}={\eps}^{12} \,=\,+1. \label{A.12} \end{equation}
 As a
consequence one has
 \begin{eqnarray}
  {\eps}^{\al \bt}
{\eps}_{\gm \delta} &=& -\dtud{\al}{\gm} \dtud{\bt}{\delta}
+\dtud{\al}{\delta} \dtud{\bt}{\gm}, \label{A.13} \\ {\eps}^{\al
\bt} {\eps}_{\bt \delta} &=& \dtud{\al}{\delta}, \label{A.14}
\end{eqnarray}
 together with the cyclic identity (indices $\bt, \gm,
\delta$)
\begin{equation}
{\eps}_{\al \bt} {\eps}_{\gm \delta} +{\eps}_{\al \delta}
{\eps}_{\bt \gm} +{\eps}_{\al \gm} {\eps}_{\delta \bt} \,=\,0.
\label{A.15} \end{equation}
 Exactly
the same definitions and identities hold if undotted indices are
replaced by dotted ones, \ie for the symbol ${\eps}_{\da
\db}=-{\eps}_{\db \da}$.

The $\eps$- symbols serve to lower and raise spinor indices. For a
two-component spinor $\psi_\al$, we define
\begin{equation}
{\psi}^{\bt} \,=\,{\eps}^{\bt \al} {\psi}_{\al}, \cem {\psi}_{\bt}
\,=\,{\eps}_{\bt \al} {\psi}^{\al}. \label{A.16}
\end{equation}
 The cyclic identity implies
\begin{equation}
{\eps}_{\al \bt} {\psi}_{\gm}+ {\eps}_{\gm \al} {\psi}_{\bt}+
{\eps}_{\bt \gm} {\psi}_{\al} \,=\,0. \label{A.17}
\end{equation}
 Again, exactly the same relations hold
for dotted indices.
 The standard convention for summation over spinor indices is
\begin{equation}
\psi \chi \, = \, \chi \psi \, =\, \psi^\al \chi_\al, \cem \psib
\chib \, = \, \chib \psib \,=\, \psib_\da \chib^\da. \label{A.20}
\end{equation}

The antisymmetric combination of a product of two Weyl spinors is
given in terms of the $\eps$-symbols as
 \begin{eqnarray}
\psi_\al \chi_\bt - \psi_\bt \chi_\al &=& +\eps_{\al
\bt}{\psi}^{\vp}{\chi}_{\vp}, \label{A.18} \\ \psib_\da \chib_\db
- \psib_\db \chib_\da &=& -\eps_{\da
\db}{\psib}_{\dv}{\chib}^{\dv}. \label{A.19} \end{eqnarray}
 Tensors $V_{\al \da}$ with a pair of undotted
and dotted spinor indices are equivalent to vectors $V_a$. The
explicit relation is defined in terms of the ${\si}$-matrices,
which carry the index structure ${\si}^{a}_{\al \da}$, \ie
\begin{equation}
V_{\al \da} \,=\,\si^a_{\al \da} V_a. \label{A.26}
\end{equation}
 They are defined as
\begin{equation}
\begin{array}{ll}
{\si}^0= \left(
\begin{array}{cc}
1 & 0 \\ 0 & 1
\end{array} \right),  \;\;\;
{\si}^1= \left(
\begin{array}{cc}
0 & 1 \\ 1 & 0
\end{array} \right),\;\;\;
{\si}^2= \left(
\begin{array}{cc}
0 & -i \\ i & 0
\end{array} \right), \;\;\;
{\si}^3= \left(
\begin{array}{cc}
1 & 0 \\ 0 & -1
\end{array} \right).
\end{array}
\label{A.27} \end{equation}
\\
We frequently use also the ${\sib}$-matrices,
\begin{equation}
{\sib}^{a \ \da \al} \,=\,{\eps}^{\da \db} {\eps}^{\al \bt}
{\si}^{a}_{\bt \db} \,=\,-({\eps}{\si}^{a} {\eps})^{\al \da},
\label{A.28} \end{equation}
with numerical entries such that
\begin{equation}
{\sib}^0 \,=\,{\si}^0, \cem {\sib}^{1,2,3} \,=\,-{\si}^{1,2,3}.
\label{A.29} \end{equation}
As a consequence of (\ref{A.28}) we have also
\begin{equation}
{({\si}^a {\eps})_{\al}}^{\da} \,=\,{({\sib}^a
{\eps})^{\da}}_{\al},\cem {({\eps} {\si}^a )^{\al}}_{\da} \,=\,
{({\eps}{\sib}^a)_{\da}}^{\al}. \label{A.30} \end{equation} These
matrices form a Clifford algebra, \ie
\begin{eqnarray} {
\left({\si}^a {\sib}^b + {\si}^b {\sib}^a \right)_{\al}}^{\bt} &=&
-2{\eta}^{ab} \dtdu{\al}{\bt}, \label{A.31} \\ { \left({\sib}^a
{\si}^b + {\sib}^b {\si}^a \right)^{\da}}_{\db} &=& -2{\eta}^{ab}
\dtud{\da}{\db}. \label{A.32} \end{eqnarray}
The products of two $\si$-matrices can be written as
\begin{eqnarray}
 {\si}^a {\sib}^b &=& - {\eta}^{ab} + 2{\si}^{ab},
\label{A.33} \\ \sib^a \si^b &=& - \eta^{ab} + 2\sib^{ab}.
\label{A.34} \end{eqnarray}
 The traceless antisymmetric
combinations appearing here are defined as
\begin{eqnarray}
 {({\si}^{ab})_{\al}}^{\bt} &=& \f{1}{4} {
\left({\si}^a {\sib}^b - {\si}^b {\sib}^a \right)_{\al}}^{\bt},
\label{A.35} \\ {({\sib}^{ab})^{\da}}_{\db} &=& \f{1}{4} {
\left({\sib}^a {\si}^b - {\sib}^b {\si}^a \right)^{\da}}_{\db} \ .
\label{A.36}
\end{eqnarray}
 They are self-dual resp. anti-self-dual, \ie
\begin{equation}
{\eps}_{abcd}\, {\si}^{cd} \,=\,-2i{\si}_{ab}, \cem {\eps}_{abcd}
\,{\sib}^{cd} \,=\,+2i{\sib}_{ab}, \label{A.37} \end{equation}
 and
satisfy (as a consequence of vanishing trace)
 \begin{eqnarray}
{({\eps} {\si}^{ab} {\eps})^{\bt}}_{\al}
 \,=\,-{({\si}^{ab})_{\al}}^{\bt} ,&& \cem
{({\eps} {\sib}^{ab} {\eps})_{\db}}^{\da}
 \,=\,-{({\sib}^{ab})^{\da}}_{\db}, \label{A.38} \\
({\eps}{\si}^{ab})^{\al \bt} \,=\,({\eps}{\si}^{ab})^{\bt \al}, &&
\cem ({\eps}{\sib}^{ab})_{\da \db} \,=\,({\eps}{\sib}^{ab})_{\db
\da}. \label{A.39} \end{eqnarray} Other useful identities
involving two $\si$-matrices are
\begin{eqnarray}
 \tr({\si}^a {\sib}^b) &=& -2{\eta}^{ab},
\label{A.40} \\ {{\si}^a}_{\al \da}\ {{\sib}_a}^{\db \bt} &=&-2 \,
\dtdu{\al}{\bt} \, \dtdu{\da}{\db}, \label{A.41}
\\ {{\si}^a}_{\al \da}\ {{\si}_a}_{\bt \db}
&=&-2{\eps}_{\al \bt} \ {\eps}_{\da \db}, \label{A.42} \\
{\sib}^{a \da \al} {{\sib}_a}^{\db \bt} &=&-2{\eps}^{\al \bt} \
{\eps}^{\da \db}, \label{A.43} \end{eqnarray}
 which may
be viewed as special cases of the "Fierz" reshuffling
\begin{eqnarray}
 {{\si}^a}_{\al \da}\ {{\si}^b}_{\bt \db}
&=&-\f {1}{2}{\eps}_{\al \bt} {\eps}_{\da \db}{\eta}^{ab} +
{\eps}_{\da \db} ({\si}^{ab} {\eps} )_{\al \bt} +{\eps}_{\al \bt}(
{\eps} {\sib}^{ab})_{\da \db} \nonumber \\ & &+({{\si}^a}_f
{\eps})_{\al \bt}({\eps} {\sib}^{bf})_{\da \db} +({{\si}^b}_f
{\eps})_{\al \bt}({\eps} {\sib}^{af})_{\da \db}. \label{A.44}
\end{eqnarray}
 As to the products of three
$\si$-matrices, useful identities are
 \begin{eqnarray}
  \left( {\si}^{ab}
{\si}^c \right)_{\al \dg} &=& \f{1}{2} \left( {\eta}^{ac}
{\eta}^{bd} -{\eta}^{bc}{\eta}^{ad} +i{\eps}^{abcd} \right)
{\si}_{d \ \al {\dg}}, \label{A.45} \\ \left( {\si}^{a}
{\sib}^{bc} \right)_{\al \dg} &=& \f{1}{2} \left( {\eta}^{ac}
{\eta}^{bd} -{\eta}^{ab} {\eta}^{cd}+i{\eps}^{abcd} \right)
{\si}_{d \ \al {\dg}}, \label{A.46} \\ \left( {\sib}^{ab} {\sib}^c
\right)^{\da \gm} &=& \f{1}{2} \left( {\eta}^{ac} {\eta}^{bd}
-{\eta}^{bc}{\eta}^{ad} -i{\eps}^{abcd} \right) {\sib}_{d}^{\da
{\gm}}, \label{A.47} \\ \left( {\sib}^{a} {\si}^{bc} \right)^{\da
\gm} &=& \f{1}{2} \left( {\eta}^{ac} {\eta}^{bd} -{\eta}^{ab}
{\eta}^{cd}-i{\eps}^{abcd} \right) {\sib}_{d}^{\da {\gm}},
\label{A.48} \end{eqnarray}
 and
\begin{eqnarray}
 \left( {\si}^{a} {\sib}^{b} {\si}^c
\right)_{\al \dg} &=& \left(-{\eta}^{ab}{\eta}^{cd}+{\eta}^{ca}
{\eta}^{bd}-{\eta}^{bc}{\eta}^{ad} +i{\eps}^{abcd}\right) {\si}_{d
\ \al {\dg}}, \label{A.49} \\ \left( {\sib}^{a} {\si}^{b}
{\sib}^{c} \right)^{\da \gm} &=&
\left(-{\eta}^{ab}{\eta}^{cd}+{\eta}^{ca} {\eta}^{bd} -{\eta}^{bc}
{\eta}^{ad} -i{\eps}^{abcd}\right) {\sib}_{d}^{\da {\gm}},
\label{A.50} \end{eqnarray}
 In explicit computations we also made
repeated use of the relations
 \begin{eqnarray}
  \si_{b
\bt \db}{({\si}^{ab})_{\al}}^{\vp} &=& - \dtdu{\bt}{\vp}\,
\si^a_{\al \db} + \f{1}{2} \dtdu{\al}{\vp}\, \si^a_{\bt \db},
\label{A.51}
\\ \si_{b \bt \db}{({\sib}^{ab})^{\dv}}_{\da} &=& +
\dtdu{\db}{\dv}\, \si^a_{\bt \da} - \f{1}{2} \dtud{\dv}{\da}\,
\si^a_{\bt \db}, \label{A.52}
\\ \sib_b^{\db \bt}{({\sib}^{ab})^{\da}}_{\dv} &=& -
\dtud{\db}{\dv}\, \sib^{a \da \bt} + \f{1}{2} \dtud{\da}{\dv}\,
\sib^{a \db \bt}, \label{A.53} \\ \sib_b^{\db
\bt}{({\si}^{ab})_{\vp}}^{\al} &=& + \dtud{\bt}{\vp}\, \sib^{a \db
\al} - \f{1}{2} \dtdu{\vp}{\al}\, \sib^{a \db \bt}, \label{A.54}
\end{eqnarray}

\begin{eqnarray}
 \tr \left( {\si}^{ab} {\si}^{cd} \right) &=&-{\f {1}{2}}
\left( {\eta}^{ac} {\eta}^{bd} -{\eta}^{ad}
{\eta}^{bc}+i{\eps}^{abcd} \right), \label{A.55} \\ \tr \left(
{\sib}^{ab} {\sib}^{cd} \right) &=&-{\f {1}{2}} \left( {\eta}^{ac}
{\eta}^{bd} -{\eta}^{ad} {\eta}^{bc}-i{\eps}^{abcd} \right),
\label{A.56} \\ \left[ {\si}^{ab},{\si}^{cd} \right] &=&
{\eta}^{ac} {\si}^{bd} -{\eta}^{ad} {\si}^{bc} -{\eta}^{bc}
{\si}^{ad} +{\eta}^{bd} {\si}^{ac}, \label{A.57} \end{eqnarray}
\begin{eqnarray}
 {\left\{ {\si}^{ab}, {\si}^{cd}
\right\}_{\al}}^{\bt} &=& \tr \left( {\si}^{ab} {\si}^{cd} \right)
\dtdu{\al}{\bt}, \label{A.58} \\ \left( {\eps} {\si}^{ab}
\right)^{\al \bt} \left( {\si}_{ab} {\eps} \right)_{\gm \delta}
&=& -\dtud{\al}{\gm} \dtud{\bt}{\delta} -\dtud{\al}{\delta}
\dtud{\bt}{\gm}, \label{A.59}
\end{eqnarray}

\begin{equation}
- \frac{i}{4!} {\eps}_{abcd} {\left( {\si}^a {\sib}^b {\si}^c
{\sib}^d \right)_{\al}}^{\bt} \,=\,\dtdu{\al}{\bt}, \cem
 \frac{i}{4!} {\eps}_{abcd} {\left( {\sib}^a {\si}^b {\sib}^c {\si}^d
\right)^{\da}}_{\db} \,=\,\dtud{\da}{\db}.  \label{A.60}
\end{equation}
Finally let us note that
\begin{equation}\oint_{lmn} \ \oint_{\da \db \dg} \ \si^l_{\al \da} (\eps
\sib^{mn})_{\db \dg} =0,
\label{cycsig}
\end{equation}
with cyclic permutations on vector and spinor indices.

\vskip .2in
 \indent In the Weyl basis the Dirac matrices are given
by
\begin{equation}\gm^a \ =
 \left(
\begin{array}{cc}
0 & \si^a \\ \sib^a & 0
\end{array} \right)~. \end{equation}
 A Majorana spinor $\Psi$
is made of a Weyl spinor $\chi_\al$ with two components, $\al =
1,2$ and of its complex conjugate $\chib^\da \, , \, \da =
\dot{1},\dot{2}$:
\begin{equation}
\Psi_M \,=\,\lp \chi_\al \atop \chib^\da \rp~, \end{equation}
\begin{equation}
\Psib_M \,=\,\left( \chi^\al , \chib_\da \right)~.
\end{equation}
 A Dirac spinor is made of two different Weyl
spinors, $\chi_\al, \vpb^\da$,
\begin{equation}
\Psi_D \, = \, \lp \chi_\al \atop \vpb^\da \rp~,\cem  \Psib_D \, =
\, \left( \vp^\al , \chib_\da \right)~.
\end{equation}
In the Lagrangian calculations we need to know conjugation rules,
\begin{eqnarray}
\lp \psi_1 \si^m \psib_2 \rp ^\dagger \,&=&\,- \lp \psib_1 \sib^m
\psi_2 \rp \,=\,+ \lp \psi_2 \si^m \psib_1 \rp,
\nn \\  \lp \psi_1
\si^{mn} \psi_2 \rp ^\dagger \,&=&\,+ \lp \psib_1 \sib^{mn}
\psib_2 \rp \,=\,-\lp \psib_2 \sib^{mn} \psib_1 \rp,
\end{eqnarray}
and some Fierz relations
\begin{eqnarray}
(\psi_1 \psi_2) (\chi_1 \chi_2) \,&=&\, -\f{1}{2}(\psi_1 \chi_1)
(\psi_2 \chi_2), \nn \\
 (\psi_1 \psi_2) (\chib_1 \chib_2) \,&=&\,
-\f{1}{2}(\psi_1 \si^m \chib_1) (\psi_2 \si_m \chib_2), \nn \\
\psi_\al \chib_\db \,&=&\,-\f{1}{2} \si^m_{\al \db}\, (\psi \si_m
\chib). \end{eqnarray}

 \subsec{Spinor Notations for Tensors \label{appA2}}

We can convert vector indices into spinor indices and vice-versa
using ${\si}$ and ${\sib}$ matrices.
 \begin{eqnarray}
V_{\al \da} &=& {\si}^a_{\al \da} V_a, \label{A.61} \\ V_a &=& -\f
{1}{2} {\sib}_a^{\da \al} V_{\al \da}. \label{A.62}
\end{eqnarray}
 So the scalar product of two vectors writes
\begin{equation}
T_a V^a \,=\,- \f {1}{2} T_{\al \da} V^{\al \da}. \label{A.63}
\end{equation}
 Tensors $T_{\al \bt}$,
$\bar{T}_{\da \db}$ with two spinor indices have the standard
decompositions
\begin{eqnarray}
 T_{\al \bt}
&=& +\eps_{\al \bt} T + T_{\sym{\al \bt}}, \label{A.21}
\\ \bar{T}_{\da \db} &=& -\eps_{\da \db} \bar{T} +
\bar{T}_{\sym{\da \db}}, \label{A.22} \end{eqnarray} with
\begin{equation}
T \,=\,\f{1}{2}{T^\al}_\al, \cem \bar{T} \,=\,
\f{1}{2}{\bar{T}_\da}{}^\da, \label{A.23} \end{equation}
 and
\begin{eqnarray}
T_{\sym{\al \bt}} &=& \f{1}{2}(T_{\al \bt} + T_{\bt \al}),
\label{A.24} \\ \bar{T}_{\sym{\da \db}} &=& \f{1}{2}(\bar{T}_{\da
\db} + \bar{T}_{\db \da}). \label{A.25}
\end{eqnarray}

 For an antisymmetric tensor with two
indices, like $F_{ba}=-F_{ab}$, in spinor notations we have
\begin{eqnarray}
 F_{  \bt \db \ \al \da} &=& {\si}^a_{\al
\da} {\si}^b_{\bt \db} F_{ba} \nonumber \\ &=& \left[ ({\si}^{ba}
{\eps})_{ \bt \al} {\eps}_{ \db \da} + ({\eps}{\sib}^{ba})_{ \db
\da} {\eps}_{ \bt \al} \right] F_{ba}. \label{A.64}
\end{eqnarray}
 Using the standard
decomposition
\begin{equation}
F_{  \bt \db \al \da} \ \equiv \ -2{\eps}_{ \bt \al} F_{\sym{ \db
\da}} + 2 {\eps}_{ \db \da} F_{\sym{ \bt \al}}, \label{A.65}
\end{equation}
 we obtain
  \begin{eqnarray}
   F_{\sym{
\bt \al}} &=& + \f {1}{2} ({\si}^{ba}{\eps})_{ \bt \al}\ F_{ba},
\label{A.66}
\\ F_{\sym{ \db \da}} &=& -\f {1}{2} ({\eps}{\sib}^{ba})_{ \db \da}\
F_{ba}, \label{A.67} \end{eqnarray}
 and vice-versa
\begin{equation}
F_{ba} \,=\,({\sib}_{ba} {\eps})^{ \db \da} F_{\sym{ \da \db}}
-({\eps}{\si}_{ba})^{ \bt \al } F_{\sym{ \al \bt}}.
\label{A.68}
\end{equation}
As a consequence, the kinetic term reads
\begin{equation}
F^{ba} F_{ba} \,=\,2 F_{\sym{ \bt \al}} \ F^{\sym{ \bt \al}}
 + 2 F_{\sym{ \db \da}} \ F^{\sym{ \db \da}}.
\end{equation}
 One often uses the dual tensor defined as
\begin{equation}
{\, {}^*F}^{dc} \,=\, \frac{1}{2} \ \eps^{dcba} \ F_{ba},
\end{equation}
 whose spinor components are
\begin{equation}
{\, {}^*F}^{\delta \dd \ \gm \dg} \,=\,2 i \eps^{\dd \dg} F^{\sym{
\delta \gm}} +2 i \eps^{\delta \gm} F^ {\sym{ \dd \dg}}.
\end{equation}
 The topological combination
${\, {}^*F}^{ba} F_{ba} $ takes the form
\begin{equation}
{\, {}^*F}^{ba} F_{ba} \,=\,2i F_{\sym{ \db \da}} \ F^{\sym{ \db
\da}} - 2i F^{\sym{ \bt \al}} \ F_{\sym{ \bt \al}}.
\end{equation}
 Along the same lines, for a symmetric tensor with
two indices, $S_{ba}=S_{ab}$, one has the decomposition
\begin{eqnarray}
 S_{ \bt \db \al \da} &\equiv& {\eps}_{ \bt
\al} {\eps}_{\db \da}S +S_{\vt{3} \sym{ \bt \al}} {}_{\sym{\db
\da}} \nonumber
\\ &=& - \f {1}{2} {\eps}_{ \bt \al} {\eps}_{\db \da}{S_b}^b
+2 ({{\si}^b}_f {\eps})_{ \bt \al}({\eps}{\sib}^{af})_{\db
\da}S_{ba}. \label{A.69} \end{eqnarray}
 Finally, for a three
index, antisymmetric tensor, say $H_{cba}$, the spinor structure
is most easily analyzed using its dual tensor, ${\, {}^*{H}}^d$,
defined as
\begin{equation}
{\, {}^*{H}}^d \,=\,\frac{1}{3!}{\eps}^{dcba}H_{cba},\cem H_{abc}
\,=\,{\eps}_{abcd} {\, {}^*{H}}^d. \label{A.70} \end{equation}

Then thanks to the spinor expression for the ${\eps}$-symbol
\begin{equation}
{\eps}_{ \delta \dd \ \gm \dg \  \bt \db \  \al \da} \,=\,4i
\left( {\eps}_{\delta \gm} {\eps}_{\bt \al} {\eps}_{\dd \db}
{\eps}_{\dg \da}- {\eps}_{\dd \dg} {\eps}_{\db \da} {\eps}_{\delta
\bt} {\eps}_{\gm \al}
 \right), \label{A.72}
\end{equation}
 one obtains
\begin{equation}
H_{ \gm \dg \ \bt \db \ \al \da} \,=\,2i \left( {\eps}_{\dg \db}
{\eps}_{ \gm \al}{\, {}^*{H}}_{\bt \da}-{\eps}_{\gm \bt}
{\eps}_{\dg \da} {\, {}^*{H}}_{\al \db} \right). \label{A.73}
\end{equation}

\newpage
\sect{  Elements of $U(1)$ superspace  \label{appB}}

As we have seen in the main text, $U(1)$ superspace provides the
underlying structure for the geometric description of the
supergravity/matterYang-Mills system. Matter fields are incorporated
through well defined specifications in the $U(1)$ gauge sector, leading to
K\"ahler superspace geometry. Very often, however, explicit
calculations are done to a large extend without taking into
account the special features of K\"ahler superspace. For this
reason we found it useful to provide a compact account of the
properties of $U(1)$ superspace.

\subsec{General Definitions  \label{appB1}}

The basic superfields are the supervielbein ${E_M}^A(z)$, the
Lorentz connection ${\phi_{MB}}^A (z)$ and the gauge potential
$A_M (z)$ for chiral $U(1)$ transformations. These superfields are
coefficients of 1-forms in superspace,
\begin{eqnarray}
 E^A \ &=& \ dz^M {E_M}^A (z), \\ \label{B.1}
{\phi_B}^A  \ &=& \ dz^M {\phi_{MB}}^A (z), \\ \label{B.2} A \ &=&
\ dz^M A_M(z). \label{B.3} \end{eqnarray}
 Torsion curvatures and
$U(1)$ field strengths are then defined as
 \begin{eqnarray}
  T^A \
&=& \ dE^A + E^B {\phi_B}^A + w(E^A) E^A A, \\ \label{B.4} {R_B}^A
\ &=& \ d {\phi_B}^A + {\phi_B}^C {\phi_C}^A, \\ \label{B.5} F \
&=& \ dA. \label{B.6} \end{eqnarray}
 The chiral $U(1)$ weights
$w(E^A)$ are given as
\begin{equation}
w(E^a) \,=\,0, \ \ \ \ w(E^\al) \,=\,1, \ \ \ \ w(E_{\da}) \, = \,
-1. \label{B.7} \end{equation}
 Torsion, Lorentz curvature and
$U(1)$ field strength are 2-forms in superspace,
\begin{eqnarray}
 T^A \ &=& \ \f{1}{2} E^B E^C {T_{CB}}^A, \\
\label{B.8} {R_B}^A \ &=& \ \f{1}{2} E^C E^D {R_{DCB}}^A, \\
\label{B.9} F \ &=& \ \f{1}{2} E^C E^D F_{DC}. \label{B.10}
\end{eqnarray}
 They satisfy Bianchi identities
\begin{equation}
{\cd}T^A - E^B {R_B}^A - w(E^A) E^A F \,=\,0. \label{B.11}
\end{equation}
 A more explicit form of the Bianchi identities is
\begin{equation}
\oint_{(DCB)} ({\cd}_D{T_{CB}}^A+{T_{DC}}^F{T_{FB}}^A-{R_{DCB}}^A
-w(E^A)F_{DC}{\delta}^{A}_{B}) \,=\,0, \label{B.12}
\end{equation}
 with the graded cyclic combination of super-indices
$D,C,B$ defined as
\begin{equation}
\oint_{(DCB)} DCB \,=\,DCB+{(-)}^{b(d+c)}BDC+{(-)}^{d(b+c)}CBD.
\label{B.13} \end{equation}
 Covariant derivatives are always
understood to be maximally covariant, unless explicitly otherwise
stated. In our present case this means covariance with respect to
both, Lorentz and $U(1)$ transformations. As an example, take the
generic 0-form superfield $\cx_A$ of chiral weight $w(\cx_A)$. Its
covariant derivative is defined as
\begin{equation}
{\cd}_B \cx_A \,=\,{E_B}^M \prt_M \cx_A-{\phi_{BA}}^C
\cx_C+w(\cx_A) A_B \cx_A, \label{B.14} \end{equation}
 with graded commutator
\begin{equation}
({\cd}_{C}, {\cd}_{B}) \cx_A \,=\,- {T_{CB}}^F {\cd}_F \cx_A
 - {R_{CBA}}^F \cx_F + w(\cx_A) F_{CB} \cx_A.
\label{B.15} \end{equation}
 The chiral weights of the various
quantities are given as
\begin{equation}
w({\cd}_A) \,=\,-w(E^A),\cem  w({T_{CB}}^A) \,=\,w(E^A) - w(E^B) -
w(E^C), \label{B.16} \end{equation}
\begin{equation}
w({R_{CBA}}^F) \,=\,- w(E^B) - w(E^C). \label{B.17}
\end{equation}

\subsec{Torsion Tensor Components   \label{appB2}}

For a discussion of the $U(1)$ superspace torsion constraints we
refer to the main text and to the original literature. Here we
content ourselves to note that all the coefficients of torsion,
curvature and $U(1)$ field strength are given in terms of the few
superfields
\begin{equation}
R \ , \ R^{\dagger} \ , \ G_a \ , \ W_{\lsym{\gm \bt \al}} \ ,
 \ W_{\lsym{\dg \db \da}}
\label{B.18} \end{equation}
 and their superspace derivatives. The
chiral weights of these superfields are determined according to
their appearance in the torsion coefficients (see below), \ie
\begin{equation}
w(R) \,=\,+2, \ \ \ \ w(R^{\dagger}) \,=\,-2, \ \ \ \ w(G_a) \, =
\,0, \label{B.19} \end{equation}
\begin{equation}
w(W_{\lsym{\gm \bt \al}}) \,=\,+1, \ \ \ \ w(W_{\lsym{\dg \db
\da}}) \,=\,-1. \label{B.20} \end{equation}
 We present the torsion tensor components in order of increasing
canonical dimension (remember that $[x]=-1$ and $[\theta]=-\f
{1}{2}$). We try to be as exhaustive as possible. In particular,
in many places we give the results in vector as well as in spinor
notation, with $ \underline{\al} \sim (\al, \da)$ defined as usual.

\begin{itemize}
\item {\em dimension $0$ }

\begin{equation}
{T_{\gm \bt}}^a \,=\,0, \cem {T^{\dg \db a}} \,=\,0, \label{B.21}
\end{equation}
\begin{equation}
{T_{\gm}}^{\db a} \,=\,-2i {(\si^a \eps)_{\gm}}^{\db}.
\label{B.22} \end{equation}

\indent

\item{\em dimension $\f{1}{2}$ }

\begin{equation}
{T_{\underline{\gm} \underline{\bt}}}^{\underline{\al}} \,=\,0,
\cem {T_{\underline{\gm} b}}^a \,=\,0 \label{B.23}
\end{equation}

\indent

\item{\em dimension $1$}
\end{itemize}

At this level appear the superfields $R, R^{\dagger}$ and $G_a$,
\ie
\begin{eqnarray}
 {T_{\gm b}}^{\al} \,=\,{\f{i}{2}}
{({\si}_c {\sib}_b)_{\gm}}^{\al}G^c\ \ \ &\leadsto&\ \ \ T_{\gm\
\bt \db \ \al} \,=\,i \eps_{\bt \al} G_{\gm \db}, \label{B.24}
\\
{T^{\dg}}_{b \da} \,=\,- {\f{i}{2}}{({\sib}_c
{\si}_b)^{\dg}}_{\da} G^c \ \ \ &\leadsto&\ \ \ T_{\dg\ \bt \db \
\da} \,=\,i \eps_{\db \da} G_{\bt \dg}, \label{B.25}
\\
T_{\gm b \da} \,=\,-i {\si}_{b \ \gm \da}R^{\dagger}\ \ \
&\leadsto& \ \ \ T_{\gm \  \bt \db\  \da} \,=\,-2i {\eps}_ {\gm
\bt}{\eps}_{\db \da} R^{\dagger}, \label{B.26}
\\
{{T^{\dg}}_b}^{\al} \,=\,-i{{\sib}_b}^{\dg \al} R\ \ \ &\leadsto&\
\ \ T_{\dg\  \bt \db\  \al} \,=\,-2i {\eps}_{\dg \db} {\eps}_{\bt
\al} R, \label{B.27}
\\
{T_{cb}}^a \,=\,0\ \ \ &\leadsto&\ \ \ T_{\gm \dg\ \bt \db\ \al
\da} \,=\,0. \label{B.28} \end{eqnarray}

\begin{itemize}
\item{\em dimension $\f{3}{2}$}
\end{itemize}

Here, the basic objects are
${T_{cb}}^{\underline{\al}}$, expressed in
terms of the Weyl spinor superfields $W_{\lsym{\gm \bt \al}}$,
$W_{\lsym{\dg \db
\da}}$ and of spinor derivatives of the superfields $R$,
$R^{\dagger}$ and $G_a$. These properties are most clearly
exhibited using spinor notation, \ie
\begin{equation}
{T_{cb}}^{\underline{\al}} \ \ \ \ \leadsto \ \ \ \ {T_{\gm \dg \
\bt \db}}^{\underline{\al}} \,=\,2{\eps}_{\dg \db} {T_{\sym{\gm
\bt}}}^{\underline{\al}} -2 {\eps}_{\gm \bt} {T_{\sym{\dg
\db}}}^{\underline{\al}} \label{B.29}
\end{equation}
 with further tensor decompositions
\begin{equation}
{T_{\sym{\gm \bt}}}{}_{\al} \, = \, W_{\lsym{\gm \bt \al}} +\f{1}{3}
({\eps}_{\al \gm} {S_{\bt}} +{\eps}_{\al \bt} {S_{\gm}}) ,
\label{B.30}
\end{equation}
\begin{equation}
{T_{\sym{\dg \db}}}{}_{\vt{3} \da} \, = \, W_{\sym{\dg \db \da}}
+\f{1}{3} ({\eps}_{\da \dg} {S_{\db}} +{\eps}_{\da \db} {S_{\dg}}).
\label{B.31}
\end{equation}
The various tensors appearing here are defined as
 \begin{eqnarray}
{S_{\bt}} &=&{T_{\sym{\gm \bt}}}^{\gm}  \ =\ +\f{1}{4} {\cd}^{\db}
G_{\bt \db}-{\cd}_{\bt}R \, = \, \f {1}{2} (T_{cb} \, \si^{cb}
\eps)_{\bt},
\\ \label{B.32}
{S_{\db}} &=&{T_{\sym{\dg
\db}}}{}^{\dg} \,=\,- \f{1}{4} {\cd}^{\bt} G_{\bt \db}+{\cd}_{\db}
R^{\dagger} = \ \f {1}{2}(T_{cb} \, \sib^{cb})_{\db},
\label{B.33}
\end{eqnarray}
and
\begin{eqnarray}
 {T_{\sym{\dg \db}}}{}_{\vt{3} \al} &=& -
\f{1}{4} \left( {\cd}_{\dg} G_{\al \db} +{\cd}_{\db} G_{\al \dg}
\right), \label{B.34}
\\ {T_{\sym{\gm \bt}}}{}_{\da} &=& + \f{1}{4} \left(
{\cd}_{\gm} G_{\bt \da} +{\cd}_{\bt} G_{\gm \da} \right).
\label{B.35} \end{eqnarray}

\subsec{Curvature and $U(1)$ Field Strength
Components\label{appB3}}

 The curvature 2-form takes its values in the Lie algebra of
the Lorentz group. Vector and spinor components are therefore
related by means of the canonical decomposition
\begin{equation}
{R \vs{1}_{ DC\ b}}^a \ \ \  \leadsto \ \ \  R_{DC\  \bt \db \ \al
\da}
 \,=\,2{\eps}_{\db \da}{ R \vs{1}}_{DC }{}_{\ \sym{\bt \al}}
-2 {\eps}_{\bt \al}{R \vs{1}}_{DC}{}_{\ \sym{\db \da}},
\label{B.36}
\end{equation}
as defined in appendix \ref{appA}.
The indices $D$ and $C$ are superspace
2-form indices. As a general feature of superspace geometry, the
components of curvature and $U(1)$ field strengths are completely
determined from the torsion components and their covariant
derivatives. We proceed again in order of increasing canonical dimension.

\begin{itemize}
\item {\em dimension $1$}
\end{itemize}

Here, the 2-form indices $D$ and $C$ are spinor
indices.
 \begin{eqnarray}
  R_{\delta \gm\  b a}&=& 8({\si}_{ba}
{\eps})_{\delta \gm} R^{\dagger}, \label{B.37} \\ {R^{\dd \dg}}_{\
ba}&=&8({\sib}_{ba}{\eps})^{\dd \dg} R. \label{B.38}
\\ {{R_{\delta}}^{\dg}}_{\ ba}&=& 2i G^c {({\si}^d
{\eps})_{\delta}}^{\dg} {\eps}_{dcba}. \label{B.39} \end{eqnarray}
In spinor notation these components become, respectively,
\begin{eqnarray}
R_{\delta \gm}{}_{\sym{\bt \al}}&=& 4({\eps}_{\delta \bt}
{\eps}_{\gm \al}
 +{\eps}_{\delta \al} {\eps}_{\gm \bt})R^{\dagger}, \\
\label{B.40} {R}{ \vs{1}}_{\delta \gm}{\vu{2}}_{\sym{\db
\da}}&=&0, \\ \label{B.41} & & \nonumber \\ R_{\dd \dg}{}_{\
\sym{\bt \al}}&=&0, \\ \label{B.42} R_{\dd \dg}{\vs{2}}_{\
\sym{\db \da}}&=& 4({\eps}_{\dd \db}{\eps}_{\dg \da}
 +{\eps}_{\dd \da}{\eps}_{\dg \db})R, \\
\label{B.43} & & \nonumber \\ R_{\delta \dg}{}_{\ \sym{\bt \al}}
&=& -{\eps}_{\delta \bt } G_{\al \dg} -{\eps}_{\delta \al} G_{\bt
\dg}, \\ \label{B.44} {R \vs{1.5}}_{\delta \dg}{}_{\ \sym{\db
\da}}&=& -{\eps}_{\dg \db} G_{\delta \da} -{\eps}_{\dg \da}
G_{\delta \db}, \label{B.45}
\end{eqnarray}

The $U(1)$ field strengths are
\begin{equation}
F_{\delta \gm} \,=\,0,\cem F^{\dd \dg} \,=\,0,\cem
{F_{\delta}}^{\dg} \,=\,3{({\si}^a{\eps})_{\delta}}^{\dg} G_a.
\label{B.46} \end{equation}

\indent

\begin{itemize}
\item {\em dimension $\f{3}{2}$}
\end{itemize}

Bianchi identities tell us directly that the relevant curvatures
are given in terms of torsion as:
 \begin{eqnarray}
  R_{\delta c \
ba} & = & i \si_{c \delta \dd} {T_{ba}}^\dd -
                        i \si_{b \delta \dd} {T_{ac}}^\dd -
                        i \si_{a \delta \dd} {T_{cb}}^\dd,
\label{B.47} \\ {} \nonumber \\ {R^{\dd}}_{c\ ba} & = & i
{\sib}_c^{\dd \delta} {T_{ba \delta}} -
                        i {\sib}_b^{\dd \delta} {T_{ac \delta}} -
                        i {\sib}_a^{\dd \delta} {T_{cb \delta}}.
\label{B.48} \end{eqnarray}
In spinor notation one obtains, respectively
\begin{eqnarray}
 R_{\delta \ \gm \dg}{}_{\ \sym{\bt \al}} &=& + i
\sum_{\bt \al}\left ({\eps}_{\delta \al} {T_{\sym{\gm
\bt}}}{}_{\dg} +{\eps}_{\delta \gm} {T_{\sym{\bt \al}}}{}_{\dg}
-{\eps}_{\delta \bt} {\eps}_{\gm \al} S_{\dg} \right),
\label{B.49} \\ {R \vs{1}}_{\delta \ \gm \dg}{}_{\ \sym{\db
\da}}&=& +4i{\eps}_{\delta \gm } W_{\lsym{\dg \db \da}} +i
\sum_{\db \da}{\eps}_{\dg \db} \left ({T_{\sym{\delta
\gm}}}{}_{\da}+ \f{1}{3}{\eps}_{\delta \gm} S_{\da}
 \right),
\label{B.50} \\ {} \nonumber \\ {} \nonumber \\ R_{\dd\  \gm
\dg}{\vs{2}}_{\ \sym{\bt \al}}&=& -4i{\eps}_{\dd \dg}W_{\lsym{\gm
\bt \al}}-i \sum_{\bt \al} {\eps}_{\gm \bt} \left({T_{\sym{\dd
\dg}}}{}_{\vt{3} \al}+ \f{1}{3} {\eps}_{\dd \dg} S_{\al} \right),
\label{B.51} \\ R_{\dd \ \gm \dg}{}_{\ \sym{\db \da}} &=& -i
\sum_{\db \da} \left ({\eps}_{\dd \da} {T_{\sym{\dg
\db}}}{}_{\vt{3} \gm} +{\eps}_{\dd \dg} {T_{\sym{\db
\da}}}{}_{\vt{3} \gm} -{\eps}_{\dd \db} {\eps}_{\dg \da} S_{\gm}
\right). \label{B.52} \end{eqnarray}
Using the explicit form of the torsion coefficients as defined in
the previous subsection, these curvatures may also be written as
\begin{eqnarray}
 R_{\delta \ \gm \dg}{}_{\ \sym{\bt \al}} &=& i
\sum_{\bt \al}\left (\f{1}{2}{\eps}_{\delta \gm} {\cd}_{\bt}G_{\al
\dg} +\f{1}{2}{\eps}_{\delta \bt} {\cd}_{\gm}G_{\al \dg}
-{\eps}_{\delta \bt} {\eps}_{\gm \al}{\cd}_{\dg} R^{\dagger}
\right), \label{B.53} \\ {R \vs{1}}_{\delta \ \gm \dg}{}_{\
\sym{\db \da}}&=& 4i{\eps}_{\delta \gm } W_{\lsym{\dg \db \da}}+i
\sum_{\db \da}{\eps}_{\dg \da} \left( \f{1}{3} {\eps}_{\delta \gm}
{\bX}_{\db}+\f{1}{2}{\cd}_{\delta} G_{\gm \db} \right),
\label{B.54} \\[5mm]
 R_{\dd\  \gm
\dg}{\vs{2}}_{\ \sym{\bt \al}}&=&- 4i{\eps}_{\dd \dg} W_{\lsym{\gm
\bt \al}}+i \sum_{\bt \al} {\eps}_{\gm \al} \left( \f{1}{3}
{\eps}_{\dd \dg} {X}_{\bt}+\f{1}{2}{\cd}_{\dd} G_{\bt \dg}
\right), \label{B.55} \\ R_{\dd \ \gm \dg}{}_{\ \sym{\db \da}} &=&
i \sum_{\db \da} \left (\f{1}{2} {\eps}_{\dd \dg}
{\cd}_{\db}G_{\gm \da} +\f{1}{2} {\eps}_{\dd \db}
{\cd}_{\dg}G_{\gm \da} -{\eps}_{\dd \db} {\eps}_{\dg
\da}{\cd}_{\gm} R \right). \label{B.56}
\end{eqnarray}
Here symmetric sums over indices $\al, \bt$ resp. $\da, \db$ are
understood in an obvious way and we have used the definitions
 \begin{eqnarray}
  X_{\bt} &=&
{\cd}_{\bt}R -{\cd}^{\db}G_{\bt \db}, \label{B.57} \\ {\bX}_{\db}
&=& {\cd}_{\db} R^{\dagger} - {\cd}^{\bt} G_{\bt \db}.
\label{B.58} \end{eqnarray}
 These superfields are naturally
identified in the $U(1)$ field strengths
 \begin{eqnarray}
F_{\delta c} &=&\f{3i}{2}{\cd}_{\delta} G_c +\f{i}{2} {\si}_{c
\delta \dd} {\bX}^{\dd}, \label{B.59} \\ {F^{\dd}}_c &=&
\f{3i}{2}{\cd}^{\dd} G_c -\f{i}{2}{{\sib}_c}^{\dd \delta}
X_{\delta}, \label{B.60} \end{eqnarray}
which, in spinor notation, read
\begin{eqnarray}
 F_{\delta \ \gm \dg}
&=&\f{3i}{2}{\cd}_{\delta} G_{\gm \dg}
                           +i{\eps}_{\delta \gm}{\bX}_{\dg},
\label{B.61} \\ F_{\dd \ \gm \dg} &=&\f{3i}{2}{\cd}_{\dd} G_{\gm
\dg}
                        +i{\eps}_{\dd \dg}X_{\gm}.
\label{B.62} \end{eqnarray}

\indent

\begin{itemize}
\item {\em dimension $2$}
\end{itemize}

The curvature tensor ${R_{dc\ b}}^a$ has the property
\begin{equation}
R_{dc \ ba}=R_{ba \ dc}. \label{.63} \end{equation}
 Its
decomposition in spinor notations is given as
 \begin{eqnarray}
R_{\delta \dd \ \gm \dg \ \bt \db \ \al \da}&=& +4{\eps}_{\dd \dg}
\left( {\eps}_{\db \da}{\chi}_{\sym{\delta \gm} \ \sym{\bt
\al}}-{\eps}_{\bt \al}{\psi \vs{2}}_{\sym{\delta \gm}} {{\vu{2}}_{
\  \sym{\db \da}}} \right) \nonumber \\ & &+4{\eps}_{\delta \gm}
\left( {\eps}_{\bt \al} {\chi}_{\sym{\dd \dg} \ \sym{\db
\da}}-{\eps}_{\db \da} {\psi} {\vu{2}}_{\sym{\dd \dg}}{\vs{3}}_{ \
\sym{\bt \al}} \right), \label{B.64}
\end{eqnarray}
 where
\begin{eqnarray}
 {\chi}_{\sym{\delta \gm} \ \sym{\bt \al}}
&=& {\chi}_{\lsym{\delta \gm \bt \al}} +({\eps}_{\delta
\bt}{\eps}_{\gm \al}
 +{\eps}_{\delta \al}{\eps}_{\gm \bt}) {\chi},
\label{B.65} \\ {\chi}_{\sym{\dd \dg} \ \sym{\db \da}} &=&
{\chi}_{\lsym{\dd \dg \db \da}} +({\eps}_{\dd \db}{\eps}_{\dg \da}
 +{\eps}_{\dd \da}{\eps}_{\dg \db}){\chi}
\label{B.66} \end{eqnarray}
 and
\begin{equation}
{\chi} \,=\,\f{1}{24}{R_{ba}}^{ \ ba}. \label{B.67}
\end{equation}
 The tensors appearing in the spinor decomposition
of the curvature are, respectively,
$$\begin{array}{ll} T_{cb}{}^{\al}, \  T_{cb}{}_{\da} & \cem
\mbox{the Rarita-Schwinger field strength}, \\
{R_{dc \ b}}^a & \cem
\mbox{the Lorentz curvature}, \\
X_{\al}, \ {\bX}^{\da} & \cem
\mbox{the $U(1)$ superfield}.
\end{array}$$
Here ${\chi}_{\lsym{\delta \gm \bt \al}}$ and ${\chi} _{\lsym{\dd
\dg \db \da}}$ describe the Weyl tensor in spinor notation,
whereas ${\psi} {\vu{1.5}}_{\sym{\dd \dg}}{\vs{3}}_{\sym{\bt
\al}}$
 and ${\chi}$
correspond respectively to the Ricci
${\cal R}_{dc} = {R_{da\ b}}^a$ tensor and to the curvature
scalar $ {\cal R} = {R_{ba}}^{ \ ba}$.
These superfields are related to the basic superfields
obtained in the preceding section in the following way
\begin{eqnarray}
 {\chi}_{\lsym{\delta \gm \bt \al}} &=&
\f{1}{4} ({\cd}_{\delta}W_{\lsym{\gm \bt \al}}
+{\cd}_{\gm}W_{\lsym{\bt \al \delta}} +{\cd}_{\bt}W_{\lsym{\al
\delta \gm}} +{\cd}_{\al}W_{\lsym{\delta \gm \bt}}), \label{B.68}
\\ {\chi}_{\lsym{\dd \dg \db \da}}&=& \f{1}{4}
({\cd}_{\dd}W_{\lsym{\dg \db \da}} +{\cd}_{\dg}W_{\lsym{\db \da
\dd}} +{\cd}_{\db}W_{\lsym{\da \dd \dg}} +{\cd}_{\da}W_{\lsym{\dd
\dg \db}}), \label{B.69} \end{eqnarray}
\begin{equation}
{\psi}{{\vs{3}}_{\sym{\delta \gm}}}{\vu{2}}_{ \ \sym{\db \da}} \ =
\ \f{1}{8}\sum_{\delta \gm} \sum_{\db \da} \left( G_{\delta \db}
G_{\gm \da} -\f{1}{2} \left[ {\cd}_{\delta},{\cd}_{\db} \right]
G_{\gm \da} \right), \label{B.70} \end{equation}
 and
\begin{eqnarray}
 {\chi}&=&-\f{1}{12} \left({\cd}^{\al}
{\cd}_{\al}R +{\cd}_{\da} {\cd}^{\da} R^{\dagger} \right)
+\f{1}{48} \left[ {\cd}^{\al},{\cd}^{\da} \right] G_{\al \da}
 \nonumber \\
& &-\f{1}{8} G^{\al \da}G_{\al \da} +2 R R^{\dagger}. \label{B.71}
\end{eqnarray}
 The $U(1)$ field strength $F_{dc}$
with canonical spinor decomposition
\begin{equation}
F_{\delta \dd \ \gm \dg}
 \,=\,2{\eps}_{\dd \dg} F_{\sym{\delta \gm}}
      -2{\eps}_{\delta \gm} F_{\sym{\dd \dg}},
\label{B.72} \end{equation}
 can be expressed as
  \begin{eqnarray}
F_{\sym{\delta \gm}} &=& +\f{1}{8} \sum_{\delta \gm}
({\cd}_{\delta} {\cd}^{\dd} G_{\gm \dd} +3i {{\cd}_{\delta}}^{\dd}
G_{\gm \dd}), \label{B.73} \\ F_{\sym{\dd \dg}} &=& -\f{1}{8}
\sum_{\dd \dg} ({\cd}_{\dd} {\cd}^{\delta} G_{\delta \dg} +3i
{{\cd}^{\delta}}_{\dd} G_{\delta \dg}). \label{B.74}
\end{eqnarray}

 \subsec{Derivative Relations  \label{appB4}}

Superspace constraints, via the Bianchi identities, imply
covariant restrictions on the basic superfields encountered in the
previous subsections. Most important are the chirality conditions
\begin{equation}
{\cd}_{\al}R^{\dagger} =0, \cem {\cd}^{\da}R=0, \label{B.75}
\end{equation}
 and
\begin{equation}
{\cd}_{\al}W_{\lsym{\dg \db \da}}=0, \cem {\cd}^{\da}W_{\lsym{\gm
\bt \al}}=0. \label{B.76}
\end{equation}
 Superfield expansions are defined in terms of
covariant derivatives. We have seen that the geometry of $U(1)$
superspace can be expressed in terms of some basic superfields and
their covariant derivatives. Conversely, this means that
tensors like $ T_{cb}{}^{\al}$, $T_{cb}{}_{\da}$, ${R_{dc \
b}}^a$, $X_{\al}$, ${\bX}^{\da}$ are located in the superfield
expansions of these basic superfields. At dimension $\tdem$ the
relevant equations are
\begin{eqnarray}
{\cd}_{\bt}R &=& - \f{1}{3} X_{\bt}
                   - \f{2}{3} (T_{cb}{\si}^{cb}{\eps})_{\bt},
\label{B.77} \\ {\cd}^{\db}R^{\dagger} &=& - \f{1}{3} {\bX}^{\db}
                   - \f{2}{3} (T_{cb}{\sib}^{cb}{\eps})^{\db},
\label{B.78} \end{eqnarray}
 and
 \begin{eqnarray}
  {\cd}_{\bt}G_a
&=& - \f{1}{2} (T_{cb}{\sib}_a{\si}^{cb}{\eps})_{\bt}
                     + \f{1}{6} (T_{cb}{\sib}^{cb}{\sib}_a{\eps})_{\bt}
                     - \f{1}{3} ({\bX}{\sib}_a{\eps})_{\bt},
\label{B.79} \\ {\cd}^{\db}G_a &=& + \f{1}{2}
(T_{cb}{\si}_a{\sib}^{cb}{\eps})^{\db}
                   - \f{1}{6} (T_{cb}{\si}^{cb}{\si}_a{\eps})^{\db}
                   + \f{1}{3} (X{\si}_a{\eps})^{\db}.
\label{B.80} \end{eqnarray}
 Note that, in order to compactify the
notation, we have suppressed a number of spinor indices. They are
easily (and without ambiguity) restored with reference to the
index structures of ${\si}$-matrices explicitly defined in
appendix \ref{appA}. In spinor notation, these relations may
equivalently be written as
\begin{eqnarray}
 {\cd}_{\bt}R &=& - \f{1}{3} X_{\bt} - \f{4}{3} S_{\bt},
\label{B.81} \\ {\cd}^{\db}R^{\dagger} &=& - \f{1}{3} X^{\db} +
\f{4}{3} S^{\db}, \label{B.82} \end{eqnarray}
 and
  \begin{eqnarray}
{\cd}_{\bt}G_{\al \da} &=& + 2 {T_{\sym{\bt \al}}}{}_{\da}
                                + \f{2}{3} {\eps}_{\bt \al} S_{\da}
                                - \f{2}{3} {\eps}_{\bt \al} {\bX}_{\da},
\label{B.83} \\ {\cd}_{\db}G_{\al \da} &=& - 2 {T_{\sym{\db
\da}}}{}_{\vt{3} \al}
                              - \f{2}{3} {\eps}_{\db \da} S_{\al}
                              - \f{2}{3} {\eps}_{\db \da} X_{\al}.
\label{B.84} \end{eqnarray}
 In the $U(1)$ gauge sector, at
dimension 2, one has
\begin{equation}
{\cd}_{\al}{\bX}_{\da} \,=\,0, \cem {\cd}_{\da}X_{\al} \,=\,0,
\label{B.85} \end{equation}
 and
\begin{equation}
{\cd}^{\al}X_{\al} \,=\,{\cd}_{\da}{\bX}^{\da}. \label{B.86}
\end{equation}
 Substituting for $X_{\al}$,
${\bX}^{\da}$ yields the equivalent equations
\begin{equation}
{\cd}^{\vp}{\cd}_{\vp}G_a \,=\,4i{\cd}_a R^{\dagger},\cem
{\cd}_{\dv}{\cd}^{\dv}G_a \,=\,-4i{\cd}_a R, \label{B.87}
\end{equation}
 and
\begin{equation}
{\cd}^{\al}{\cd}_{\al}R - {\cd}_{\da}{\cd}^{\da}R^{\dagger} \, =
\, 4i{\cd}_a G^a.
\label{B.88} \end{equation}
 The orthogonal
combination is given as
\begin{equation}
{\cd}^2R + {\cdb}^2 R^{\dagger} \, = \, - \f{2}{3} {R_{ba}}^{ \
ba} - \f{2}{3} {\cd}^{\al}X_{\al} + 4G^aG_a + 32RR^{\dagger}.
\label{B.89}
\end{equation}
 As a consequence of the chirality conditions, the mixed second
spinor derivatives on $R$, $R^{\dagger}$ are
 \begin{eqnarray}
{\cd}_{\da}{\cd}_{\al}R &=& -2i{\cd}_{\al \da}R -6G_{\al \da}R,
\label{B.90} \\ {\cd}_{\al}{\cd}_{\da}R^{\dagger} &=&
-2i{\cd}_{\al \da}R^{\dagger} +6G_{\al \da}R^{\dagger}.
\label{B.91} \end{eqnarray}
The relation
\begin{eqnarray}
 [ \cd_\bt , \cd_\db ] G_{\al \da} &=& - 4 \
{\psi}_{{\sym{\bt \al}}
  \ {\sym{\db \da}}}+ 2 G_{\bt \db} G_{\al \da}
   + 4 \left( \eps_{\bt \al} F_{\sym{\db \da}}
    + \eps_{\db \da} F_{\sym{ \bt \al}} \right) \nonumber \\
&&+ 2i \eps_{ \bt \al} {\cd^{\vp}} _\db  G_{\vp \da} -2i \eps_{
\db \da} {\cd_{\bt}} ^\dv  G_{\al \dv} \nonumber
\\ && + \eps_{\bt \al} \eps_{\db \da}  \left( 8 R R^{\dagger}
 +2G^c G_c - \frac{2}{3} \cd^\vp  X_\vp -4 \chi \right),
\end{eqnarray}
may be equivalently written as
  \begin{eqnarray}
   [ \cd_\al, \cdb_\da] G_a \; &=&
\;-({\si}_a)_{\al \da}(4R \rd + G^b G_b +\f{1}{6} \car ) \nonumber
\\ && + ({\si}^b)_{\al \da} \left( \car_{ba} +2 G_a G_b
+\varepsilon_{abcd} \cd^c G^d \right). \end{eqnarray}
As to the Weyl spinor superfields, their nontrivial spinor
derivatives are determined to be
 \begin{eqnarray}
{\cd}_{\delta}W_{\lsym{\gm \bt \al}} &=& {\chi}_{\lsym{\delta \gm
\bt \al}} +\f{1}{4}{\eps}_{\delta \gm}{\cd}^{\vp}W_{\lsym{\vp \bt
\al}} +\f{1}{4}{\eps}_{\delta \bt}{\cd}^{\vp}W_{\lsym{\vp \al
\gm}} +\f{1}{4}{\eps}_{\delta \al}{\cd}^{\vp}W_{\lsym{\vp \gm
\bt}}, \label{B.92}
\\ {\cd}_{\dd}W_{\lsym{\dg \db \da}} &=& {\chi}_{\lsym{\dd \dg \db
\da}} +\f{1}{4}{\eps}_{\dd \dg}{\cd}^{\dv}W_{\lsym{\dv \db \da}}
+\f{1}{4}{\eps}_{\dd \db}{\cd}^{\dv}W_{\lsym{\dv \da \dg}}
+\f{1}{4}{\eps}_{\dd \da}{\cd}^{\dv}W_{\lsym{\dv \dg \db}},
\label{B.93} \end{eqnarray} with
\begin{eqnarray}
 {\cd}^{\vp}W_{\lsym{\vp \bt \al}}
&=& -\f{1}{6} \sum_{\bt \al} \left( {\cd}_{\bt}{\cd}^{\dv} G_{\al
\dv}+3i {{\cd}_{\bt}}^{\dv}G_{\al \dv} \right) = - \f{4}{3}
F_{\sym{\bt \al}}, \label{B.94} \\ {\cd}^{\dv}W_{\lsym{\dv \db
\da}} &=& +\f{1}{6} \sum_{\db \da}\left(
{\cd}_{\db}{\cd}^{\vp}G_{\vp \da} +3i{{\cd}^{\vp}}_{\db}G_{\vp
\da} \right) = -  \f{4}{3} F_{\sym{\db \da}}. \label{B.95}
\end{eqnarray}
Observe that these relations may
also be identified in the more compact identity
\begin{equation}
{\cd}^{\al}T_{cb \ \al}+{\cd}_{\da}{T_{cb}}^{\da} =0. \label{B.96}
\end{equation}

\subsec{Yang-Mills in $U(1)$ Superspace  \label{appB5}}

As in section \ref{RS3}, the Yang-Mills
connection and its curvature are Lie algebra valued forms in
$U(1)$ superspace,
\begin{eqnarray}
 {\ca} \ &=& \ E^A
{\ca}_A^{(r)}{\bf T}_{(r)} \,=\,{\ca}^{(r)}{\bf T}_{(r)},
\label{B.97} \\ {\cf} \ &=& \ {\f {1}{2}} E^A E^B
{{\cf}}^{(r)}_{BA}{\bf T}_{(r)}\,=\, {\cf}^{(r)}{\bf T}_{(r)},
 \label{B.98} \end{eqnarray}
  with $\cf = d \ca + \ca \ca$, or
\begin{equation}
{\cf}^{(r)} \,=\,d{\ca}^{(r)}
                  + \f{i}{2}{\ca}^{(p)}{\ca}^{(q)}{c_{(p)(q)}}^{(r)}.
\label{B.102} \end{equation}
 The Bianchi identities are
\begin{equation}
{\cd}{\cf} \, =  \, d{\cf} -{\ca}{\cf} +{\cf}{\ca} \,=\,0,
\label{B.103} \end{equation}
 \ie
\begin{equation}
{\cd}{\cf}^{(r)} \, = \, d{\cf}^{(r)} -
i{\ca}^{(p)}{\cf}^{(q)}{c_{(p)(q)}}^{(r)}
 \,=\,0.
\label{B.104} \end{equation}
 More explicitly, decomposing on the
covariant superspace basis this 3-form, we obtain
\begin{equation}
\oint_{(CBA)} ({\cd}_C{\cf}_{BA} + {T_{CB}}^F{\cf}_{FA}) \,=\,0.
\label{B.105} \end{equation}
 In the discussion of the Yang-Mills
Bianchi identities the complete structure of $U(1)$ superspace as
presented in this appendix must be taken into account, derivatives
are now covariant with respect Lorentz, chiral $U(1)$ and
Yang-Mills gauge transformations.
 The covariant constraints\footnote{The explicit solution of the
constraints, as explained in section \ref{RS3}, in particular the
construction of the chiral and antichiral basis, carries
straightforwardly over to $U(1)$ superspace.}
\begin{equation}
{\cf}^{{\da}{\db}} \,=\,0, \cem {\cf}_{\bt \al} \,=\,0 \cem
{{\cf}_{\bt}}^{\da} \,=\,0, \label{B.106} \end{equation} together
with the Bianchi identities restrict the form of the remaining
components of the Yang-Mills field strength such that
\begin{eqnarray}
 {\cf}_{{\bt} a} &=&+i({\si}_a)_{\bt
\db}{\cw}^{\db}, \label{B.107} \\ {{\cf}^{\db}}_a
&=&-i({\sib}_a)^{\db \bt} {\cw}_{\bt}, \label{B.108} \\
{\cf}_{ba}&=&{\f {1}{2}}({\eps}{\si}_{ba})^{{\bt}{\al}}
{\cd}_{\al}{\cw}_{\bt} +{\f {1}{2}}({\sib}_{ba}
{\eps})^{{\db}{\da}} {\cd}_{\da} {\cw}_{\db}. \label{B.109}
\end{eqnarray}
 The Yang-Mills superfields
\begin{equation}
{\cw}_{\al} \,=\,{\cw}^{(r)}_{\al}{\bf T}_{(r)}, \cem {\cw}^{\da}
\,=\,{\cw}^{(r) \da}{\bf T}_{(r)}, \label{B.110}
\end{equation}
 with respective chiral weights, $+1$
and $-1$, are subject to the reduced set of Bianchi identities
\begin{equation}
{\cd}_{\al}{\cw}^{\da} \,=\,0, \cem {\cd}^{\da} {\cw}_{\al} \,=\,
0, \label{B.112} \end{equation}
\begin{equation}
{\cd}^{\al} {\cw}_{\al} \,=\,{\cd}_{\da} {\cw}^{\da}.
\label{B.113} \end{equation}

We also define the D-term superfield ${\bf D}^{(r)}$ as
\begin{equation}
{\bf D}^{(r)} \,=\,-\f{1}{2}{\cd}^{\al}{\cw}_{\al}^{(r)},
\label{B.114} \end{equation}
 with vanishing chiral weight, $
w({\bf D}^{(r)}) =  0 $.  In spinor notation the components of the
field strength are given as
 \begin{eqnarray}
  {\cf}_{{\db}\
{\al}{\da}} &=& 2i{\eps}_{\db \da}{\cw}_{\al}, \label{B.116}
\\ {\cf}_{{\bt}\ {\al}{\da}} &=& 2i{\eps}_{\bt
\al}{\cw}_{\da}, \label{B.117} \end{eqnarray}
 and
\begin{equation}
{\cf}_{\bt \db \ \al \da} \,=\,2{\eps}_{\db \da}{\cf}_{\sym{\bt
\al}} -2{\eps}_{\bt \al}{\cf}_{\sym{\db \da}}, \label{B.118}
\end{equation}
 with
\begin{eqnarray}
 {\cf}_{\sym{\bt \al}} &=&
-\f{1}{4}({\cd}_{\bt}{\cw}_{\al}+{\cd}_{\al}{\cw}_{\bt}),
\label{B.119} \\ {\cf}_{\sym{\db \da}} &=&
+\f{1}{4}({\cd}_{\db}{\cw}_{\da}+{\cd}_{\da}{\cw}_{\db}).
\label{B.120} \end{eqnarray}
 Conversely, the nontrivial spinor
derivatives of the Yang-Mills superfields are given as
\begin{eqnarray}
 \cd_{\bt}\cw_{\al}^{(r)} &=&
-(\si^{ba}\eps)_{\bt \al}\cf_{ba}^{(r)}
                                  -\eps_{\bt \al}{\bf D}^{(r)},
\label{B.121} \\ \cd_{\db}\cw_{\da}^{(r)} &=&
-(\eps\sib^{ba})_{\db \da}\cf_{ba}^{(r)}
                                  +\eps_{\db \da}{\bf D}^{(r)},
\label{B.122} \end{eqnarray}
and those of the D-term superfield are
\begin{eqnarray}
 \cd_\al {\bf D}^{(r)} &=& i\si^a_{\al \da} \cd_a
\cw^{(r) \da}, \label{B.123} \\ \cd^\da {\bf D}^{(r)} &=& i\sib^{a
\da \al} \cd_a \cw^{(r)}_ \al. \label{B.124} \end{eqnarray} The
covariant derivative appearing here is defined as
\begin{equation}
\cd_A {\bf D}^{(r)}  \,=\, {E_A}^M \partial_{M} {\bf D}^{(r)}
                            -i\ca_A^{(p)} {\bf D}^{(q)}
                             {c_{(p)(q)}}^{(r)}.
\label{B.125} \end{equation}
Recall that the graded commutator of two
covariant derivatives is
\begin{equation}
(\cd_B, \cd_A){\bf D}^{(r)} \,=\,-{T_{BA}}^F \cd_F {\bf D}^{(r)}
-i\cf_{BA}^{(p)} {\bf D}^{(q)} {c_{(p)(q)}}^{(r)}. \label{B.126}
\end{equation}
 In the case of the Yang-Mills superfields additional terms
appear due to their non-trivial Lorentz and $U(1)$ structures:
\begin{eqnarray}
 (\cd_C, \cd_B) \cw_\al^{(r)} &=& -{T_{CB}}^F
\cd_F \cw_\al^{(r)} -i\cf_{CB}^{(p)} \cw_\al^{(q)}
{c_{(p)(q)}}^{(r)} \nonumber \\ & & -{R_{CB \ \al}}^\vp
\cw_\vp^{(r)} + F_{CB} \cw_\al^{(r)}, \label{B.127} \\ (\cd_C,
\cd_B) \cw^{(r) \da} &=& -{T_{CB}}^F \cd_F \cw^{(r) \da}
-i\cf_{CB}^{(p)} \cw^{(q) \da} {c_{(p)(q)}}^{(r)} \nonumber \\ & &
-{{R_{CB}}^\da}_\dv \cw^{(r) \dv} - F_{CB} \cw^{(r) \da}.
\label{B.128} \end{eqnarray}
 In the evaluation of (\ref{B.123}),
(\ref{B.124}) these relations are used in combination with
(\ref{B.112}), (\ref{B.113}). Further useful relations are
\begin{eqnarray}
 \cd^2 \cw_\al^{(r)} &=& 4i
\si^a_{\al \da} \cd_a \cw^{(r) \da} + 12 R^{\dagger}
\cw_\al^{(r)}, \label{B.129} \\ \cdb^2 \cw^{(r) \da} &=& 4i
\sib^{a \da \al} \cd_a \cw_\al^{(r)} + 12 R \cw^{(r) \da}.
\label{B.130}
\end{eqnarray}

\newpage
\sect{  Gauged Isometries \label{appC}}

In the general supergravity/matter/Yang-Mills system the chiral matter
superfields para\-metrize a \ka manifold. These structures are quite well
understood in the geometric framework of \ka superspace. In general, from
the point of view of differential geometry, \ka manifolds admit
non-linear isometry transformations, which can be gauged using suitable
Yang-Mills potentials.

This appendix provides a description of gauged isometries compatible with
superspace. Of course, the relevant language makes use of superfields. In
a first subsection we develop the general formalism on a manifold
parametrized by complex superfields, not yet necessarily subject to
chirality conditions. The second subsection shows how \ka superspace can
be modified to take care of gauged isometries. The resulting geometric
structure is called {\em isometric \ka superspace}. In the third subsection
we derive the supergravity transformations in this context and in the
fourth and last subsection we establish the relation with Yang-Mills
transformations of the matter superfields, which correspond to linear
isometry transformations.

\subsec{Isometries and Superfields  \label{appC1}}

As a starting point we consider a complex manifold spanned by complex
superfields $\phi^k$ and their complex conjugates $\phib^\bk$.
Following \cite{Bag83,Bag85} we define infinitesimal variations
\begin{equation}
 {\delta}{\phi}^k  =
-{\al}^{(r)}V_{(r)}{\phi}^k, \cem {\delta}{\phib}^{\bk} =
-{\al}^{(r)}{\bV}_{(r)}{\phib}^{\bk}, \label{C.42}
\end{equation}
 of generators $V_{(r)}$ and ${\bV}_{(r)}$
which depend holomorphically resp. anti-holomorphically on the
superfield coordinates
 \begin{equation}
 V_{(r)}= {V_{(r)}}^k(\phi) \, \frac{\prt}{\prt{\phi}^k}, \cem
  {\bV}_{(r)} ={{\bV}_{(r)}}{}^{{\bk}}(\phib)
\, \frac{\prt}{\prt{\phib}^{\bk}}, \label{C.44} \end{equation}
 and which satisfy commutation relations
  \begin{eqnarray}
\left[V_{(r)},V_{(s)}\right] &=& {c_{(r)(s)}}^{(t)} V_{(t)},
\label{C.45} \\ \left[{\bV}_{(r)},{\bV}_{(s)} \right] &=&
{c_{(r)(s)}}^{(t)} {\bV}_{(t)}, \label{C.46} \\
\left[V_{(r)},{\bV}_{(s)}\right] &=&0. \label{C.47} \end{eqnarray}
 In
addition to holomorphy properties, solution of the Killing
equations of the hermitean metric implies the appearance of
Killing potentials, $G_{(r)}(\phi,{\phib})$, such that
\begin{equation}
g_{k{\bk}}{{\bV}_{(r)}}{}^{{\bk}} = + i{{\prt}G_{(r)} \over
{\prt}{\phi}^k}, \cem g_{k{\bk}}{V_{(r)}}^k =
-i\frac{{\prt}G_{(r)}}{{\prt}{\phib}^{\bk}}. \label{C.49}
\end{equation}
 In the case of
K\"ahler geometry, \ie
\begin{equation}
g_{k{\bk}} \,=\,\frac{{\prt}^2 K(\phi,{\phib})}{{\prt}{\phi}^k
\, {\prt}{\phib}^{\bk}}, \label{C.50} \end{equation}
 these equations in turn
are solved in terms of holomorphic resp. anti-holomorphic
functions $F_{(r)}(\phi)$ resp. ${{\bF}}_{(r)}({\phib})$ \ - which
one might call Killing pre-potentials -  such that
\begin{equation}
G_{(r)} \,=\,\f {i}{2}(V_{(r)}-{\bV}_{(r)})K -\f
{i}{2}(F_{(r)}-{{\bF}}_{(r)}), \label{C.51} \end{equation}
 and
\begin{equation}
(V_{(r)}+{\bV}_{(r)})K \,=\,F_{(r)}+{{\bF}}_{(r)}. \label{C.52}
\end{equation}
 As a consequence of the commutation relations for
$V_{(r)},{\bV}_{(r)}$, the pre-potentials $F_{(r)}$ and
${{\bF}}_{(r)}$ satisfy consistency conditions
\begin{eqnarray}
V_{(r)}F_{(s)}-V_{(s)}F_{(r)} &=& {c_{(r)(s)}}^{(t)}F_{(t)}+i
C_{(r)(s)}, \label{C.53} \\
{\bV}_{(r)}{{\bF}}_{(s)}-{\bV}_{(s)}{{\bF}}_{(r)}
&=&{c_{(r)(s)}}^{(t)}{{\bF}}_{(t)}-i C_{(r)(s)}, \label{C.54}
\end{eqnarray}
 with antisymmetric separation constants
\begin{equation}
C_{(r)(s)} \,=\,- C_{(s)(r)}. \label{C.55} \end{equation}
 Moreover, multiplying
eqs. (\ref{C.49}), which define the Killing potential
$G_{(r)}$, appropriately with ${V_{(r)}}^k$ resp. ${{\bV}_{(r)}}{}^{\bk}$
one obtains
\begin{equation}
V_{(r)}G_{(s)}+{\bV}_{(s)}G_{(r)} \,=\,0. \label{C.56}
\end{equation}
 Other useful relations in this context are
\begin{eqnarray}
 V_{(r)}G_{(s)}-V_{(s)}G_{(r)} &=&
{c_{(r)(s)}}^{(t)}G_{(t)}+ C_{(r)(s)}, \label{C.57} \\
{\bV}_{(r)}G_{(s)}-{\bV}_{(s)}G_{(r)}
&=&{c_{(r)(s)}}^{(t)}G_{(t)}+ C_{(r)(s)}, \label{C.58} \\
(V_{(r)}+{\bV}_{(r)})G_{(s)}&=&{c_{(r)(s)}}^{(t)}G_{(t)}+
C_{(r)(s)}. \label{C.59} \end{eqnarray}
 In the following we shall
restrict ourselves to cases where it is possible to take
\begin{equation}
C_{(r)(s)} \,=\,0, \label{C.60} \end{equation}
 and discuss gauged
isometries, \ie variations of ${\phi}^k$ and ${\phib}^{\bk}$ where
the parameters ${\al}^{(r)}$ are unconstrained real superfields.
Covariant derivatives are then constructed with the help of
superfield gauge potentials which are 1-forms in superspace
\begin{equation}
{\ca}^{(r)} \,=\,E^A {{\ca}_A}^{(r)}, \label{C.61} \end{equation}
subject to gauge variations
\begin{equation}
{\delta}{\ca}^{(r)} \,=\,{\al}^{(p)}{\ca}^{(q)}{c_{(p)(q)}}^{(r)}
-id{\al}^{(r)}. \label{C.62} \end{equation}
 The covariant exterior derivatives of
the matter superfields are defined as
 \begin{eqnarray}
  {\cd}{\phi}^k &=& \left( d+i{\ca}^{(r)} V_{(r)} \right){\phi}^k,
\label{C.63} \\
{\cd}{\phib}^{\bk} &=&
\left(d+i{\ca}^{(r)} {\bV}_{(r)} \right){\phib}^{\bk}.
\label{C.64}
\end{eqnarray}
 By construction, they change covariantly under
gauged isometries, \ie
\begin{eqnarray}
 {\delta}
{\cd}{\phi}^k&=&-{\al}^{(r)}{{\prt}{V_{(r)}}^k \over
{\prt}{\phi}^l}{\cd}{\phi}^l, \label{C.65}\\ {\delta}
{\cd}{\phib}^{\bk}&=&-{\al}^{(r)}\frac{{\prt}{{\bV}_{(r)}}{}^{\bk}}
{{\prt}{\phib}^{\bar{l}}}{\cd}{\phib}^{\bar{l}}. \label{C.66}
\end{eqnarray}
 Of course, the covariant exterior derivative is no
longer nilpotent, its square being related to the field
strength
\begin{equation}
{\cf}^{(r)} \,=\,d{\ca}^{(r)}+\f {i}{2}{\ca}^{(p)}{\ca}^{(q)}
{c_{(p)(q)}}^{(r)}, \label{C.67} \end{equation}
 such that
 \begin{eqnarray}
  {\cd}{\cd}{\phi}^k
&=&i{\cf}^{(r)}V_{(r)}{\phi}^k, \label{C.68} \\
{\cd}{\cd}{\phib}^{\bk} &=&i{\cf}^{(r)}{\bV}_{(r)}{\phib}^{\bk}.
\label{C.69}
\end{eqnarray}
 In a somewhat more explicit notation, \ie
\begin{equation}
{\cd}{\phi}^k \,=\,E^A{\cd}_A{\phi}^k, \cem {\cd}{\phib}^{\bk} \ =
\ E^A{\cd}_A{\phib}^{\bk}, \label{C.70} \end{equation}
 and
\begin{equation}
{\cf}^{(r)} \,=\,{\f {1}{2}}E^A E^B {{\cf}_{BA}}^{(r)},
\label{C.71} \end{equation}
 this yields the graded commutation
relations
\begin{eqnarray}
 ({\cd}_B,{\cd}_A){\phi}^k &=&-{T_{BA}}^C
{\cd}_C{\phi}^k +i{{\cf}_{BA}}^{(r)}V_{(r)}{\phi}^k, \label{C.72}
\\ ({\cd}_B,{\cd}_A){\phib}^{\bk} &=&-{T_{BA}}^C
{\cd}_C{\phib}^{\bk} +i{{\cf}_{BA}}^{(r)}{\bV}_{(r)}{\phib}^{\bk}.
\label{C.73} \end{eqnarray}

\subsec{Isometric \ka Superspace  \label{appC1a}}

The composite K\"ahler gauge potential was defined in terms of chiral
matter superfields as a 1-form in superspace such that
\begin{equation}
A \,=\,\f {1}{4} (K_k d{\phi}^k -K_{\bar{k}}d{\phib}^{\bar{k}})
+\f {i}{8}E^a(12G_a + {{\sib}}_a^{{\da} \al}g_{k{\bar{k}}}
{\cd}_\al {\phi}^k {\cd}_{\da}{\phib}^{\bar{k}}).
\label{kalcon}
\end{equation}
Consider now the spinor
derivatives to be covariant with respect to gauged isometries, as defined
above, rendering the last term invariant. However, the term
\begin{equation}
{\Delta} \,=\,K_kd{\phi}^k-K_{\bk}d{\phib}^{\bk},
\label{C.74}
\end{equation}
changes under gauged isometry transformations as
\begin{equation}
{\delta}{\Delta} \,=\,-2id \ima ({\al}^{(r)}F_{(r)})
+2i(d{\al}^{(r)})G_{(r)}.
\label{C.75}
\end{equation}
This can be verified
using the relations presented so far.
Interestingly enough the first term has the form of a gauge
transformation, it closely resembles a K\"ahler
transformation. As to the second term, it is easy to see that it
corresponds to
\begin{equation}
{\delta}({\ca}^{(r)}G_{(r)}) \,=\,-i(d{\al}^{(r)})G_{(r)}.
\label{C.76} \end{equation}
 Therefore the combination
\begin{equation}
{\widetilde{\Delta}} \,=\,{\Delta} + 2 {\ca}^{(r)}G_{(r)},
\label{C.77} \end{equation}
 transforms as a gauge field, both
under gauged isometries and under K\"ahler transformations, \ie
\begin{equation}
{\delta}{\widetilde{\Delta}} \,=\,2i\ d[ \ima
(F-{\al^{(r)}}F_{(r)})]. \label{C.78} \end{equation}

 This is completely in line with our understanding of supergravity/matter
couplings, \ie gauged isometries can be reconciled with the
structure of K\"ahler superspace provided we replace ${\Delta}$ by
${\widetilde{\Delta}}$ and require that the frame of superspace
changes under a gauged isometry as well such that
\begin{equation}
{\delta}E^A \,=\,- \f {i}{2}w(E^A)E^A  \ima (-{\al^{(r)}}F_{(r)}).
\label{C.79}
\end{equation}
This leads us to the definition of {\em isometric \ka superspace},
with a modified composite gauge potential
 \begin{eqnarray}
  {\frak{A}}&=&\f
{1}{4}K_k d{\phi}^k-\f {1}{4}K_{\bk} d{\phib}^{\bk} +{\f
{1}{2}}{\ca}^{(r)}G_{(r)} \nonumber \\ & &+\f {i}{8}E^a(
12G_a+{\sib}_a^{\da\al}g_{k{\bk}}{\cd}_{\al}
{\phi}^k{\cd}_{\da}{\phib}^{\bk}),
\label{C.81}
\end{eqnarray}
in the $U(1)$ sector, giving rise to the torsion 2-form
\begin{equation}
{\frak{T}}^A \,=\,dE^A +E^B{{\phi}_B}^A +w(E^A)E^A
{\frak{A}},
\label{C.80}
\end{equation}
invariant under \ka transformations and gauged isometries.
Gauged isometries appear in the
structure group of superspace via (\ref{C.79}) in close analogy
with K\"ahler transformations. Covariance with respect to these
transformations is obtained with the help of the modified gauge
potential defined in (\ref{C.81}) and the usual rules of K\"ahler
superspace. Furthermore, following the definitions (\ref{C.63}),
(\ref{C.64}), the matter superfields are defined to be
{\em{covariantly chiral}}, \ie
 \begin{eqnarray}
{\cd}_{\al}{\phib}^{\bk} &=& \left( {E_{\al}}^M{\prt}_M
+i{\ca}_{\al}^{(r)} \, {\bV}_{(r)} \right) {\phib}^{\bk} \,=\,0,
\label{C.82} \\
{\cd}^{\da}{\phi}^k&=& \left( E^{{\da}M}{\prt}_M
+i{\ca}^{(r){\da}} \, V_{(r)} \right) {\phi}^k \,=\,0.
\label{C.83}
\end{eqnarray}
 Likewise, in the definition of ${\frak{A}}$, -cf.
(\ref{C.81}), one has
 \begin{eqnarray}
{\cd}_{\al}{\phi}^k &=& \left({E_{\al}}^M{\prt}_M +i{\ca}_{\al}^{(r)} \,
V_{(r)}\right) {\phi}^k,
\label{C.84} \\
{\cd}^{\da}{\phib}^{\bk}&=&\left(E^{{\da}M}{\prt}_M +i{\ca}^{(r){\da}}
\,{\bV}_{(r)}\right){\phib}^{\bk}.
\label{C.85}
\end{eqnarray}
 The superspace
geometry we have established here describes supergravity and
matter and accounts consistently for K\"ahler transformations and
for gauged isometries of the K\"ahler metric (of which Yang-Mills
symmetries are a particular case).

\indent

The field strength superfields $X_{\al},{\bX}^{\da}$, already
discussed in the ungauged case, receive now additional
contributions (hereafter we shall denote them ${\frak{X}}_{\al}$
and ${\bar{\frak{X}}}^{\da}$), involving the Yang-Mills field
strength, ${\cf^{(r)}}$, and the Killing potential, $G_{(r)}$. To
see this, apply the exterior derivative to ${\widetilde{\Delta}}$
to obtain
\begin{equation}
d{\widetilde{\Delta}} \, = \,
2g_{k{\bk}}{\cd}{\phi}^k{\cd}{\phib}^{\bk}+2{\cf}^{(r)}G_{(r)}.
\label{C.86} \end{equation}
 Due to
\begin{equation}
{\frak{A}} \,=\,\f {1}{4}{\widetilde{\Delta}}+\f {i}{8}E^a(12G_a+
{\sib}_a^{{\da}{\al}}g_{k{\bk}}{\cd}_{\al}{\phi}^k
{\cd}_{\da}{\phib}^{\bk}), \label{C.87} \end{equation}
 the
relation between $d{\widetilde{\Delta}}$ and ${\frak{F}} =
d{\frak{A}}$ is obvious. As before, the superfields
${\frak{X}}_{\al}$ and ${\bar{\frak{X}}}^{\da}$ are identified
in the field strengths ${{{\frak{F}}^\db}_{ a}}$ and
${{\frak{F}}_{\bt a}}$ as
 \begin{eqnarray}
{\frak{X}}_\al \ & = & -\f {i}{2}g_{k{\bar{k}}} {\si}^a_{\al{\da}}
{\cd}_a{\phi}^k {\cd}^{\da} {\phib}^{\bar{k}} +\f
{1}{2}g_{k{\bar{k}}}{\cd}_\al {\phi}^k \, {\bF}^{\bar{k}}
+{\cw}_{\al}^{(r)}G_{(r)}, \label{C.88} \\
{\bar{\frak{X}}}^{\da}
& = &- \f {i}{2}g_{k {\bar{k}}} {{\sib}}^{a{\da} \al}
{\cd}_a{\phib}^{\bar{k}}{\cd}_{\al}{\phi}^k + \f {1}{2}g_{k
{\bar{k}}}{\cd}^{\da}{\phib}^{\bar{k}}F^k +{{\cw}}^{(r){\da}}
G_{(r)}. \label{C.89} \end{eqnarray}
 In distinction
to the ungauged case all derivatives are now fully covariant with
respect to gauged isometries. $F^k$ and ${\bF}^{\bk}$ are still
defined as
\begin{equation}
F^k \,=\,- \f {1}{4}{\cd}^{\al}{\cd}_{\al}{\phi}^k, \cem
{\bF}^{\bk} \,=\,- \f {1}{4}{\cd}_{\da}{\cd}^{\da}{\phib}^{\bk},
\label{aux}
\end{equation}
  but
the covariant derivatives of ${\cd}_\al \phi^k$ and $\cd^{\da}
{\phib}^{\bk}$ appearing in this definition contain now new terms
which take into account the gauged isometries, explicitly
 \begin{eqnarray}
  {\cd}_B {\cd}_\al \phi^k
&=& {E_B}^M \partial_M {\cd}_\al \phi^k - {\phi_{B\al}}{}^\vp
{\cd}_\vp \phi^k + i {\ca}^{(r)}_B \frac{\partial
{V_{(r)}}^k}{\partial \phi^l} {\cd}_\al \phi^l \nonumber \\ & & -
{\frak{A}}_B {\cd}_\al \phi^k + {\Gamma^k}_{ij} {\cd}_B
\phi^i{\cd}_\al \phi^j , \label{C.90}\\ {\cd}_B {\cd}^{\da}
\phib^{\bk} &=& {E_B}^M
\partial_M {\cd}^{\da} \phib^{\bk} - {{\phi_B}^{\da}}{}_{\dv}
{\cd}^{\dv}\phib^{\bk}+ i {\ca}^{(r)}_B \frac{\partial
{{\bV}_{(r)}}{}^{\bk}}{\partial {\phib}^{\bl}} {\cd}^{\da}
{\phib}^{\bl} \nonumber \\ & &  + {\frak{A}}_B {\cd}^{\da}
\phib^{\bk}+ {\Gamma^{\bk}}_{\bi\bj} {\cd}_B
\phib^{\bi}{\cd}^{\da}\phib^{\bj}. \label{C.91} \end{eqnarray}
The Yang-Mills superfields appearing in
(\ref{C.88}), (\ref{C.89}) are identified in the field strength
${\cf}^{(r)}$, \ie
\begin{equation}
{\cf}^{(r)}_{\ {\bt} \, a} \,=\,
i{\si}_{a{\bt}{\db}}{{\cw}}^{(r) {\db}}, \cem
{\cf}^{(r){\db}}{}_a \, = \,
i{\sib}_a^{{\db}{\bt}}{\cw}^{(r)}_{\bt}, \label{C.92}
\end{equation}
 and satisfy the relations
(\ref{B.112}) - (\ref{B.113}). Since the Yang-Mills gauge potentials
are now defined in the framework of K\"ahler superspace geometry,
all the chiral weights and therefore the transformation laws under
K\"ahler transformations and gauged isometries are determined and
should be taken into account in the definition of covariant
derivatives.

The relevant quantity in the construction of the component field
action is the K\"ahler D-term, defined as the lowest component of the
superfield ${\cd}^{\al}{\frak{X}}_{\al}$. The geometric
construction presented so far has the great advantage that full
invariance is automatically insured. The explicit form of the
D-term superfield is
 \begin{eqnarray}
  -{\f{1}{2}}{\cd}^{\al}{\frak{X}}_{\al}&=&
-g_{k{\bk}}{\eta}^{ab}{\cd}_b {\phi}^k {\cd}_a {\phib}^{\bk}
\nonumber \\ & &-\f {i}{4}g_{k{\bk}}{\si}^a_{{\al}{\da}}\left(
{\cd}^{\al} {\phi}^k {\cd}_a {\cd}^{\da}{\phib}^{\bk} +{\cd}^{\da}
{\phib}^{\bk} {\cd}_a {\cd}^{\al} {\phi}^k \right) \nonumber \\ &
&+g_{k{\bk}}F^k {\bF}^{\bk} +\f
{1}{16}R_{j{\bj}k{\bk}}{\cd}^{\al}{\phi}^k {\cd}_{\al} {\phi}^j
{\cd}_{\da} {\phib}^{\bk}{\cd}^{\da}{\phib}^{\bj} \nonumber \\ &
&-ig_{k{\bk}}{{\bV}_{(r)}}{}^{\bk} {\cw}_{\al}^{(r)} {\cd}^{\al}
{\phi}^k
-ig_{k{\bk}}{V_{(r)}}^k{{\cw}}^{(r)}_{\da}{\cd}^{\da}{\phib}^{\bk}
\nonumber \\ & &-{\f {1}{2}} \left( {\cd}^{\al}{\cw}_{\al}^{(r)}
\right) G_{(r)}. \label{C.96}
\end{eqnarray}

The discussion of this section shows that gauged isometries allow
for a very suggestive description in the framework of K\"ahler
superspace geometry. The results presented here in superfield form
are particularly useful to extract component field expressions in
a constructive and concise way as illustrated in section
\ref{CPN}, where we fully develop Lagrangians in component
fields.\\ \indent
 So far we have mainly dealt with matter
superfields, which play the role of coordinates of the K\"ahler
manifold, and with their covariant differentials.
It will be useful to consider the more general case
of a generic superfield, ${\bf U}^k$, of transformation law
\begin{equation}
\delta {\bf U}^k \,=\,- \al^{(r)} \ \frac{\delta
{V_{(r)}}^k}{\delta \phi^l} {\bf U}^l. \label{C.97} \end{equation}
For simplicity we assume ${\bf U}^k$ to be a superfield (0-form) of
vanishing chiral weight and scalar with respect to Lorentz
transformations. The exterior covariant derivative is then defined
as
\begin{equation}
{\cd} {\bf U}^k \,=\,d {\bf U}^k +  i \ca^{(r)} \frac{\prt
{V_{(r)}}^k}{\prt \phi^l} {\bf U}^l +
 {\Gamma^k}_{lm} \, {\cd} \phi^m \, {\bf U}^l,
\label{C.98} \end{equation}
 with
\begin{equation}
{\cd} {\bf U}^k \,=\,E^A \cd_A {\bf U}^k. \label{C.99}
\end{equation}
Note that, as a consequence of the chirality of the matter
superfields, the Levi-Civita term is absent in $\cd^{\da} {\bf U
}^k$.

The graded commutator of two such covariant derivatives is
obtained in taking the covariant exterior derivative of the one
form ${\cd} {\bf U}^k$, \ie
 \begin{equation}
  {\cd}{\cd} {\bf U}^k \, = \, i
\cf^{(r)} \left(
\frac{\prt {V_{(r)}}^k}{\prt \phi^l}
{\bf U}^l
+ {V_{(r)}}^m {\Gamma^k}_{lm} {\bf U}^l \right)
- g^{k \bk} \, R_{m \bk l \bj} \, {\cd} \phib^{\bj} \, {\cd} \phi^m \,
{\bf U}^l.
\label{C.100} \end{equation}
 Decomposing the left hand term
according to
\begin{equation}
{\cd}{\cd} {\bf U}^k \,=\,E^A E^B (\cd_B \cd_A {\bf U}^k + \f
{1}{2} \ {T_{BA}}^C \cd_C {\bf U}^k),
\label{C.101} \end{equation}
 allows
to read off the graded commutator of two covariant derivatives of
${\bf U }^k$ to be
 \begin{eqnarray}
  (\cd_B , \cd_A) {\bf U}^k&=&
- {T_{BA}}^C \ \cd_C {\bf U}^k
+ i {\cf_{BA}}^{(r)} \left(  \frac{\prt {V_{(r)}}^k}{\prt
\phi^l} {\bf U}^l + {V_{(r)}}^m {\Gamma^k}_{lm} {\bf U}^l \right)
\nonumber \\ & &
+ g^{k \bk} R_{m \bk l \bj} \, {\bf U}^l \left( \cd_B
\phib^{\bj} \, \cd_A \phi^m - (-)^{ab} \cd_A \phib^{\bj} \, \cd_B
\phi^m \right).
\label{C.102}
\end{eqnarray}

We have considered ${\bf U}^k$ as a superfield inert under Lorentz
and K\"ahler transformations. The spinor
derivative $\cd_\al \phi^k$ of a chiral superfield $\phi^k$ will
transform in the same manner as ${\bf U}^k$ under gauged
isometries but will pick up additional contributions from Lorentz
and K\"ahler transformations.

\subsec{Supergravity Transformations  \label{appC2}}

We have constructed a superspace geometry in terms of the basic
geometric objects
\begin{itemize}
\item $E^A=dz^M {E_M}^A \hs{1.4cm}$        frame of superspace,
\item ${\phi}^k, {\phib}^{\bk}\hs{3.1cm}$  chiral matter superfields,
\item ${\ca}^{(r)}=dz^M {{\ca}_M}^{(r)}\cem$  Yang-Mills potential.
\end{itemize}
The chiral matter superfields take their values in a K\"ahler
manifold and we have seen that superspace geometry and K\"ahler
geometry are intimately related. In order to describe gauged
isometries of the superfield K\"ahler metric we have introduced
the corresponding Yang-Mills potential. Infinitesimal variations
of parameters
\begin{itemize}
\item $\xi_M \hs{1.3cm}$   superspace diffeomorphisms,
\item ${{\Lambda}_B}^A\cem$ Lorentz transformations,
\item ${\al}^{(r)}\hs{1.1cm}$ Yang-Mills transformations,
\end{itemize}
change the basic geometric objects such that
\begin{eqnarray}
E^A &\mapsto& E^A + \delta E^A,
\\  \label{C.103}
{\phi}^{k} &\mapsto&{\phi}^k + \delta {\phi}^k,
\\ \label{C.104}
{\phib}^\bk &\mapsto& {\phib}^{\bk} + \delta {\phib}^\bk,
\\ \label{C.105}
{\ca}^{(r)} &\mapsto& {\ca}^{(r)} + \delta {\ca}^{(r)},
\label{C.106}
\end{eqnarray}
 with
  \begin{eqnarray}
\delta E^A &=& L_{\xi} E^A +E^B
{{\Lambda}_B}^A-\f {i}{2}w(E^A) E^A  \ima  \left(F({\phi})
-{\al}^{(r)}F_{(r)}({\phi}) \right),
\\ \label{C.107}
\delta {\phi}^k &=& L_{\xi}{\phi}^k -{\al}^{(r)}
{V_{(r)}}^k({\phi}),
\\[1mm] \label{C.108}
\delta {\phib}^{\bk} &=&
L_{\xi}{\phib}^{\bk}-{\al}^{(r)}{{\bV}_{(r)}}{}^{\bk}({\phib}),
\\[1mm] \label{C.109}
\delta {\ca}^{(r)} &=& L_{\xi}{\ca}^{(r)} -id{\al}^{(r)} +
{\al}^{(p)} {\ca}^{(q)} {c_{(p)(q)}}^{(r)}. \label{C.110}
\end{eqnarray}
 Here, the Lie derivative in superspace is defined
as
\begin{equation}
L_{\xi} \,=\,{\iota}_{\xi}d +d {\iota}_{\xi}. \label{C.111}
\end{equation}
 Remarkably enough, K\"ahler transformations and
gauged isometries appear in a well defined way in the structure
group of superspace. In the next step we wish to express these
transformation laws as much as possible in terms of covariant
objects - torsion, field strength and covariant derivatives -
which were defined earlier as
\begin{eqnarray}
 {\frak T}^A &=& {\cd}E^A = dE^A +E^B{\phi_B}^A +w(E^A) E^A {\frak{A}},
\\
\label{C.112}
{\cd}{\phi}^k &=& d{\phi}^k +i{\ca}^{(r)}
{V_{(r)}}^k({\phi}),
\\[1mm] \label{C.113}
{\cd}{\phib}^{\bk} &=&
d{\phib}^{\bk} +i{\ca}^{(r)} {{\bV}_{(r)}}{}^{\bk}({\phib}), \\
\label{C.114} {\cf}^{(r)}&=& d{\ca}^{(r)}+\f
{i}{2}{\ca}^{(p)}{\ca}^{(q)} {c_{(p)(q)}}^{(r)}. \label{C.115}
\end{eqnarray}
 Straightforward substitution yields
 \begin{eqnarray}
  \delta
E^A &=& {\cd}{\xi}^A+{\iota}_{\xi}{\frak T}^A + E^B \left( {\Lambda_B}^A
-\iota_{\xi}{\phi_B}^A \right) \nonumber \\ & &-w(E^A)E^A \left[
\iota_{\xi}{\frak{A}}+\f {i}{2} \ima ( F-\al^{(r)}F_{(r)})\right],
\label{C.116}
\\ \label{C.116a} {\delta} {\phi}^k &=&
{\iota}_{\xi}{\cd}{\phi}^k-({\al}^{(r)}
+i{\iota}_{\xi}{\ca}^{(r)}){V_{(r)}}^k({\phi}),
\\[2mm] \label{C.117}
{\delta} {\phib}^{\bk} &=&
{\iota}_{\xi}{\cd}{\phib}^{\bk}-({\al}^{(r)}
+i{\iota}_{\xi}{\ca}^{(r)}){{\bV}_{(r)}}{}^{\bk}({\phib}), \\[2mm]
\label{C.118}
\delta {\ca}^{(r)}&=&\iota_{\xi}{\cf}^{(r)}+(\al^{(p)}+i
\iota_{\xi}{\ca}^{(p)}){\ca}^{(q)} {c_{(p)(q)}}^{(r)}
-id(\al^{(r)}+i\iota_{\xi}{\ca}^{(r)}). \label{C.119}
\end{eqnarray}
 Supergravity transformation ${\delta}_{\WZ}$ are
then defined as certain combinations of superspace diffeomorphisms
and field dependent compensating Lorentz and gauged
isometry transformations, namely
 \begin{eqnarray}
  {\Lambda_B}^A
&=& {\iota}_{\xi} {\phi_B}^A,
\\[1mm] \label{C.120}
{\al}^{(r)} &=&
-i\iota_{\xi}{\ca}^{(r)}. \label{C.121} \end{eqnarray}
Taking into account the explicit form of ${\frak{A}}$, -cf. (\ref{C.81}),
we obtain
\begin{eqnarray}
 {\delta}_{\WZ}\,E^A &=& {\cd}{\xi}^A +{\iota}_{\xi}{\frak T}^A
- \f {1}{4}w(E^A)E^A (K_k {\iota}_{\xi}{\cd}{\phi}^k -K_{\bk}
{\iota}_{\xi}{\cd}{\phib}^{\bk})
 \nonumber \\
& &-\f {i}{8}w(E^A)E^A{\xi}^b\left(12G_b +
 {{\sib}}_b^{{\da} \al}g_{k{\bk}}{\cd}_{\al}{\phi}^k
{\cd}_{\da}{\phib}^{\bk} \right),
\\[2mm] \label{C.122}
{\delta}_{\WZ}\, {\phi}^k &=& {\iota}_{\xi}{\cd}{\phi}^k,
\\[1mm] \label{C.123}
{\delta}_{\WZ}\,{\phib}^{\bk}&=&
{\iota}_{\xi}{\cd}{\phib}^{\bk},
\\[1mm] \label{C.124}
{\delta}_{\WZ}\,{\ca}^{(r)} &=& {\iota}_{\xi}{\cf}^{(r)}.
\label{C.125} \end{eqnarray}
 Recall that the last term in the transformation law of $E^A$ is spurious
in that it could be absorbed in covariant redefinitions of the
first two terms. The interior product of ${\xi}^M$ with torsion
and Yang-Mills field strength is defined as
 \begin{eqnarray}
{\iota}_{\xi}{\frak T}^A &=& E^B {\xi}^C {{\frak T}_{CB}}^A, \\
\label{C.126} {\iota}_{\xi} {\cf}^{(r)} &=& E^A {\xi}^B {{\cf}_{BA}}^{(r)}.
\label{C.127} \end{eqnarray}
 For later convenience we consider
also generic superfields $\Phi$ and ${\bf U}^k$ of transformation
laws
\begin{eqnarray}
 {\delta}  \, {\Phi} &=& L_{\xi}d{\Phi}
-\f{i}{2}w(\Phi){\Phi} \, \ima  \left( F(\phi)
-{\al}^{(r)}F_{(r)}(\phi) \right),
\\ \label{C.128}
{\delta}  \, {\bf U}^k &=& L_{\xi}{\bf U}^k -{\al}^{(r)} \frac{\prt
{V_{(r)}}^k}{\prt \phi^l} {\bf U}^l \label{C.129}
\end{eqnarray}
 and covariant derivatives
  \begin{eqnarray}
{\cd}{\Phi} &=& d{\Phi} + w({\Phi}){\Phi}{\frak{A}},
\\[1mm] \label{C.130}
{\cd}{\bf U}^k &=& d{\bf U}^k +i {\ca}^{(r)}
\frac{\prt {V_{(r)}}^k}{\prt \phi^l} {\bf U}^l +{{\Gamma}^k}_{lm}
{\cd} {\phi}^l {\bf U}^m. \label{131} \end{eqnarray}
 Straightforward
substitution allows to derive the supergravity transformations
\begin{eqnarray}
 {\delta}_{\WZ}\, {\Phi} &=&{\iota}_{\xi}{\cd}{\Phi} -\f
{1}{4}w({\Phi}){\Phi} (K_k {\iota}_{\xi}{\cd}{\phi}^k
-K_{\bar{k}}{\iota}_{\xi}{\cd}{\phib}^{\bar{k}})
\nonumber \\ & &
-\f {i}{8}w({\Phi}){\Phi}{\xi}^a(12G_a +
{{\sib}}_a^{{\da} \al}g_{k{\bar{k}}}{\cd}_\al {\phi}^k
{\cd}_{\da}{\phib}^{\bar{k}}),
\\ \label{132}
\delta_{\WZ}\, {\bf U}^k &=&{\iota}_{\xi}{\cd}{\bf U}^k + {{\Gamma
}^k}_{lm}{\iota}_{\xi} {\cd}{\phi}^l {\bf U}^m. \label{C.133}
\end{eqnarray}
 The supergravity transformations presented so far
at the full superfield level will provide the basic building
blocks for the derivation of supersymmetry transformations of the
component fields. We will also use these supergravity
transformations in the more explicit form
 \begin{eqnarray}
{\delta}_{\WZ}\,{E_M}^A &=& {\cd}_M {\xi}^A +{E_M}^B {\xi}^C
{{\frak T}_{CB}}^A
\nonumber \\ & &- \f {1}{4}w(E^A){E_M}^A
{\xi}^B(K_k{\cd}_B {\phi}^k -K_{\bar{k}} {\cd}_B{\phib}^{\bar{k}})
\nonumber \\ & &
-\f {i}{8}w(E^A){E_M}^A{\xi}^b (12G_b +
{{\sib}}_b^{{\da} \al}g_{k{\bar{k}}} {\cd}_\al {\phi}^k
{\cd}_{\da}{\phib}^{\bar{k}}),
\label{C.134} \\[2mm]
{\delta}_{\WZ}\,{\phi}^k &=&{\xi}^A{\cd}_A{\phi}^k, \\[1mm]
\label{C.135} {\delta}_{\WZ}\,{\phib}^{\bk} &=& {\xi}^A {\cd}_A
{\phib}^{\bk},
\\[1mm] \label{C.136}
{\delta}_{\WZ}\,{\Phi}&=&
{\xi}^A{\cd}_A {\Phi} -\f {1}{4}w({\Phi}){\Phi}\ {\xi}^A
(K_k{\cd}_A{\phi}^k -K_{\bar{k}}{\cd}_A{\phib}^{\bar{k}})
\nonumber \\ & &
-\f {i}{8}w({\Phi}){\Phi}{\xi}^b(12G_b +
{{\sib}}_b^{{\da} \al} g_{k{\bar{k}}}{\cd}_\al {\phi}^k
{\cd}_{\da}{\phib}^{\bar{k}}),
\\[2mm] \label{C.137}
{\delta}_{\WZ}\,{\bf U}^k &=& {\xi}^A{\cd}_A {\bf
U}^k+{{\Gamma}^k}_{lm}{\xi}^A {\cd}_A {\phi}^l {\bf U}^m.
\label{C.138} \end{eqnarray}
 Observe the presence of the terms
\begin{equation}
K_k
{\iota}_{\xi}{\cd}{\phi}^k-K_{\bk}{\iota}_{\xi}{\cd}{\phib}^{\bk}
\,=\,{\xi}^A(K_k{\cd}_A{\phi}^k
-K_{\bar{k}}{\cd}_A{\phib}^{\bar{k}}). \label{C.139}
\end{equation}
 The corresponding gauge transformations are field
dependent K\"ahler transformations and isometries, there is no
free parameter which could compensate these terms unlike the case
of Lorentz and Yang-Mills transformations.

\subsec{The Yang-Mills Case  \label{appC3}}

 Let us consider the situation where the gauged isometries
reduce to the standard Yang-Mills transformations. This
corresponds to the case where the isometries act linearly on
the fields such that
\begin{eqnarray}
 {V_{(r)}}^k &=& V_{(r)} \phi^k \, = \,
+i \lp{{\bf T}_{(r)}}\phi \rp^k
\label{C.159} \\
{\bV_{(r)}}{}^\bk &=& \bV_{(r)} \phib^\bk \, = \,
-i\lp \phib {\bf
T}_{(r)}{}\rp^{\bk} . \label{C.160} \end{eqnarray}
 where the
${\bf T}_{(r)}$,
are in a suitable matrix
representation of the generators of the gauge group considered, with
commutation relations
\begin{equation}
\left[{\bf T}_{(r)},{\bf T}_{(s)}\right] = i {c_{(r)(s)}}^{(t)}
{\bf T}_{(t)}, \end{equation}
implied by those of the $V_{(r)}$'s.
Using the notation $\ca = \ca^{(r)} {{\bf T}_{(r)}}$,
the covariant derivatives of the matter superfields take the form
\begin{equation}
{\cd} \phi^k \, = \,
\left( d \phi -\ca\phi \right)^k, \cem
{\cd} \phib^\bk \, = \,
\lp d \phib +\phib \ca \rp^\bk.
\end{equation}
 Next we can determine the Killing potential using
(\ref{C.51}), (\ref{C.52}). Since the K\"ahler potential is
invariant under gauge transformations, (\ref{C.52}) tells us
\begin{equation}
K_k \lp {\bf T}_{(r)}\phi \rp^k -K_\bk \lp \phib {\bf T}_{(r)}
\rp^\bk \, = \, 0 \, = \, F_{(r)} (\phi) + {\bar{F}}_{(r)}(\phib),
\end{equation}
implying that $F_{(r)} (\phi)$ and
${\bar{F}}_{(r)}(\phib)$ are just constants, which can safely be
set to zero. The real Killing potential $G_{(r)}$ then becomes
\begin{equation}
G_{(r)} \, = \,
+ \f{i}{2}\lp K_k {V_{(r)}}^k - K_\bk {\bV_{(r)}}{}^\bk \rp \, = \,
-\f{1}{2} \left[ K_k \lp{{\bf
T}_{(r)}}{\phi} \rp^k  + K_\bk \lp \phib{{{\bf T}}_{(r)}}\rp^{\bk}
\right].
    \label{C.161}
\end{equation}
Using this information, together with the vanishing of the Killing
pre-potentials, in the combination
$\widetilde{\Delta} =\Delta + 2 \ca^{(r)} G_{(r)}$, we obtain
 \begin{eqnarray}
  \Delta + 2
\ca^{(r)} G_{(r)}  &=& K_k \lp d \phi^k +i \ca^{(r)} {V_{(r)}}^k
\rp
 - K_\bk \lp d \phib^\bk +i \ca^{(r)} {\bV_{(r)}}{}^\bk \rp \\ \nonumber
 &=& K_k {\cd} \phi^k - K_\bk {\cd} \phib^\bk. \end{eqnarray}
As a consequence, we recover the K\"ahler connection
$A = \frak A $ of section
\ref{GRA32}, given as
\begin{equation}
A \, = \, \f{1}{4}K_k {\cd}{\phi}^k-\f {1}{4}K_{\bk} {\cd}{\phib}^{\bk}
  + \f{i}{8}E^a( 12G_a+{\sib}_a^{\da\al}g_{k{\bk}}{\cd}_{\al}
{\phi}^k{\cd}_{\da}{\phib}^{\bk}),
\label{kelkon}
\end{equation}
with Yang-Mills covariant derivatives everywhere.

Finally, the supergravity transformations are directly read off from the
previous discussions, eqs. (\ref{C.116}) - (\ref{C.119}) and
(\ref{C.134}) - (\ref{C.138}).

\newpage
\sect{  Superfield equations of motion  \label{appD}}

Given the geometric formulation of supersymmetric theories it is
desirable to have a superfield action principle, in the sense that the
variation of suitable superspace densities gives rise to superfield
equations of motion.

On the other hand, the geometric descriptions of
supersymmetric theories are characterized by covariant constraints
(torsion constraints for supergravity, field strength constraints for
Yang-Mills, 2-form and 3-form gauge theories and chirality constraints for
matter superfields). As a consequence, the basic building blocks
initially  used in the geometric construction (frame of superspace,
Lorentz, Yang-Mills, 2-form and 3-form gauge potentials, and chiral
superfields) are no longer the fundamental objects - they are given in
terms of unconstrained pre-potentials which arise from the explicit
solution of the superspace constraint equations.

A possible way to formulate a superfield action principle is therefore to
write superfield densities in terms of the unconstrained pre-potentials and
to vary them accordingly \cite{GGRS83}.
This approach is particularly useful in the context of supergraph
perturbation theory.

Another possibility \cite{WZ78a}, more closely related to superspace
geometry, and which will be pursued here, is to solve directly the
variational version of the constraint equations. In this way one determines
directly the variations of the basic geometric objects in terms of
unconstrained entities. In this (equivalent) formulation, superspace
densities are written in the usual way and the relation to component field
formalism is quite transparent.

In this appendix we derive, as an example, the superfield equations of
motion for the complete supergravity/matter/Yang-Mills system in the
presence of gauged isometries. In the first two subsections, we work in
generic $U(1)$ superspace, defining and solving the variational constraint
equations in the first subsection and discussing superspace densities and
integration by parts in the second one. The variational equations
pertaining to isometric superspace are treated in subsection \ref{appD3}.
In subsection \ref{appD4} we derive the superfield equations of motion
for the complete supergravity/matter/Yang-Mills system.

\subsec{Integration by Parts in $U(1)$ Superspace
\label{appD2}}

The superfield action principle for supergravity proposed by Wess and Zumino
\cite{WZ78a}, \cite{Zum79a} is a generalization of usual gravity. In
general relativity, especially when coupled to spinor fields, densities are
constructed by means of the determinant of the vierbein, or frame. The
corresponding basic superspace object is $E$, the superdeterminant of the
frame $E_M{}^A$ in superspace. In general, a supersymmetric action will be
given as the product of $E$ with some suitable covariant superfield,
integrated over superspace, \ie over space-time and the anticommuting
spinor coordinates. In the derivation of superfield equations of motion,
integration by parts in superspace will be used systematically.
This means that expressions like
\begin{equation}
\int_\ast \! \! E \, \cd_\al v^\al , \cem
\int_\ast \! \! E \, \cd^{\da} v_{\da}, \cem
\int_\ast \! \! E \, \cd_a v^a,
\label{D.113}
\end{equation}
with $v^A$ some generic, covariant superfield of chiral weight
$w(v^A)$, should be related to pure superspace surface terms. The asterisk
indicates that integration is understood over full superspace, \ie
anticommuting coordinates {\em and} space-time.

In order to explain the mechanism of integration by parts in some more
detail let us recall first some definitions. The exterior covariant
derivative
${\cd} v^A = dz^M \cd_M v^A$ being given as
\begin{equation}
{\cd} v^A \, = \, d v^A + v^B {\phi_B}^A + w(v^A) v^A A,
\label{D.114}
\end{equation}
we identify the 1-form coefficients
\begin{equation}
\cd_M v^A \, = \, \prt_M v^A
+ (-)^{mb} v^B {\phi_{MB}}^A + w(v^A) A_M v^A.
\label{D.116}
\end{equation}
Another crucial ingredient is the torsion 2-form
$T^A = \frac{1}{2} dz^M dz^N {T_{NM}}^A$ defined as
\begin{equation}
T^A \,=\,{\cd}E^A \,=\,dE^A + E^B {\phi_B}^A + w (E^A) E^A A.
\label{D.119}
\end{equation}
Its components
\begin{equation}
{T_{NM}}^A \, = \, \cd_N {E_M}^A - (-)^{mn} \cd_M {E_N}^A,
\label{D.121}
\end{equation}
are given in terms of the covariant derivatives
\begin{equation}
\cd_N {E_M}^A \, = \, \prt_N {E_M}^A + (-)^{(m+b)n}
{E_M}^B {\phi_{NB}}^A + w (E^A) A_N {E_M}^A.
\label{D.122}
\end{equation}
It is a matter of straightforward calculation to establish the
superspace identity \cite{Zum79a}
\[
\prt_M (E v^A {E_A}^M) (-)^m
 \,=\,E \left[ \prt_M v^A + v^N (\prt_N {E_M}^A
-(-)^{mn} \prt_M {E_N}^A) \right] {E_A}^M (-)^m.
\]
Covariantizing the derivatives, this identity takes the form
\begin{eqnarray}
\prt_M (E v^A {E_A}^M) (-)^m &=& E \, \cd_{\! A} v^A (-)^a
+ E \, v^B {T_{BA}}^A (-)^a
\nonumber \\ &&
+ E \left( w (E^A) - w( v^A) \right)v^A A_A.
\label{D.124}
\end{eqnarray}
This is the central point in the discussion of integration by
parts in superspace. Observe that so far we did not make any use
of torsion constraints. Taking into account the explicit form of
the torsion coefficients in $U(1)$ superspace, one shows that the
only non-vanishing contributions to the torsion term are
\begin{equation}
{T_{b \, \al}}^\al \,=\,+i \, G_b, \cem
{{T_b}^{\, \da}}_{\da} \,=\,-i \, G_b,
\label{D.125} \end{equation}
which add to zero in the supertrace. The
torsion term is therefore absent.
If, in addition, we require
\begin{equation}
w(v^A) \,=\,w(E^A),
\label{D.126}
\end{equation}
we obtain
\begin{equation}
\prt_M (E \, v^A {E_A}^M) (-)^m \ =\  E \, \cd_{\! A} v^A (-)^a.
\label{D.127}
\end{equation}
This establishes the relation alluded to above, identifying expressions
like (\ref{D.113}) as pure superspace surface terms.
This relation will be frequently used
in the derivation of superfield equations of motion.

\subsec{Variational Equations in $U(1)$ Superspace
\label{appD1}}

We first introduce as basic
variables the variations of the vielbein, Lorentz and $U(1)$
connections modulo the effects of superspace diffeomorphisms and structure
group transformations. Subsequently we present a concise and systematic
analysis of the consequences of the constraints of $U(1)$ superspace for
these variables.

\indent

\begin{itemize}

\item{\bf Basic definitions}

Consider the infinitesimal variations
 \begin{eqnarray}
  \delta E^A &=&
H^A, \\ \label{D.1} \delta {\phi_B}^A &=& {\Omega_B}^A, \\
\label{D.2} \delta A &=& \omega, \label{D.3} \end{eqnarray}
 of the
frame, Lorentz and U(1) gauge potential. These superspace 1-forms
are pa\-ra\-me\-tri\-zed in such a way that
 \begin{eqnarray}
  H^A
\,=\,E^B {H_B}^A, & & \cem {H_B}^A \,=\,{E_B}^M \delta {E_M}^A,
\label{D.4} \\ {\Omega_B}^A \,=\,E^C {\Omega_{CB}}^A, & & \cem
{\Omega_{CB}}^A \,=\,{E_C}^M \delta {\phi_{MB}}^A, \label{D.5} \\
\omega \,=\,E^A \omega_A, & & \cem \omega_A \,=\,{E_A}^M \delta
A_M. \label{D.6} \end{eqnarray}
 As a consequence of these
definitions the variations of torsion, curvature and $U(1)$ field
strength become
\begin{eqnarray}
 \delta T^A &=& {\cd}H^A + E^B {\Omega_B}^A + w (E^A) E^A \omega,
\label{D.7}
\\ \delta {R_B}^A &=& {\cd} {\Omega_B}^A, \label{D.8} \\ \delta F
&=& d \omega. \label{D.9} \end{eqnarray}
 Here, ${\cd}$ denotes the
covariant exterior derivative in $U(1)$ superspace. It is
straightforward to work out the explicit expressions for the
coefficients of these 2-forms in superspace. The torsion
variational equations
\begin{eqnarray}
  \delta {T_{CB}}^A & = & \cd_C {H_B}^A
- (-)^{cb} \cd_B {H_C}^A + {T_{CB}}^F {H_F}^A \nonumber \\ && -
{H_C}^F {T_{FB}}^A + (-)^{cb} {H_B}^F {T_{FC}}^A \nonumber \\ && +
{\Omega_{CB}}^A - (-)^{cb} {\Omega_{BC}}^A \nonumber \\ && +
w(E^A) (\dtdu{B}{A} \omega_C - (-)^{cb} \dtdu{C}{A} \omega_B),
\label{D.10}
\end{eqnarray}
are of particular importance.
The vielbein and gauge potential variations must leave the torsion
constraints invariant. This determines the unconstrained
variational superfields. The corresponding
variations of curvature and $U(1)$ field strength are
\begin{eqnarray}
 \delta {R_{DCB}}^A & = & \cd_D {\Omega_{CB}}^A
-(-)^{dc} \cd_C {\Omega_{DB}}^A + {T_{DC}}^F {\Omega_{FB}}^A
\nonumber \\ && - {H_D}^F {R_{FCB}}^A + (-)^{dc} {H_C}^F
{R_{FDB}}^A, \label{D.11} \end{eqnarray}
 \begin{eqnarray}
  \delta
F_{DC} & = & \cd_D \omega_{C} -(-)^{dc} \cd_C \omega_{D} +
{T_{DC}}^F \omega_{F} \nonumber \\ && - {H_D}^F F_{FC} + (-)^{dc}
{H_C}^F F_{FD}. \label{D.12} \end{eqnarray}
Observe that the variational superfields are determined modulo
diffeomorphisms and structure group transformations, \ie up to
redefinitions of the form
\begin{eqnarray}
\unddt \, H^A &=& {\cal L}_\xi E^A + E^B {\chi_B}^A + w(E^A)E^A \rho,
\label{D.13} \\
\unddt \, {\Omega_B}^A &=& -{\cd} {\chi_B}^A + \iota_\xi {R_B}^A,
\label{D.14} \\
\unddt \, \omega &=& - d \rho + \iota_\xi F.
\label{D.15}
\end{eqnarray}
As a consequence, the variational equations change as
 \begin{eqnarray}
 \unddt \, \delta T^A &=& \cl_\xi T^A + T^B {\chi_B}^A + w(T^A) T^A \rho,
\label{D.16} \\
 \unddt \, \delta {R_B}^A &=& \cl_\xi {R_B}^A + {R_B}^C {\chi_C}^A -
{\chi_B}^C {R_C}^A, \label{D.17} \\
 \unddt \, \delta F &=& \cl_\xi F .
\label{D.18}
\end{eqnarray}
The covariant Lie derivative appearing here is given as
\begin{equation}
\cl_\xi \,=\,\iota_\xi {\cd} + {\cd} \iota_\xi. \label{D.19}
\end{equation}
Using $\iota_\xi E^A \,=\,\xi^A$,
the variation of ${H_B}^A$ reads
\begin{equation}
\unddt \, {H_B}^A \,=\,\xi^C {T_{CB}}^A + {\cd}_B \xi^A + {\chi_B}^A
+ w(E^A) \, \delta_B^A \, \rho.
\label{D.21}
\end{equation}
 Similarly,
\begin{eqnarray}
 \unddt \, {\Omega_{CB}}^A &=& - \cd_C {\chi_B}^A + \xi^D {R_{DCB}}^A,
\label{D.22} \\
 \unddt \, \omega_A &=& - \cd_A \rho + \xi^B F_{BA}.
\label{D.23} \end{eqnarray}
Clearly, the variational equations of the
torsion constraints are invariant under these redefinitions.

\indent

\item{\bf Torsion constraints I.}

In a first step we consider the variational equations of the
torsions
\begin{equation}
{T_{\gm \bt}}^a \,=\,0, \cem T^{\dg \db a} \,=\,0,
\label{D.24}
\end{equation}
\begin{equation}
{T_\gm}^{\db a} \, =\,  -2i {(\si^a \eps)_{\gm}}^{\db},
\label{D.25}
\end{equation}
\begin{equation}
 T_{\gm \bt \da} \, = \, 0, \cem T^{\dg \db \al} \,=\,0.
\label{D.26}
\end{equation}
 From (\ref{D.10}) we read off the explicit equations
 \begin{eqnarray}
  \delta {T_{\gm \bt}}^a &=&
\sum_{\gm \bt} \left( \cd_\gm {H_\bt}^a - H_{\gm \dv}
T^{\dv}{}_\bt{}^a \right),
\label{D.27} \\
\delta T_{\gm \bt \da}
&=& \sum_{\gm \bt} \left( \cd_\gm H_{\bt \da} - {H_{\gm}}^f T_{f
\bt \da} \right). \label{D.28}
\end{eqnarray}
 The pure gauge solution of (\ref{D.27}), (\ref{D.28}) is
\begin{eqnarray}
 {H_\bt}^a & = & \cd_\bt \bar{\Xi}^a +
\bar{\Xi}_{\dv} T^{\dv}{}_\bt{}^a,
\label{D.29} \\
H_{\bt \da} & = & \cd_\bt \bar{\Xi}_{\da} + \bar{\Xi}^f T_{f \bt \da}.
\label{D.30} \end{eqnarray}
 Likewise, the complex conjugate equations are solved by
 \begin{eqnarray}
  H^{\db a} & = &
\cd^{\db} \Xi^a + \Xi^{\vp} \, {T_{\vp}}^{\db a},
\label{D.31} \\
H^{\db \al} & = & \cd^{\db} \Xi^\al + \Xi^f \, {T_f}^{\db \al}.
\label{D.32} \end{eqnarray}
 Finally, making
use of the invariance of the variational equations under
redefinitions of the form (\ref{D.21}) we arrive at
\begin{eqnarray}
 {H_\bt}^a \,=\,\cd_\bt \cv^a, && \hs{0.5cm} H^{\db
a} \,=\,- \cd^{\db} \cv^a, \label{D.33} \\ H_{\bt \da} \, = \,
-\cv^c T_{\bt c \da} \,=\,i R^{\dagger} \cv_{\bt \da}, &&
\hs{0.5cm} H^{\db \al} \,=\,\cv^c{T^{\db}}_c{}^\al \ =\ -i R
\cv^{\al \db} .
\label{D.34}
\end{eqnarray}
 It remains to
discuss the variation of (\ref{D.25}),
 \begin{eqnarray}
  \delta
{T_\gm}^{\db a} & = & \cd_\gm H^{\db a} + \cd^{\db} {H_\gm}^a +
{T_\gm}^{\db f} {H_f}^a \nonumber \\ && - {H_\gm}^\vp {T_\vp}^{\db
a} - {H^\db}_{\dv} {T_\gm}^{\dv a}.
\label{D.35}
\end{eqnarray}
 We eliminate the
traceless parts of ${H_\bt}^\al, {H^{\db}}_{\da}$ by suitably
choosing ${\chi_\bt}^\al, {\chi^{\db}}_{\da}$ in (\ref{D.21}) to
arrive at
\begin{eqnarray}
 {H_\bt}^\al & = & \f {1}{2} \ \dtdu{\bt}{\al} H,
\label{D.36} \\
{H^{\db}}_{\da} & = & \f {1}{2} \ \dtud{\db}{\da} \bar{H}.
\label{D.37}
\end{eqnarray}
As a consequence, (\ref{D.35}) becomes
\begin{equation}
{T_\gm}^{\db f} {H_f}^a - \f {1}{2} (H+\bar{H}) {T_\gm}^{\db a} -
\left[\cd_\gm, \cd^{\db} \right] \cv^a \,=\,0,
\label{D.38}
\end{equation}
 showing that ${H_b}^a$ is completely determined as
a function of the unconstrained superfields $H + \bar{H}$ and
$\cv^a$. In spinor notation this equation reads
\begin{equation}
H_{\bt \db \ \al \da} \,=\,- \eps_{\bt \al} \eps_{\db \da}
(H+\bar{H})
- \f {i}{2} \left[\cd_\bt, \cd_{\db} \right] \cv_{\al \da}.
\label{D.39} \end{equation}
The supertrace of ${H_B}^A$ is now given as
\begin{equation}
{H_A}^A (-)^a \,=\,H + \bar{H}
+ \f {i}{4} \left[\cd^\al, \cd^{\da} \right]
\cv_{\al \da}. \label{D.40} \end{equation}
 Observe
that we did not make use of the redefinitions which correspond
to the chiral $U(1)$ in (\ref{D.21}).

\indent

\item{\bf Torsion constraints II.}

The variations of the torsions
\begin{equation}
{T_{\gm b}}^a \,=\,0, \cem {T_\gm}^{\db}{}_{\da} \,=\,0, \cem
{T_{\gm \bt}}^\al \,=\,0, \label{D.41}
\end{equation}
 give rise to the equations
\begin{equation}
\cd_\gm {H_b}^a - \cd_b {H_\gm}^a + {\Omega_{\gm b}}^a
 + {T_{\gm b}}^F {H_F}^a
 - {H_\gm}^{\underline{\vp}} {T_{\underline{\vp} b}}^a
 + H_{b \dv} {T^{\dv}}_\gm{}^a \,=\,0,
\label{D.42}
\end{equation}
\begin{equation}
\cd_\gm {H^{\db}}_{\da} + \cd^{\db} H_{\gm \da}
 + {{T_\gm}^{\db f}} H_{f \da}
 - {H_\gm}^f {T_f}^{\db}{}_{\da} - H^{\db f} T_{f \gm \da}
 + {\Omega_\gm}^{\db}{}_{\da} - \dtud{\db}{\da}\,  \omega_\gm \,=\,0,
\\[3mm]
\label{D.43}
\end{equation}
\begin{equation}
\sum_{\gm \bt} \left(\cd_\gm {H_\bt}^\al
 - {H_\gm}^f {T_{f\bt}}^\al
 + {\Omega_{\gm \bt}}^\al
 + \dtdu{\bt}{\al}\, \omega_\gm \right) \,=\,0.
\label{D.44} \end{equation}

These relations serve
to express the variations ${\Omega_{\gm b}}^a, H_{b \da}$ and
$\omega_\al$ in terms of the so far unconstrained superfields $H,
\bar{H}$ and $\cv^a$. In this context it is
convenient to define
 \begin{eqnarray}
  {\chi_b}^a  &=&  {H_b}^a -
\cd_b \cv^a, \label{D.45} \\ \chi_{b \da}  &=&  H_{b \da} - \cv^c
T_{cb \da}, \label{D.46} \\ {\chi_\bt}^\al  &=& {H_\bt}^\al +
\cv^c {T_{\bt c}}^\al, \label{D.47} \\ {\chi^{\db}}_{\da} &=&
{H^{\db}}_{\da} + \cv^c {T^{\db}}_{c \da}, \label{D.48} \\
{\Pi_{\gm b}}^a  &=&  {\Omega_{\gm b}}^a + \cv^d {R_{\gm d b}}^a,
\label{D.49} \\ \Sigma_\gm  &=& \omega_\gm + \cv^d F_{\gm d}
\label{D.50}
\end{eqnarray}
 and
to write (\ref{D.42}) - (\ref{D.44}) in the form
\begin{equation}
{\Pi_{\gm b}}^a  +  \cd_\gm {\chi_b}^a + \chi_{b \dv} {T_\gm}^{\dv
a} - 2T_{\gm b \dv} \cd^{\dv} \cv^a \,=\, 0, \label{D.51}
\end{equation}
\begin{equation}
{\Pi_\gm}^{\db}{}_{\da}  -  \dtud{\db}{\da} \Sigma_\gm
 + {T_\gm}^{\db d} \ \chi_{d \da} + \cd_\gm {\chi^{\db}}_{\da}
 - 2(\cd^{\db}\cv^d) T_{\gm d \da} \,=\,0, \\[2mm]
\label{D.52}
\end{equation}
\begin{equation}
{\Pi_{\gm \bt}}^\al + {\Pi_{\bt \gm}}^{\al}
 + \dtdu{\bt}{\al} \Sigma_\gm + \dtdu{\gm}{\al} \Sigma_\bt
 + \cd_\gm {\chi_\bt}^\al + \cd_\bt {\chi_\gm}^\al \,=\,0.
\label{D.53}
\end{equation}

The first of these equations allows to determine both ${\Pi_{\gm b}}^a$
and $\chi_{b \da}$. This is most easily seen in
spinor notation, where (\ref{D.51}) takes the form
\begin{equation}
2 \eps_{\db \da} \Pi_{\vt{2}\gm}{}_{\sym{\bt \al}}
 - 2 \eps_{\bt \al} \Pi_\gm{}_{\sym{\db \da}}
 - 4i \eps_{\gm \al} \chi_{\bt \db \, \da}
 + \cd_\gm \chi_{\bt \db \, \al \da}
 + 4i \eps_{\gm\bt} R^\dagger \cd_{\db} \cv_{\al \da} \,=\,0.
\label{D.54} \end{equation}
Taking into account
\begin{equation}
\chi_{\bt \db \ \al \da} \, = \, - \eps_{\bt \al} \eps_{\db \da} (H +
\bar{H}) -i \cd_\bt \cd_{\db} \cv_{\al \da}
 - i G_{\al \db} \cv_{\bt \da} + i G_{\bt \da} \cv_{\al
\db} ,
\label{D.55}
\end{equation}
 we obtain
\begin{eqnarray}
 \Pi_{\vt{3}\gm}{}_{\sym{{\db} {\da}}} & = & -
\f {i}{4} \cd_\gm \sum_{\db \da} \left(\cd^\vp \cd_{\db} \cv_{\vp
\da} - {G^\vp}_{\db} \cv_{\vp \da}\right), \label{D.56}
\\ \Pi_{\vt{2}\gm}{}_{\sym{\bt \al}} & = & \f {i}{4}
\cd_\gm \sum_{\bt \al} {G_\bt} ^{\dv} \cv_{\al \dv} \nonumber
\\ && - \sum_{\bt \al} \eps_{\gm \bt} \cd_\al \left(\f
{1}{2} (H+ \bar{H})
 + \f {i}{4} \ \cd^\vp \cd^{\dv} \cv_{\vp \dv} \right),
\label{D.57} \end{eqnarray}
 as well as
\begin{equation}
8i \, \chi_{\bt \db \, \da} \,=\,- 4 \eps_{\db \da} \cd_\bt (H+
\bar{H}) + 2i\cd_\bt \cd^\vp \cd_{\da} \cv_{\vp \db} -8i R^\dagger
\cd_{\db} \cv_{\bt \da}. \label{D.58}
\end{equation}
 This exhausts the information contained in
(\ref{D.51}). Substituting these results reduces (\ref{D.52})
simply to
\begin{equation}
\Sigma_\gm \,=\,-\cd_\gm \left(H + \f {1}{2} \bar{H} + \f {i}{4}
\cd^\vp \cd^\dv\cv_{\vp \dv} - \f {i}{2} \cv^aG_a \right)
\label{D.59} \end{equation}
 and (\ref{D.53}) is then
identically satisfied.

\indent

\item{\bf Torsion constraints III.}

As to the complex conjugate torsions,
\begin{equation}
{T^{\dg}}_b{}^a \,=\,0, \cem {T^{\dg}}_\bt{}^\al \,=\,0, \cem
{T^{\dg \db}}_{\da} \,=\,0 , \label{D.60} \end{equation}
 the variational equations read
\begin{equation}
\cd^{\dg} {H_b}^a - \cd_b H^{\dg a} + {{\Omega^{\dg} }_b}^a
 + {{T^{\dg}}_b}^{\underline{\vp}} {H_{\underline{\vp}}}^a
+ {H_b}^\vp {T_\vp}^{\dg a} \,=\,0,
\label{D.61}
\end{equation}
\begin{equation}
\cd^{\dg} {H_\bt}^\al + \cd_\bt H^{\dg \al}
 + {T_\bt}^{\dg f}{H_f}^\al
 - H^{\dg f} {T_{f \bt}}^\al
 - {H_{\bt}}^f {T_f}^{\dg \al}
 + {{\Omega^{\dg}}_\bt}^\al + \dtdu{\bt}{\al} \omega^{\dg} \,=\,0,
\\[2mm]
\label{D.62}
 \end{equation}
\begin{equation}
\sum_{\dg \db} \left( \cd^{\dg} {H^{\db}}_{\da}
 - H^{\dg f} {T_f}^{\db}{}_{\da}
 + {\Omega^{\dg \db}}_{\da}
 - \dtud{\db}{\da} \, \omega^{\dg} \right) \,=\,0 .
\label{D.63} \end{equation}

In this sector it is convenient to define
 \begin{eqnarray}
{{\ovch}_b}{}^a &=& {H_b}^a + \cd_b \cv^a, \label{D.64} \\
{{\ovch}_b}{}^\al &=& {H_b}^\al + \cv^c {T_{cb}}^\al, \label{D.65}
\\ {{\ovch}_\bt}{}^\al  &=&  {H_\bt}^\al - \cv^c
{T_{\bt c}}^\al, \label{D.66} \\ {{\ovch}^{\db}}{}_{\da}  &=&
{H^{\db}}_{\da} - \cv^c {T^{\db}}_{c \da}, \label{D.67} \\
{{{\oPi}^{\dg}}_b}{}^{a}  &=& {\Omega^{\dg}}_b{}^a - \cv^d
{R^{\dg}}_{db}{}^a, \label{D.68} \\ \bar{\Sigma}^{\dg}  &=&
\omega^{\dg} - \cv^d {F^{\dg}}_d . \label{D.69} \end{eqnarray}
 With these notations and after some
manipulations involving superspace Bianchi identities,
(\ref{D.61}) - (\ref{D.63}) can be written as
\begin{equation}
{{\oPi^{\ \dg}}{}_b}{}^{a}
+ \cd^{\dg} \, {{\ovch}_b}^{\ a}
+ {{{\ovch}}_b}^{\ \vp} {T_\vp}^{\dg a}
+ 2 \, {{T^{\dg}}_b}^\vp\cd_\vp \, \cv^a \,=\,0, \\[1mm]
\label{D.70}
\end{equation}
\begin{equation}
\oPi^{\, \dg}{}_\bt{}^\al
 + \dtdu{\bt}{\al} \bar{\Sigma}^{\dg}
 + {T_\bt}^{\dg d} \; {{\ovch}_d}^{\ \al}
 + \cd^{\dg} \, {{{\ovch}}_\bt}^{\ \al}
 + 2 \, (\cd_\bt \cv^d) {{T^{\dg}}_d}^\al \,=\,0, \\[1mm]
\label{D.71}
\end{equation}
\begin{equation}
{\oPi^{\, \dg \, \db}}{}_{\da} + {{\oPi}^{\, \db \, \dg}}_{\ \da}
 - \dtud{\db}{\da} \bar{\Sigma}^{\dg}
 - \dtud{\dg}{\da} \bar{\Sigma}^{\db}
 + \cd^{\dg} {{{\ovch}}^{\db}}_{\ \da}
 + \cd^{\db} {{{\ovch}}^{\dg}}_{\ \da} \,=\,0.
\label{D.72} \end{equation}

As before we employ spinor notation. (\ref{D.70}) becomes
\begin{equation}
2 \eps_{\db \da} {\oPi}_\dg{}_{\sym{\bt \al}}
 - 2 \eps_{\bt \al} {\oPi}_\dg{}_{\sym{\db \da}}
 + 4 i \eps_{\dg \db} {\ovch}_{\bt \db \al}
 + \cd _{\dg} {\ovch}_{\bt \db \ \al \da}
 + 4i \eps_{\dg \db} R \cd_\bt \cv_{\al \da} \,=\,0,
\label{D.73}
\end{equation}
 with
\begin{equation}
   {\ovch}_{\bt
\db \ \al \da} \, = \, - \eps_{\bt \al} \eps_{\db \da} (H+ \bar{H})
+ i \ \cd_{\db} \cd_{\bt} \cv_{\al \da}  + i \
G_{\al \db} \cv_{\bt \da} - i \ G_{\bt \da} \cv_{\al \db} .
\label{D.74}
\end{equation}
 From
(\ref{D.73}) we obtain
\begin{eqnarray}
 {\oPi}_\dg{}_{\sym{\bt \al}} & = &
 - \f {i}{4} \ \cd_{\dg} \sum_{\bt \al}
\left(\cd^{\dv} \cd_\bt \cv_{\al \dv}
 + {G_\bt}^{\dv} \cv_{\al \dv} \right),
\label{D.75} \\
{\oPi}_\dg{}_{\sym{\db \da}} & = &
 - \f {i}{4} \cd_{\dg} \sum_{\db \da} {G^\vp}_{\db} \cv_{\vp \da}
\nonumber \\ && + \sum_{\db \da} \eps_{\dg \db} \cd_{\da} \left(\f
{1}{2} (H+\bar{H}) - \f {i}{4} \cd^{\dv} \cd^\vp \cv_{\vp \dv}
\right),
 \label{D.76}
\end{eqnarray}
 as well as
\begin{equation}
8i {\ovch}_{\bt \db \, \al} \,=\,4 \eps_{\bt \al} \cd_{\db} (H+
\bar{H}) + 2i\cd_{\db} \cd^{\dv} \cd_\al \cv_{\bt \dv} + 8i R
\cd_\bt \cv_{\al \db} \, .
\label{D.77}
\end{equation}
 Equation (\ref{D.71}) then yields
\begin{equation}
\bar{\Sigma}^{\dg} \,=\,\cd^{\dg} \left( \bar{H} + \f {1}{2} H
 - \f {i}{4} \cd^{\dv} \cd^\vp \cv_{\vp \dv}
 - \f {i}{2} \cv^a G_a \right)
\label{D.78} \end{equation}
 and (\ref{D.72}) is identically satisfied.

\indent
This concludes our discussion of torsion constraints at dim = 0
and dim = $\f {1}{2}$ in $U(1)$ superspace. We have found that the
vielbein and connection variations are described in terms of the
independent unconstrained superfields $H$, $\bar{H}$ and
$\cv^a$. The torsion coefficients at dim =1 can then be used to
determine the variations of the covariant superfields $R,
R^\dagger$ and $G_a$. For our present purpose it is sufficient to
work out $\delta R$ and $\delta R^\dagger$ (which are
most conveniently obtained in using the corresponding curvature
equations)
 \begin{eqnarray}
\delta R & = &
- \left(\cv^a \cd_a + \bar{H} -i \cv^a G_a \right) R
\nonumber \\ &&
- \f {1}{8} \cd_{\da} \cd^{\da} \left( H + \bar{H}
- \f {i}{2} \cd^{\dv} \cd^\vp \cv_{\vp \dv} \right),
\label{D.79}
\\ \ & \ & \ \nonumber
\\ \delta R^\dagger & = & + \left(\cv^a \cd_a - H + i \cv^a G_a
\right) R^\dagger
\nonumber \\ &&
- \f {1}{8} \cd^\al \cd_\al
\left(H+ \bar{H} + \f {i}{2} \cd^\vp \cd^{\dv} \cv_{\vp \dv}
\right).
\label{D.80}
\end{eqnarray}

\indent

\item{\bf Chiral $U(1)$ gauge sector}

The solutions of the constraints
\begin{equation}
F_{\bt \al} \,=\,0 , \cem F^{\db \da} \,=\,0, \label{D.81}
\end{equation}
 in the $(\dem,\dem)$-basis are parametrized in
terms of a pre-potential $K$ (which, later on will be specialized
to the K\"ahler potential) such that
 \begin{eqnarray}
  A_\al & = & +\f {1}{4} {E_\al}^M \prt_M K,
\label{D.82} \\
A^{\da} & = & - \f {1}{4} E^{\da M} \prt_M K.
\label{D.83}
\end{eqnarray}
Using $\delta A = \om$, the variation of these equations gives
\begin{eqnarray}
\omega_{\al} - {H_\al}^B \left( A_B - \f {1}{4} \cd_B K \right)
- \f {1}{4} \cd_\al \delta K &=& 0,
\label{D.84} \\
\omega^{\da} - H^{\da B} \left( A_B + \f {1}{4} \cd_B K \right)
- \f {1}{4} \cd^{\da} \delta K &=& 0.
\label{D.85}
\end{eqnarray}
 Taking into
account our solution for ${H_A}^B$ leads to
 \begin{eqnarray}
\Sigma_\al & = &
+ \cd_\al \left(\f {1}{4} \delta K + \cv^b A_b
- \f{1}{4} \cv^b \cd_b K \right),
\label{D.86} \\
\bar{\Sigma}^{\da} & = &
- \cd^{\da} \left(\f {1}{4} \delta K + \cv^b A_b
+ \f {1}{4} \cv^b \cd_b K \right).
\label{D.87}
\end{eqnarray}
Finally, comparing with (\ref{D.59}) and (\ref{D.78}), we arrive
at the chirality conditions
\begin{eqnarray}
&&\! \! \! \! \!  \cd_\al \left(H + \f {1}{2} \bar{H} + \f {1}{4}
\delta K + \f {i}{4} \cd^\vp \cd^{\dv} \cv_{\vp \dv} + \cv^a (A_a
- \f {i}{2} G_a) - \f {1}{4}  \cv^a \cd_a K \right) = 0,
\label{D.88} \nonumber \\[0.5ex]  &&
\\ &&\! \! \! \! \! \cd^{\da} \left(\bar{H} + \f {1}{2} H + \f {1}{4} \delta K - \f
{i}{4}\cd^{\dv} \cd^\vp \cv_{\vp \dv} + \cv^a (A_a - \f {i}{2}
G_a) + \f {1}{4} \cv^a \cd_a K \right) = 0. \nn \\[0.5ex]
&&\label{D.89}
\end{eqnarray}
 These chirality constraints in turn are solved with the help
of chiral projection operators acting on unconstrained superfields
$U, \bar{U}$ and we obtain
 \begin{eqnarray}
  H + \bar{H} & = & -
\f {1}{3} \delta K
- \f {i}{6} \left[\cd^\al, \cd^{\da}\right] \cv_{\al \da}
- \f {4}{3} \cv^a (A_a - \f {i}{2} G_a)  \nonumber \\ && - \f
{2}{3} (\cd^\al \cd_\al - 8 R^\dagger) \bar{U}
   - \f {2}{3}(\cd_{\da} \cd^{\da} - 8R)U, \label{D.90} \\
&& \nonumber \\ H - \bar{H} & = & 2 \cd^a \cv_a + \cv^a \cd_a K
\nonumber \\ && - 2 (\cd^\al \cd_\al - 8 R^\dagger) \bar{U}
   + 2(\cd_{\da} \cd^{\da} - 8R)U. \label{D.91}
\end{eqnarray}
 In conclusion, the combinations $H- \bar{H}$
and $H + \bar{H} +\f {1}{3} \delta K$ of variational superfields
are given in terms of unconstrained superfields $U, \bar{U}$ and
$\cv_a$.

\indent

\item{\bf Yang-Mills sector}

We parametrize the variation of the Yang-Mills gauge potential in
$U(1)$ superspace such that
\begin{equation}
\delta A^{(r)} \, = \, \Gamma^{(r)} \,=\,E^A {\Gamma_A}^{(r)}.
\label{D.92} \end{equation}
 The Yang-Mills field strength,
$\cf^{(r)} \,=\,\f {1}{2} E^A E^B {\cf_{BA}}^{(r)}$, defined as
\begin{equation}
\cf^{(r)} \,=\,d A^{(r)} + \f {i}{2} A^{(p)} A^{(q)}
{f_{(p)(q)}}^{(r)},
\label{D.93}
\end{equation}
 changes under these variations as
\begin{equation}
\delta \cf^{(r)} \,=\,d\Gamma^{(r)} + i\Gamma^{(p)} A^{(q)}
{f_{(p)(q)}}^{(r)} \,=\,{\cd}\Gamma^{(r)}.
 \label{D.94}
\end{equation}
The variational equations of its coefficients are
\begin{eqnarray}
 \delta {\cf_{BA}}^{(r)} & = & \cd_B
{\Gamma_A}^{(r)} - (-)^{ab} \cd_A {\Gamma_B}^{(r)} + {T_{BA}}^F
{\Gamma_F}^{(r)} \nonumber \\ && - {H_B}^F {\cf_{FA}}^{(r)} +
(-)^{ab} {H_A}^F {\cf_{FB}}^{(r)} . \label{D.96} \end{eqnarray}
 As
in the gravitational case, we are only interested in infinitesimal
variations modulo ordinary gauge variations $\eps^{(r)}$, given as
\begin{eqnarray}
 \unddt \, \Gamma^{(r)} &=& d \eps^{(r)}
+ i \eps^{(p)} A^{(q)} {f_{(p)(q)}}^{(r)} = {\cd}{\eps}^{(r)},
\label{D.97} \\
\unddt \, \delta \cf^{(r)} &=& i \eps^{(p)}
\cf^{(q)}{f_{(p)(q)}}^{(r)}.
 \label{D.98}
\end{eqnarray}
 The solution of the variational equations of the constraints
\begin{equation}
\delta {\cf_{\bt \al}}^{(r)} =0 , \cem \delta \cf^{\db \da (r)} \
= \ 0, \label{D.99} \end{equation}
 is expressed in
terms of an unconstrained superfield $\Sigma^{(r)}$ such that
\begin{eqnarray}
 {\Gamma_\al}^{(r)} &=& + \cd_\al \Sigma^{(r)}
+ \cv^f {\cf_{f \al}}^{(r)}, \label{D.100} \\ \Gamma^{\da (r)} &=&
-\cd^{\da} \Sigma^{(r)} - \cv^f {\cf_f}^{\da (r)}. \label{D.101}
\end{eqnarray}
 The constraint
\begin{equation}
\delta {\cf_\bt}^{\da (r)} \,=\,0,
 \label{D.102}
\end{equation}
serves to express the vector component ${\Gamma_a}^{(r)}$ in terms
of $\Sigma^{(r)}$ as well. It is convenient to define
 \begin{eqnarray}
  {\Lambda_a}^{(r)} & = &
{\Gamma_a}^{(r)} - \cv^b {\cf_{ba}}^{(r)} - \cd_a\Sigma^{(r)},
\label{D.103} \\
{\bar{\Lambda}_{a}}{}^{(r)} & = &
{\Gamma_a}^{(r)} + \cv^b {\cf_{ba}}^{(r)} + \cd_a \Sigma^{(r)}.
\label{D.104}
\end{eqnarray}
 Accordingly, the solution of
(\ref{D.102}) can be written in two ways,
 \begin{eqnarray}
{\Lambda_{\al \da}}^{(r)} & = & i \cd_\al {\Gamma_{\da}}^{(r)} + i
(\cd_{\da} \cv^b) {\cf_{b \al}}^{(r)}, \label{D.105} \\
{\bar{\Lambda}_{\al \da}}{}^{(r)} & = & i \cd_{\da}
{\Gamma_{\al}}^{(r)} - i (\cd_{\al} \cv^b) {\cf_{b \da}}^{(r)}.
\label{D.106} \end{eqnarray}
 The variations of the
covariant Yang-Mills superfields ${\cw_\al}^{(r)}$, $\cw^{\da (r)}$
are obtained from
$\delta \cf^\db{}_a{}^{(r)}$,
$\delta \cf_{\bt a}{}^{(r)}$ to be
\begin{eqnarray}
 \delta
{\cw_\al}^{(r)} & = & - \cv^b \cd_b {\cw_\al}^{(r)} - (\bar{H} +
\f {1}{2} H) {\cw_\al}^{(r)} + i \Sigma^{(s)} {\cw_\al}^{(t)}
{f_{(t)(s)}}^{(r)} \nonumber \\ && + \f {i}{2} (\cd^{\dv} \cd_\al
\cv_{\vp \dv}) \cw^{\vp (r)} -\f {i}{2} \cv_{\al \dv} G^{\vp \dv}
{\cw_{\vp}}^{(r)} \nonumber
\\ && - \f {1}{4} (\cd_{\dv} \cd^{\dv} - 8R)
{\Gamma_\al}^{(r)}, \label{D.107} \\ && \nonumber \\ \delta
{\cw_{\da}}^{(r)} & = & + \cv^b \cd_b {\cw_\da}^{(r)} - (H + \f
{1}{2} \bar{H}){\cw_{\da}}^{(r)} - i \Sigma^{(s)} \cw_{\da}^{(t)}
{f_{(t)(s)}}^{(r)} \nonumber \\ && - {\f {i}{2}} (\cd^{\vp}
\cd_{\da} \cv_{\vp \dv}) \cw^{\dv (r)} -{\f {i}{2}} \cv_{\vp \da}
G^{\vp \dv} {\cw_{\dv}}^{(r)} \nonumber \\ && +{ \f {1}{4}}
(\cd^{\vp} \cd_{\vp} - 8R^\dagger) {\Gamma_{\da}}^{(r)}.
\label{D.108}
\end{eqnarray}

\end{itemize}

\subsec{Superspace Densities
\label{appD2bis}}

As a first application of the previous discussion, we consider the
superfield action
\begin{equation} \int_\ast \! \! E. \label{D.128} \end{equation}
Recalling that the asterisk denotes integration over space-time and
anticommuting coordinates, this superspace integral might be called the
volume of superspace. It serves to generalize the D-term construction
of invariant actions to local supersymmetry. Taking into account
(\ref{D.40}), the variation of the superdeterminant
\begin{equation}
\dt E \, = \, E H_A{}^A (-)^a,
\end{equation}
gives rise to
\begin{equation}
 \delta \! \! \int_\ast \! \! E \, = \, \int_\ast \! \! E (H + \bar{H}),
\label{D.130}
\end{equation}
with superspace surface terms neglected after integration by
parts. Observe that in generic $U(1)$ superspace, the superfield
$H + \bar H$, as given in (\ref{D.90}), contains $\dt K$, the
variation of the $U(1)$ pre-potential as an independent
unconstrained variable. As a consequence, the superfield
equations of motion would imply the volume of superspace to
vanish. Therefore, the action (\ref{D.128}) is not very useful in
$U(1)$ superspace. However, when specified to pure Wess-Zumino
superspace (resp. \ka superspace), $\dt K$ will be subject to
constraints and the same action will provide the pure
supergravity (resp. supergravity/matter) action.

 Another
useful concept in constructing superfield actions is the chiral
density. It serves to generalize the F-term construction of
invariant actions to the case of local supersymmetry. As a
starting point consider the superspace action
\begin{equation}
\int_\ast \! \frac{E}{R} \, \cs ,
\label{D.129}
\end{equation}
with $\cs $ some generic chiral superfield of weight $w(\cs) =2 \ $ to
ensure invariance under $U(1)$ transformations. Using the relation
\begin{equation}
\cs \, = \, \left( \cd_{\! \da} \cd^\da - 8 R \right) \, \Sigma (\cs),
\end{equation}
expressing the chiral superfield in terms of the unconstrained superfield
$\Sigma (\cs)$, together with integration by parts yields
\begin{equation}
\int_\ast \! \frac{E}{R} \, \cs \, = \,
           -8 \! \! \int_\ast \! \! E \, \Sigma (\cs).
\label{D.129a}
\end{equation}
This shows that integrating the chiral superfield $\cs$ using the chiral
density is the same as integrating its pre-potential $\Sigma (\cs)$ using
the complete volume density. Note that adding a linear superfield to
$\Sigma (\cs)$ does not change $\cs$. This is coherent with relation
(\ref{D.129a}), because the superspace integral of a linear superfield
vanishes (this, in turn, is due to the fact that a linear superfield can be
expressed in terms of spinor derivatives of unconstrained pre-potentials).

In spite of the equivalence established in (\ref{D.129a}), it is very
often quite useful to work with the chiral density expression
(\ref{D.129}), and its complex conjugate
\begin{equation}
\int_\ast \! \frac{E}{\rd} \, \bar \cs ,
\label{D.129b}
\end{equation}
with chiral weight $w(\bar \cs)=-2$ assigned to $\bar \cs$. Taking into
account (\ref{D.40}), as well as (\ref{D.79}) and (\ref{D.80}) we find
\begin{eqnarray}
\delta \! \! \int_\ast \! \frac{E}{R} \cs &=&
\int_\ast \! \frac{E}{R}
\left( \,
(\delta \cs + \cv^a \cd_{\! a} \cs)
+ (H+ 2\bar{H}  - i \cv^a G_{\! a}) \, \cs \,
\right),
\label{D.131} \\[2mm]
\delta \! \! \int_\ast \! \frac{E}{R^\dagger} \bar \cs  &=&
\int_\ast \! \frac{E}{R^\dagger} \left( \,
(\delta \bar \cs - \cv^a \cd_{\! a} \bar \cs)
+ (\bar{H}+2H - i \cv^a G_{\! a}) \, \bar \cs \,
\right),
\label{D.132}
\end{eqnarray}
with $H$ and $\bar H$ determined in (\ref{D.90}), (\ref{D.91}).

\subsec{Variational Equations in \ka Superspace
\label{appD3}}

So far, in this appendix, we worked in the framework of $U(1)$
superspace. Supergravity/matter coupling is obtained in suitably
specializing the $U(1)$ sector. We will present here the general case,
where chiral superfields parametrize a \ka manifold with gauged isometries.
The relevant geometric framework is isometric \ka superspace as defined in
appendix \ref{appC1a}.

After a discussion of the variational equations for chiral
superfields and a summary of the properties of covariant isometric
superspace derivatives, we will solve the variational equations
pertaining to isometric superspace, thus identifying the fundamental
variables relevant for the derivation of superfield equations of motions for
the complete supergravity/matter/Yang-Mills system.

\begin{itemize}

\item{\bf Chirality conditions}

The variational equations corresponding to the chirality conditions can be
treated along the same lines as the constraint equations discussed earlier.
We will first describe in some detail the procedure for the superfield
$\phi^k$ and give the results for $\phib^{\, \bar{k}}$ afterwards.

In (\ref{C.63}), the covariant derivative
${\cd}{\phi}^k = E^A{\cd}_A{\phi}^k$ has been defined as
\begin{equation}
{\cd}{\phi}^k \, = \, \left( d+i{\ca}^{(r)} V_{(r)} \right){\phi}^k.
\end{equation}
Its variation in terms of $\dt {\phi}^k$ and
$\dt {\ca}^{(r)} = \Gamma^{(r)}$ is given as
\begin{equation}
\dt \, {\cd}{\phi}^k \, = \, {\cd} \, \dt {\phi}^k
+ i\Gamma^{(r)} V_{(r)}{}^k (\phi),
\end{equation}
with the definition
\begin{equation}
{\cd} \, \dt {\phi}^k \, = \, d \, \dt {\phi}^k
+ i{\ca}^{(r)}{{\prt}{V_{(r)}}^k \over {\prt}{\phi}^l} \, \dt {\phi}^l.
\end{equation}
Using
\begin{equation}
\dt \, {\cd}{\phi}^k \, = \, E^A \dt \, {\cd}_{\! A} {\phi}^k
                          + E^A H_A{}^B {\cd}_B {\phi}^k,
\end{equation}
the variational equation for ${\cd}_{\! A} {\phi}^k$ becomes
\begin{equation}
 \dt \, {\cd}_{\! A} {\phi}^k \, = \,
{\cd}_{\! A} \, \dt {\phi}^k
+ i\, \Gamma_{\! A}{}^{(r)} V_{(r)}{}^k (\phi)
- H_A{}^B {\cd}_{\! B} {\phi}^k.
\end{equation}
We are now in a position to study the consequences of the chirality
condition $\cd^{\da} \phi^k = 0$, \ie to determine the variations
$\delta \phi^k$ of chirally constrained matter
superfields in terms of unconstrained variational superfields.
This is achieved in taking the $\da$ component of the previous
equation,
\begin{equation}
 \dt \, {\cd}^\da {\phi}^k \, = \, 0 \, = \,
{\cd}^\da \, \dt {\phi}^k
+ i\, \Gamma^{\da (r)} V_{(r)}{}^k (\phi)
- H^\da{}^B {\cd}_{\! B} {\phi}^k,
\end{equation}
and making use of (\ref{D.101}), \ie
\begin{equation}
\Gamma^{\da (r)} \, = \,
-\cd^{\da} \Sigma^{(r)} - \cv^b {\cf_b}^{\da (r)},
\end{equation}
in the second term. Taking into account (\ref{D.33}) and (\ref{D.34})
allows to write the third term in the form
\begin{equation}
- H^\da{}^B {\cd}_{\! B} {\phi}^k \, = \,
-{\cd}^\da \! \left( \cv^b {\cd}_{\! b} {\phi}^k \right)
+ \cv^b \left[ {\cd}^\da, {\cd}_{\! b} \right] {\phi}^k
+ \cv^b \, T^\da{}_b{}^\vp \, \cd_\vp {\phi}^k.
\end{equation}
Finally, substituting (\ref{C.72}) for the commutator, gives rise to the
chirality condition
\begin{equation}
\cd^{\da} \eta^k \, = \, 0,
\label{D.140}
\end{equation}
with
\begin{equation}
\eta^k \, = \, \delta \phi^k + \cv^b \cd_b \phi^k
-i \Sigma^{(r)} {V_{(r)}}^k.
\label{D.142}
\end{equation}
The corresponding expressions for $\delta \phib^{\bar{k}}$ are obtained in
complete analogy. There, the chirality condition
\begin{equation}
 \cd_\al \etab^{\bar{k}} \, = \, 0,
\label{D.139}
\end{equation}
is obtained for the combination
\begin{equation}
 \etab^{\bar{k}} \, = \, \delta \phib^{\bar{k}}
- \cv^b \cd_b \phib^{\bar{k}}
+i \Sigma^{(r)} {{\bV}_{(r)}}{}^{{\bk}}.
\label{D.141}
\end{equation}
The chirality conditions are solved in terms of
unconstrained superfields ${\vpb}^{\bk}$ and $\vp^k$,
\ie
\begin{eqnarray}
 \etab^{\bk} &=& (\cd^\al \cd_\al - 8R^\dagger) {\vpb}^{\bk},
\label{D.145} \\
\eta^k &=& (\cd_{\da} \cd^{\da} - 8R) \vp^k.
\label{D.146}
\end{eqnarray}

\indent

\item{\bf Covariant superspace derivatives and gauged isometries}

Let ${\bf U}^k$ be some generic $p$-form in superspace, undergoing
non-linear transformations
\begin{equation}
\delta {\bf U}^k \,=\,- \al^{(r)} \frac{\prt {V_{(r)}}^k}{\prt
\phi^l} {\bf U}^l.
\label{D.147}
\end{equation}
For simplicity, we suppose that ${\bf U}^k$ is inert under Lorentz and \ka
transformations.  The exterior
covariant derivative of this $p$-form is
\begin{equation}
{\cd} {\bf U}^k \,=\,d {\bf U}^k + (-)^p i \ca^{(r)} \frac{\prt
{V_{(r)}}^k}{\prt \phi^l} {\bf U}^l + (-)^p {\Gamma^k}_{lm} {\cd}
\phi^m {\bf U}^l,
\label{D.148}
\end{equation}
with ${\Gamma^k}_{lm}$ defined as in (\ref{levicivicom}).
 In verifying the covariant
transformation law of (\ref{D.148}) it is convenient to use
identities such as
\begin{eqnarray}
 (V_{(r)} + \bV_{(r)}) g_{k \bk} + \frac{\prt
{V_{(r)}}^m}{\prt \phi^k} g_{m \bk} + \frac{\prt
{\bV_{(r)}}{}^{\bl}}{\prt \phib^{\bk}} g_{k \bl} \ =\ 0,
\label{D.152} \\ (V_{(r)} + \bV_{(r)}) g^{l \bl} + \frac{\prt
{V_{(r)}}^l}{\prt \phi^k} g^{k \bl} - \frac{\prt
{\bV_{(r)}}{}^{\bl}}{\prt \phib^{\bk}} g^{l \bk} \,=\,0,
\label{D.153} \end{eqnarray}
 and
\begin{equation}
(V_{(r)} + \bV_{(r)} ) {\Gamma^l}_{mn} \,=\,\frac{\prt
{V_{(r)}}^l}{\prt \phi^k}{\Gamma^k}_{mn}
 - \frac{\prt {V_{(r)}}^k}{\prt \phi^m} {\Gamma^l}_{kn}
 - \frac{\prt {V_{(r)}}^k}{\prt \phi^n} {\Gamma^l}_{mk}
 - \frac{\prt^2 {V_{(r)}}^l}{\prt \phi^m \prt\phi^n}.
\label{D.154}
\end{equation}
In the case
$p=0$,  ${\bf U}^k$ is a superfield and its covariant derivative
is given as
\begin{equation}
{\cd} {\bf U}^k \,=\,E^A \cd_A {\bf U}^k.
\label{D.155}
\end{equation}
The graded commutator of two such covariant
derivatives is obtained by taking the covariant
exterior derivative of (\ref{D.155}), using (\ref{D.148}) for $p=1$.
The
result is
\begin{equation}
{\cd}{\cd} {\bf U}^k \,=\,i \cf^{(r)} \left( \frac{\prt
{V_{(r)}}^k}{\prt \phi^l} {\bf U}^l +{V_{(r)}}^m {\Gamma^k}_{lm}
{\bf U}^l \right) -g^{k \bl} R_{m \bl l \bk} {\cd} \phib^{\bk}
{\cd} \phi^m {\bf U}^l.
\label{D.156}
\end{equation}
 Decomposing
\begin{equation}
{\cd}{\cd} {\bf U}^k \,=\,E^A E^B \left(\cd_B \cd_A {\bf U}^k + \f
{1}{2} \ {T_{BA}}^C \cd_C {\bf U}^k \right), \label{D.157}
 \end{equation}
we find
\begin{eqnarray}
 (\cd_B , \cd_A) {\bf U}^k &=&-
{T_{BA}}^C \ \cd_C {\bf U}^k + i {\cf_{BA}}^{(r)} \left(
\frac{\prt {V_{(r)}}^k}{\prt \phi^l} {\bf U}^l + {V_{(r)}}^m
{\Gamma^k}_{lm} {\bf U}^l \right) \nonumber \\
 && + g^{k \bl} R_{m \bl l \bk} \ {\bf U}^l
\left( \cd_B \phib^{\bk} \cd_A \phi^m - (-)^{ab}
\cd_A \ \phib^{\bk} \ \cd_B \phi^m \right).
\label{D.158}
\end{eqnarray}
The spinor derivative
$\cd_\al \phi^k$ of a chiral superfield $\phi^k$ transforms in the
same manner as ${\bf U}^k$ under gauged isometries but picks up
additional contributions from Lorentz and K\"ahler
transformations. Taking into account these modifications, we have
\begin{equation}
F^k \,=\,- \f {1}{4} \cd^\al \cd_\al \phi^k, \label{D.159}
\end{equation}
 and
\begin{equation}
\cd_\al F^k \,=\,-2 R^\dagger \cd_\al \phi^k. \label{D.160}
\end{equation}

\indent

\item{\bf Variations in isometric \ka superspace}

As we have shown in appendix \ref{appC1a}, gauged isometries can be
included in the geometric description in replacing the generic $U(1)$
connection by the composite connection
\begin{equation}
\frak{A} \,=\, \f{1}{4}  \widetilde{\Delta}
+ \f {i}{8} E^a \left(12 \, G_a
+ {\sib}_a^{\da \al} g_{k \bk} \cd_\al \phi^k \cd_{\da} \phib^{\bk}
 \right),
\label{D.166}
\end{equation}
 with
\begin{equation}
\widetilde{\Delta} \,=\, K_k d \phi^k - K_{\bk} d \phib^{\bk}
                         + 2 \, \ca^{(r)} G_{(r)}.
\label{D.167}
\end{equation}
The resulting geometric structure in superspace is called isometric \ka
superspace. As a consequence of the particular form of the composite
connection, the variational equations in the $U(1)$ sector will furnish
additional information.

Recall that the field strength $\frak F = d \frak A$ satisfies the same
constraints as that of the generic $U(1)$ connection. For this reason the
generic $U(1)$ pre-potential $K$ will be replaced by a field dependent
quantity. In standard \ka superspace, this is just the superfield \ka
potential. In the presence of gauged isometries, the dependence on the
matter sector and the Yang-Mills sector involved in the gauging of
isometries will be quite intricate.

Fortunately enough, in the investigation of the variational equations, the
know\-ledge of the explicit form of the composite pre-potential can be
circumvented in considering directly the variations in terms of $\frak A$.

The relevant object in this analysis is the variation of
$\widetilde{\Delta}$, which may be written as
\begin{equation}
\delta \widetilde{\Delta} \, = \,
d \left(K_k \delta \phi^k
- K_{\bk} \delta \phib^{\bk} \right)
+ 2 \, g_{k \bk} {\cd} \phi^k \delta \phib^{\bk}
- 2 \, g_{k \bk} {\cd} \phib^{\bk} \delta \phi^k
+ 2 \, \Gamma^{(r)} G_{(r)}.
\label{D.168}
\end{equation}
We parametrize
\begin{equation}
\delta \widetilde{\Delta} \, = \, E^A {B}_A,
\label{D.169}
\end{equation}
and consider the spinor coefficient
\begin{equation}
{B}_\al \, = \,
{E_\al}^M \prt_M
\left(K_k \delta \phi^k
- K_{\bk} \delta\phib^{\bk} \right)
+ 2 \, g_{k \bk} \cd_\al \phi^k \delta \phib^{\bk}
+ 2 \, \Gamma^{(r)}_\al G_{(r)}.
\label{D.170}
\end{equation}
Taking into
account the explicit expression for $\Gamma_\al^{(r)}$, -cf.
(\ref{D.100}), we obtain
\begin{eqnarray}
{B}_\al & = & {E_\al}^M \prt_M
\left( K_k \delta \phi^k - K_{\bk} \delta \phib^{\bk}
+ 2 \, \Sigma^{(r)} G_{(r)} \right)
\nonumber \\ &&
+ 2 \, g_{k \bk} \cd_\al \phi^k \etab^{\bk}
+ 2 \, \cv^b \left( {\cf_{b \al}}^{(r)} G_{(r)}
+ g_{k \bk} \cd_b \phib^{\bk} \cd_\al \phi^k \right).
\label{D.171}
\end{eqnarray}
Remember that our aim is to determine
$\dt \frak A = \om$, -cf. (\ref{D.3}),
with the definition $\Sigma_\al \,=\, \omega_\al - \cv^b {\frak F}_{b \al}$,
-cf. (\ref{D.50}).
To this end we have to add the variation of the second term in
(\ref{D.166}) to arrive at
\begin{eqnarray}
\Sigma_\al & = &
\f{1}{4} {E_\al}^M \prt_M \left(
K_k \delta \phi^k - K_{\bk} \delta \phib^{\bk}
+ 2 \Sigma^{(r)} G_{(r)} \right.
\nonumber \\ && \left.
+ 6i \cv^b G_b
+ \f {i}{2} \cv^b {\sib_b}^{\ \da \al} g_{k
\bk} \cd_\al\phi^k \cd_{\da} \phib^{\bk} \right)
+ \f {1}{2} g_{k \bk} \cd_\al \phi^k \etab^{\bk}.
\label{D.172}
\end{eqnarray}
An explicit
calculation shows that the last term in this equation can be
written as a total spinor derivative as well, namely
\begin{equation}
\f {1}{2} g_{k \bk} \cd_\al \phi^k \etab^{\bk} \,=\,{E_\al}^M
\prt_M \left( 2 \vpb^{\bk} g_{k \bk} F^k - g_{k \bk}\cd^\vp \phi^k
\cd_{\vp} \vpb^{\bk} \right). \label{D.174}
\end{equation}
 This leads then to
\begin{eqnarray}
\Sigma_\al & = & \f {1}{4} {E_\al}^M \prt_M
\left(
K_k \delta \phi^k - K_{\bk} \delta \phib^{\bk}
+ 2 \Sigma^{(r)} G_{(r)}
\right. \nonumber \\ &&
+ 6i \cv^b G_b
+ \f {i}{2} \cv^b {\sib_b}^{\ \da \al} g_{k \bk} \cd_\al \phi^k \cd_{\da}
\phib^{\bk}
\nonumber \\ && \left.
+ 8 \vpb^{\bk} g_{k \bk} F^k -
g_{k \bk} \cd^\vp \phi^k \cd_\vp \vpb^{\bk} \right).
\label{D.175}
\end{eqnarray}
This relation summarizes the consequences of the variational
equations in the $U(1)$ sector which arise from the fact that
$\frak A$ is a composite connection, dependent on the \ka and
Yang-Mills sector. On the other hand, in the analysis of the
consequences of the torsion constraints, -cf. (\ref{D.59}), the superfield
$\Sigma_\al$ had been given in terms of the, up to this point,
unconstrained superfields $H$ and $\bar H$, \ie
\begin{equation}
\Sigma_\al \, = \, - {E_\al}^M \prt_M
\left(
H + \f {1}{2} \bar{H}
+ \f {i}{4}\cd^\vp \cd^{\dv} \cv_{\vp \dv}
- \f {i}{2} {\cv}^a G_a
\right)
\label{D.164}
\end{equation}
Comparing the expressions in
(\ref{D.164}) and (\ref{D.175}) leads to a chirality condition
which is solved in terms of an unconstrained variational
superfield $\cz$ such that
\begin{eqnarray}
H + \f {1}{2} \bar{H} &=&
- \f {i}{4} \cd^\vp \cd^{\dv} \cv_{\vp \dv}
- \f {1}{4} \left( K_k \delta \phi^k
- K_{\bk} \delta \phib^{\bk}
+ 2 \Sigma^{(r)} G_{(r)} \right)
\nonumber \\ &&
- i \cv^b G_b
- \f {i}{8}
\cv^b {\sib_b}^{\ \da \al} g_{k \bk} \cd_\al \phi^k \cd_{\da} \phib^{\bk}
\nonumber \\ &&
- 2 \, \vpb^{\bk} g_{k \bk} F^k
+ g_{k \bk} \cd^\vp \phi^k \cd_\vp \vpb^{\bk}
+ (\cd^\vp \cd_\vp -
8R^\dagger)\cz.
\label{D.176}
\end{eqnarray}
 Performing the corresponding analysis for the
complex conjugate sector leads to
\begin{eqnarray}
\bar{H} + \f{1}{2} H &=&
+ \f {i}{4} \cd^{\dv} \cd^\vp \cv_{\vp \dv}
+ \f {1}{4} \left( K_k \delta \phi^k
- K_{\bk} \delta \phib^{\bk}
- 2 \Sigma^{(r)} G_{(r)} \right)
\nonumber \\ &&
- i \cv^b G_b
- \f {i}{8}
\cv^b {\sib_b}^{\ \da \al} g_{k \bk} \cd_\al \phi^k \cd_{\da} \phib^{\bk}
\nonumber \\ &&
+ 2 \vp^k g_{k \bk} {\bF}^{\bk}
- g_{k \bk} \cd_{\dv} \phib^{\bk} \cd^{\dv} \vp^k
+ (\cd_{\dv}
\cd^{\dv} - 8R) \cz^\dagger.
\label{D.177}
\end{eqnarray}
 This completes our discussion of the variational
equations of superspace constraints. The basic variational
superfields are $\cv_a$ and $\cz, \cz^\dagger$ for supergravity,
$\vp^k$ and $\vpb^{\bk}$ for chiral matter superfields and
$\Sigma^{(r)}$ for the Yang-Mills sector. Recall that the variations
$\dt \phi^k$, $\dt \phib^\bk$ are expressed in terms of $\cv_a$, $\vp^k$
and $\vpb^{\bk}$ according to (\ref{D.142}), (\ref{D.141}) and
(\ref{D.145}), (\ref{D.146}). Observe that in the standard Yang-Mills case,
\ie no gauged isometries, the results (\ref{D.176}), (\ref{D.177}) should
reproduce those derived from (\ref{D.90}), (\ref{D.91}) with $\dt K$
evaluated directly as a function of chiral superfields.

\end{itemize}

\subsec{Variation of the Action Functionals \label{appD4}}

We are now in a position to derive the superspace equations of motion for
the complete supergravity/matter/Yang-Mills system. The full action
\begin{equation}
\ca \,=\,\ASM + \AYM + \ASPOT, \label{D.194}
\end{equation}
consists of three separately supersymmetric and \ka invariant pieces. It
remains to perform the superfield variations and write down the equations
of motion.

\begin{itemize}

\item{\bf Variation of $\ASM \ $}

The kinetic action for the supergravity+matter system is given as
\begin{equation}
\ASM \,=\,-3 \! \int_\ast \! \! E. \label{D.178} \end{equation}
 This is the
form of the prototype action (\ref{D.128}) discussed earlier. In its
variation, cf. (\ref{D.130}),
\begin{equation}
\delta \ASM \,=\,-3 \int_\ast \! \! E (H + \bar{H}), \label{D.179}
\end{equation}
$H + \bar{H}$ is given as the sum of (\ref{D.176}) and (\ref{D.177}), \ie
\begin{eqnarray}
\f {3}{2} (H + \bar{H}) & = &
\f {i}{4} \left[\cd^{\dv}, \cd^\vp \right] \cv_{\vp \dv}
- \Sigma^{(r)} G_{(r)}
- 2i \cv^b G_b
- \f {i}{4} {\sib_b}^{\ \da \al}
g_{k \bk} \cd_{\al} \phi^k \cd_{\da} \phib^{\bk}
\nonumber \\[2mm] &&
- 2 \vpb^{\bk} g_{k \bk} {F}^k + g_{k \bk} \cd^\vp
\phi^k \cd_{\vp} \vpb^{\bk}
+ 2 \vp^k g_{k \bk}
{\bF}^{\bk} - g_{k \bk} \cd_{\dv} \phib^{\bk} \cd^{\dv} \vp^k
\nonumber \\[2.5mm] &&
+ (\cd^\vp \cd_\vp - 8R^\dagger)\cz + (\cd_{\dv}
\cd^{\dv} -8R) \cz^\dagger. \label{D.180}
\end{eqnarray}
Substituting, integrating by parts and neglecting superspace surface terms
gives rise to
\begin{eqnarray}
\delta \ASM & = &
 4i \! \int_\ast \! \! E \, \cv^b (G_b + \f {1}{8}
{\sib_b}^{\ \da \al} g_{k \bk} \cd_\al \phi^k \cd_{\da}
\phib^{\bk})
\nonumber \\ &&
+ 16 \! \int_\ast \! \! E \, \cz R^\dagger
+ 16 \! \int_\ast \! \! E \, \cz^\dagger R
\nonumber \\ &&
+ 4 \! \int_\ast \! \! E \, {F}^k g_{k \bk} \vpb^{\bk}
- 4 \! \int_\ast \! \! E \, \vp^k g_{k \bk} {\bF}^{\bk}
\nonumber \\ &&
+ 2 \! \int_\ast \! \! E \, \Sigma^{(r)} G_{(r)}.
\label{D.181}
\end{eqnarray}

\indent

\item{\bf Variation of $\AYM$}

The Yang-Mills action of (\ref{GRA.241}) is obtained from the
prototype action (\ref{D.129}) in identifying $\cs$ with
\begin{equation}
\cs_{\mathrm{Yang-Mills}} \,=\,
\f {1}{8} f_{(r)(s)}(\phi)\cw^{(r) \al}\cw^{(s)}_\al,
\label{D.182}
\end{equation}
and accordingly for $\bar \cs$.
The function $f_{(r)(s)}(\phi)$ of the chiral
matter superfields is required to satisfy
\begin{equation}
V_{(p)} f_{(r)(s)} (\phi) \,=\,
{f_{(p)(r)}}^{(q)} f_{(q)(s)} (\phi)
+ {f_{(p)(s)}}^{(q)} f_{(r)(q)} (\phi),
\label{D.184}
\end{equation}
assuring  that $\cs_{\mathrm{Yang-Mills}}$ is indeed a chiral
superfield of weight $w(\cs_{\mathrm{Yang-Mills}}) =2$. Then, taking into
account the variations of $\cw^{(r)}_\al$ and $\phi^k$ as determined in
this appendix, working out $\delta \cs_{\mathrm{Yang-Mills}}$, substituting
in the general variation given in (\ref{D.131}) and neglecting
superspace surface terms yields, as an intermediate result
\begin{equation}
\delta \! \int_\ast \! \frac{E}{R} \, \cs_{\mathrm{Yang-Mills}} \,=\,
\f {1}{8} \int_\ast \! \frac{E}{R} \, \eta^k \,
 \frac{\prt f_{(r)(s)}}{\prt \phi^k} \, \cw^{(r)\al} \cw^{(s)}_\al
+\f {1}{2} \int_\ast \! \! E \, f_{(r)(s)} \cw^{(r) \al} \,
\Gamma^{(s)}_\al.
\label{D.185}
\end{equation}
 Using furthermore the
explicit form of $\eta^k$ and ${\Gamma_\al}^{(r)}$ gives rise to
\begin{eqnarray}
\delta \! \int_\ast \! \frac{E}{R} \, \cs_{\mathrm{Yang-Mills}} &=&
- \f {1}{2} \int_\ast \! \! E
\, \Sigma^{(r)} \left(
f_{(r)(s)}  \cd^\al \cw^{(s)}_\al
+ \frac{\prt f_{(r)(s)}}{\prt \phi^k} \cd_\al \phi^k \cw^{(s)\al}
\right) \nonumber \\[2mm] &&
 - \f{i}{2} \int_\ast \! \! E \, \cv_b \,
 {\cw^{(r)}}^\al\,{\si^b}_{ \al \da}\, {{\cw}^{(s)}}^{\da}\,f_{(r)(s)}
(\phi)
\nonumber \\[2mm] &&
- \int_\ast \! \! E \, \vp^k \,
\frac{\prt f_{(r)(s)}}{\prt \phi^k} \, \cw^{(r)\al} \cw^{(s)}_\al.
\label{D.186}
\end{eqnarray}
Observe that in the variation of the full Yang-Mills action
\begin{equation}
\AYM \,=\, \rea \int_\ast \! \frac{E}{R} \, \cs_{\mathrm{Yang-Mills}},
\label{D.183}
\end{equation}
we have to take into account the complex conjugate term as well.

\indent

\item{\bf Variation of $\ASPOT$}

The action for the superpotential, -cf. (\ref{GRA.242}), is a special
case of prototype action as well, in this case we identify $\cs$ with
\begin{equation}
\cs_{\mathrm{superpotential}} \,=\,\f {1}{2} e^{K(\phi, \phib)/2}
W(\phi).
\label{D.187}
\end{equation}
In the presence of gauged isometries the condition
\begin{equation}
V_{(r)} W + F_{(r)} W \,=\, 0,
\label{D.189}
\end{equation}
ensures that $\cs_{\mathrm{superpotential}}$ is indeed
a chiral superfield of weight $w(\cs_{\mathrm{Superpotential}}) = 2$.
An explicit calculation shows that the variation of the superpotential term
is given as
\begin{equation}
\delta \left( \f {1}{2} \int_\ast \! \frac{E}{R} e^{K/2} W \right)
\, = \,
-8 \! \int_\ast \! \! E \, \cz^\dagger \, e^{K/2} W
-4 \! \int_\ast \! \! E \, \vp^k \, e^{K/2} \left( W_k + K_k W \right) .
\label{D.192}
\end{equation}
For the complete superpotential action
\begin{equation}
\ASPOT \,=\,\rea \int_\ast \! \frac{E}{R} \cs_{\mathrm{superpotential}},
\label{D.188}
\end{equation}
we have to take into account the complex conjugate term as well.
\indent

\item{\bf The superfield equations of motion}

In order to find the superfield equations of motion of the
complete action
\begin{equation}
\ca \,=\,\ASM + \AYM + \ASPOT, 
\end{equation}
we simply identify the factors of the various
variational superfields.
From the coefficient of $\cz^\dagger$ we obtain
\begin{equation}
R - \f {1}{2} e^{K/2} W \,=\,0 . \label{D.195}
\end{equation}
 The superfield equation corresponding to $\cv^b$
reads
\begin{equation}
G_b + \f {1}{8} {\sib_b}^{\ \da \al} g_{k \bk} \cd_\al \phi^k
\cd_{\da} \phib^{\bk} - \f {1}{8} {\sib_b}^{\ \da \al} (f+
\bar{f})_{(r)(s)} \cw^{(r)}_\al \cw^{(s)}_{\da} \,=\,0.
\label{D.196} \end{equation}
 Matter and Yang-Mills variations, respectively, give
rise to the equations of motion
\begin{equation}
4 \, g_{k \bk} {\bF}^{\bk}
+ \frac{\prt f_{(r)(s)}}{\prt \phi^k}
\, \cw^{(r) \al} \cw^{(s)}_\al
+ 4 \, e^{K/2} \left( W_k + K_k W \right) \,=\,0,
\label{D.197}
\end{equation}
 and
\begin{equation}
\f {1}{2} f_{(r)(s)} \cd^\al \cw^{(s)}_\al - \f {1}{2} \frac{\prt
f_{(r)(s)}}{\prt \phi^k} \cd_\al {\phi}^k \cw^{(s)\al}
 - G_{(r)} + h.c. \,=\,0.
\label{D.198} \end{equation}

\end{itemize}

\newpage
\sect{Linear multiplet component field formalism}
\label{appE}
The discussion of the linear superfield formalism in section
\ref{2F} was mainly in terms of superfields. As component field
expressions are notoriously heavy in notations and size we have
deferred their presentation to the present appendix.
We display here the
complete component field action for the particular kinetic potential
$K=K_0 (\phi, \phib) + \al \log L $
of (\ref{3.25})
and discuss shortly the effective anomaly
cancellation mechanism in terms of component fields. This appendix is
designed as a complement to section \ref{2F}.

\subsec{List of Component Fields
\label{appE1}}

Component fields have been defined in various places in the main
text. For the sake of clarity we give here a complete list
of the component fields which will
appear in the Lagrangian below.

\begin{itemize}

\item
In the {\em supergravity sector} we have
\[ e_m{}^a, \cem \psi_m{}^\al, \cem \psib_{m \, \da},
\cem M, \cem \overline M, \cem b_a, \]
the vierbein and the Rarita-Schwinger fields as dynamical variables
and a complex scalar and a real vector as auxiliary fields.

\item
The {\em matter sector} is described in terms of
\[ A^k, \cem \bar A {}^{\bar k}, \cem \chi_\al^k, \cem
\chib^{\bar k \da}, \cem F^k, \cem \bar F {}^{\bar k}, \]
a set of complex scalars and of Majorana spinors as physical fields,
together with
another set of complex scalars as auxiliary fields, indices
$k$ and $\bar k$ referring to the \ka variety.

\item
The{\em Yang-Mills sector} contains
\[ {\aym}_m, \cem \la^\al, \cem \lab_\da, \cem {\bf D}, \]
the gauge potential, the gaugino Majorana spinor and a real scalar
auxiliary field, all Lie algebra valued with matricial generators
${\bf T}_{(r)}$ in a suitable representation.

\item
The {\em linear multiplet} consists of
\[ b_{mn}, \cem L, \cem \Lambda_\al, \cem \bar \Lambda{}^\da, \]
an antisymmetric tensor gauge field, a real scalar and a Majorana
spinor; it does not contain auxiliary fields. We should stress that in
the actual component field Lagrangian given below the Majorana spinor
always appears in the combination $\vp_\al = L^{-1} \La_\al$
and $\vpb^\da = L^{-1} \Lab^\da$.

\end{itemize}
When derived from superspace, the component field Lagrangian contains a
number of compact building blocks, which
arise in a natural manner and gather complicated component field expressions
in a concise way. The same structures appear in the derivation of
supergravity transformations.
Examples of this mechanism are the spin
connection, as defined in (\ref{CPN.15}) and (\ref{CPN.20}),
super-covariant field strength or curvature tensors like the curvature
scalar in (\ref{CPN.26}), the projection ${R_{ab}}^{ab}{\loco}$ in
(\ref{CPN.27}), or the field strength
$T_{cb}{}^\al$, $T_{cb}{}_\da$ in (\ref{CPN.23}), (\ref{CPN.24}). Other
important building blocks which arise naturally are the super-covariant
component field derivatives and the composite \ka connection. This
has already been described in section \ref{CPN}, for the general
supergravity/matter/Yang-Mills system, but is even more dramatic in the
presence of linear multiplets. For the sake of illustration we will
discuss two examples of super-covariant component field derivatives and
the construction of the explicit form of the composite of the \ka
connection in the presence of a linear multiplet (coupled to Chern-Simons
forms).

\subsec{Construction of Supercovariant Derivatives}
\label{appE2}

It might be instructive and useful to review shortly how
the super-covariant component field derivatives are derived
from superspace.  To be definite we shall discuss here as representative
examples the super-covariant derivatives of $A^k$ and $\chi_\al^k$.

Let us begin with $A^k$. The starting point is the superspace covariant
exterior derivative
\begin{equation}
D \phi^k \, = \, d \phi^k - {\ca}^{(r)} \lp {\bf T}_{(r)} \phi \rp^k.
\end{equation}
Using the double bar projection as introduced in section \ref{CPN} one finds
\begin{equation}
D \phi^k \doubar \, = \,
dx^m \lp \prt_m A^k -  i {\aym}_m^{(r)} \lp {\bf T}_{(r)} \phi \rp^k \rp,
\end{equation}
suggesting the definition
\begin{equation}
\cd_m A^k \, = \,
 \prt_m A^k -  i {\aym}_m^{(r)} \lp {\bf T}_{(r)} \phi \rp^k,
\end{equation}
for the component field covariant space-time derivative. On the other hand,
double bar projection in terms of covariant differentials gives
\begin{equation}
D \phi^k \doubar \, = \, dx^m
\lp
e_m{}^a \cd_a \phi^k \loco
+ \frac{1}{\sqrt{2}} \psi_m{}^\al \chi_\al^k
+ \frac{1}{\sqrt{2}} \psib_m{}_\da \chib^k{}^\da
\rp.
\end{equation}
The object $\cd_a \phi^k \loco$ is called the super-covariant space-time
derivative of $A^k$, explicitly given as
\begin{equation}
e_m{}^a \cd_a \phi^k \loco  \, = \, \cd_m A^k
- \frac{1}{\sqrt{2}} \psi_m{}^\al \chi_\al^k
- \frac{1}{\sqrt{2}} \psib_m{}_\da \chib^k{}^\da.
\end{equation}

The analogous construction for $\chi_\al^k$ is slightly more involved.
Here
the starting point is the exterior covariant derivative
\begin{equation}
D \, \cd_\al \phi^k \, = \, d \cd_\al \phi^k
     -\phi_\al{}^\bt \cd_\bt \phi^k - A \, \cd_\al \phi^k
     - {\ca}^{(r)} \lp {\bf T}_{(r)} \cd_\al \phi \rp^k
     + \Gamma^k{}_{lj} D \phi^j \cd_\al \phi^l,
\end{equation}
which upon double bar projection gives rise to
\[
D \, \cd_\al \phi^k \doubar \, = \, \sqrt{2} dx^m
\lp
\prt_m \chi_\al^k -\om_m{}_\al{}^\bt \chi_\bt^k -  A_m \chi_\al^k
- i {\aym}_m^{(r)} \lp {\bf T}_{(r)} \chi_\al \rp^k
+ \Gamma^k{}_{lj} \cd_m A^j \chi_\al^l
\rp,
\]
with $\cd_m A^j$ defined above. This suggests to define
\begin{equation}
\cd_m \chi_\al^k  \, = \,
\prt_m \chi_\al^k -\om_m{}_\al{}^\bt \chi_\bt^k -  A_m \chi_\al^k
- i {\aym}_m^{(r)} \lp {\bf T}_{(r)} \chi_\al \rp^k
+ \Gamma^k{}_{lj} \cd_m A^j \chi_\al^l.
\end{equation}
The double bar projection on covariant differentials yields now
\begin{equation}
D \cd_\al \phi^k \doubar \, = \, dx^m
\lp
e_m{}^a \cd_a \cd_\al \phi^k \loco
+ \frac{1}{2} \psi_m{}^\bt \cd_\bt \cd_\al \phi^k \loco
+ \frac{1}{2} \psib_m{}_\db \cd^\db \cd_\al \phi^k \loco
\rp.
\end{equation}
Here, the quantity $\cd_a \cd_\al \phi^k \loco$ is called the
super-covariant component field derivative of $\chi_\al^k$. However, the
two remaining terms still need some workout. Whereas the second term
involves the auxiliary field
$F^k$, the third term gives rise to the super-covariant
component field derivative $\cd_a\phi^k \loco$, just derived above. As a
result one recovers the same form as in (\ref{CPN.56}), \ie
\begin{equation}
{\cd}_a \cd^{\al} \phi^i {\loco} \, = \,
{e_a}^m \left(
\sqrt{2} {\cd}_m {\chi}^{\al i}
- {\psi_m}^{\al} F^i + i (\psib_m \sib^n)^{\al} (\cd_n A^i
- \frac{1}{\sqrt{2}} {\psi_n}^{\varphi} \chi_{\varphi}^i)
\right).
\label{CPNE.56}
\end{equation}
Observe however that this expression is different from (\ref{CPN.56a}),
because now the composite \ka connection $A_m$ contains additional
terms due to the linear superfield dependence of the kinetic potential.

\subsec{The Composite $U_K(1)$ Connection}
\label{appE3}

Let us first recall the identification of the spinor and vector components of
the $U_K(1)$ gauge potential in terms of the kinetic potential $K$, adapted
to the present situation, where $K$ depends on a linear superfield as well.
The relevant equations are generalizations of (\ref{GRA.146}), which read now
\begin{equation}
 A_\al \, = \, \frac{1}{4} {E_\al}^M \prt_M K(\phi, \phib, L), \cem
 A^\da \, = \, - \frac{1}{4} E^\da{}^M \prt_M K(\phi, \phib, L),
\label{GRAE.146a}
\end{equation}
\begin{equation}
 A_{\al \da}- \frac{3i}{2}G_{\al \da} \, = \,
\frac{i}{2} \lp \cd_\al  A_\da + \cd_\da  A_\al \rp.
\label{GRAE.146}
\end{equation}
The important point to notice here is that the entities which are known
{\em a priori} are the covariant components $ A_\al$,
$ A^\da$ and ${ A}_a$.
As a consequence, the space-time component ${ A}_m$ identified in
(\ref{CPN.21}), \ie $ A \doubar = dx^m { A}_m (x)$,
must be evaluated from the expression
\begin{equation}
{ A}_m(x) \,=\,{e_m}^a { A}_a{\loco}
+ \f{1}{2} {\psi_m}^\al { A}_\al{\loco}
+ \f{1}{2} \psib_{m \da}\, { A}^\da{\loco}.
\label{eq:3.31}
\end{equation}
Taking into account the linear multiplet couplings, section \ref{2F},
we obtain
\begin{eqnarray}
{ A}_m{\loco} + \frac{i}{2}\, {e_m}^a b_a &=&
\f{1}{4} \, K_k \, \cd_m A^k
 - \f{1}{4} K_\bk \, \cd_m \bA^\bk
+ \f{i}{4} \, g_{k \bk} \; \chi^k \sigma_m \chib^\bk
\nn \\ &&
+ \f{i \al}{6} \, {e_m}^a b_a + \f{i \al}{4} \kl \, {}^*h_m
- \f{i \al}{4} \kl \tr(\la \sigma_m \lab)
- \f{i \al}{8} \vp \sigma_m \vpb
\nn \\ &&
- \f{\al}{8} \lp \psi_n \sigma_m \sib^n \vp
- \psib_n \sib_m \sigma^n \vpb \rp
- \f{\al}{8} \, {\vep}_{mnpq} \, \psi^n \sigma^p \psib^q.
\label{eq:3.32}
\end{eqnarray}
Compared to the pure \ka  superspace construction, (\ref{CPN.40}),
\ie without linear multiplets, a number of new terms appear.
In particular, the dual field strength of the antisymmetric tensor gauge
field,
\begin{equation}
{}^* h^k = \f{1}{3!} \vep^{klmn} h_{lmn},
\label{defdualh}
\end{equation}
with $h_{lmn}$ identified in (\ref{Hbb}), is given as
\begin{equation}
{}^* h^k \, = \, \f{1}{3!} \vep^{klmn}
\lp
3 \, \prt_n b_{ml}
+ k ({\aym}_{l} \prt_m {\aym}_{n}
- \f{2i}{3} {\aym}_{l} {\aym}_m {\aym}_{n})
\rp.
\label{dualh}
\end{equation}
Instead of keeping all these terms encoded in
the component field definitions of the covariant derivatives, we only
retain the combination
\begin{equation}
v_m \,=\,\f{1}{4} K_k \, \cd_m A^k - \f{1}{4} K_\bk \, \cd_m \bA^\bk
 + \f{i}{4} g_{k \bk} \; \chi^k \si_m \chib^\bk,
\label{eq:3.33}
\end{equation}
in these definitions.
This renders the component field action more complicated, but shows
explicitly the various couplings related to the linear multiplet. The
corresponding covariant derivatives will be denoted $\nabla_m$, they coincide
with those defined in section \ref{CPN}.

\subsection
{Genesis of the factor $L K_L - 3$}
\label{app3/a}

The chiral supergravity superfield $R$ and its spinor derivatives are
essential building blocks in the construction of supersymmetric actions
and the derivation of supersymmetry transformations. A detailed knowledge
of $\cd_{\! \al} R$ and $\cd^\al \cd_{\! \al} R$ is crucial for the
construction of supersymmetric component field actions. In section
\ref{2F3} we have pointed out modifications to the normalization of the
Einstein term in the linear superfield formalism.

We will explain here in some detail the superspace mechanism
which underlies these modifications. To be definite we shall consider the
superfield $R$. Its spinor derivative is given as
\begin{equation}
-3 \, \cd_{\! \al} R \, = \, X_\al + 4 S_\al,
\label{dalR}
\end{equation}
as a consequence of the Bianchi identities, see (\ref{B.81}). The
superfield $S_\al$, as defined in (\ref{B.32}),
is related to the torsion
$T_{cb}{}^\al$, while $X_\al$ is given in (\ref{2F.31}),
\begin{equation}
X_\al  \, = \,
-\f {1}{8}({\cdb}^2 -8R) \, {\cd}_{\al}K(\phi,\phib, L).
\end{equation}
Although straightforward, it will be instructive to illustrate in detail
the appearance of the term $L K_L \, \cd_{\! \al} R$ in $X_\al$, in
successively applying the spinor derivatives. In a first step, we write
\[
-8 \, X_\al \, = \,
\cdb^2 \left( K_k \cd_\al \phi^k \right)
+ \cdb^2 \left( K_L \cd_\al L \right).
\]
It is clear that the linearity condition will arise from the second term,
evaluation of the spinor derivatives yields
\[
\cdb^2 \left( K_L \cd_\al L \right) \, = \,
\cd_\da \left( \cd^\da\! K_L \, \cd_\al L \right)
+ \left( \cd_\da K_L \right) \cd^\da  \cd_\al L
+ K_L \left[ \cdb^2 , \cd_\al \right] L
+ K_L  \, \cd_\al \cdb^2 \! L.
\]
At this point the modified linearity condition (\ref{mlc1})
\[
\proki L \, = \, 2k \, \tr \lp \cw^\vp \cw_\vp \rp,
\]
must be used to arrive at
\[
K_L \, \cd_\al \cdb^2 \! L \, = \, 8 L K_L \, \cd_\al R
+ 8 R \, K_L \cd_\al L + 2k \, \cd_\al \tr \lp \cw^\vp \cw_\vp \rp.
\]
In this way we recover (\ref{2FXmod}) in the form
\begin{equation}
X_\al \, = \, - L K_L \, \cd_\al R + Y_\al,
\end{equation}
with $Y_\al$ determined from the string of equations above. Combining
this with (\ref{dalR}) gives rise to
\begin{equation}
(L K_L - 3) \, \cd_\al R \, = \, Y_\al + 4 S_\al,
\label{dalRbis}
\end{equation}
identifying $\cd_\al R$ in terms of other, already known, superfields.
When projected to lowest superfield components, $S_\al \loco$ will
contain the super-covariant field strength of the gravitino. As to $Y_\al
\loco$, one has to go through the various terms and identify properly the
component field expressions. This is straightforward, but rather lengthy,
and will not be done here.

\subsec{Supersymmetry Transformations}
\label{appE3/4}

One of the advantages of superspace geometry is that supersymmetry
transformations are defined geometrically. We have outlined in
detail how this mechanism works in the case of
supergravity/matter coupled to Yang-Mills in section \ref{CPN3},
based on the general formalism developed in appendix \ref{appC2}.
Deriving supersymmetry transformations for component fields
amounts to a bookkeeping activity in the sense that one has to
apply a set of well-defined rules to extract component field
properties from superspace.

The emphasis will be rather on the method of
derivation of the component field transformations than their explicit
gestalt (which is often quite lengthy and not very illuminating).

Here we will discuss supersymmetry
transformations for component fields in the linear superfield formalism,
based on the general notion of supergravity transformations extended to
2-form geometry. This will allow to derive the component field
transformations for the linear multiplet, coupled to the
supergravity/matter/Yang-Mills system.

At the same time, the presence of the linear superfield $L$ in the
kinetic potential $K(\phi, \phib, L)$, which replaces the \ka potential,
will modify the supersymmetry transformations in the supergravity, matter
and Yang-Mills sectors.

We will discuss here, sector by sector, how these modifications are
induced from superspace geometry, before turning to the derivation of the
supersymmetry transformations of the linear multiplet component fields.

\begin{itemize}

\item
{\bf Matter and Yang-Mills multiplets}

The supersymmetry transformations of component fields in the case
of the general supergravity/matter/Yang-Mills system have been
derived in section \ref{CPN}. The transformations of $A^k$,
$\chi_\al^k$, $F^k$ are given in (\ref{CPN.92}) - (\ref{CPN.94}),
those of $\bA^k$, $\chib^k$, $\bF^k$ in (\ref{CPN.98}) -
(\ref{CPN.100}), whereas those of the Yang-Mills multiplet
$\aym_m$, $\la^\al$, $\lab_\da$, $\bf D$ are given in
(\ref{CPN.102}) - (\ref{CPN.105}).

In the linear superfield formalism,
the general structure of these transformation laws remains unchanged.
The modifications caused by the linear field dependence of the kinetic
potential $K(\phi, \phib, L)$ occur in two ways. First of all, whenever a
covariant space-time derivative acts on a component of non vanishing
chiral weight, it should be written in terms of the new composite $U(1)$
connection (\ref{eq:3.32}) instead of (\ref{CPN.40}).

The second source of modifications is the term $\iota_\xi A= \xi^A A_A$,
see (\ref{CPN.68}), in the generic case of a component with non vanishing
chiral weight. As $A_\al$ and $A^\da$ are now given in terms of the
kinetic potential rather then the \ka potential, new terms appear. This
amounts in replacing everywhere the combination
$K_k \xi \chi^k - K_\bk \xib \chib^\bk$ by
\begin{equation}
K_k \xi \chi^k - K_\bk \xib \chib^\bk
+ \frac{\al}{\sqrt{2}} \, K_L (\xi \vp - \xib \vpb).
\end{equation}
In this way the supergravity transformations of matter and Yang-Mills
fields are adapted to the linear superfield formalism.

\item
{\bf Supergravity multiplet}

The mechanism just pointed out will occur for the gravitino supergravity
transformations and the scalar auxiliary fields as well.
Geometrically, the
starting point for deriving supersymmetry transformations of the vierbein
$e_m{}^a$ and the gravitino $\psi_m{}^\al$, $\psib_m{}_\da$
is the general superspace equation (\ref{CPN.67})
\begin{equation}
 \delta {E_M}^A  \, = \,
\cd_M \xi^A + {E_M}^B \xi^C {T_{CB}}^A
-w(E^A) {E_M}^A \, \xi^C \! A_C  ,
\label{CPN.67bis}
\end{equation}
derived in section \ref{CPN3}. This relation is still valid in the linear
superfield formalism. What kind of modifications arise for the component
fields? Consider first the case of the vierbein $e_m{}^a$.
Choosing $M=m$ and $A=a$ in (\ref{CPN.67bis}) and projecting to lowest
components reproduces the supersymmetry transformation (\ref{CPN.74}).
No dependence on the linear multiplet appears, the supersymmetry
transformation for $e_m{}^a$ remains unchanged.

What happens in the case of the gravitino? Taking $M=m$ and $A=\al$ gives
rise to
\begin{equation}
\frac{1}{2} \dt \psi_m{}^\al \, = \,
\cd_m \xi^\al
+ e_m{}^b \xi^\gamma T_{ \gm b}{}^\al \loco
+e_m{}^b \xi_\dg T^\dg{}_b{}^\al \loco
- \psi_m{}^\al \left(
\xi^\gm  A_\gm \loco + \xib_\dg  A^\dg \loco \right).
\label{deltapsi}
\end{equation}
Clearly, the torsion terms are expressed in terms of the supergravity
auxiliary fields as before, no modification. However, in the covariant
derivative of $\xi^\al$ -cf. (\ref{CPN.77})- the composite \ka connection
${ A}_m \loco$ is now given by (\ref{eq:3.32}) instead of
(\ref{CPN.40}).  Moreover, in the
last term, the linear superfield dependence must be taken into account,
giving rise to the second type of modification pointed out before.
It is then an easy exercise to write down explicitly all the terms in the
supersymmetry transformation of the gravitino in the linear superfield
formalism, the result should be compared to (\ref{CPN.75}) and
(\ref{CPN.76}).

Let us next turn to the auxiliary fields $M$, $\ovM$ and $b_a$. As we
point out now, the situation is more intricate in this case. To be
definite we concentrate on
$M = -6 R
\loco$. Its generic supersymmetry transformation  -cf.
(\ref{CPN.82})- reads
\begin{equation}
\dt M \, = \, - 6 \, \xi^\al \cd_\al R \loco
- \frac{1}{\sqrt{2}} \, M \left(
K_k \xi \chi^k - K_\bk \xib \chib^\bk
+ \frac{\al}{\sqrt{2}} \, K_L (\xi \vp - \xib \vpb) \right).
\end{equation}
As to the lowest component of $\cd_\al R$ we should take into account
the discussion in the previous subsection, in particular (\ref{dalRbis}).
As a result, we find
\begin{eqnarray}
 \, \dt M &=& \frac{1}{(\al - 3)} \left(
2 \, \xi^\al {({\si}^{cb} \eps)}_{\al\varphi}{T_{cb}}^\varphi \loco
+ \xi^\al Y_\al \loco \right)
\nn \\ &&
- \frac{1}{\sqrt{2}} \, M \left(
K_k \xi \chi^k - K_\bk \xib \chib^\bk
+ \frac{\al}{\sqrt{2}} \, K_L (\xi \vp - \xib \vpb) \right).
\label{deltaM}
\end{eqnarray}
This is a very compact form of a quite complicated expression. First of
all the super-covariant field strength ${T_{cb}}^\varphi \loco$ of the
gravitino is given in (\ref{CPN.23}). Here, the covariant derivative
(\ref{CPN.17}) must now be written in terms of the composite \ka
connection constructed in (\ref{eq:3.32}).
As to $Y_\al \loco$, its superfield form is to be determined from the
string of equations of the preceding subsection and then projected to
lowest components with carefully paying attention to $U(1)$ covariant
space-time derivatives. The procedure is straightforward, but a bit
lengthy and so is the result, which will not be presented here. Note,
however, that the same quantity  $Y_\al \loco$ appears in the variation of
$b_a$ as well.

\item
{\bf Linear multiplet}

The linear multiplet and its couplings to the
supergravity/matter/Yang-Mills system, including Chern-Simons forms,
is described in the framework of 2-form geometry in superspace.
In order to extract the supergravity transformations of the antisymmetric
tensor we have to extend the notion of supergravity transformations to
this geometric structure as well.

Recall that invariance of the 3-form field strength
$H = dB +k Q$ under Yang-Mills gauge transformations of the Chern-Simons
form $Q = \tr ( \ca \cf - 1/3 \, \ca \ca \ca )$ is achieved in assigning
a compensating Yang-Mills transformation to
the 2-form gauge potential, in addition to superspace diffeomorphisms and
1-form gauge transformations $\bt = dz^M \bt_M$, such that
\begin{equation}
\dt_{tot} B \, = \, L_\xi B + d \bt + i k \tr \left( \alym \, d
\ca \right),
\end{equation}
with $\alym = \al^{(r)} {\bf T}_{(r)}$. In the first term, we
explicit the Lie-derivative, and use $\iota_\xi d B = \iota_\xi H
- \iota_\xi Q$ with
\begin{equation}
\iota_\xi Q \, = \, \tr \left(\ca \, \iota_\xi \cf \right)
+ \tr \left( (\iota_\xi \ca) \, d \ca \right),
\end{equation}
to arrive at
\begin{equation}
\dt \, B \, = \, \iota_\xi H - k \tr \left(\ca \, \iota_\xi \cf
\right)  +d (\bt + \iota_\xi B) + i k \tr \left( (\alym +
\iota_\xi \ca)\, d \ca \right).
\end{equation}
Supergravity transformations, along the same lines of reasoning as in
appendix \ref{appC2} are then defined as
\begin{equation}
{\delta}_{\WZ}\, B \, = \, \iota_\xi H - k \tr \left(\ca \, \iota_\xi\cf
\right) ,
\label{deltaB}
\end{equation}
\ie a combination of superspace diffeomorphisms and field dependent
compensating Yang-Mills and 1-form gauge transformations of parameters
\begin{equation}
\alym \, = \, - \iota_\xi \ca, \cem \bt \, = \, - \iota_\xi B.
\end{equation}
The supergravity transformation of the antisymmetric tensor gauge field
$b_{mn} (x)$ is then obtained from (\ref{deltaB}) in applying
systematically the double bar projection, which yields
\begin{eqnarray}
dx^m dx^n \, \frac{1}{2} {\delta}_{\WZ}\, b_{nm} &=& dx^m dx^n  \left[
\xi \si_{nm} \Lambda
+ \xib \sib_{nm} \bar \Lambda
- i L \, \psi_n \si_m \xib
- i L \, \psib_n \sib_m \xi \right.\nn \\
&& \hspace{16mm} \left.
+ i k \tr \left(
{\aym}_m (\xi \si_n \lab + \xib \sib_n \la )
\right)
\right].
\end{eqnarray}
Supergravity transformations of $L(x)$ and $\La_\al$, $\Lab^\da$ are
obtained in the usual way, applying spinor derivatives to the superfields
$L$ and $\cd_\al L$, $\cd^\da L$. As to $L(x)$ it is immediate to find
$\dt L = \xi \La + \xib \Lab$. The case of $\La_\al$ is slightly more
interesting, let us outline the general procedure to obtain its
supergravity transformation. The starting point is the superfield
equation,
$ \, \dt \cd_\al L = \xi^\bt \cd_\bt \cd_\al L
+ \xi_\db \cd^\db \cd_\al L \, $
written in the form
\begin{equation}
\dt \cd_\al L
\, = \,
- \frac{1}{2} \xi_\al \cd^2 L
+ \frac{1}{2} \xi_\db \left\{ \cd_\al , \cd^\db \right\} L
- \frac{1}{2} \xi_\db \left[ \cd_\al , \cd^\db \right] L.
\end{equation}
Using the modified linearity condition (\ref{mlc1}) and substituting for
the commutator (\ref{commrel}) gives rise to
\begin{eqnarray}
\dt \cd_\al L &=&
i\xib_\da (\sib^a\eps)^\da{}_\al \, \cd_a L
- \frac{1}{6} \xib_\da (\sib_d\eps)^\da{}_\al \eps^{dcba}H_{cba}
\nn \\[1mm]
&&
- 4 \xi_\al \rd L
+ 2 \xib_\da (\sib^a\eps)^\da{}_\al \, G_a L
\nn \\[1mm]
&& - k \, \xi_\al \tr \left( \cw_\da \cw^\da \right) +2k \,
\xib_\da \tr \left( \cw _\al \cw^\da \right).
 \end{eqnarray}
The supergravity transformation of $\La_\al$ is then obtained after
projecting to lowest superfield components with special care to the
super-covariant component derivative $\cd_a L \loco$ and field strength
$H_{cba} \loco$.

\end{itemize}

\subsec{Component Field Lagrangian - I}
\label{appE4}

We display here the complete component field Lagrangian for the example of
section \ref{2F}, \ie a special kinetic function of the form
\begin{equation}
K(\phi, \phib, L) \, = \, K_0(\phi, \phib) + \al \log L.
\end{equation}
Requiring a canonical normalization function $N=1$
gives rise to a
subsidiary function
\begin{equation}
F(\phi, \phib, L) \, = \, 1 - \frac{\al}{3} + L \, V(\phi, \phib),
\label{specialF}
\end{equation}
with arbitrary linear potential $V(\phi, \phib)$.
The component field action is then derived from the generic
procedure of section \ref{CPN4}, for the chiral superfield
(\ref{genact}) in section \ref{NHGC}, \ie
\begin{equation}
{\bf r} \, = \, -\frac{1}{8} \proki F(\phi, \phib, L), \cem
{\bf {\bar r}} \, = \, -\frac{1}{8} \prokib F(\phi, \phib, L),
\end{equation}
with $F$ given by (\ref{specialF}). Working through all the
necessary steps leads then to the Lagrangian
\begin{eqnarray}
&&\f{1}{e} \cl = - \f{1}{2} {\cal R}
          + \f{1}{2} \vep^{mnpq} \lp \psib_m \sib_n \nabla_p \psi_q
                                     - \psi_m \sigma_n \nabla_p \psib_q \rp
\nonumber \\[1mm]
      & & -\lp K_{k \bk}-3LV_{k \bk} \rp  \nabla_m A^k \, \nabla^m \! \bA^\bk
 - \f{i}{2}\lp K_{k \bk}-3LV_{k \bk} \rp
\lp \chi^k \sigma^m \nabla_m \chib^\bk
+ \chib^\bk \sib^m \nabla_m \chi^k \rp
\nonumber \\[1mm]
      & & + \f{\al}{4 L^2} {\, {}^* \! h}^m {\, {}^* \!h}_m
          - \f{\al}{4L^2}  \prt^m \! L \, \prt_m L
          - \f{i \al}{4} \lp \vp \sigma^m \nabla_m \vpb
     + \vpb \sib^m \nabla_m \vp \rp
\nonumber \\[2mm]
      & & + \f{k}{4}\lp\f{\al}{L}-3V \rp {\fym}_{mn}^{(r)}\, {\fym}^{mn}_{(r)}
+ \f{i k}{2} \lp\f{\al}{L}-3V \rp \lp \la^{(r)} \sigma^m \nabla_m
\lab_{(r)} + \lab^{(r)} \sib^m \nabla_m \la_{(r)} \rp
 \nonumber \\ [0.5ex] & &
 -\f{3i}{2} (V_k\nabla_m A^k -V_\bk\nabla_m\bA^\bk) \, {}^* \!h^m
 + \f{\al(\al-4)}{8L}{\, {}^* \!h}^m (\vp \sigma_m \vpb)
\nonumber \\ [0.5ex]
 && \nonumber \\ [0.5ex]
 & & + \f{1}{9} (\al-3) M \bM - \f{1}{9}
(\al-3) b_a b^a +\lp K_{k \bk}-3LV_{k \bk} \rp F^k \bar{F}^\bk
 - \f{ k}{2}\lp\f{\al}{L}-3V \rp \,{{\bf D}}^{(r)} {{\bf D}}_{(r)}
\nonumber \\ [0.5ex]
& &  + \f{1}{6} (\al-3) \left[ \lp K_{k \bk} -3LV_{k \bk} \rp
(\chi^k \sigma^a \chib^\bk)
 + \f{\al}{2} (\vp \sigma^a \vpb)
 + k \lp\f{\al}{L}-3V \rp (\la^{(r)} \sigma^a \lab_{(r)}) \right] b_a
\nonumber \\ [0.5ex]
 & & - \f{1}{2}  \left[
(K_k-3LV_k) (T_{(r)} A)^k + (\bA T_{(r)})^\bk (K_\bk-3LV_\bk
)\right. \nonumber \\ [0.5ex]
 && \hs{7mm} \ \ \left. -3i\sqrt{2} k( V_k\,\chi^k\la_{(r)} - V_\bk\,
 \chib^\bk\lab_{(r)} )
  + i \al \kl (\lab_{(r)} \vpb - \la_{(r)} \vp)  \right]{{\bf D}}^{(r)}
   \nonumber\\ [0.5ex]&&
    + \f{3L}{2}  \l[ \, \s2 V_{k\bk}\,\vpb\chib^\bk
+\cv_{k\bl\bk}\,\chib^\bl\chib^\bk
             -\kl V_k\,\la^{(r)}\la_{(r)} \, \r]\,F^k
\nonumber\\ [0.5ex]&&
 + \f{3L}{2}  \l[ \, \s2 V_{k\bk}\,\vp\chi^k +\cv_{\bk
l k}\,\chi^l\chi^k
 -\kl V_\bk\,\lab^{(r)}\lab_{(r)} \, \r]\,\bF^\bk
  \nonumber\\ [0.5ex]&&
 \nonumber \\ [0.5ex] & &
 + \, \l[\f{\al}{4L}
 \lp K_{k\bk}-3LV_{k\bk}\rp
 +\f{3}{2} V_{k\bk} \r]\, {}^* \!h^m(\chi^k\sigma_m\chib^\bk)
 +\f{\al k}{4L} \left[\f{\al -2}{L}-3V \right] \, {}^* \!h^m
 \lp \la^{(r)}\sigma_m\lab_{(r)}\rp
 \nonumber\\ [0.5ex] && - \f{i3 L}{2}  \lp \s2 V_{k\bk}\,
 \vp\sigma^m\chib^\bk
 +\cv_{k \bk \,l}\,\chi^l\sigma^m\chib^\bk \, \rp \nabla_m A^k \nn \\ &&
-\f{i3 L}{2} \lp\, \s2 V_{k\bk}\,\vpb\sib^m\chi^k
 +\cv_{\bk \,k\bl}\,\chib^\bl\sib^m\chi^k \rp \, \nabla_m
 \bA^\bk \nn \\ [0.5ex] &&
 + \left[ i \sqrt{2} \lp K_{k \bk} -3LV_{k\bk}\rp
 ( \chib^\bk \lab^{(r)}) -3iLV_k (\vpb \lab^{(r)})\right](T_{(r)}A)^k
     \nonumber \\ [0.5ex]&&
 - \left[ i \sqrt{2} \lp K_{k \bk} -3LV_{k\bk}\rp
 ( \chi^k \la^{(r)})-3iLV_\bk (\vp \la^{(r)})
  \right](\bA T_{(r)})^\bk
  \nonumber\\ [0.5ex]
  &&+ \f{1}{4} \lp R_{k \bk l \bl}- 3 L  \cv_{k\bk l\bl}  \rp (\chi^k
\chi^l) (\chib^\bk\chib^\bl) + \f{\al}{8}(\al-3)(\vp\vp)(\vpb\vpb)
 \nonumber\\ [0.5ex] &&
  -\f{\al}{8}  \lp K_{k \bk} -3LV_{k \bk} \rp\lp \chi^k \sigma^m \chib^\bk \rp
 \lp \vp \sigma_m \vpb \rp \nonumber\\ [0.5ex]&&
 + \f{ \al k}{8}\left[3 V- \f{2}{L} (\al-2)\right]
  (\la^{(r)} \sigma^m \lab_{(r)})(\vp \sigma_m \vpb) \nonumber \\ [0.5ex]
&& - \f{\al k^2}{4 L}\lp\f{\al}{L}-3V -\f{1}{L}\rp
 (\la^{(r)} \sigma^m \lab_{(r)})(\la^{(s)} \sigma_m \lab_{(s)})
 \nonumber \\ [0.5ex]
 &&-\f{k}{4} \l[\f{\al
 }{L} \lp K_{k \bk} -3LV_{k \bk} \rp +6
 V_{k\bk} \r](\la^{(r)} \sigma^m \lab_{(r)})(\chi^k \sigma_m \chib^\bk)
  \nonumber \\ [0.5ex]&&
 +\f{i3k}{2} ( V_k\nabla_m A^k
-V_\bk\nabla_m\bA^\bk )(\la^{(r)}\sigma^m\lab_{(r)} )
 \nonumber \\ [0.5ex]&&
 - \f{3 L }{2 \s2}
\l[\cv_{k\bl\bk}(\chib^\bl\chib^\bk)(\chi^k \vp) + \cv_{\bk
lk}(\chi^l\chi^k)(\chib^\bk \vpb)\r] \nonumber\\ [0.5ex] &&+
\f{3k}{4}\, \cv_{kl}(\chi^k\chi^l)(\la^{(r)}\la_{(r)}) + \f{3}{4}
k\, \cv_{\bk\bl}(\chib^\bk\chib^\bl)(\lab^{(r)}\lab_{(r)})
\nonumber \\ [0.5ex]&&
 - \f{\al k}{4 L}\left[ (\lab^{(r)} \lab_{(r)}) (\vpb \vpb)
  + (\la^{(r)} \la_{(r)}) (\vp \vp) \right]
 - \f{\al k^2}{4 L^2}(\la^{(r)} \la_{(r)})(\lab^{(s)} \lab_{(s)})
\nonumber \\ [0.5ex] &&
 -\left[ \f{\al k}{2L}  \lp \vp \sigma^{mn} \la_{(r)}
+\vpb \sib^{mn} \lab_{(r)} \rp + \f{3 k}{\s2}  \lp
V_k\,\chi^k\sigma^{mn}\la_{(r)}
  +V_\bk\,\chib^\bk\sib^{mn}\lab_{(r)} \rp \right]\, {\fym}_{mn}^{(r)}
 \nonumber \\ [0.5ex]
  & &- \f{ik}{2} \lp \f{\al}{L}-3V\rp \lp {\fym}^{(r)\, mn}
 + i\,\, {}^* \!{\fym}^{(r)\, mn} \rp
 (\psi_m \sigma_n \lab_{(r)}) \nonumber \\ [0.5ex]
     & &   - \f{ik}{2}\lp \f{\al}{L} -3V \rp
\lp {\fym}^{(r)\, mn} -i\, \, {}^* \!{\fym}^{ (r)\,mn} \rp
(\psib_m \sib_n \la_{(r)}) \nonumber \\[0.5ex] &&
  -\f{1}{2} \lp K_k -3LV_k \rp (T_{(r)}A)^k (\psib_m \sib^m \la^{(r)})
 +\f{1}{2}  \lp K_\bk -3LV_\bk \rp (\bA T_{(r)})^\bk
  (\psi_m \sigma^m \lab^{(r)})\nonumber \\ [0.5ex]
  & &   - \f{1}{\sqrt{2}}\lp K_{k \bk}-3LV_{k \bk} \rp
  \left[ (\psi_n \sigma^m \sib^n \chi^k) \nabla_m \bA^\bk
  + (\psib_n \sib^m \sigma^n \chib^\bk) \nabla_m A^k \right]
  \nonumber \\ [0.5ex] &&
 -\f{\al}{4L} (\psi_n \sigma^m \sib^n \vp
 + \psib_n \sib^m \sigma^n \vpb)\, \prt_m L+ \f{i \al}{4L}
 (\psi_n \sigma^m \sib^n \vp - \psib_n \sib^m \sigma^n \vpb)\,{ {}^* \!h}_m
  \nonumber \\
  & & + \f{i \al k}{8 L}
\left[ (\psib_m \sib^m \vp)(\la^{(r)} \la_{(r)})
 + (\psi_m \sigma^m \vpb)(\lab^{(r)} \lab_{(r)}) \right]
\nonumber \\
   & & - \f{i \al}{16} (\al-4) \left[ (\psi_m \sigma^m \vpb)(\vp \vp)
    + (\psib_m \sib^m \vp)(\vpb \vpb) \right]
   \nonumber \\ [0.5ex]
 && - \f{i3 k}{4 \s2}\, V_k\,(3\psi^m\chi^k
 +2\psi_n\sigma^{nm}\chi^k)\,
(\la^{(r)}\sigma_m\lab_{(r)})   \nonumber\\ [0.5ex]
 && + \f{i3 k}{4 \s2}
 \,  V_\bk\,(3\psib^m\chib^\bk +2\psib_n\sib^{nm}\chib^\bk)
\, (\la^{(r)}\sigma_m\lab_{(r)}) \nonumber\\ [0.5ex] && + \f{i3 k}{4
\s2}  \l[ \, V_k(\psib_m\sib^m\chi^k)(\la^{(r)}\la_{(r)})
              +V_\bk(\psi_m\sigma^m\chib^\bk)(\lab^{(r)}\lab_{(r)}) \, \r]
\nonumber \\ [0.5ex]
     & & - \f{i \al k}{4 L}  \lp \psi^m
     \vp -\psib^m \vpb \rp (\la^{(r)} \sigma_m \lab_{(r)})
     \nonumber \\ [0.5ex]
& &  +\f{i \al k}{8} \lp \f{\al}{L}-3V -\f{1}{L} \rp
  \lp \psi_n \sigma^m \sib^n \vp
  -\psib_n \sib^m \sigma^n\vpb \rp (\la^{(r)} \sigma_m \lab_{(r)})
  \nonumber  \\ [0.5ex]
& & + \f{1}{4}\lp K_{k \bk}-3LV_{k\bk}\rp
 \lp \psi_n \sigma^m \psib^n
 - \f{i}{2} (\al-4) \vep^{mnpq} \psi_m \sigma_n \psib_p \rp
 (\chi^k \sigma_q \chib^\bk)
  \nonumber\\ [0.5ex]
 & & +\f{1}{16} \lp\f{\al}{L}-3V\rp \left[
(3\psi_n\psi^n + 2\psi_n\sigma^{nm}\psi_m)(\la^{(r)}\la_{(r)})\right.
\nn \\[0.5ex]
&& \left. \hs{32mm}
+(3\psib_n\psib^n+2\psib_n\sib^{nm}\psib_m)(\lab^{(r)}\lab_{(r)})\right]
\nonumber\\ [0.5ex]
 & & + \f{k}{8}\lp \f{\al}{L}-3V\rp \lp {\oint}_{npq}g^{mn} g^{pq}
 - i(\al-1) \vep^{mnpq} \rp (\psi_m \sigma_n \psib_p)(\la^{(r)} \sigma_q\lab_{(r)})
 \nonumber\\ [0.5ex]
  & & +\f{3ikV}{4}\vep^{mnpq}(\psi_m \sigma_n
  \psib_p)(\la^{(r)} \sigma_q \lab_{(r)})
   - \f{i \al}{4L}\vep^{mnpq}
  \, (\psi_m \sigma_n \psib_p)\, {{}^* \! h}_q
 \nonumber \\ [0.5ex] &&
 + \f{\al}{8} \lp g^{mn} g^{pq} + g^{mq} g^{np}
             - \f{i}{2} (\al-4) \vep^{mnpq} \rp (\psi_m \sigma_n \psib_p)
(\vp \sigma_q \vpb)\nonumber \\ [0.5ex]
     & & + \l[\f{i \al}{8}\lp K_{k \bk}-3LV_{k\bk}\rp
     +\f{3i}{4}LV_{k\bk} \r] (\psi_n \sigma^m \sib^n \vp - \psib_n \sib^m \sigma^n \vpb)
(\chi^k \sigma_m \chib^\bk)\nonumber \\ [0.5ex]
    & & + \f{\al}{8}\l[ (\psi_m \psi^m) (\vp \vp)+
     (\psib_m \psib^m) (\vpb \vpb)\r]
+ \f{\al}{8} \vep^{mnpq} (\psi_m \sigma_n \psib_p)
           (\psi_s \sigma_q \sib^s \vp - \psib_s \sib_q \sigma^s \vpb)
\nonumber \\ [0.5ex]
    & & - \f{\al}{16} \vep^{mnpq} \vep_{mrst}
           (\psi_n \sigma_p \psib_q) (\psi^r \sigma^s \psib^t).
\end{eqnarray}
Recall that  the first term in this expression, the curvature scalar $\car$,
is defined in (\ref{CPN.26}). As mentioned above, the
covariant derivatives $\nabla_m$ coincide with those
defined in section \ref{CPN}. For the sake of completeness, we recall here
the explicit expressions.
The nabla derivatives of the
Rarita-Schwinger field are given in (\ref{CPN.17}), (\ref{CPN.18}),
\begin{eqnarray}
\nabla_n {\psi_m}^\al & = & \prt_n {\psi_m}^\al
+ {\psi_m}^\beta {\om_{n \beta}}^\al + {\psi_m}^\al v_n,
\label{CPNE.17} \\
\nabla_n {\psib_{m \da}} & = & \prt_n {\psib_{m \da}}
+ {\psib_{m \db}} {{\om_{n}}^{\db}}_{\da} - \psib_{m \da} v_n,
\label{CPNE.18}
\end{eqnarray}
whereas (\ref{CPN.31}), (\ref{CPN.31}) define those of the
matter complex scalars:
\begin{equation}
\nabla_m A^i \, = \, \prt_m A^i
-i {\aym}^{(r)}_m \lp {{\bf T}_{(r)}} A \rp^i, \cem
\nabla_m \bA^{\bj} \, = \, \prt_m \bA^{\bj}
+i {\aym}^{(r)}_m \lp \bA {{\bf T}_{(r)}}\rp^{\bj}.
\label{CPNE.32}
\end{equation}
The derivatives for the spinors in the matter sector are,
(\ref{CPN.60}), (\ref{CPN.61}),
\begin{eqnarray}
\nabla_m \chi_\al^i & = & \prt_m \chi_\al^i
- {\om_{m \, \al}}^\varphi {\chi}_{\varphi}^i
-i {\aym}^{(r)}_m  \lp{{\bf T}_{(r)}} \chi_\al \rp^i
-v_m \chi_\al^i
+\chi^j_\al\ {\Gamma^i}_{jk} \nabla_m A^k,
\label{CPNE.33}
\\[2mm]
\nabla_m \chib^{\da \bj} & = & \prt_m \chib^{\da \bj}
 - {{\om_{m}}^\da}_{\dv} \chib^{\dv \bj}
+i {\aym}^{(r)}_m \lp \chib^{\da} \, {{\bf T}_{(r)}} \rp^{\bj}
+ v_m \chib^{\da \bj}
+ \chib^{\da\bi} \ {\Gamma^\bj}_{\bi\bk} \nabla_m {\bA}^{\bk},
\label{CPNE.34}
\end{eqnarray}
whereas those of the Majorana spinor of the linear multiplet are
given as
\begin{equation}
\nabla_m \varphi_\al^i \, = \, \prt_m \varphi_\al^i
- {\om_{m \, \al}}^\varphi {\varphi}_{\varphi}^i
-v_m \varphi_\al^i, \cem
\nabla_m \vpb^{\da \bj} \, = \, \prt_m \vpb^{\da \bj}
- {{\om_{m}}^\da}_{\dv} \vpb^{\dv \bj}
+ v_m \vpb^{\da \bj}.
\end{equation}
Finally, the gaugino covariant derivatives, (\ref{CPN.153}),
(\ref{CPN.154}), are
\begin{eqnarray}
\nabla_m \la^{(r)}_\al \
 &=&
\partial_m \la^{(r)}_\al
-{\om_{m \, \al}}^\varphi \la^{(r)}_\varphi
- {\aym}^{(t)}_m \la^{(s)}_\al {c_{(s)(t)}}^{(r)}
+ v_m \la^{(r)}_\al ,
\label{CPNE.38} \\[2mm]
\nabla_m \lab^{(r)\da} &=&
\partial_m \lab^{(r)\da} -{\om_m}^{\da}{}_{\dv} \lab^{(r)\dv}
 - {\aym}^{(t)}_m \lab^{(s)\da} {c_{(s)(t)}}^{(r)}
- v_m \lab^{(r)\da}.
\label{CPNE.39}
\end{eqnarray}
As to the field strength tensors,
${}^* h^k = \f{1}{3!} \vep^{klmn} h_{lmn}$ is given above in
(\ref{dualh}). The Yang-Mills field strength, defined in (\ref{CPN.66}),
reads
\begin{equation}
{\fym}_{mn}^{(r)} \,=\,\prt_m {\aym}_n^{(r)} - \prt_n {\aym}_m^{(r)}
+ {\aym}_m^{(s)} {\aym}_n^{(t)} c_{{(s)} {(t)}}{}^{(r)},
\label{CPNE.66}
\end{equation}
with dual $2 \, {}^* \!{\fym}^{(r)\, kl} =  \vep^{klmn} {\fym}_{mn}^{(r)}$.

As to the manifold of the matter scalar fields, the basic
objects are the kinetic potential $K$ and the linear potential $V$.
Subscripts attached to these objects denote derivatives with respect to the
complex scalars. In particular, the \ka metric $g_{k \bar k} = K_{k \bar k}$
is defined in (\ref{Kamet}), and its inverse shows up in the Levi-Civita
symbols
\begin{equation}
{\Gamma^k}_{ij} \,=\,g^{k\bl}g_{i\bl,j} \, , \cem
{\Gamma^{\bk}}_{{\bi}{\bj}} \,=\,g^{l{\bk}}g_{l{\bi},{\bj}}\,.
\label{graE.161}
\end{equation}
The curvature tensor is given as (\ref{GRA2.165})
\begin{equation}
R_{k{\bk} j{\bj}} \,=\,g_{k{\bk},j{\bj}}
- g^{l{\bar{l}}} g_{k{\bar{l}},j} \, g_{l{\bk},{\bj}}\, .
\label{GRAE.165}
\end{equation}
As to the derivatives of the linear potential we have introduced the
covariant objects
\begin{equation}
\cv_{i  j} \, = \,V_{i j}-{{\Gm_i}^k}_j \, V_{k}, \cem
\cv_{\bi  \bj} \, = \,V_{\bi \bj}-{{\Gm_\bi}^\bk}_\bj \, V_{\bk},
\end{equation}
\begin{equation}
\cv_{i \bk\,  j} \, = \,V_{i \bk\, j}-{{\Gm_i}^k}_j \, V_{k \bk},
\cem
\cv_{\bi \, k  \bj} \, = \,V_{\bi\, k \bj}
-{{\Gm_\bi}^\bk}_\bj \, V_{k\bk},
\end{equation}
as well as
\begin{equation}
\cv_{i \bi j \bj} \, = \,
V_{i \bi j \bj}+ {{\Gm_i}^k}_j \, {{\Gm_\bi}^\bk}_\bj\, V_{k \bk}
- {{\Gm_i}^k}_j \, V_{\bi k \bj} - {{\Gm_\bi}^\bk}_\bj\, V_{i \bk j}.
\end{equation}

Before turning to a discussion of the auxiliary field sector we shortly
discuss the effective transformations $V \mapsto V + H + \bar H$. Observe
that any term containing either $V$ itself or derivatives $V_k$, $V_{\bar k}$
or $V_{kl}$, $V_{\bar k \bar l}$ changes under such transformations. Of
particular interest is the term
\[ (V_k\nabla_m A^k -V_\bk\nabla_m\bA^\bk) \, {}^* \!h^m, \]
which transforms into
\[ (H - \bar H) \, {\fym}_{mn}^{(r)}\, {}^* \! {\fym}^{mn}_{(r)}, \]
after integration by parts. On the other hand, the Yang-Mills
kinetic term gives rise to
\[ (H + \bar H) \, {\fym}_{mn}^{(r)}\, {\fym}^{mn}_{(r)}. \]

Finally, we have to comment on the structure of the auxiliary field sector.
Collecting in $\cl_{\mathrm{aux}}$ all the terms containing auxiliary fields,
that is components $M$,
$\overline M$,
$b_a$,
$F^k$, $\bar F^{\bar k}$ and ${\bf D}_{(r)}$, we diagonalize in terms of
new, hatted auxiliary fields which have trivial equations of motion.
As a result, the auxiliary sector of the Lagrangian takes the form
\begin{eqnarray}
&&e^{-1} \cl_{\mathrm{aux}}\,=\, + \f{1}{9} (\al-3) M  \bM -
\f{1}{9} (\al-3)\widehat b_a \,\widehat b^a
 \nonumber\\ [0.5ex] &&
 +\lp K_{k \bk}-3LV_{k \bk} \rp {\widehat F}^k \, {\widehat {\bar F}}{}^\bk
 - \f{ k}{2}\lp\f{\al}{L}-3V \rp \,{\widehat{\bf D}}^{(r)}\, {\widehat{\bf D}}_{(r)}
 \nonumber\\ [0.5ex] &&
 - \left[
\f{9L^2}{4} \cv_{\bj \, k \bl} {\lp K_{k \bk}-3LV_{k\bk}\rp}^{-1}
\cv_{j \bk l} +\f{\al-3}{8}  \lp K_{j \bj}-3LV_{j\bj}\rp \lp K_{l
\bl}-3LV_{l\bl}\rp \right]\lp \chi^j \chi^l \rp \lp \chib^\bj
\chib^\bl \rp
 \nonumber \\ [0.5ex]& &
 -\f{9k^2}{4}  V_k {\lp K_{k
\bk}-3LV_{k\bk}\rp}^{-1} V_\bk \,(\la^{(r)}\la_{(r)})
(\lab^{(s)}\lab_{(s)})
 \nonumber\\ [0.5ex] &&
 -\f{9 L^2}{2} V_{k \bj}\,{\lp K_{k
\bk}-3LV_{k\bk}\rp}^{-1}\, V_{j \bk}\, ( \vp \chi^j)\,(\vpb
\chib^\bj)\nonumber \\ [0.5ex]& & +\f{9kL}{4} V_\bk  {\lp K_{k
\bk}-3LV_{k\bk}\rp}^{-1} \left[ \cv_{\bj\, k \bl} (\chib^\bj
\chib^\bl) + \sqrt{2} V_{k \bj}(\vpb \chib^\bj) \right]
(\lab^{(s)}\lab_{(s)})
\nonumber\\ [0.5ex] &&
+\f{9kL}{4} V_k  {\lp K_{k \bk}-3LV_{k\bk}\rp}^{-1} \left[
\cv_{j\, \bk l}\, (\chi^j \chi^l) + \sqrt{2} V_{\bk j}\,(\vp
\chi^j) \right](\la^{(s)}\la_{(s)}) \nonumber
\\ [0.5ex] &&
-\f{9 \sqrt{2}L^2}{4}  \cv_{\bj\, k \bl} {\lp K_{k
\bk}-3LV_{k\bk}\rp}^{-1}  V_{j \bk}\,(\chib^\bj \chib^\bl)\,(
\chi^j \vp) \nonumber\\[0.5ex] &&
 -\f{9 \sqrt{2}L^2}{4}
\cv_{j \bk l}  {\lp K_{k \bk}-3LV_{k\bk}\rp}^{-1}  V_{k
\bj}(\chi^j \chi^l)\,( \chib^\bj \vpb)\nonumber\\[0.5ex] &&
+\f{(\al-3) k^2}{16} \lp \f{\al}{L} -3V \rp^2 (\la^{(r)}\sigma^m
\lab_{(r)})(\la^{(s)}\sigma_m \lab_{(s)})-\f{\al^2 (\al-3)}{32}(\vp
\vp) (\vpb \vpb) \nonumber\\[0.5ex] && +\f{(\al-3)}{16}{\lp K_{k
\bk}-3LV_{k\bk}\rp} \lp \chi^k \sigma_m \chib^\bk \rp \left[2k \lp
\f{\al}{L} -3V \rp (\la^{(r)}\sigma^m \lab_{(r)})+ \al (\vp \sigma^m
\vpb) \right] \nonumber\\[0.5ex] && +\f{\al
(\al-3)k}{16}\lp\f{\al}{L} -3V \rp (\vp \sigma^m \vpb)(
\la^{(r)}\sigma_m \lab_{(r)})
 \nonumber\\[0.5ex] && +\f{1}{8k}{\lp\f{\al}{L} -3V \rp}^{-1}
 \left[(K_k-3LV_k) (T_{(r)} A)^k + (\bA T_{(r)})^\bk (K_\bk-3LV_\bk
)\right]^2 \nonumber \\ [0.5ex]
 &&  +\f{9k}{8}{\lp \f{\al}{L} -3V \rp}^{-1}
 \left[ V_k\,V_j (\chi^k \chi^j)(\la^{(r)} \la_{(r)})
  + V_\bk\, V_\bj (\chib^\bk \chib^\bj)(
  \lab^{(r)}\lab_{(r)})\right.
 \nn \\ && \hs{34mm}
 \left. -2 V_k \, V_\bk ( \chi^k \sigma^m \chib^\bk)( \la^{(r)} \sigma_m \lab_{(r)})
 \right] \nn \\[0.5ex] &&
  +\f{\al^2 k}{16L^2}{\lp \f{\al}{L} -3V \rp}^{-1}\left[
  (\lab^{(r)} \lab_{(r)})( \vpb \vpb)+ (\la^{(r)} \la_{(r)})
  ( \vp  \vp)-2( \la^{(r)} \sigma^m \lab_{(r)})(\vp \sigma_m \vpb) \right]
 \nonumber \\ [0.5ex]&&
 -\f{3 i}{2 \s2}{\lp \f{\al}{L} -3V \rp}^{-1}
 \left[(K_k-3LV_k) (T_{(r)} A)^k + (\bA T_{(r)})^\bk (K_\bk-3LV_\bk
)\right]\nn \\ &&\hs{32mm} \times \left[ V_j\,\chi^j\la^{(r)} -
V_\bj\,\chib^\bj \lab^{(r)} \right]
  \nonumber \\ [0.7ex]&&
 +\f{i \al}{4L} {\lp \f{\al}{L} -3V \rp}^{-1}
  \left[(K_k-3LV_k) (T_{(r)} A)^k + (\bA T_{(r)})^\bk (K_\bk-3LV_\bk
)\right]\nn \\ && \hs{29mm} \times\left[
  \lab^{(r)} \vpb - \la^{(r)} \vp  \right]
 \nonumber \\ [0.7ex]&&
+\f{3  k \al}{4 \s2 L} {\lp \f{\al}{L} -3V \rp}^{-1} \left[
V_k\,(\chi^k \vp)( \la^{(r)}\la_{(r)}) + V_\bk\,(\chib^\bk \vpb) (
\lab^{(r)} \lab_{(r)})
 \right]\nonumber \\ [0.5ex]&&
 - \f{3  k \al}{4 \s2 L} {\lp \f{\al}{L} -3V \rp}^{-1} \lp
 V_k\, \chi^k \sigma^m\vpb -  V_\bk \, \chib^\bk \sib^m\vp  \rp
 ( \la^{(r)} \sigma_m \lab_{(r)}).
\end{eqnarray}
Clearly, the role of effective transformations after elimination of the
auxiliary fields deserves further study.

\subsec{Component Field Lagrangian - II}
\label{appE5}

We can merge these new contributions into the Lagrangian and
eliminate trivially the auxiliary fields; this will yield a huge
expression which we simplify somehow by making the following changes
\begin{itemize}
\item {Change the \ka metric in the Lagrangian:}
\end{itemize}
Consider the \ka potential
\begin{eqnarray}\label{Khat}
  \hhK(\phi, \phib, L) &=& K(\phi,\phib,L)-3 L V(\phi, \phib) \nn
\\ &=& K_0(\phi,\phib)+{\al}\log L -3 L V(\phi, \phib),
\end{eqnarray}
we promote $\hhK_{k \bk} = K_{k \bk} -3L V_{k \bk}$ to a metric
denoted $G_{k \bk}$ and define symbols and tensors in this new
scheme. For instance:
\begin{equation}
  \hhK_{j \bj k} \,=\, g_{l \bj}\, {{\Gm_j}^l}_k -3L \cv_{j \bj\, k},
\end{equation}
so that we can define new Christoffel symbols
\begin{equation}
  {\widehat{\Gm}}_j{}^{l}{}_k \,\equiv G^{l \bj}\, \hhK_{j \bj\, k}
  \,=\, {{\Gm_j}^l}_k -3L G^{l \bj}\, \cv_{j \bj\, k},
\end{equation}
\begin{equation}
  {\widehat{\Gm}}_\bj{}^{\bl}{}_\bk \,\equiv G^{j \bl}\, \hhK_{\bj\,j \bk}
  \,=\, {{\Gm_\bj}^\bl}_\bk -3L G^{j \bl}\, \cv_{\bj \, \bk}
\end{equation}
and a curvature tensor
\begin{eqnarray}
  {\widehat R}_{j \bj\, k \bk} &\equiv& {\hhK}_{j \bj\, k \bk}-G_{l \bl}
 \, {\widehat{\Gm}}_j{}^{l}{}_k
  \,{\widehat{\Gm}}_{\bj}{}^{\bl}{}_\bk \nn \\
  &=& {R}_{j \bj\, k \bk} -3L \cv_{j \bj\, k \bk}-9 L^2\, \cv_{j \bl\,
  k}\,
  G^{l \bl}\, \cv_{\bj\, l \bk}.
\end{eqnarray}
We can then define the corresponding "hat" covariant derivatives
like ${\widehat {\cv}}_{i j}, {\widehat {\cv}}_{i \bk j}$, etc.

 Finally let us note that
\begin{equation}
  \f{\al}{L} -3 V \equiv \hhK_L,
\end{equation}
and that Yang-Mills invariance of $\hhK$ tells us
 \begin{equation}
\hhK_k (T_{(r)} A)^k \, =\, (\bA T_{(r)})^\bk \hhK_\bk,
\end{equation}
which again simplifies the expression of the Lagrangian. With the
new metric in the \ka connection we define new covariant
derivatives
 \begin{eqnarray}
{\widehat {\nabla}}_m A^i &\equiv&
  \cd_m
A^i  =  \prt_m A^i -i {\aym}^{(r)}_m \lp {{\bf T}_{(r)}} A
\rp^i,\nn
\\ {\widehat {\nabla}}_m \bar{A}^{\bj}&\equiv& \cd_m \bar{A}^{\bj}
 =  \prt_m \bar{A}^{\bj} +i {\aym}^{(r)}_m \lp \bar{A} {{\bf
T}_{(r)}}\rp^{\bj},\nn \\
  {\widehat {\nabla}}_m \chi_\alpha^i & = & \prt_m
\chi_\alpha^i - {\omega_{m \alpha}}^\varphi \chi_{\varphi}^i
 - i {\aym}^{(r)}_m \lp{{\bf T}_{(r)}} \chi_\alpha \rp^i
 +\chi^j_\alpha \ {\widehat{\Gamma}}^{i}_{j\,k} {\cal
D}_m A^k
  \nonumber \\ &&-\f{1}{4} ( \hhK_j \cd_m A^j -  \hhK_{\bj} \cd_m \bar{A}^{\bj})
\chi_\alpha^i
 - \f{i}{4} G_{j\bk} (\chi^j \si_m\chib^\bk)\ \chi^i_\alpha ,
 \\ {\widehat {\nabla}}_m \chib^{\bj \da} & = & \prt_m \chib^{\bj
\da} - {{\omega_{m}}^{\da}}_{\dot{\varphi}} \chib^{\bj
\dot{\varphi}} + i {\aym}^{(r)}_m \lp  \chib^ \da{{\bf T}_{(r)}}
\rp^\bj+ \chib^{\da\bi}\ {\widehat{\Gamma}}^{\bj}_{\bi \bk} {\cal
D}_m {\bar{A}}^{\bk}
 \nonumber \\ &&+ \f{1}{4} ( \hhK_k \cd_m A^k -  \hhK_{\bk} \cd_m
\bar{A}^{\bk}) \chib^{\bj \da}
 +\f{i}{4} G_{j\bk} (\chi^j \si_m\chib^\bk)\ \chib^{\bj\da}
, \end{eqnarray}
  \begin{eqnarray}
{\widehat{\nabla}}_n \psi^\alpha_m & = & \prt_n \psi^\alpha _m +
\psi_m^\beta {\omega_{n \beta}}^\alpha \nonumber
\\ && + {\psi_m}^\alpha \left(\f{1}{4} \hhK_i \cd_n A^i - \f{1}{4}
\hhK_{\bj}
 \cd_n A^{\bj}  + \f{i}{4} G_{i \bj} \chi^i \sigma_n \chib^{\bj} \right),
 \\ {\widehat{\nabla}}_n \psib_{m \da} & = & \prt_n \psib_{m \da}
+ \psib_{m \db} {{\omega_n}^{\db}}_{\da} \nonumber
\\ && - \psib_{m \da} \left(\f{1}{4} \hhK_i \cd_n A^i - \f{1}{4}
\hhK_{\bj} \cd_n A^{\bj}  + \f{i}{4} G_{i \bj} \chi^i \sigma_m
\chib^{\bj} \right)  \end{eqnarray}
\begin{eqnarray}
 {\widehat{\nabla}}_m \lambda_\alpha^{(r)}  \ & = &
\prt_m \lambda_\alpha^{(r)} - {\omega_{m \alpha}}^\varphi
\lambda_{\varphi}^{(r)}-
 {\aym}^{(t)}_m   {c_{(s)(t)}}^{(r)} \lambda^{(s)}_\alpha
 \nonumber \\ && + \f{1}{4} (\hhK_j \cd_m A^j -  \hhK_{\bj} \cd_m
\bar{A}^{\bj}) \lambda_\alpha^{(r)} + \f{i}{4} G_{j\bk} (\chi^j
\si_m\chib^\bk)\ \lambda^{(r)}_\alpha, \nn \\ {\widehat{\nabla}}_m
\lab^{{(r)}\da} & = & \prt_m \lab^{{(r)} \da} -
{{\omega_{m}}^{\da}}_{\dot{\varphi}} \lab^{{(r)} \dot{\varphi}} -
{\aym}_m^{(t)}{c_{(s)(t)}}^{(r)} \lab^{{(s)} \da} \nonumber \\ &&
- \f{1}{4} ( \hhK_k \prt_m A^k -  \hhK_{\bk} \prt_m \bar{A}^{\bk})
\lab^{{(r)} \da}
 -\f{i}{4} G_{j\bk} (\chi^j \si_m\chib^\bk)\
\lab^{{(r)}\da}. \end{eqnarray}
\begin{itemize}
\item { Make a shift on ${\, {}^* \!h}_m$}
\end{itemize}
\begin{eqnarray}
{\, {}^* \!h}_m \mapsto {\, {}^* \!h}_m  &+&\f{iL}{2}
\left[\vep^{mnpq}  (\psi_n \si_p \psib_q)-i\, G_{k \bk} (\chi^k
\si_m \chib^\bk)\right.\nn \\ &&\left.\hs{8mm} -\f{i \al}{2} (\vp
\si_m \vpb)+ik\lp\f{2}{L}- \hhK_L \rp(\la^{(r)} \si_m
\lab_{(r)})\right].
\end{eqnarray}
Putting everything together
 this gives rise the new Lagrangian
\begin{eqnarray} &&\f{1}{e} \cl = - \f{1}{2} {\cal R}
          + \f{1}{2} \vep^{mnpq} \lp \psib_m \sib_n {\hat \nabla}_p \psi_q
          - \psi_m \si_n {\hat \nabla}_p \psib_q \rp
\nonumber \\ [0.5ex]
      & & -G_{k \bk}  {\hat \nabla}_m A^k {\hat \nabla}^m \bA^\bk
 - \f{i}{2}G_{k \bk}  \lp \chi^k \si^m {\hat \nabla}_m \chib^\bk
 + \chib^\bk \sib^m {\hat \nabla}_m \chi^k \rp
\nonumber \\ [0.5ex]
      & & + \f{\al}{4 L^2} {\, {}^* \! h}^m {\, {}^* \!h}_m
          - \f{\al}{4L^2}\,  \prt^m L \, \prt_m L
          - \f{i \al}{4} \lp \vp \si^m {\hat \nabla}_m \vpb
     + \vpb \sib^m {\hat \nabla}_m \vp \rp
\nonumber \\ [0.5ex]
      & & + \f{k}{4}\hhK_L \, {\fym}_{mn}^{(r)}\, {\fym}^{mn}_{(r)}
  + \f{i k}{2} \hhK_L \lp \la^{(r)} \si^m {\hat \nabla}_m \lab_{(r)}
   + \lab^{(r)} \sib^m {\hat \nabla}_m \la_{(r)} \rp
 \nonumber \\ [0.5ex] & &
 \nonumber \\ [0.5ex] & & +\f{1}{8k}{\lp {\hhK_L} \rp}^{-1}
 \left[\hhK_k (T_{(r)} A)^k + (\bA T_{(r)})^\bk \hhK_\bk\right]^2
 \nonumber \\ [0.5ex]&&
 -\f{3}{2}i {\,
{}^* \!h^m} ( V_k{\hat \nabla}_m A^k -V_\bk{\hat
\nabla}_m\bA^\bk)-\f{\al}{2L}{}^* \!h^m \, (\vp \si_m \vpb)
 +\f{3}{2} V_{k\bk} \, {}^* \!h^m \,(\chi^k\si_m\chib^\bk)
 \nonumber \\ [0.5ex]& &- \left[\f{3 k}{\s2}
  \lp  V_k\,\chi^k\si^{mn}\la_{(r)}
  +V_\bk\,\chib^\bk\sib^{mn}\lab_{(r)}  \rp
  + \f{\al k}{2L} \lp \vp \si^{mn} \la_{(r)}
+\vpb \sib^{mn} \lab_{(r)} \rp \right]  {\fym}^{(r)}_{mn}
 \nonumber\\ [0.5ex] &&
 +i(T_{(r)}A)^k \left[\lp  \sqrt{2} G_{k
 \bk}+\f{3}{\s2}\f{\hhK_k}{\hhK_L} \, V_\bk \rp
     ( \chib^\bk \lab^{(r)})-\lp 3LV_k -\f{\al}{2L}
    \f{\hhK_k}{\hhK_L}\rp
      (\vpb \lab^{(r)})\right]\nonumber \\ [0.5ex]&&
 -i(\bA T_{(r)})^\bk \left[ \lp \sqrt{2} G_{k \bk}+
\f{3}{\s2}\f{\hhK_\bk}{\hhK_L} \, V_k \rp ( \chi^k \la^{(r)})-\lp
3LV_\bk-\f{\al}{2L}\f{\hhK_\bk}{\hhK_L} \rp (\vp \la^{(r)})
  \right]\nonumber \\ [0.5ex]  & &
  + \f{i3 L}{\s2}  \left[  V_{k\bk}\,(\chib^\bk\sib^m \vp)
{\hat \nabla}_m A^k + V_{k\bk}\,(\chi^k\si^m \vpb)\,{\hat
\nabla}_m \bA^\bk
 \right]
 \nn \\[0.5ex] &&
+ \f{1}{4} \lp{\widehat R}_{k \bk l \bl} +\f{3}{2} G_{k \bk} G_{l
\bl} \rp  (\chi^k \chi^l) (\chib^\bk\chib^\bl) - \f{\al
(\al+12)}{32} (\vp \vp) (\vpb \vpb) \nonumber\\ [0.5ex] && -
\f{k^2}{4}\lp \f{\al}{ L^2}  +9 V_k \,{G^{k \bk}}\, V_\bk
\rp(\la^{(r)} \la_{(r)})(\lab^{(s)} \lab_{(s)})
 \nonumber \\ [0.5ex] &&- \f{3 k^2}{16} \, {\hhK_L}^2 \,
 (\la^{(r)} \si^m \lab_{(r)})\,(\la^{(s)} \si_m \lab_{(s)})
 \nonumber \\ [0.5ex]&&
+ \f{3 k }{4} \lp {\widehat \cv}_{kl}+\f{3}{2}{\hhK_L}^{-1} V_k
V_l \rp (\chi^k\chi^l)(\la^{(r)}\la_{(r)}) \nn \\ && + \f{3 k}{4}
\lp {\widehat\cv}_{\bk\bl} +\f{3}{2}{\hhK_L}^{-1} V_\bk\, V_\bl
\rp
   (\chib^\bk\chib^\bl)(\lab^{(r)}\lab_{(r)})  \nonumber \\ [0.5ex]
&& - \f{\al k}{4 L}\lp 1- \f{\al}{4 L {\hhK_L}}\rp
 \left[ (\lab^{(r)} \lab_{(r)}) (\vpb \vpb)
  + (\la^{(r)} \la_{(r)}) (\vp \vp) \right]
\nonumber \\ [0.5ex] &&  -\f{\al k}{16 L} \hhK_L\lp 1+ \f{2
\al}{{\hhK_L}^2 L^2}\rp
  (\la^{(r)} \si^m \lab_{(r)})(\vp \si_m \vpb) \nonumber \\ [0.5ex]
&& + \left[-\f{\al}{16}G_{k \bk}
  +\f{9 L^2}{4}  V_{j \bk}\, {G^{j\bj}}\, V_{k \bj} \right]
 \lp \chi^k \si^m \chib^\bk \rp
 \lp \vp \si_m \vpb \rp   \nonumber\\ [0.5ex]&&
 -\f{3k}{8} \lp {\hhK_L}\, G_{k \bk}\,+ 6{\hhK_L}^{-1} V_k\, V_\bk
\rp
  (\chi^k \si^m \chib^\bk)(\la^{(r)} \si_m \lab_{(r)})
  \nonumber \\ [0.5ex]&&
+\f{3 k}{4\sqrt{2}L}\lp 6 L^2 V_\bk\,  {G^{k \bk}} \,V_{k \bj} +
\al  {\hhK_L}^{-1} V_\bj \rp( \chib^\bj \vpb)
(\lab^{(s)}\lab_{(s)})\nn
\\ && +\f{3k}{4\sqrt{2}L}\lp 6 L^2 V_k \, {G^{k \bk}}\, V_{\bk j}+
\al {\hhK_L}^{-1} V_j\rp (\chi^j \vp)(\la^{(s)}\la_{(s)})
\nonumber\\ [0.5ex] && -\f{3 k \al}{4  \sqrt{2}L} {\lp {\hhK_L}
\rp}^{-1} \left[ V_k\,(\chi^k \si_m \vpb) - V_\bk\,(\chib^\bk
\sib_m \vp ) \right]
 (\la^{(r)}\si^m \lab_{(r)})
 \nonumber \\ [0.5ex]&& - \f{3  L}{2\s2}\left[
{\widehat {\cv}}_{\bk\, j\, \bl}\,(\chib^\bl\chib^\bk)\, (\chi^j
\vp) +\,{\widehat {\cv}}_{k \bj\, l}\,(\chi^l\chi^k)\,(\chib^\bj
 \vpb)\right]
 \nonumber\\ [0.5ex] && \nonumber\\ [0.5ex] &&
   -\f{1}{2} \, \hhK_k \, (T_{(r)}A)^k
\,  (\psib_m \sib^m \la^{(r)}) +\f{1}{2} \, \hhK_\bk \,(\bA
T_{(r)})^\bk \,(\psi_m \si^m \lab^{(r)})\nonumber \\ [0.5ex]  & &
 - \f{1}{\sqrt{2}}G_{k \bk}
  \left[ (\psi_n \si^m \sib^n \chi^k) {\hat \nabla}_m \bA^\bk
  + (\psib_n \sib^m \si^n \chib^\bk) {\hat \nabla}_m A^k \right]
  \nonumber \\ [0.5ex] &&
 -\f{\al}{4L} (\psi_n \si^m \sib^n \vp
 + \psib_n \sib^m \si^n \vpb) \prt_m L+ \f{i \al}{4L}
 (\psi_n \si^m \sib^n \vp - \psib_n \sib^m \si^n \vpb){\, {}^* \!h}_m
  \nonumber\\[0.5ex]
  & & + \f{i \al k}{8 L}
\left[ (\psib_m \sib^m \vp)(\la^{(r)} \la_{(r)}) + (\psi_m \si^m
\vpb)(\lab^{(r)} \lab_{(r)}) \right] \nonumber \\[0.5ex] & & -
\f{i \al}{16} (\al-4) (\vp \si^m \vpb)
 \lp\,\psi_m  \vp \,+\,\psib_m \vpb \, \rp
    \nonumber\\ [0.5ex]&&
 - \f{i 3 k}{4\s2}
   V_k(3 \chi^k\psi^m +2\chi^k \si^{mn}\psi_n)(\la^{(r)}\si_m\lab_{(r)})
   \nonumber\\ [0.5ex]&& + \f{i 3 k}{4\s2}
   V_\bk(3\chib^\bk \psib^m +2\chib^\bk \sib^{mn}\psib_n)(\la^{(r)}\si_m\lab_{(r)})
 \nonumber\\ [0.5ex]&& - \f{i 3 k}{4\s2} \l[ \, V_k(\chi^k \si^m \psib_m)(\la^{(r)}\la_{(r)})
 +V_\bk(\chib^\bk \sib^m \psi_m)(\lab^{(r)}\lab_{(r)}) \, \r]
\nonumber \\ [0.5ex] & & - \f{i \al k}{4 L}  \lp
 \psi^m \vp -\psib^m \vpb \rp(\la^{(r)} \si_m \lab_{(r)}) \nonumber \\ [0.5ex]
& &  +\f{i \al k}{8L}  \lp \psi_n \si^m \sib^n \vp
  -\psib_n \sib^m \si^n\vpb \rp(\la^{(r)} \si_m \lab_{(r)})
  \nonumber  \\ [0.5ex] & &
   +\f{3i}{4}LV_{k\bk}(\chi^k \si_m \chib^\bk)
   (\psi_n \si^m \sib^n \vp - \psib_n \sib^m \si^n \vpb)
\nonumber \\ [0.5ex]& & -\f{i \al (\al+4)}{32} (\psi_n \si^m
\sib^n \vp - \psib_n \sib^m \si^n \vpb)(\vp \si_m \vpb ) \nonumber
\\ [0.5ex]& & + \f{1}{4}G_{k \bk} (\chi^k \si_m \chib^\bk)
 \lp \psi_n \si^m \psib^n
 - \f{i}{2} \vep^{mnpq} \psi_n \si_p \psib_q \rp
 \nonumber\\ [0.5ex]& & - \f{ik}{2}  \hhK_L \left[ \lp {\fym}^{(r)\, mn}
 + i\,\, {}^* \!{\fym}^{(r)\, mn} \rp
 (\psi_m \si_n \lab_{(r)})+
\lp {\fym}^{(r)\, mn} -i\, \, {}^* \!{\fym}^{ (r)\,mn} \rp
(\psib_m \sib_n \la_{(r)}) \right]\nonumber
 \\ [0.5ex] &&+\f{1}{16 L} \hhK_L \l[
(3\psi_n\psi^n + 2\psi_n\si^{nm}\psi_m)(\la^{(r)}\la_{(r)}) +
(3\psib_n\psib^n+2\psib_n\sib^{nm}\psib_m)(\lab^{(r)}\lab_{(r)})
\r]\nonumber
\\ [0.5ex] & & + \f{k}{8}\hhK_L
            \lp {\oint}_{\!npq}g^{mn} g^{pq}
 - i \vep^{mnpq} \rp (\psi_m \si_n \psib_p)(\la^{(r)}
 \si_q \lab_{(r)})\nonumber\\ [0.5ex]
  & &  + \f{\al}{8}\l[ (\psi_m \psi^m) (\vp \vp)+
     (\psib_m \psib^m) (\vpb \vpb)\r] \nonumber \\ [0.5ex] &&
 + \f{\al}{8} \lp g^{mn} g^{pq} + g^{mq} g^{np} \rp
  (\psi_m \si_n \psib_p)(\vp \si_q \vpb).
\end{eqnarray}

\newpage
\sect{ 3-form gauge potential and Chern-Simons forms \label{appF}}

The analogy between Chern-Simons forms and 3-form gauge potentials
will be employed to determine the Chern-Simons superfield
(\ref{CSsfi}). To this end we present first the explicit solution
of the 4-form constraints in terms of an unconstrained superfield.
Already important by itself, in the description of constrained
chiral multiplets -cf. section \ref{3F}- this analysis underlies
the explicit construction of the Chern-Simons superfield. After
some general remarks and definitions concerning Chern-Simons forms
in superspace, the Chern-Simons superfield is determined as the
counterpart of the pre-potential of the 3-form.

 \subsec{Explicit Solution of the Constraints \label{appF1}}

As shown in the main text, the constraints
 \begin{equation}
 \Si_{\unddt \, \undgm \, \undbt \, A} \
= \ 0, \end{equation} allow to express all the coefficients of the
4-form field strength in terms of the constrained chiral fields
$Y, \ovY$. The Bianchi identities in the presence of the
constraints are summarized in the chirality conditions together
with the additional constraint (\ref{con3f}). Alternatively, as we
will explain now, the explicit solution of the superspace
constraints allows us to determine the unconstrained pre-potential
of the constrained superfield. An important ingredient in this
procedure will be the use of the gauge freedom of the 3-form
potential, $C$, parametrized by a 2-form $\La$,
 \begin{equation}
 {}^\La {C}_{CBA} ={C}_{CBA} + {\oint}_{CBA} \lp \cd_C
 \La_{BA} + {T_{CB}}^F \La_{FA} \rp.  \end{equation}
 As usual ${\oint}_{CBA}$ denotes the graded sum $CBA
 +(-)^{c(b+a)}BAC+(-)^{a(b+c)}ACB$.
 In a first step consider
\begin{equation}
\Si_{\dt \, \gm \, \bt \, A} \,=\,0, \end{equation}
 which we satisfy with
\begin{equation}
C{}_{\gm \bt A} \,=\,\cd_A U_{\gm \bt} +
    \oint_{\gm \bt} \lp \cd_\gm U_{\bt A} + T_{A \gm}{}^F U_{F \bt} \rp,
\end{equation}
 and the complex conjugate
  \begin{equation}
  \Si^{\dd \, \dg \,
\db}{}_A \,=\,0, \end{equation}
    by
\begin{equation}
C{}^{\dg \db}{}_A \,=\,\cd_A V^{\dg \db} +
    \oint^{\dg \db} \lp \cd^\dg V^\db{}_A + T_A{}^{\dg \, F} V_F{}^\db \rp.
\end{equation}

Since the pre-potentials $U_{\bt A}$ and $V^\db{}_A$ should
reproduce the gauge transformations of the gauge potentials
$C{}_{\gm \bt A}$ and $C{}^{\dg \db}{}_A$ we assign
\begin{equation}
U_{\bt A} \ \mapsto \ {}^\La U_{\bt A} \,=\, U_{\bt A} + \La_{\bt
A}, \end{equation}
 and
\begin{equation}
V^\db{}_A \ \mapsto \ {}^\La V^\db{}_A \,=\, V^\db{}_A +
\La^\db{}_A, \end{equation}
 as gauge transformation
laws for the pre-potentials. On the other hand, the so-called {\em
pre-gauge transformations} are defined as the zero-modes of the
gauge potentials themselves, that is transformations which leave
$C{}_{\gm \bt A}$ and $C{}^{\dg \db}{}_A$ invariant. They are
given as
 \begin{equation}
 U_{\bt A} \ \mapsto \ U_{\bt A}
      + \cd_\bt \chi_A - (-)^a \cd_A \chi_\bt + T_{\bt A}{}^F \chi_F, \end{equation}
and
 \begin{equation}
 V^\db{}_A \ \mapsto \ V^\db{}_A
      + \cd^\db \psi_A - (-)^a \cd_A \psi^\db + T^\db{}_A{}^F \psi_F. \end{equation}

We parametrize the pre-potentials now as follows:
 \begin{eqnarray}
U_\bt{}^\da &=& W_\bt{}^\da + T_\bt{}^{\da \, f} K_f, \\
V^\db{}_\al &=& W_\al{}^\db - T_\al{}^{\db \, f} K_f,
\end{eqnarray}
 and
\begin{eqnarray}
 U_{\bt \, a} &=& W_{\bt \, a} - \cd_\bt K_a, \\ V^\db{}_a &=&
W^\db{}_a + \cd^\db K_a. \end{eqnarray}
 Explicit substitution
shows that the $K_a$ terms drop out in $C{}_{\gm \bt A}$ and
$C{}^{\dg \db}{}_A$. Denoting furthermore
\begin{equation}
U_{\bt \al} \,=\,W_{\bt \al}, \cem {\rm and} \cem V^{\db \da} \ =
\ W^{\db \da}, \end{equation}
 we arrive at
 \begin{eqnarray}
C{}_{\gm \bt A} &=& \cd_A W_{\gm \bt} +
    \oint_{\gm \bt} \lp \cd_\gm W_{\bt A} + T_{A \gm}{}^F W_{F \bt} \rp, \\
C{}^{\dg \db}{}_A &=& \cd_A W^{\dg \db} +
    \oint^{\dg \db} \lp \cd^\dg W^\db{}_A + T_A{}^{\dg \, F} W_F{}^\db \rp,
\end{eqnarray}
 \ie a pure gauge form for the coefficients $C{}_{\gm \bt
A}$ and $C{}^{\dg \db}{}_A$ with the 2-form gauge parameter $\La$
replaced by the pre-potential 2-form
 \begin{equation}
 W \,=\,
\f12 E^A E^B W_{BA}, \cem {\rm with} \cem W_{ba} \,=\,0.
\end{equation}
 We take advantage of this fact to perform a
redefinition of the 3-form gauge potentials, which has the form of
a gauge transformation,
\begin{equation}
\hC \ := \ {}^{-W} C \,=\,C - dW. \end{equation}
 This
leaves the field strength invariant and leads in particular to
\begin{equation}
\hC{}_{\gm \bt A} \,=\,0, \cem \mbox{and} \cem \hC{}^{\dg
\db}{}_A \,=\,0, \end{equation}
 whereas the coefficient
$C{}_\gm{}^\db{}_a$ is replaced by
\begin{equation}
\hC{}_\gm{}^\db{}_a \,=\,C{}_\gm{}^\db{}_a
     - \cd_\gm W^\db{}_a - \cd^\db W_{\gm a} - \cd_a W_\gm{}^\db.
\end{equation}

We define the tensor decomposition
\begin{equation}
\hC{}_\gm{}^\db{}_a \,=\,
     T_\gm{}^\db{}^f \lp \eta_{fa} \, \Om + \hW_{[fa]} + {\tilde{\Om}}_{(fa)} \rp,
\end{equation}
 where $\hW_{[fa]}$ is antisymmetric and ${\tilde{\Om}}_{(fa)}$
symmetric and traceless, and perform another redefinition which
has again the form of a gauge transformation, this time of
parameter
\begin{equation}
\hW \,=\,\f12 E^a E^b \hW_{[ba]},
\end{equation}
 such that
 \begin{equation}
 \Om \ := \ {}^{-\hW}
\hC \,=\,\hC - d \hW. \end{equation}
 Note that this
reparametrization leaves $\hC{}_{\gm \bt A}$ and $\hC{}^{\dg
\db}{}_A$ untouched, they remain zero.

Let us summarize the preceding discussion: we started out with the
3-form gauge potential $C$. The constraints on its field strength
led us to introduce pre-potentials. By means of pre-potential
dependent redefinitions of $C$, which have the form of gauge
transformations (and which, therefore, leave the field strength
invariant), we arrived at the representation of the 3-form gauge
potential in terms of $\Om$, with the particularly nice properties
\begin{equation}
\Om_{\gm \bt A} \,=\,0, \cem \Om^{\dg \db}{}_A \,=\,0,
\end{equation}
 and
\begin{equation}
\Om_\gm{}^\db{}_a \,=\,
     T_\gm{}^\db{}^f \lp \eta_{fa} \, \Om + {\tilde{\Om}}_{(fa)} \rp,
\end{equation}
 Clearly, in this representation, calculations simplify
considerably. We shall therefore, from now on, pursue the solution
of the constraints in terms of $\Om$ and turn to the equation
\begin{equation}
\Si_{\dt \gm}{}^{\db \da} \,=\,0 \,=\,
      \oint_{\dt \gm}^{\db \da} T_\dt{}^{\dg \, f} \Om_f{}^\db{}_\al,
\end{equation}
 which tells us simply that ${\tilde{\Om}}_{(ba)}$ is zero.
Hence,
\begin{equation}
\Om_\gm{}^\db{}_a \,=\,T_\gm{}^\db{}_a \, \Om. \end{equation}

We turn next to the constraints
\begin{equation}
\Si_\dt{}^{\dg \db}{}_a \,=\,0 \,=\,\oint^{\dg \db}
   \lp \cd^\dg \Om_\dt{}^\db{}_a + T_\dt{}^{\dg \, f} \Om_f{}^\db{}_a \rp,
\end{equation}
 and
\begin{equation}
\Si^\dd{}_{\gm \bt a} \,=\,0 \,=\,\oint_{\gm \bt}
   \lp \cd_\gm \Om^\dd{}_{\bt a} + T_\gm{}^{\dd \, f} \Om_{f \bt a} \rp,
\end{equation}
 which, after some straightforward spinor index gymnastics
give rise to
 \begin{eqnarray}
  \Om_{\gm \, ba} &=& 2
(\si_{ba})_\gm{}^\vp \, \cd_\vp \Om, \\ \Om^\dg \, {}_{ba} &=& 2
(\sib_{ba})^\dg{}_\dv \, \cd^\dv \Om. \end{eqnarray}

This completes the discussion of the solution of the constraints,
we discuss next the consequences of this solution for the
remaining components in $\Si$ \ie $\Si_{\unddt \, \undgm \, ba}$,
$\Si_{\unddt \, cba}$ and $\Si_{dcba}$. As a first step we
consider
\begin{equation}
\Si_{\dt \, \gm \, ba} \,=\,\oint_{\dt \gm} \lp \cd_\dt \Om_{\gm
\, ba} - T_{\dt b \dv} \Om_\gm{}^\dv{}_a + T_{\dt a
\dv}\Om_\gm{}^\dv{}_b \rp, \end{equation}
 and
\begin{equation}
\Si^{\dd \dg}{}_{ba} \,=\,\oint^{\dd \dg}
   \lp \cd^\dd \Om^\dg{}_{ba}
   - T^\dd{}_b{}^\vp \Om_\vp{}^\dg{}_a + T^\dd{}_a{}^\vp \Om_\vp{}^\dg{}_b \rp.
\end{equation}

Substituting for the 3-form gauge potentials as determined so far,
and making appropriate use of the supergravity Bianchi identities
yields
\begin{equation}
\Si_{\dt \, \gm \, ba} \,=\,-2 (\si_{ba} \eps)_{\dt \gm} \ \prokib
\Om, \end{equation}
 and
\begin{equation}
\Si^{\dd \dg}{}_{ba} \,=\,-2 (\sib_{ba}\eps)^{\dd \dg} \ \proki
\Om.
\end{equation}
 The appearance of the chiral projection operators suggests to
define
\begin{eqnarray}
 \ovY &=& -4 \prokib \Om, \\ Y &=& -4 \proki \Om.
\end{eqnarray}
The gauge invariant superfields $Y$ and $\ovY$ have chirality
properties
\begin{equation}
 \cd_\al \ovY \,=\,0, \cem \cd^\da Y \,=\,
0,
\end{equation}
 and we obtain
 \begin{eqnarray}
  \Si_{\dt \gm \, ba}
&=& \f12 (\si_{ba}\eps)_{\dt \gm} \, \ovY, \\ \Si^{\dd
\dg}{\,}_{ba} &=& \f12 (\sib_{ba}\eps)^{\dd \dg} \, Y.
\end{eqnarray}

In the next step we observe that, due to the information extracted
so far from the solution of the constraints, the field strength
\begin{equation}
\Si_\dt{}^\dg{\,}_{ba} \,=\,T_\dt{}^{\dg \, c} \, \Si_{cba},
\end{equation}
 is determined such that $\Si_{cba}$ is totally
antisymmetric in its three vector indices. As, in its explicit
definition a linear term appears (due to the constant torsion
term), \ie
\begin{equation}
\Si_\dt{}^\dg{\,}_{ba} \,=\,T_\dt{}^{\dg \, c} \Om_{cba} +
            {\rm \ derivative \ and \ other \ torsion \ terms},
\end{equation}
 we can absorb $\Si_{cba}$ in a modified 3-form gauge potential
\begin{equation}
\undOm_{cba} \,=\,\Om_{cba} - \Si_{cba}, \end{equation}
 such that
the corresponding modified field strength vanishes, \ie
\begin{equation}
\undSi_\dt{}^\dg{\,}_{ba} \,=\,0. \end{equation}
 The outcome  of
this discussion is then the relation
\begin{equation}
\lp [\cd_\al, \cd_\da] - 4 G_{\al \da} \rp \Om
             \,=\, -\f13 \si_{d\, \al \da} \, \eps^{dcba} \undOm_{cba},
\end{equation}
 which identifies $\undOm_{cba}$ in the superfield expansion of
the unconstrained pre-potential $\Om$.

Working, from now on, in terms of the modified quantities, the
remaining coefficients, at canonical dimensions 3/2 and 2, \ie
${\undSi}_{\,\unddt \, cba}$ and ${\undSi}_{\,dcba}$,
respectively, are quite straightforwardly obtained in terms of
spinor derivatives of the basic gauge invariant superfields $Y$
and $\ovY$. To be more precise, at dimension 3/2 one obtains
 \begin{eqnarray}
  \undSi_{\,\dt
\, cba} &=&
         -\f{1}{16} \si^d_{\dt \dd} \, \eps_{dcba} \, \cd^\dd \, \ovY,
 \\ \undSi^{\,\dd} {}_{cba} &=&
  +\f{1}{16} \sib^{d \, \dd \dt} \, \eps_{dcba} \, \cd_\dt\, Y,
\end{eqnarray}
 and the Bianchi identity at dimension 2 takes the simple form
\begin{equation}
\lp \cd^2 - 24 \rd \rp Y - \lp \cdb^2 - 24 R \rp \ovY \,=\,
         \f{8i}{3} \eps^{dcba} \undSi_{\,dcba}.
\end{equation}

As to the gauge structure of the 3-form gauge potential we note
that in the transition from $C$ to $\Om$, the original 2-form
gauge transformations have disappeared, $\Om$ is invariant under
those. In exchange, however, as already mentioned earlier, $\Om$
transforms under so-called pre-gauge transformations which, in
turn, leave $C$ unchanged. As a result, the residual pre-gauge
transformations of the unconstrained pre-potential superfield,
\begin{equation}
 \Om \ \mapsto \ \Om' \,=\,\Om + \la,
\end{equation}
 are parametrized in terms of a linear superfield
$\la$ which satisfies
\begin{equation}
\prokib \,\la \,=\,0, \cem \proki \, \la \,=\,0. \end{equation}
 In turn, $\la$ can be expressed in terms of an unconstrained
superfield, as we know from the explicit solution of the
superspace constraints of the 2-form gauge potential, actually
defining the linear superfield geometrically. In other words, the
pre-gauge transformations should respect the particular form of
the coefficients of the 3-form $\Om$.

\subsec{Chern-Simons Forms in Superspace  \label{appF2}}

 Under gauge transformations the
Chern-Simons 3-forms change by the exterior derivative of a
2-form, which depends on the gauge parameter and the gauge
potential. Due to this property one may view the Chern-Simons form
as a special case of a generic 3-form gauge potential -cf. the
preceding subsection. This point of view is particularly useful
for the supersymmetric case. To be as clear as possible we first
recall some general properties of Chern-Simons forms in
superspace.

To begin with we consider two gauge potentials $\ca_0$ and $\ca_1$
in superspace. Their field strength squared invariants are related
through
\begin{equation}
\tr \lp \cf_0 \cf_0 \rp - \tr \lp \cf_1 \cf_1 \rp \,=\,
                                        d \cq \lp \ca_0, \ca_1 \rp.
\end{equation}
 This is the superspace version of the Chern-Simons formula,
where
\begin{equation}
\cf_0 \,=\,d \ca_0 + \ca_0 \ca_0, \cem \cf_1 \,=\,d \ca_1 + \ca_1
\ca_1. \end{equation}
 On the right appears the superspace
Chern-Simons form,
\begin{equation}
\cq \lp \ca_0, \ca_1 \rp \,=\,
          2 \int_0^1 dt \ \tr \left\{ \lp \ca_0 - \ca_1 \rp \cf_t \right\},
\end{equation}
 where
  \begin{equation}
  \cf_t \,=\,d \ca_t + \ca_t \ca_t, \end{equation}
is the field strength for the interpolating gauge
potential
\begin{equation}
 \ca_t \,=\,(1-t) \ca_0 + t \ca_1.
\end{equation}
 The Chern-Simons form is antisymmetric in its
arguments, \ie
 \begin{equation}
 \cq \lp \ca_0, \ca_1 \rp \,=\,-
\cq \lp \ca_1, \ca_0 \rp.
\end{equation}
 In the particular case $\ca_0=\ca$, $\ca_1=0$, one
obtains
\begin{equation}
\cq \lp \ca \rp \ := \ \cq \lp \ca, 0 \rp \,=\,
                  \tr \lp \ca \cf- \f{1}{3} \ca \ca \ca \rp.
\end{equation}
 We shall also make use of the identity
\begin{equation}
\cq \lp \ca_0, \ca_1 \rp + \cq \lp \ca_1, \ca_2 \rp + \cq \lp
\ca_2, \ca_0 \rp \,=\,d \chi \lp \ca_0, \ca_1, \ca_2 \rp,
\end{equation} with
\begin{equation}
\chi \lp \ca_0, \ca_1, \ca_2 \rp
    \,=\,\tr \lp \ca_0 \ca_1 + \ca_1 \ca_2 + \ca_2 \ca_0 \rp.
\end{equation}
This last relation (the so-called {\em triangular equation}) is
particularly useful for the determination of the gauge
transformation of the Chern-Simons form. The argument goes as
follows: first of all, using the definition given above, one
observes that
\begin{equation}
\cq \lp {}^{\gym} \ca, 0 \rp \,=\, \cq \lp \ca, d{\gym} \,
{\gym}^{-1} \rp. \end{equation}
 Combining
this with the triangular equation for the special choices
\begin{equation}
\ca_0 \,=\,0, \cem \ca_1 \,=\,\ca, \cem \ca_2 \,=\,d{\gym} \,
{\gym}^{-1}, \end{equation}
 one obtains
\begin{equation}
\cq \lp 0, \ca \rp + \cq \lp {}^{\gym} \ca, 0 \rp + \cq \lp
d{\gym} \, {\gym}^{-1}, 0 \rp
   \,=\,d \, \tr \lp \ca \,d{\gym} \, {\gym}^{-1} \rp,
\end{equation}
 or, using the antisymmetry property
\begin{equation}
\cq \lp {}^{\gym} \ca \rp - \cq \lp \ca \rp \,=\,
   d \, \tr \lp \ca \,d{\gym} \, {\gym}^{-1} \rp - \cq \lp d{\gym} \, {\gym}^{-1} \rp.
\end{equation}
The last term in this equation is an exact differential form
in superspace as well, it can be written as
\begin{equation}
\cq \lp d{\gym} \, {\gym}^{-1} \rp \,=\,d \si, \end{equation}
 where
the 2-form $\si$ is defined as
\begin{equation}
\si \,=\,\int_0^1 dt \ \tr \lp
          \prt_t {\gym}_t {\gym}_t^{-1} \, d{\gym}_t {\gym}
          _t^{-1} \, d{\gym}_t {\gym}_t^{-1} \rp,
\end{equation}
 with the interpolating group element ${\gym}_t$ parametrized
such that for $t \in  [\,0,1]$
 \begin{equation}
 {\gym}_0 \,=\,1,
\cem {\gym}_1 \,=\,{\gym}. \end{equation}

This shows that the gauge transformation of the Chern-Simons form,
which is a 3-form in superspace, is given as the exterior
derivative of a 2-form,
\begin{equation}
\cq \lp {}^{\gym} \ca \rp - \cq \lp \ca \rp \,=\,d \Dt({\gym},
\ca), \label{Qgtf}
\end{equation}
 with $\Dt = \chi-\si$.

The discussion so far was quite general and valid for some generic
gauge potential. It does not only apply to the Yang-Mills case but
to gravitational Chern-Simons forms as well.

\subsec{The Chern-Simons Superfield  \label{appF3}}

We specialize here to the Yang-Mills case, \ie we shall now take
into account the covariant constraints on the field strength,
which define supersymmetric Yang-Mills theory. It is the purpose
of the present subsection to elucidate the relation between the
unconstrained pre-potential, which arises in the constrained
3-form geometry, and the Chern-Simons superfield. Moreover, based
on this observation and on the preceding subsections we present a
geometric construction of the explicit form of the Yang-Mills
Chern-Simons superfield in terms of the unconstrained
pre-potential of supersymmetric Yang-Mills theory.

In this construction of the Chern-Simons superfield we will
combine the knowledge acquired in the discussion of the 3-form
gauge potential with the special features of Yang-Mills theory in
superspace. Recall that the Chern-Simons superfield $\Om^{\ym}$ is
identified in the relations
\begin{eqnarray}
 \tr \lp \cw_\da \cw^\da \rp &=&
\f12 \prokib \Om^{\ym}, \\ \tr \lp \cw^\al \cw_\al \rp &=& \f12
\proki \Om^{\ym}. \end{eqnarray}
 The appearance of one and the same
superfield under the projectors reflects the fact that the gaugino
superfields $\cw_\undal$ are not only subject to the chirality
constraints (\ref{RS.120}) but satisfy the additional
condition (\ref{RS.121}). It is for this reason that the
Chern-Simons form can be so neatly embedded in the geometry of the
3-form.
As explained in section \ref{2F2} the terms on the left
hand side are located in the superspace 4-form
 \begin{equation}
 \Si^{\ym} \,=\,\tr (\cf \cf). \end{equation}
Of course, the constraints on the Yang-Mills field strength induce
special properties on the 4-form coefficients, in particular
\begin{equation}
{\Si^{\ym}}_{\unddt \, \undgm \, \undbt \, A} \ = \ 0,
\end{equation}
 which is just the same tensor structure as
the constraints on the field strength of the 3-form gauge
potential. Therefore {\em{the Chern-Simons geometry can be
regarded as a special case of that of the 3-form gauge
potential}}. Keeping in mind this fact we obtain
\begin{eqnarray}
 {\Si^{\ym}}_{\dt \gm \ ba} & = & \f{1}{2}
(\si_{ba} \eps)_{\dt \gm} \ovY^{\ym}, \\ {{\Si^{\ym}}^{\dd
\dg}}_{\ ba} & = & \f{1}{2} ({\sib}_{ba} \eps)^{\dd {\dg}}
Y^{\ym},
\end{eqnarray}
 with
 \begin{eqnarray}
 Y^{\ym} &=& - 8 \, \tr (\cw^\al \cw_\al), \\ \ovY^{\ym} &=& - 8 \,
\tr (\cw_\da \cw^\da). \end{eqnarray}

These facts imply the existence and provide a method for the
explicit construction of the Chern-Simons superfield: comparison
of these equations with those obtained earlier in the 3-form
geometry clearly suggests that the Chern-Simons superfield
$\Om^{\ym}$ will be the analogue of the unconstrained
pre-potential superfield $\Om$ of the 3-form. In order to
establish this correspondence in full detail we translate the
procedure developed in the case of the 3-form geometry to the
Chern-Simons form (in the following we shall omit the ${}^{\ym}$
superscript). The starting point for the explicit construction of
the Chern-Simons superfield is the relation
\begin{equation}
\tr(\cf \cf) \,=\,d \cq(\ca). \end{equation}

In the 3-form geometry we know unambiguously the exact location of
the pre-potential in superspace geometry. Since we have identified
Chern-Simons as a special case of the 3-form, it is now rather
straightforward to identify the Chern-Simons superfield following
the same strategy. To this end we recall that the pre-potential
was identified after certain field dependent redefinitions which
had the form of a gauge transformation, simplifying considerably
the form of the potentials. For instance, the new potentials had
the property
\begin{equation}
 \Om_{\gm \bt A} \,=\,0, \cem \Om^{\dg
\db}{}_A \,=\,0.
\end{equation}
 Note, en passant, that these redefinitions are not
compulsory for the identification of the unconstrained
pre-potential. They make, however, the derivation a good deal more
transparent. Can these features be reproduced in the Chern-Simons
framework? To answer this question we exploit a particularity of
Yang-Mills in superspace, namely the existence of different types
of gauge potentials corresponding to the different possible types
of gauge transformations as described in subsection \ref{RS.32}.
These gauge potentials are superspace 1-forms denoted by $\ca$,
$\ca(0) = \lgoa$ and $\ca(1) = \lgoab$, with gauge transformations
parametrized in terms of real, chiral and antichiral superfields,
respectively. Moreover, the chiral and antichiral bases are
related by a redefinition which has the form of a gauge
transformation involving the pre-potential superfield $\cw$
\begin{equation}
 \lgoa \,=\,\cw^{-1} \lgoab \ \cw - \cw^{-1} d
\cw \,=\,{}^\cw \lgoab. \end{equation}
 Writing the superspace
Chern-Simons form in terms of $\lgoa$ shows immediately
that
\begin{equation}
\cq^{\dg \db}{}_A (\lgoa) \,=\, 0, \end{equation}
 due to $\lgoa^\da=0$, but
\begin{equation}
\cq_{\gm \bt A} (\lgoa) \ \neq \ 0. \end{equation}
 Of course,
in the antichiral basis, things are just the other way round,
there we have
\begin{equation}
\cq_{\gm \bt A} (\lgoab)\,=\,0.\label{F.316}
\end{equation}
On the other hand, due to the relation between $\lgoa$ and
$\lgoab$ and the transformation law of the Chern-Simons form
(\ref{Qgtf}) we have
\begin{equation}
\cq (\lgoa) - \cq (\lgoab) \,=\,d \Dt (\cw, \lgoab),\label{F.317}
\end{equation}
 where now the
group element, $\gym$, is replaced by the pre-potential superfield
$\cw$. In some more detail, in $\Dt = \chi - \si$, we have
\begin{equation}
\chi \,=\,\chi (0, \lgoab, \Upsilon) \,=\,\tr (\lgoab \,
\Upsilon),
\end{equation}
 where
\begin{equation}
\Upsilon \,=\,d \cw \, \cw^{-1} \,=\,E^A \Upsilon_A,
\end{equation}
 has zero field strength
\begin{equation}
d \Upsilon + \Upsilon\Upsilon \,=\,0. \end{equation} The
coefficients of the 2-form, $\chi$, are given as
\begin{equation}
\chi_{BA} \,=\,\tr \lp \Upsilon_B \, \lgoab_A - (-)^{ab}
\Upsilon_A \, \lgoab_B \rp. \end{equation} For $\si$, we define
the interpolating pre-potential $\cw_t$
\begin{equation}
\Upsilon_t \,=\,d\cw_t \, \cw_t^{-1},
\end{equation}
such that
\begin{equation}
 \si_{BA} \,=\,\int_0^1 dt \,
\tr \lp \prt_t \cw_t
  \, \cw_t^{-1} (\Upsilon_{t \, B}, \Upsilon_{t \, A}) \rp.
\end{equation}
 Consider now
\begin{equation}
\cq_{\gm \bt A} (\lgoa) \,=\,\cd_A \Dt_{\gm \bt}
  + \oint_{\gm \bt} \lp \cd_\gm \Dt_{\bt A}
  - (-)^a T_{\gm A}{}^F \Dt_{F \bt} \rp,
\end{equation}
following from (\ref{F.317}), and (\ref{F.316}) and perform a
redefinition
\begin{equation}
{\hat{\cq}} \ := \ \cq(\lgoa) - d \La, \end{equation}
 which leaves $\tr(\cf \cf)$ invariant. We then determine the 2-form $\La$ in terms of the coefficients of the
2-form $\Dt$ such that
\begin{equation}
{\hat{\cq}}_{\gm \bt A}\, = \, 0, \end{equation}
 and maintain, at the same time,
\begin{equation}
{\hat{\cq}}^{\dg \db}{}_A  \,=\,0.
\end{equation}
 This is achieved with the identification
\begin{equation}
\La_{\bt A} \,=\,\Dt_{\bt A}, \cem
    \La^\db {}_a \,=\,-\f{i}{2} \cd^\db \Dt_a, \cem
    \La_{\db \da} \,=\,0.
\end{equation}
 For later convenience, we put also
\begin{equation}
\La_{ba} \,=\,\f{i}{2} \lp \cd_b \Dt_a - \cd_a \Dt_b \rp.
\end{equation}
 Here $\Dt_a$ is identified using spinor notation such that
\begin{equation}
\Dt_\gm {}^\db \,=\,- \f{i}{2} T_\gm{} ^{\db \, a} \Dt_a.
\end{equation}
 We have, of course, to perform this redefinition on
all the other coefficients, in particular
\begin{equation}
{\hat{\cq}}_\gm{}^\db{}_a \,=\,\cq_\gm{}^\db{}_a (\lgoa) - \cd^\db
\Xi_{\gm a}. \end{equation}
 In the derivation of this
equation one uses the anticommutation relation of spinor
derivatives and suitable supergravity Bianchi identities together
with the definition
\begin{equation}
\Xi_{\gm a} \,=\,\Dt_{\gm a} + \f{i}{2} \cd_\gm \Dt_a.
\end{equation}
 We parametrize
\begin{equation}
{\hat{\cq}}_\gm{}^\db{}_a \,=\,
       T_\gm{}^\db{}_a \Om^{\ym} + T_\gm{}^{\db \, b}
       {\hat{\cq}^{\ym}}_{[ba]},
\end{equation}
 where we can now identify the explicit form of the
Chern-Simons superfield
\begin{equation}
\Om^{\ym} \,=\, \cq (\lgoa) - \f{i}{16} \cd^\da \Xi^\al{}_{\al
\da}.
\end{equation}
 The first term is obtained from the spinor
contraction of
\begin{equation}
\cq_\gm{}^\db{}_a (\lgoa) \,=\,\tr \lp \lgoa_\gm
\cf^\db{}_a(\lgoa) \rp \,=\,-i (\sib_a \eps)^\db{}_\bt \, \tr \lp
\lgoa_\gm \cw^\bt (\lgoa) \rp,
\end{equation}
 \ie
\begin{equation}
\cq(\lgoa) \,=\,\f{i}{16} \cq^{\al \da}{}_{\al \da} (\lgoa)
   \,=\,- \f{1}{4} \tr \lp \lgoa^\al \cw_\al (\lgoa) \rp.
\end{equation}
 It remains to read off the explicit form of the second term
from the definitions above.

In closing we note that a more symmetrical form of the
Chern-Simons superfield may be obtained in exploiting the relation
\begin{equation}
\cq_\gm{}^\db{}_a (\lgoa) - \cq_\gm{}^\db{}_a (\lgoab) \,=\,
       \cd_\gm \Xi^\db{}_a + \cd^\db \Xi_{\gm a}
      + T_\gm{}^{\db \, b} \lp \Dt_{ba} + \f{i}{2} (\cd_b \Dt_a - \cd_a \Dt_b) \rp,
\end{equation}
 with
 \begin{equation}
 \Xi^\db{}_a \,=\,\Dt^\db{}_a + \f{i}{2}
\cd^\db \Dt_a. \end{equation}

Observe that different appearances of the Chern-Simons superfields
should be equivalent modulo linear superfields. To establish the
explicit relation of the Chern-Simons superfield presented here
and that given in \cite{CFV87} is left as an exercise. So far, we
have dealt with the superspace Chern-Simons form alone; when
coupled to the linear multiplet the modified field strength is
\begin{equation}
H^{\ym} \,=\, H+ k\, \cq^{\ym},
\end{equation}
with $H=dB$. In the preceding discussion we have split $\cq^{\ym}$
\begin{equation}
\cq^{\ym}\,=\, {\hat{\cq}}^{\ym} + d \La^{\ym},
\end{equation}
such that ${\hat{\cq}}^{\ym}$ has the same vanishing components as
$H$. Defining  $\cah^{\ym} = H^{\ym} -{\hat{\cq}}^{\ym}$ and
$\cb^{\ym} = B +\La^{\ym}$ leads to
\begin{equation}
\cah^{\ym} \,=\, d \cb^{\ym}.
\end{equation}
Although $\cah^{\ym}$ is no longer invariant under Yang-Mills
gauge transformations, it has the same constraints as $H$.
Therefore the solution of the modified linearity conditions can be
obtained by the same procedure as employed in the case without
Chern-Simons forms.

\newpage
\addcontentsline{toc}{section}{\bf  REFERENCES}
\bibliography{Refs}
\bibliographystyle{plain}
\end{document}